\documentclass[epjc3]{svjour3}
\usepackage{amsmath}  

\usepackage{url}

\usepackage{amssymb}  
\usepackage{float} 
\usepackage{xcolor} 
\usepackage{graphicx} 
\usepackage{cite} 
\usepackage[normalem]{ulem}
\usepackage{bbm}
\usepackage{dsfont}
\usepackage[mathscr]{euscript}
\usepackage{ntheorem}

\newcommand{\ringzeta}{{  \mathring \zeta}}

\newcommand{\redc}{}
\newcommand{\barh}{\blue{ h^{\mathrm{reg}}}}
\newcommand{\hBo}{\blue{ h^{\mathrm{Bo}}}}

\newcommand{\Vol}{{V}}

\newcommand{\pmass}{{+m^2}}

\newcommand{\omthree}[1]{\overset{\mbox{\tiny (-3)}}{#1}}

\newcommand{\omtwo}[1]{\overset{\mbox{\tiny (2)}}{#1}}
\newcommand{\omone}[1]{\overset{\mbox{\tiny (1)}}{#1}}
\newcommand{\ozero}[1]{\overset{\mbox{\tiny (0)}}{#1}}

\newcommand{\othreeone}[1]{\overset{\mbox{\tiny (-3,1)}}{#1}}
\newcommand{\otwo}[1]{\overset{\mbox{\tiny (-2)}}{#1}}
\newcommand{\othree}[1]{\overset{\mbox{\tiny (-3)}}{#1}}
\newcommand{\ofour}[1]{\overset{\mbox{\tiny (-4)}}{#1}}
\newcommand{\oone}[1]{\overset{\mbox{\tiny (-1)}}{#1}}
\newcommand{\ok}[1]{\overset{\mbox{\tiny (-k)}}{#1}}

\newcommand{\hdpot}{\blue{\hat \chi}}
\newcommand{\hrotpot}{\blue{\check \chi}}

\newcommand{\TB}{\blue{\mathrm{TB}}}

\newcommand{\breh}{\blue{\breve{h}}}

\newcommand{\IntPsi}{{\mathring \Ipsi}}

  \newcommand{\mypi}{{\pi}}
  \newcommand{\tpi}{{\tilde\mypi}}
  \newcommand{\tphi}{{\tilde\phi}}
  \newcommand{\bmetric}{{b}}

\newcommand{\R}{\mathbb{R}}

\newcommand{\purple}{}

\newcommand{\Ichi}{{\mu}}
\newcommand{\Ipsi}{{\lambda}}

\newcommand{\lapseTB}{\blue{N}}

\newcommand{\deltaU}{{ {\zhTB}_{u}{}} }

\newcommand{\ud}[2]{^{#1}_{\phantom{#1 } #2}}

\newcommand{\zhTB}{\blue{\check h}}

\newcommand{\zhTBW}{\blue{\gamma}}
\newcommand{\zzhTBW}{\blue{\mathring{\zhTBW}}}

\newcommand{\zh}{{\zzhTBW }}

 \newcommand{\NullQ}{{Q}}

\newcommand{\mcN}{{\mycal N}}

\newcommand{\sectionofScri}%
{{ \,\,\,\,\mathring{\!\!\!\!\mcN}}}

\newcommand{\Bo}{\Bobo}
\newcommand{\Bobo}{{}}

\newcommand{\ourU}{\mathbb U}%

\newcommand{\Kp}{{\mathfrak{g}}}
\newcommand{\Bgamma}{{B}}

\newcommand{\backg}{{b}}
 \newcommand{\dlambda}[1]{{{d #1}\over {d\lambda}}}

 \newcommand{\tomega}{{\tilde \omega}}
 \newcommand{\hyp}{{\mycal S}}
\newcommand{\mcL}{{\mycal L}}

\newcommand{\rbo}{r_{\Bo}}
\newcommand{\rBo}{\rbo}

\newcommand{\scri}{{\mycal I}}%
\newcommand{\scrip}{\scri^{+}}%
\newcommand{\ringh}{{  \zzhTBW }}%

\newcommand{\znabla}{ {\mathring{\nabla}}}

\newcommand{\eq}[1]{(\ref{#1})}

\newcommand{\og}{{\overline{g}}}

\newcommand{\noalpha}{{{\varphi}}}
\newcommand{\nobeta}{{{\psi}}}
\newcommand{\nosigma}{{{\chi}}}

\newcommand{\noomega}{{w}}

\newcommand{\eeal}[1]{\label{#1}\end{eqnarray}}

\newtheorem{Proposition}[theorem]{\sc Proposition\rm}
\newtheorem{Remark}[theorem]{\sc Remark\rm}

\DeclareFontFamily{OT1}{rsfs}{}
\DeclareFontShape{OT1}{rsfs}{m}{n}{ <-7> rsfs5 <7-10> rsfs7 <10-> rsfs10}{}
\DeclareMathAlphabet{\mycal}{OT1}{rsfs}{m}{n}

\newcommand{\mmr}[1]{\mnotex{{\bf mm:} {\color{violet} #1}}}

\newcommand{\blue}[1]{{\color{blue} #1}}

\definecolor{applegreen}{rgb}{0.55, 0.71, 0.0}
\definecolor{armygreen}{rgb}{0.29, 0.33, 0.13}

\newcommand{\green}[1]{{\color{caribbeangreen} #1}}

\definecolor{caribbeangreen}{rgb}{0.0, 0.8, 0.6}

\newcounter{mnotecount}[section]

\renewcommand{\themnotecount}{\thesection.\arabic{mnotecount}}

\newcommand{\mnotex}[1]
{\protect{\stepcounter{mnotecount}}$^{\mbox{\footnotesize
$
\bullet$\themnotecount}}$ \marginpar{
\raggedright\tiny\em
$\!\!\!\!\!\!\,\bullet$\themnotecount: #1} }

\newcommand{\ptc}[1]{\mnotex{{\bf ptc:} {  #1}}}
\newcommand{\ptcrr}[1]{\mnotex{{\bf ptc:} {  #1}}}

\newcommand{\bel}[1]{\begin{equation}\label{#1}}
\newcommand{\bea}{\begin{eqnarray}}
\newcommand{\bean}{\begin{eqnarray}\nonumber}
\newcommand{\beal}[1]{\begin{eqnarray}\label{#1}}
\newcommand{\eea}{\end{eqnarray}}

\newcommand{\nn}{\nonumber}
\newcommand{\Eq}[1]{Equation~\eq{#1}}

\newcommand{\qedskip}{\hfill $\Box$\medskip}
\newcommand{\be}{\begin{equation}}
\newcommand{\eeq}{\end{equation}}
\newcommand{\ee}{\end{equation}}
\newcommand{\beqa}{\begin{eqnarray}}
\newcommand{\eeqa}{\end{eqnarray}}
\newcommand{\beqan}{\begin{eqnarray*}}
\newcommand{\eeqan}{\end{eqnarray*}}
\newcommand{\ba}{\begin{array}}
\newcommand{\ea}{\end{array}}

\newcommand{\const}{\mbox{\rm const}} 

\newcommand{\ptcheck}[1]{\ptc{checked on #1}}

\newcommand{\spaceD}{{ D}}
\newcommand{\zspaceD}{ {\mathring D}}

\newcommand{\mcC}{\mycal C}
\newcommand{\mcE}{\mycal E}
\newcommand{\mcH}{\mycal H}

\newcommand{\gone}{\overset{\mbox{\tiny (-1)}}{g}}

\newcommand{\gmone}{\overset{\mbox{\tiny (1)}}{g}}
\newcommand{\gmtwo}{\overset{\mbox{\tiny (2)}}{g}}

\newcommand{\gzero}{\overset{\mbox{\tiny (0)}}{g}}

\newcommand{\nobarg}{\blue{g}}

\newcommand{\wmcL}{{\,\,\,\widetilde{\!\!\!\mcL}}}
\newcommand{\wmcH}{{\,\,\,\widetilde{\!\!\!\mcH}}}

\newcommand{\Lie}{{\mathcal{L}}}

\newcommand{\jhbr}[1]{\mnotex{{\bf jh:} {\color{blue} #1}}}

\newcommand{\TSphi}{\blue{\phi}}
\newcommand{\TSpi}{\blue{\pi}}
\newcommand{\TSa}{\blue{a}}
\newcommand{\TSb}{\blue{b}}
\newcommand{\TStphi}{\blue{{\tilde\TSphi}}}
\newcommand{\TStpi}{\blue{{\tilde\TSpi}}}

\newcommand{\tscomr}[1]{\mnotex{{\bf ts:} {  #1}}}

\newcommand{\TSxip}{\blue{\zeta}}
\newcommand{\TSxi}{\blue{\xi}}
\newcommand{\TSr}{\blue{r}}
\newcommand{\TSu}{\blue{u}}
\newcommand{\TSoxi}{\blue{\mathring \TSxi}}
\newcommand{\TSoLie}{\blue{\mathring \Lie}}
\newcommand{\TSzlap}{\blue{\Delta_{\ringh}}}
\newcommand{\TSgreen}{}
\newcommand{\TSred}{}
\newcommand{\TSlambden}{\blue{\Pi}}
\newcommand{\TSblue}{}
\newcommand{\TSgrn}{}

\newcommand{\dgbar}[1]{\delta_{#1} \overline{g}}
\newcommand{\tshbar}{\overline{h}}

\def\d{\partial}

\def\beq{\begin{eqnarray}}
\def\eeq{\end{eqnarray}}

\def\a{\alpha}
\def\b{\beta}
\def\c{\gamma}
\def\d{\delta}
\def\e{\epsilon}
\def\f{\sigma}

\def\be{\begin{equation}}
\def\ee{\end{equation}}
\def\bea{\begin{eqnarray}}
\def\eea{\end{eqnarray}}
{\catcode `\@=11 \global\let\AddToReset=\@addtoreset}
\AddToReset{equation}{section}

\renewcommand{\ptcrr}[1]{}

\renewcommand{\mmr}[1]{}
\renewcommand{\tscomr}[1]{}
 
\renewcommand{\og}{{g}}
\renewcommand{\ptcheck}[1]{}
\renewcommand{\jhbr}[1]{}
\renewcommand{\blue}[1]{#1}
\renewcommand{\green}[1]{#1}

\journalname{arXiv version}
\begin{document}

\title{On the canonical energy of weak gravitational fields  with   a cosmological constant\thanksref{tit} $\Lambda\in\R$
}

\author{ P.T. Chru\'sciel \thanksref{addr1a,addr1b}
    \and
     Sk J. Hoque \thanksref{addr2,jh}
     \and
      M. Maliborski \thanksref{addr1a,mm}
      \and
       T. Smo{\l}ka \thanksref{addr4}
}

\thankstext{tit}{Preprint UWThPh-2020-27}
\thankstext{jh}{co-author of Sections \ref{s14VI21.1},
\ref{C26X20.2}-\ref{C26X20.4}}
\thankstext{mm}{co-author of Sections \ref{s14VI21.1} and \ref{C26X20.3}}
\institute{Faculty of Physics, University of Vienna \label{addr1a}
     \and  Institut de Hautes \'Etudes Scientifiques, Bures-sur-Yvette \label{addr1b}
     \and
     Institute of Theoretical Physics, Faculty of Mathematics and Physics, Charles University, Prague \label{addr2}
    \and Department of Mathematical Methods in Physics, Faculty of Physics, University of Warsaw
       \label{addr4}
}
\date{}
\maketitle

\abstract{
    We analyse the canonical energy  of vacuum linearised gravitational fields on light cones on a de Sitter, Minkowski, and Anti de Sitter backgrounds in  Bondi gauge. We derive the associated asymptotic symmetries.  When $\Lambda>0$ the energy diverges, but a renormalised formula with well defined flux is obtained.  We show that the renormalised energy in the asymptotically off-diagonal gauge  coincides with the quadratisation of the generalisation of the Trautman-Bondi mass proposed in \cite{ChIfsits}.
}

\tableofcontents

\section{Introduction}
\label{s14VI21.1}

\ptcrr{should be repeated in an arbitrary FRW metric}
A question of current interest is the amount of energy that can be radiated by a gravitating system in the presence of a positive cosmological constant. This problem has been addressed in~\cite{ChIfsits} using an approach based on the characteristic constraint equations and involving Bondi coordinates. The analysis there showed ambiguities in the resulting expression. The question then arises, whether some insight into the problem at hand could be gained by considering linearised gravity on the de Sitter background. The aim of this work is to carry out this project.

We start, in Section~\ref{C26X20.1} with a general analysis of the canonical energy in linearised Lagrangian theories. The main point is to derive a formula for the canonical energy including all boundary terms, which are usually neglected and which play a key role in general relativity. {The results presented in this section are essentially known~\cite{KijowskiTulczyjew,WaldZoupas}, but a coherent and systematic presentation for linearised field theories, keeping track of all terms in the integrals, does not exist in the literature. One of the main results in this section is Proposition~\ref{P10IX18.1}, which does not seem to have appeared in the literature in this generality.
}

In Section~\ref{C26X20.2} we show how to put a linearised gravitational field into Bondi gauge, analyse the small-$r$ behaviour of the fields and derive the freedom remaining. { This gives a unified treatment for all $\Lambda \in \R$ of asymptotic symmetries \emph{\`a la Bondi-Metzner-Sachs} in the linearised regime,  which leads to a clear view of the differences arising from the sign of the cosmological constant and from the boundary conditions imposed. We note that the current approach to asymptotic symmetries, based on characteristic initial data and their evolution, gives a perspective different from the one based e.g. on Fefferman-Graham expansions, as it introduces naturally a foliation of the conformal boundary $\scri$ by spheres obtained by  intersecting $\scri$ with the light cones.}

In Section~\ref{C26X20.3} we analyse the large-$r$ behaviour of vacuum metric perturbation in the Bondi gauge  and   show explicitly how the formalism works for a class of linearised solutions of the vacuum Einstein equations discovered by Blanchet and Damour~\cite{BlanchetDamour}.
{Our asymptotic conditions on the linearised perturbations of the metric are modelled on the asymptotic behaviour of the full solutions of the Einstein vacuum field equations with positive cosmological constant and with smooth initial Cauchy data on $S^3$, as derived by Friedrich in~\cite{Friedrich}. Here some comments might be in order. In~\cite{Friedrich} Friedrich shows that small perturbations of de Sitter Cauchy data on $S^3$ lead to vacuum spacetimes with smooth conformal boundaries at infinity. He isolates a set of data on the spacelike boundary at infinity which parameterise uniquely all vacuum spacetimes with a positive cosmological constant and with smooth conformal completions at infinity. His analysis carries over without difficulties to the linearised equations; a convenient way to proceed is to linearise the equations in~\cite{TimConformal}. The results of Friedrich provide a rigorous justification of the asymptotic expansions proposed by Starobinsky~\cite{Starobinsky:1982mr}, revisited later from a more general perspective in~\cite{Fefferman:2007rka}. The readers familiar with the Fefferman-Graham expansions can view the results in this Section as a translation of these expansions to characteristic Cauchy data in the linearised setting.

It should be emphasised that requiring asymptotic conditions  more stringent than ours will lead to non-generic fields, in the sense that generic smooth changes of initial data on a Cauchy surface in de Sitter spacetime will lead to solutions which do not satisfy the more stringent conditions. And allowing less stringent conditions will lead to linearised metric fields which are not smoothly conformally extendable along, and  in a neighborhood of, the initial light cone. In other words, our asymptotic conditions are optimal for linearised fields which are smoothly conformally extendable.
}

In Section~\ref{C26X20.4} we apply the formalism developed so far to derive our formula for the canonical energy of vacuum linearised metric perturbations on a de Sitter, Minkowski, and
Anti de Sitter backgrounds.

In the asymptotically flat case we recover the linearised-theory version of the usual Trautman-Bondi mass. Similarly when $\Lambda\le 0$  and the old-fashioned { \cite{Ashtekar:1984zz,Hawking:1983mx,AshtekarDas,CJL}}
  strong decay conditions are imposed on the linearised field  we obtain the linearised version of the usual, conserved, Anti de Sitter mass.

 When $\Lambda >0$, {
  or when $\Lambda<0$ but the asymptotic conditions are relaxed to the linearised version of not-conformally-flat-$\scri$,} we obtain an expression for the canonical mass on light cones truncated to radius $R$ which diverges when $R$ tends to infinity.
It turns out  that the divergent part of the energy has a dynamics of its own, which allows one to introduce a renormalised energy which satisfies a well defined flux formula.
This formula, and its flux, are the main results of this work, to be found in Section~\ref{s9XI20.1}.
We show in Section~\ref{s1IX20.1} that the renormalised energy coincides with a quadratisation of a generalisation of the Trautman-Bondi mass proposed in \cite{ChIfsits} when an ``asymptotically off-diagonal gauge" is used. On the other hand, the ``asymptotically block-diagonal gauge'' leads to a formula for the renormalised energy which differs from the previous one by a  boundary term determined in Section~\ref{s18II21.1}.
{
We find several natural candidates for a renormalised canonical energy:  $ \delta^2 m_\TB(\mcC_{u })$,
$\hat E_{c,I}$ and $\hat E_{c,II}$; yet another one, $\delta^2 E^{(\Lambda)}$, motivated by holography, has been proposed in \cite{compere2019lambda}; cf.\ Section~\ref{s23II21/1a}. 
We find that
the only one which is invariant under asymptotic symmetries is $ \delta^2 m_\TB(\mcC_{u })$, which coincides with $\hat E_{c,I}$ in the gauge where the Bondi-transformed linearised metric is as regular as possible at the origin.

Most of the results derived here have been summarised in \cite{ChHS}. Further results on, or related to, this problem  can be found in~\cite{AshtekarBK,AshtekarBongaKesavanI,AshtekarBongaKesavanIII,%
KolanowskiLewandowski,Saw,Bishop,Date:2015kma,Date:2016uzr,Hoque:2018byx,%
SzabadosTod,SzabadosTod2,Kolanowski:2021hwo}.

We note that there exists a rich literature on energy when $\Lambda<0$ { \cite{AshtekarDas,deHaro:2000xn,Barzegar:2017ijz,ChruscielSimon,Papadimitriou:2005ii,Hollands:2005wt,JJCYKADS},
and while essentially all our results apply to either sign of cosmological constant, including $\Lambda=0$, we will not carry-out a systematic comparison of our results to these,} since the main concern of  this paper is the case $\Lambda>0$, which is much less understood so far. }
\ptcrr{Jaeger's diploma might be relevant, but not really clear \url{http://www.theorie.physik.uni-goettingen.de/forschung/qft/theses/dipl/Jaeger.pdf}
}

\bigskip

\flushbottom
\noindent{\sc Acknowledgements:}  Useful discussions with P.~Aichelburg, T.~Damour, G.~Date, J.~Jezierski,  M.~Kolanowski,   P.~Krtou\v{s} and J.~Lewandowski are acknowledged.
JH is grateful to the Erwin Schr\"odinger Institute,
and University of Vienna for hospitality  during part of work on this paper.
His research was supported in part
by  the Czech Science Foundation Grant 19-01850S, and by
the  Max-Planck partner group project ``Quantum Black Holes" between CMI, Chennai
and AEI, Potsdam. TS  acknowledges the hospitality of the University of Vienna during part of work on this project and  financial support from the COST Action CA16104 GWverse. His work was supported by the University of Warsaw Integrated Development Programme (ZIP), co-financed by the European Social Fund within the framework of Operational Programme Knowledge Education Development 2016-2020, action 3.5.
PTC acknowledges the friendly hospitality of the Institut de Hautes \'Etudes Scientifiques during part of work on this project. His  research was   supported
by the Austrian Research Fund (FWF), Project  P 29517-N27 and
by the Polish National Center of Science (NCN) under grant 2016/21/B/ST1/00940.

\section{The energy of linearised fields
\\
{\small\em by PTC and TS}
}
\label{C26X20.1}

Our aim in this section is to establish that the canonical energy of the linearised theory
can be calculated, up-to-divergence, by means of the ``presymplectic current'' (using the Lee-Wald terminology) of the original theory; see Proposition~\ref{P10IX18.1} below. This  is well known, but we revisit the proof as we will need the formula for the divergence term.

We further show  gauge-independence up-to-boundary term of the canonical energy of the linearised theory. This is also well known (cf., e.g., \cite{CrnkovicWitten,Friedman1978,WaldZoupas}), and closely related to the above, but here again our focus is on the explicit form of the boundary term.

\subsection{General formalism}
 \label{ss11IX18.1}

We consider a first-order Lagrangian field theory for a collection of fields $\phi \equiv (\phi^A)$, where $A$ runs over a finite set. We write
$$
 \partial \phi \equiv \left(\phi^A{}_\mu \right)
  := \left(
   \partial_\mu \phi^A
    \right)
    \equiv \left( \frac{\partial \phi^A}{\partial x^\mu}
     \right)
 \,.
$$
Given a Lagrangian  density $\mcL(\phi, \partial \phi, \cdot)$, where $\cdot $ denotes  background fields (which might or might not be present), the field equations are
\begin{equation}\label{5IX18.1}
 \mcE_A:=   \partial_\mu
   \left(
    \frac{\partial \mcL}{\partial \phi^A{}_\mu}
     \right) - \frac{\partial \mcL}{\partial \phi^A}
      =0
   \,.
\end{equation}

For $\lambda \in I$, where $I$ is an open interval containing $0$, let $\phi(\lambda)\equiv \left(\phi^A(\lambda)\right)$ be a one-parameter family of fields differentiable with respect to $\lambda$, set
\begin{equation}\label{5IX18.2}
  \tphi \equiv (\tphi^A) :=
    \frac{d \phi^A}{d\lambda}
  \,.
\end{equation}
Then $\tphi$ satisfies the set of equations
\begin{equation}\label{5IX18.3}
  \partial_\mu
   \left(
    \frac{\partial^2  \mcL}{\partial \phi^A{}_\mu\partial \phi^B{}_\nu }
     \partial _\nu \tphi ^B
     +
    \frac{\partial^2  \mcL}{\partial \phi^A{}_\mu\partial \phi^B  }
      \tphi ^B
      \right)
     =
    \frac{\partial^2  \mcL}{ \partial \phi^A\partial \phi^B{}_\mu  }
     \partial_\mu \tphi ^B
      +
    \frac{\partial^2  \mcL}{ \partial \phi^A \partial \phi^B  }
      \tphi ^B
      + \dlambda{\mcE_A}
   \,.
\end{equation}

To continue, it is convenient to introduce some notation. We set
\begin{eqnarray}
 &
 \displaystyle
  \mypi_A{}^\mu  :=
    \frac{\partial  \mcL}{\partial \phi^A{}_\mu }
     \,,
     \quad
  \mypi_A  :=
    \frac{\partial   \mcL}{\partial \phi^A  }
     \,,
   &
\\
 &
 \displaystyle
  \mypi_A{}^\mu{}_B{}^\nu :=
    \frac{\partial^2  \mcL}{\partial \phi^A{}_\mu\partial \phi^B{}_\nu }
     \,,
     \
  \mypi_A{}^\mu{}_B  :=
    \frac{\partial^2  \mcL}{\partial \phi^A{}_\mu\partial \phi^B  }
     \,,
     \
  \mypi_{A B}  :=
    \frac{\partial^2  \mcL}{ \partial \phi^A \partial \phi^B  }
   \,.
   &
  \label{5IX18.4}
\end{eqnarray}
In this notation, \eq{5IX18.3} 
can be rewritten somewhat more concisely as
\begin{equation}\label{5IX18.5}
  \partial_\mu
  \left(
 \mypi_A{}^\mu{}_B{}^\nu
     \partial _\nu \tphi ^B
     +
    \mypi_A{}^\mu{}_B
      \tphi ^B
      \right)
   =
    \left(
    \mypi_B{}^\mu{}_A
     \partial_\mu \tphi ^B
      +
    \mypi_{A B}
      \tphi ^B
       \right)
       +  \dlambda{\mcE_A}
   \,.
\end{equation}
\emph{If $\dlambda{\mcE_A}$ has been prescribed} (e.g., equal to zero, in which case the fields $\tphi^A$ satisfy the linearised field equations),
or is known  and is $\tphi$--independent, this set of equations can be derived from a Lagrangian density $\wmcL$ given by
\begin{equation}\label{5IX18.6}
  \wmcL = \frac 12
 \mypi_A{}^\mu{}_B{}^\nu
     \partial _\mu \tphi ^A   \partial _\nu \tphi ^B
     +
    \mypi_A{}^\mu{}_B
      \partial _\mu \tphi ^A  \tphi ^B
      + \frac 12
    \mypi_{A B}
    \tphi^A
      \tphi ^B + \dlambda{\mcE_A} \tphi^A
   \,.
\end{equation}
(To avoid ambiguities: we see that \eqref{5IX18.6} with $\dlambda{\mcE_A} \equiv 0$ provides the Lagrangian for the linearised field equations at $\phi(\lambda)|_{\lambda=0}$, regardless of whether or not $\phi(\lambda)|_{\lambda=0}$ itself satisfies any field equations, though typically one would be interested in situations where $\mcE_A({\phi(0)})=0$.)

We remark that the field $\phi:=\phi(\lambda)|_{\lambda = 0}$ plays a role of a background field in $\wmcL$, so even if we assumed that $\mcL$ depends only upon $\phi$ and $\partial \phi$, we end up with a Lagrangian density where background dependence has to be taken into consideration.

The following identities  are useful, assuming   $ \dlambda{\mcE_A}=0$:
\begin{eqnarray}
 \label{6IX18.10}
  \tpi_A{}^\mu(\tphi) &:=&  \frac{\partial \wmcL}{\partial \tphi ^A{}_\mu}
  =
 \mypi_A{}^\mu{}_B{}^\nu
        \partial _\nu \tphi ^B
     +
    \mypi_A{}^\mu{}_B
        \tphi ^B
        = \frac{d \big(\frac{\partial \mcL}{\partial \phi_A{}^\mu } \big)}{d\lambda}
        = \frac{d \mypi _A{}^\mu}{d\lambda}
  \,,
  \phantom{xxxx}
   \\
  \tpi_B(\tphi) &:=&  \frac{\partial \wmcL}{\partial \tphi ^B }
  =  \mypi_A{}^\mu{}_B
      \partial _\mu \tphi ^A
      +
    \mypi_{A B}
    \tphi^A
        = \frac{d \big(\frac{\partial \mcL}{\partial \phi_B} \big)}{d\lambda}
   = \frac{d \mypi _B }{d\lambda}
   \,.
 \label{6IX18.11}
\end{eqnarray}

From now on, we consider a theory which satisfies the following:

\begin{itemize}
  \item[H1.] $\mcL$ is a scalar density.
  \item[H2.] There exists a notion of derivation with respect to a family of vector fields $X$, which we will denote by $\Lie_X$, which coincides with the usual Lie derivative on vector densities, and which  we will  call  \emph{Lie derivative} regardless of whether or not this is the usual Lie derivative on the remaining fields, such that the following holds:
      \begin{enumerate}
        \item[a)] $\Lie_X$ preserves the type of a field, thus $\Lie_X$ of a scalar density is a scalar density, etc.;
        \item[b)] the field $\pi_A{}^\mu \Lie_X \phi^A$ is a vector density;
        \item[c)] in a coordinate system in which $X={\partial_0}$ we have $\Lie_X= {\partial_0}$;
        \item[d)] $\Lie_X$ satisfies the Leibnitz rule.
      \end{enumerate}
      \label{H18XI19.1}
\end{itemize}

The above holds if the fields $\phi^A$ are tensor fields and $\mcL$ is of the form $\sqrt{|\det g|} L$, where $L$ is a scalar, with $\Lie$ the standard Lie derivative.

Let us denote by $\wmcH$ the Hamiltonian density vector (called ``Noether current'' in \cite{WaldLee})
associated with the Lagrangian density $\wmcL$ and   $X$:
\begin{eqnarray}
\nonumber
  \wmcH^\mu[X]
   & := &
     \frac{\partial \wmcL}{\partial \tphi^A{}_\mu} \Lie_X \tphi^A - X^\mu \wmcL
\\
  & =
  &
 \left(
  \mypi_A{}^\mu{}_B{}^\nu
       \partial _\nu \tphi ^B
     +
    \mypi_A{}^\mu{}_B
      \tphi ^B
    \right)
     \Lie_X \tphi^A - X^\mu \wmcL
   \,.     \label{5IX18.7}
\end{eqnarray}
%
%
Let $\mcH^\mu[X]$ be the corresponding vector density associated with the original field $\phi$:
\begin{eqnarray}
\nonumber
  \mcH^\mu[X]
   & := &
     \frac{\partial \mcL}{\partial \phi^A{}_\mu} \Lie_X \phi^A - X^\mu \mcL
\\
  & =
  &
  \mypi_A{}^\mu
     \Lie_X \phi^A - X^\mu \mcL
   \,.     \label{5IX18.7+}
\end{eqnarray}

We wish to calculate the variation of $\mcH^\mu$. Typically one assumes that the background fields, if any, are $\lambda$-independent. This might, however, not be the case for some variations, e.g.\ if the variations correspond to coordinate transformation which do not leave the background invariant. In order to allow for such situations
let us denote by
$$
 \psi:= (\psi^I)
$$
the collection of all background fields; if $\mcL$ depends both upon a background and some   derivatives thereof,  we include the derivatives of the background as part of components of $\psi$.
For completeness we will carry out the usual calculation of $\frac{d \mcH^\mu[X]}{d\lambda}$. For this, recall
the formula for the Lie derivative of a vector density $Z$:
\begin{equation}\label{20XI19.1}
  \Lie_X Z^\mu = \partial_\sigma(X^\sigma Z^\mu) - Z^\sigma \partial_\sigma X^\mu
  \,.
\end{equation}
Keeping in mind that $\mypi_A{}^\mu$ is a tensor density by H1), and our remaining hypotheses H2.a)-H2.d), we are led to the following identity:
\begin{equation}\label{24IX18.11}
  \partial_\sigma ( X^\sigma \mypi_A{}^\mu \dlambda{\phi^A}) =
  \Lie_X \mypi_A{} ^\mu \dlambda{\phi^A}
  +
 \mypi_A{} ^\mu  \Lie_X \dlambda{\phi^A}
 +
  \partial_\sigma  X^\mu \mypi_A{}^\sigma \dlambda{\phi^A}
 \,.
\end{equation}

We are ready now to calculate the variation of $\mcH$:
\begin{eqnarray}
\nonumber
 \lefteqn{
 \frac{d \mcH^\mu[X]}{d\lambda}
 =
    \Lie_X \phi^A \frac{d \mypi_A{}^\mu}{d\lambda} +
 \mypi_A{}^\mu \Lie_X \frac{d \phi^A}{d\lambda}
 +
    \underbrace{
    \mypi_A{}^\mu \Lie_{\frac{dX}{d\lambda}} \phi^A
   - \frac{dX^\mu}{d\lambda} \mcL
   }_{= \mcH^\mu[ \frac{dX}{d\lambda}]}
    }
    &&
\\
 &&
  - X^\mu\Big( \mypi_A{}^\sigma \partial_\sigma \dlambda{\phi^A}
   +\underbrace{\frac{\partial \mcL}{ \partial \phi^A}
          }_{=\partial_\sigma \mypi_A{}^\sigma - \mcE^A }  \dlambda{\phi^A} +
  \frac{\partial \mcL}{ \partial \psi^I}{\dlambda{\psi^I}}
   \Big)
 \nonumber
\\
 &
 = &
    \Lie_X \phi^A \frac{d \mypi_A{}^\mu}{d\lambda} -
    \Lie_X \mypi_A{}^\mu \frac{d \phi^A}{d\lambda}
    +
    \big(\Lie_X \mypi_A{}^\mu -\partial_\sigma (X^\mu \mypi_A{}^\sigma) \big)  \frac{d \phi^A}{d\lambda}
     \nonumber
\\
  &
  &
  %
   + \mypi_A{}^{\mu}\Lie_X  \frac{d \phi^A}{d\lambda}
 +
  \partial_\sigma  X^\mu \mypi_A{}^\sigma \dlambda{\phi^A}
  - X^\mu \mypi_A{}^\sigma \partial_\sigma \dlambda{\phi^A}
     \nonumber
\\
  &
  &
   +
    \mcH^\mu[ \frac{dX}{d\lambda}]
   + X^\mu \Big(\mcE_A \frac{d \phi^A}{d\lambda}
     -
  \frac{\partial \mcL}{ \partial \psi^I}{\dlambda{\psi^I}}
   \Big)
 \nonumber
\\
 &
 = &
    \Lie_X \phi^A \frac{d \mypi_A{}^\mu}{d\lambda} -
    \Lie_X \mypi_A{}^\mu \frac{d \phi^A}{d\lambda}
   + 2 \partial_{{\sigma}} \left(
    X^{[{{\sigma}}} \mypi_A{}^{\mu]} \frac{d \phi^A}{d\lambda}
     \right)
     \nonumber
\\
  &
  &
   +
    \mcH^\mu[ \frac{dX}{d\lambda}]
   + X^\mu \Big(\mcE_A \frac{d \phi^A}{d\lambda}
     -
  \frac{\partial \mcL}{ \partial \psi^I}{\dlambda{\psi^I}}
   \Big)
   \,,     \label{6IX18.4}
\end{eqnarray}
(One should  keep in mind that, when the background is not invariant under the flow of $X$, there might be a contribution from the background  when calculating   $\Lie_X \mypi_A{}^\mu$.)
When integrated over a   compact hypersurface $\hyp$ with boundary,
\begin{eqnarray}
 \mcH  [\hyp,X]    :=
 \int_\hyp  \mcH^\mu[X]  dS_\mu
   \,,
     \label{13IX18.6asdf}
\end{eqnarray}
 \eq{6IX18.4} leads to the usual field-theoretical version of the generating formula of Hamilton:  Indeed, for solutions of the field equations and for $\lambda$-independent vector fields $X$ and background fields  the last line in \eq{6IX18.4} vanishes and, using the notation
%
\begin{eqnarray}
 &
 \delta \phi^A := \frac{d \phi^A}{d\lambda}
  \,,
  \quad
   \delta \mypi_A{}^{\mu} :=  \frac{d\mypi_A{}^{\mu}}{d\lambda}
   \,,
   &
\\
 &
    dS_\mu = \partial_\mu \rfloor dx^{0}\wedge \cdots \wedge dx^n\,,
   \quad
   dS_{\mu\nu} = -\partial_\mu \rfloor dS_\nu
   \,,
   &
    \label{21XI19.1}
\end{eqnarray}
%
%
after integration of \eq{6IX18.4} over $\hyp$ one obtains
 \ptcrr{there is an integration with forms here, but one has to integrate form densities here instead, which might affect the sign; see email on 5 XII 2019 to Jahanur for the calculation at the level of forms, which justifies the sign here ; for $X=\partial_u$ this is just the integral of the radial component of the divergence of a vector field; checked that this is identical to 5.21 of \cite{CJK} on 31VIII20}
\begin{eqnarray}
\nonumber
 \lefteqn{
 \delta  \mcH  [\hyp,X]    :=
 \int_\hyp \frac{d \mcH^\mu[X]}{d\lambda} dS_\mu
  }
  &&
\\
  &  = &
\int_\hyp
 \big(
    \Lie_X \phi^A  \delta  \mypi_A{}^\mu  -
    \Lie_X \mypi_A{}^\mu \delta \phi^A
    \big)
     dS_\mu
      {-}
        \int_{\partial \hyp} X^{[{{\sigma}}} \mypi_A{}^{\mu]} \delta \phi^A
       \, dS_{{{\sigma}} \mu}
   \,,
    \phantom{xxxxx}
     \label{13IX18.6}
\end{eqnarray}
%
where the boundary term might or might not vanish depending upon the boundary conditions satisfied by the fields at hand. These terms do not vanish, and play a key role for the problems at hand in this work.

Given two one-parameter families  of fields $\phi^A(\lambda )$ and $\phi^A( \tau)$, the ``presymplectic current'' $\omega^\mu$ is defined as
\begin{equation}\label{7IX18.1}
  \omega^\mu\Big(\frac{d \phi }{d\lambda},
  \frac {d \phi}{d\tau}\Big)
   = \frac{d \phi ^A }{ d\tau}
  \frac{d \mypi_A{}^\mu }{d\lambda}-
  \frac{d \phi^A }{d\lambda}
  \frac{d \mypi_A{}^\mu }{d\tau}
   \equiv
   \frac{d \phi ^A }{ d\tau}
   \tpi_A{}^\mu \left(\frac{d \phi} {d\lambda}\right)
   -
  \frac{d \phi^A }{d\lambda}
  \tpi_A{}^\mu\left(\frac{d \phi }{d\tau}
    \right)
   \,,
\end{equation}
with a similar definition for $\tilde \omega^\mu$:
\begin{equation}\label{10IX18.1}
  \tomega^\mu\Big(\frac{d \tphi }{d{{\sigma}}},
  \frac {d \tphi}{d\tau}\Big)
   = \frac{d \tphi^A }{ d\tau}
  \frac{d \tpi_A{}^\mu }{d{{\sigma}}}-
  \frac{d \tphi^A }{d{{\sigma}}}
  \frac{d \tpi_A{}^\mu }{d\tau}
   =
   \frac{d \tphi ^A }{ d\tau}
   \tpi_A{}^\mu \left(\frac{d \tphi} {d\sigma}\right)
   -
  \frac{d \tphi^A }{d\sigma}
  \tpi_A{}^\mu\left(\frac{d \tphi }{d\tau}
    \right)
   \,,
\end{equation}
where in the last equation we have used linearity of $\tpi_A{}^\mu$ in its argument.
Strictly speaking, $\tpi_A{}^\mu \left(\frac{d \tphi} {d\sigma}\right)$ in \eqref{10IX18.1} should be written as $(\tpi_A{}^\mu)_*\left(\frac{d \tphi} {d\sigma}\right)$, where  $(\tpi_A{}^\mu)_*$ is the map tangent to the linear map
$$
 \frac{d \phi} {d\lambda} \mapsto  \tpi_A{}^\mu \left( \frac{d \phi } {d\lambda} \right)
  \,,
$$
but we will stick to the notation $\tpi_A{}^\mu$.

While $\tomega^\mu$ and $\omega^\mu$ look identical, one should keep in mind that they are not defined on the same spaces: the arguments of $\omega^\mu$ are sections of the bundle tangent to the bundle of fields, while the arguments of $\tomega^\mu$ are sections of the bundle of tangents to the tangents. The difference is, however, somewhat esoteric in any case.

\subsubsection{The divergence of the presymplectic  current}

We wish to calculate the  divergence of the pre-symplectic current  (\ref{7IX18.1}). Consider thus, as before, a two-parameter family of fields $\phi^{A}(\lambda, \tau)$.  Assuming that the variations and the coordinate derivatives commute  we have:
\begin{eqnarray}
\partial_{\mu} \left( \frac{d \phi ^A }{ d\tau}
	\frac{d \mypi_A{}^\mu }{d\lambda} \right)
	&=&\partial_{\mu}\frac{d \phi ^A }{ d\tau}
	\frac{d \mypi_A{}^\mu }{d\lambda}+\frac{d \phi ^A }{ d\tau}
	 	\partial_{\mu} \frac{d \mypi_A{}^\mu }{d\lambda} \nonumber \\
		&=&\partial_{\mu}\frac{d \phi ^A }{ d\tau} \frac{d \mypi_A{}^\mu }{d\lambda}+\frac{d \phi ^A }{ d\tau}  \frac{d  }{d\lambda} \left(\mcE_A+\mypi_{A}\right) \nonumber \\
		&=&\frac{d}{d \lambda} \left[\mypi_A{}^{\mu} \partial_{\mu}\frac{d \phi^A}{d \tau} +\left(\mcE_A+\mypi_{A}\right) \frac{d \phi^A}{d 	\tau}  \right] \nonumber \\
			& &	-\mypi_A{}^{\mu} \partial_{\mu}\frac{d^2 \phi^A}{d 	\tau d \lambda} -\left(\mcE_A+\mypi_{A}\right) \frac{d^2 \phi^A}{d 	\tau d \lambda } \nonumber\\
				&=&\frac{d^2 \mcL}{d \lambda \d \tau} +\frac{d}{d \lambda } \left[\mcE_A \frac{d \phi^A}{d 	\tau} \right] \nonumber \\
				& &	-\mypi_A{}^{\mu} \partial_{\mu}\frac{d^2 \phi^A}{d 	\tau d \lambda} -\left(\mcE_A+\mypi_{A}\right) \frac{d^2 \phi^A}{d 	\tau d \lambda } \, . \;
\end{eqnarray}
Changing the order of $\tau$ and $\lambda$ leads to
\begin{eqnarray}
\partial_{\mu} \omega^{\mu}
 &\equiv&
  \partial_{\mu} \left( \frac{d \phi ^A }{ d\tau}
	\frac{d \mypi_A{}^\mu }{d\lambda}-
	\frac{d \phi^A }{d\lambda}
	\frac{d \mypi_A{}^\mu }{d\tau} \right) \nonumber \\
	&=&\frac{d \mcE_A}{d \lambda } \frac{d \phi^A}{d 	\tau} -\frac{d \mcE_A}{d \tau } \frac{d \phi^A}{d 	\lambda} \, .
\end{eqnarray}
We conclude that the divergence will vanish  if the linearised field equations
\begin{equation}\label{21XI19.99}
  \frac{d \mcE_A}{d \lambda }= 0 = \frac{d \mcE_A}{d \tau }
  \,.
\end{equation} are satisfied.

 \subsubsection{The canonical energy of the linearised theory and the presymplectic form}
  \label{ss13IX18.1}

Calculating directly from the definition \eq{5IX18.7+} we have
\begin{eqnarray}
	\frac{d \mcH^\mu[X]}{d\lambda  }
 &
	=
 &
	\tpi_A{}^\mu \Lie_X \phi^A
	+\mypi_A{}^\mu \Lie_X \tphi^A
	- X^\mu \left(
		\mypi_A{}^{{\sigma}} \partial_{{\sigma}} \tphi ^A
		+ \mypi_A \tphi^A
		+\frac{\partial \mcL}{ \partial \psi^I}{\dlambda{\psi^I}}
	\right)
 \nonumber
\\
	 & &
	+\underbrace{\mypi_A{}^\mu \Lie_{\frac{d X}{d \lambda}} \phi^A
	-\frac{d X^{\mu}}{d \lambda} \mcL}_{\mcH^\mu[\frac{d X}{d \lambda}]}
 \,.
\label{2XII.pt2}
\end{eqnarray}
Comparing \eqref{2XII.pt2} with \eqref{6IX18.4} we obtain
\begin{eqnarray}
 0
&=&	
		-\pi_A{}^\mu
		\Lie_X \tphi^A
		-
		\Lie_X \pi_A{}^\mu  \tphi^A
	+ 2 \partial_{{\sigma}} \Big(
	X^{[\sigma}
	\pi_A{}^{\mu]}  \tphi^A
	\Big)
	\nonumber
\\
 &&
	 + X^{\mu}\big( \mypi_{A}{}^{\sigma} \partial_\sigma \tphi^A
    + (\mypi_A + \mcE_A)\tphi^A
        \big)
 \,.
  \label{25II20.1}
\end{eqnarray}
This is true for all fields $\phi$, $\tphi$ and $X$, regardless of whether or not the fields satisfy any equations.

We will differentiate \eqref{25II20.1} with respect to $\lambda$. Before doing this, we note first that a replacement in \eqref{25II20.1} of $X$ by $dX/d\lambda$ gives the identity
\begin{eqnarray}
 0
&=&	
		-\pi_A{}^\mu
		\Lie_{\frac{dX}{d\lambda}} \tphi^A
		-
		\Lie_{\frac{dX}{d\lambda}} \pi_A{}^\mu  \tphi^A
	+ 2 \partial_{\sigma} \Big(
	{\frac{dX}{d\lambda}}^{[\sigma}
	\pi_A{}^{\mu]}  \tphi^A
	\Big)
	\nonumber
\\
 &&
	 + {\frac{dX}{d\lambda}}^{\mu}\big( \mypi_{A}{}^{\sigma} \partial_\sigma \tphi^A
    + (\mypi_A + \mcE_A)\tphi^A
        \big)
 \,.
  \label{25II20.2}
\end{eqnarray}
Similarly, replacing $\tphi$ by $d\tphi/d\lambda$ in \eqref{25II20.1} gives
\begin{eqnarray}
 0
&=&	
		-\pi_A{}^\mu
		\Lie_{X} \frac{d\tphi^A}{d\lambda}
		-
		\Lie_{X} \pi_A{}^\mu  \frac{d\tphi^A}{d\lambda}
	+ 2 \partial_{\sigma} \Big(
	X^{[\sigma}
	\pi_A{}^{\mu]}  \frac{d\tphi^A}{d\lambda}
	\Big)
	\nonumber
\\
 &&
	 + {X}^{\mu}\big( \mypi_{A}{}^{\sigma} \partial_\sigma \frac{d\tphi^A}{d\lambda}
    + (\mypi_A + \mcE_A)\frac{d\tphi^A}{d\lambda}
        \big)
 \,.
  \label{25II20.3}
\end{eqnarray}
Differentiating \eqref{25II20.1} with respect to $\lambda$, after taking into account \eqref{25II20.2} and \eqref{25II20.3} one is led to
\begin{eqnarray}
 0
&=&	
		-\tpi_A{}^\mu
		\Lie_X \tphi^A
		-
		\Lie_X \tpi_A{}^\mu  \tphi^A
	+ 2 \partial_{{\sigma}} \Big(
	X^{[\sigma}
	\tpi_A{}^{\mu]}  \tphi^A
	\Big)
	\nonumber
\\
 &&
	 + X^{\mu}\Big( \tpi_{A}{}^{\sigma} \partial_\sigma \tphi^A
    + \big(\tpi_A + \frac{d\mcE_A}{d\lambda}\big)\tphi^A
        \Big)
 \,.
  \label{25II20.4}
\end{eqnarray}
\ptcheck{4VII20}
Adding  this  to twice the right-hand side of \eqref{5IX18.7} one obtains
\begin{eqnarray}
2 \wmcH [X]
&=&	
\underbrace{
	\Lie_X \tphi^A
	\,
	\tpi_A{}^\mu
	-
	\Lie_X \tpi_A{}^\mu  \tphi^A
}_{\equiv \, \omega^\mu(\tphi, \Lie_X \tphi)}
+ 2 \partial_{{\sigma}} \Big(
X^{[{{\sigma}}}
\tpi_A{}^{\mu]}  \tphi^A
\Big)
-  X^\mu \frac{d \mcE_A}{d \lambda} \tphi^A
\,.
\nonumber
\end{eqnarray}	
This leads us to the following  (compare~\cite[Appendix]{Wald:Iyer}); note the we are not assuming that the fields $\phi$ at which we are linearising satisfy any equations,
 nor that the background structures (if any) are invariant under the flow of $X$:

\begin{Proposition}
 \label{P10IX18.1}
Consider a solution $\tphi$ of the linearised field equations and assume that the vector field $X$ is independent of the fields considered.
The  Hamiltonian current
\begin{eqnarray}
\wmcH^\mu[X]
 & := &
    \tpi^\mu{}_ A\Lie_X \tphi^A - X^\mu \wmcL
        \label{6IX18.5-}
\end{eqnarray}
of the linearised theory can be rewritten as
\begin{eqnarray}
\wmcH^\mu[X]
 &
 = &
  \frac 12
  \omega^\mu(\tphi, \Lie_X \tphi)
   +   \partial_{{\sigma}} \Big(
    X^{[{{\sigma}}}
     \tpi_A{}^{\mu]}  \tphi^A
     \Big)
   \,.
        \label{6IX18.5}
\end{eqnarray}
Here $\wmcL$ is the Lagrangian density for the linearised equations, with $\tpi^A{}_\mu = \partial \wmcL/ \partial (\partial _\mu \tphi^A)$, and $\omega^\mu$ is the presymplectic  current \eq{7IX18.1}.
\end{Proposition}

In view of the above, the Hamiltonian $\wmcH(\hyp,X)$ for the linearised theory associated with a hypersurface $\hyp$  reads
\begin{eqnarray}
\nonumber
\wmcH [\hyp, X]
 & := &
   \int_\hyp
    \big( \tpi^A{}_\mu \Lie_X \tphi^A - X^\mu \wmcL
    \big)
    dS_\mu
\\
 &
 = &
  \frac 12
  \left( \int_\hyp
  \omega^\mu(\tphi, \Lie_X \tphi)
       \, dS_\mu
   {-}
     \int_{\partial \hyp }
    X^{[{{\sigma}}}
     \tpi_A{}^{\mu]}  \tphi^A
     dS_{{{\sigma}}\mu}
      \right)
   \,.
      \phantom{xxxxxxx}
        \label{6IX18.5+}
\end{eqnarray}
%
When $X$ is a time-translation, one often identifies the numerical value of \eq{6IX18.5+} with
the energy of the field contained in $\hyp$. We will use this terminology, momentarily ignoring all the
delicate issues associated with the boundary conditions satisfied by the fields, to which we will return in due course.

\subsubsection{Energy flux}
 \label{ss13IX18.2}

We wish to derive a formula for the flux of energy across $\partial \hyp$. For this, define
\begin{equation}\label{7IX18.2}
 \phi ( \tau)= \Phi_\tau[Y]\big(\phi \big)
  \,,
\end{equation}
where we use the symbol $\Phi_\tau[Y]$ to  denote both the flow of a vector field $Y$ and its action  on our field.
 We consider a family of fields obtaining by flowing along $Y$, and the resulting variational identity.
We will require that $X$ commutes with $Y$, that the background is invariant under the flow of $Y$, and that all the $\phi(\tau)$'s are solutions of the field equations.
\tscomr{Agree,~\eq{6IX18.4} requires $[X,Y]=0$ . }
Equation~\eq{6IX18.4} with $\lambda$ replaced by $\tau$ reads
\begin{eqnarray}
\nonumber
 \frac{d \mcH^\mu[X]}{d\tau}
  &
 =
  &
    \Lie_X \phi^A \frac{d \mypi_A{}^\mu}{d\tau}  -
    \Lie_X \mypi_A{}^\mu \frac{d \phi^A}{d\tau}
   + 2 \partial_{{\sigma}} \left(
    X^{[{{\sigma}}} \mypi_A{}^{\mu]} \frac{d \phi^A}{d\tau}
     \right)
\\
  &
 =
  &
    \Lie_X \phi^A
    \Lie_Y  \mypi_A{}^\mu  -
    \Lie_X \mypi_A{}^\mu \Lie_Y \phi^A
   + 2 \partial_{{\sigma}} \left(
    X^{[{{\sigma}}} \mypi_A{}^{\mu]} \Lie_Y \phi^A
     \right)
   \,.
   \phantom{xxxxxxx}
     \label{9IX18.1}
\end{eqnarray}
Taking $Y=X$, Equation~\eq{9IX18.1} becomes
\begin{eqnarray}
 \frac{d \mcH^\mu[X]}{d\tau}
  &
 =
  & 2 \partial_{{\sigma}} \left(
    X^{[{{\sigma}}} \mypi_A{}^{\mu]} \Lie_X\phi^A
     \right)
   \,.    \label{6IX18.4+}
\end{eqnarray}
This is  the field-theoretical analogue of the statement that a Hamiltonian in mechanics is conserved along its flow, except that here one needs  to take into account the boundary term. Indeed, given a hypersurface $\hyp$ set
$$
 \hyp_\tau := \Phi_\tau[X](\hyp)
$$
and define
\begin{equation}\label{11IX18.13}
  \mcH(\hyp_\tau, X) := \int_{\hyp_\tau} \mcH^\mu dS_\mu\,.
\end{equation}
It follows from \eq{6IX18.4+} that
%
\begin{equation}\label{11IX18.14}
  \frac{d\mcH (\hyp_\tau, X)}{d\tau}
   =
    {-}
      \int_{\partial \hyp_\tau}
    X^{[{{\sigma}}} \mypi_A{}^{\mu]} \Lie_X\phi^A dS_{{{\sigma}} \mu}
    \,,
\end{equation}
%
so that the integrand of \eq{11IX18.14} represents the flux of energy through $\partial \hyp$ when $\hyp$ is dragged along the flow of $X$.

\subsubsection{The divergence of the Noether current}
 \label{ss18XI19.1}

An important consequence of the hypotheses H1.-H2., p.~\pageref{H18XI19.1}, is the identity
\begin{equation}\label{18XI19.1}
  \partial_\mu \mcH^\mu[X]=  \mcE_A   \Lie_X \phi^A
  \,.
\end{equation}
Note that the right-hand side is zero if a) either the field equations $\mcE^A=0$ hold, or b) the solution is stationary in the sense that $\Lie_X \phi = 0$.

The identity is easiest to establish by going to coordinates in which $\Lie_X={\partial_0}$, so that
\begin{eqnarray}\label{18XI19.2}
  \partial_\mu \mcH^\mu[X]
   & = &
    \partial_\mu (\mypi_A{}^\mu {\partial_0} \phi^A - \delta^\mu_{0} \mcL)
    \nonumber
\\
 &  =  &
  ( \partial_\mu  \mypi_A{}^\mu )  {\partial_0} \phi^A
     + \mypi_A{}^\mu  \partial_\mu  {\partial_0} \phi^A  - {\partial_0}\mcL
      \nonumber
\\
 &  =  &
  ( \mypi_A  + \mcE_A)  \partial_{0} \phi^A
     + \mypi_A{}^\mu  \partial_\mu  \partial_{0} \phi^A
       -\frac{ \partial \mcL}{\partial \phi^A} \partial_{0} \phi^A
      -\frac{ \partial \mcL}{\partial \phi^A_\mu } \partial_{0} \partial_\mu \phi^A
 \nonumber
\\
 & = &
 \mcE_A   \partial_{0} \phi^A
  \,,
\end{eqnarray}
which is the same as \eqref{18XI19.1}.

Formula \eqref{18XI19.1} provides an alternative derivation of  \eqref{11IX18.14}, as follows: Let $X$ be a vector field everywhere transversal to a hypersurface $\hyp$ with boundary $\partial \hyp$. Let, as before, $\hyp_\tau$ be obtained by flowing $\hyp$ with the vector field $X$ for a time $\tau$. Let us denote by $T_\tau$ the hypersurface obtained by flowing the boundary of $\hyp$ from time zero to time $\tau$:
$$
 T_\tau =\cup_{s\in[0,\tau]} \phi_s[X](\partial \hyp)
 \,.
$$
Supposing that the right-hand side of \eqref{18XI19.1} vanishes, and applying Stokes' theorem on the set bounded by $\hyp$, $\hyp_\tau$ and $T_\tau$ we obtain
%
\begin{equation}\label{18XI19.3}
  H[\hyp_\tau,X]= H[\hyp_{0},X] + \int_{T_\tau}\mcH^\mu dS_\mu
  \,.
\end{equation}
Differentiating with respect to $\tau$ one obtains
%
\begin{equation}\label{18XI19.4}
  \frac{dH[\hyp_\tau,X]}{d\tau}=
   {-}
   \int_{\partial \hyp} X^{[\nu} \mcH^{\mu]} dS_{\nu\mu}
  \,.
\end{equation}
This coincides with \eqref{11IX18.14}, as the term $X^\mu \mcL$ in the definition of $\mcH^\mu$ drops out from the integral after antisymmetrisation.

\subsubsection{The energy flux revisited}
 \label{ss22XI19.1}

In the case of theory of fields linearised   around a solution, there exists yet  another way of computing the flux of energy across $\partial \hyp$, as follows:%
\footnote{The calculation here is due to J.Hoque.}

Since the divergence of $\omega^\mu$ vanishes for solutions of field equations, the calculation leading to \eqref{18XI19.4} gives
\begin{eqnarray}
  \frac{d}{d\tau} \int_\hyp
\omega^\mu(\tphi, \Lie_X \tphi)
       \, dS_\mu
    =
    -
     \int_{\partial \hyp }
   X^{[{{\sigma}}}
     \omega{}^{\mu]}(\tphi, \Lie_X \tphi)\,
     dS_{{{\sigma}}\mu}
   \,.
\end{eqnarray}
Calculating the $\tau$-derivative of  $\wmcH [\hyp, X]$ as given by \eqref{6IX18.5+}  we obtain
\begin{eqnarray}
\nonumber
 \frac{d \wmcH[\hyp_\tau, X]}{d\tau}
  &
 =
  &
  \frac 12
  \frac{d}{d\tau} \int_\hyp
\omega^\mu(\tphi, \Lie_X \tphi)
       \, dS_\mu
   {-}
     \frac 12\int_{\partial \hyp }
   \Lie_X \left( X^{[{{\sigma}}}
     \tpi_A{}^{\mu]}  \tphi^A\right)
     dS_{{{\sigma}}\mu}
  \\
      &
      =
      &
     -\frac 12\int_{\partial \hyp }
   X^{[{{\sigma}}}
     \omega{}^{\mu]}(\tphi, \Lie_X \tphi)\,
     dS_{{{\sigma}}\mu}
     \nonumber
\\
 && -
     \frac 12\int_{\partial \hyp }\left(
    X^{[{{\sigma}}}
     \Lie_X\tpi_A{}^{\mu]}  \tphi^A
    +  X^{[{{\sigma}}}
     \tpi_A{}^{\mu]} \Lie_X \tphi^A
     \right)
     dS_{{{\sigma}}\mu}
   \,.
\end{eqnarray}
Inserting the definition $\omega^{\mu}(\tphi, \Lie_X \tphi):= \Lie_X \tphi^{A}\tpi_A{}^{\mu}-
\tphi^{A} \Lie_X \tpi_{A}{}^{\mu}$, one term cancels out and another gets doubled, resulting in the energy flux equal to
\begin{equation}\label{11IX18.14asdf}
  \frac{d\wmcH (\hyp_\tau, X)}{d\tau}
   =
    -
      \int_{\partial \hyp_\tau}
    X^{[{{\sigma}}} \tpi_A{}^{\mu]} \Lie_X\tphi^A dS_{{{\sigma}} \mu}
    \,,
\end{equation}
%
recovering again  \eqref{11IX18.14}.

\subsection{Scalar fields on de Sitter spacetime} \label{s22XII19.1}

\tscomr{Independent calculations done 7VII2020. Files scalar field calculations TS(2).jpg on SVN. Energy and flux here and on the JPGs are the same.}

We apply the formalism to a linear scalar field in Minkowski spacetime and in de Sitter spacetime. In our signature the Lagrangian reads
\begin{equation}\label{22XII19.11}
  \mcL =- \frac 12  \sqrt{|-\det g|}
   \big(
    g^{\mu\nu}\partial_\mu \phi \, \partial_\nu \phi
      \pmass \phi^2
    \big)
  \,,
\end{equation}
for a constant $m$.
The theory coincides with its linearisation and we will therefore not make a distinction between the fields $\varphi$ and $\tilde \varphi$.
The canonical energy-momentum current $\mcH^\mu$ equals
\begin{equation}\label{22XII19.12}
  \mcH^\mu [X] = -  \sqrt{|-\det g|}
   \Big( \nabla^\mu \phi \,\Lie_X \phi -
   \frac 12
  \big(\nabla^\alpha \phi \nabla_\alpha \phi \pmass \phi^2
   \big)
    X^\mu
    \Big)
  \,.
\end{equation}

We consider simultaneously the Minkowski space-time and the de Sitter space-time in Bondi coordinates, in which the metric takes the form
\be
{\nobarg}\equiv {\nobarg}_{\a \b} dx^\a dx^\b = \epsilon \lapseTB^2 du^2-2du \, dr
 + r^2
  \underbrace{(d\theta^2+\sin^2 \theta d\phi^2)}_{=:\zzhTBW }
   \,,
   \label{8VII20.11}
\ee
with $\Lambda \ge 0$,
$$
\lapseTB := \sqrt{|
(1-\alpha^2 r^2)|}
\,,
 \quad
 \alpha \in \big\{0,\sqrt{\frac{\Lambda}{3}}
  \big\}
 \,,
 \quad
 \epsilon \in \{\pm 1\}
 \,,
$$
with  $\epsilon$ equal to one if $1-\alpha^2 r^2<0$, and minus one otherwise. Hence
$$
g^{\alpha\beta}\partial_\alpha\partial_\beta = -2 \partial_u\partial_r -  \epsilon \lapseTB^2 (\partial_r)^2
 + r^{-2}\zzhTBW^{AB}\partial_A\partial_B
 \,,
$$
and
\begin{equation}\label{27XII19.11}
  \nabla \phi = - \partial_r \phi \,\partial_u  - (\partial_u \phi + \epsilon \lapseTB^2 \partial_r \phi)\partial_r + r^{-2}\zzhTBW^{AB}\partial_A \phi \, \partial_B
  \,.
\end{equation}
(Starting from a more usual form of the de Sitter metric,
\begin{equation}\label{16I20.1}
  g =   -(1-\alpha^2 r^2)dt^2 + \frac{dr^2}{1-\alpha^2 r^2}
 + r^2
  (d\theta^2+\sin^2 \theta d\phi^2)
   \,,
\end{equation}
the form \eqref{8VII20.11} can be obtained, for $\alpha r >1$,  by introducing a coordinate $u$ through the formula
\begin{equation}\label{16I20.3}
  du : = dt - \frac{dr}{1-\alpha^2 r^2} \equiv d
   \Big(
    t + \frac{1}{2 \alpha} \ln \big(\frac{\alpha r -1}{\alpha r + 1} \big)
    \Big)
  \,;
\end{equation}
cf., e.g., \cite{FischerDeSitter}.)

We denote by $\mcC_{u}$ the light cone of constant $u$, and by  $\mcC_{u,R}$ its truncation in which the Bondi coordinate $r$  ranges from zero to  $ R$. We wish to calculate the canonical energy associated with the Killing vector field $X=\partial_u$ and contained in  $\mcC_{u,R}$.
Letting
\begin{equation}\label{14VII20.1a}
   d\mu_{\mcC} = \sqrt{\det g_{AB}} \; dr\wedge dx^2\wedge dx^3
   \ \mbox{and} \
     d\mu_{\zzhTBW} = \sqrt{\det \zzhTBW_{AB}} \;  dx^2\wedge dx^3
\end{equation}
we find
\begin{eqnarray}
 \nonumber
 \lefteqn{
    E_c[\phi,\mcC_{u,R}]  :=
   \int_{\mcC_{u,R}}  \mcH^\mu [\partial_u] dS_\mu
   =
   \int_{\mcC_{u,R}}  \mcH^u [\partial_u] dS_u
   }
   &&
\\
 && =
      \frac 12  \int_{\mcC_{u,R}} (
      \nabla^\alpha \phi \nabla_\alpha \phi - 2\nabla^u\phi \, \partial_u \phi
      +m^2 \phi^2
      )
       d\mu_{\mcC}
        \nonumber
\\
 &&
  =
    \frac 12 \int_{\mcC_{u,R}}
      \big(
     g^{AB}\partial_A \phi \, \partial_B \phi
      -\epsilon \lapseTB^2 ( \partial_r \phi)^2
      +m^2 \phi^2
      \big) r^2
       \, dr \, d\mu_{\zzhTBW}
        \,.
    \label{22XII19.112}
\end{eqnarray}
We see that $(\partial_r\phi)^2$ gives a positive contribution to the energy integral in the region where $g_{uu} = \epsilon \lapseTB^2$ is negative, and a negative contribution otherwise.

The presymplectic current is defined as
\begin{equation}\label{15VI20.1}
  \omega^\mu (\delta_1 \phi, \delta_2 \phi) =
   ( -\delta_2 \phi \nabla^\mu \delta_1 \phi +
  \delta_1 \phi \nabla^\mu \delta_2 \phi
   ) \sqrt{|\det g |}
   \,,
\end{equation}
with obviously vanishing divergence on solutions of field equations:
$$
 \nabla_\mu \omega^\mu \equiv \partial_\mu \omega^\mu = 0
  \,.
$$
The $u$-component of the presymplectic current on $\mcC_u$ reads
 \ptcheck{9VI20, sign crosschecked with JJ}
\begin{equation}\label{27XII19.2}
  \omega^u(\delta_1\phi, \delta_2\phi) =
   \big( \delta_2 \phi \, \partial_r \delta_1 \phi
   -  \delta_1 \phi \, \partial_r \delta_2 \phi
   \big) \sqrt{|\det g|}
   \,.
\end{equation}
When $X=\partial_u$, this results in the following volume integrand in \eqref{6IX18.5}
\begin{equation}\label{27XII19.3}
 \frac{1}{2 } \omega^u( \phi, \partial_u\phi) =
   \frac 12 \big( \partial_u \phi \, \partial_r \phi
   -    \phi \, \partial_r \partial_u \phi
   \big) \sqrt{|\det g|}
   \,,
\end{equation}
while the boundary integrand equals
\begin{equation}\label{27XII19.4}
   \partial_{{\sigma}} \Big(
    X^{[{{\sigma}}}
     \tpi^{\mu]}  \phi
     \Big) = \frac 1 2
      \partial_{i} \Big(
   \phi \nabla^i \phi\sqrt{|\det g|}
     \Big)
   \,.
\end{equation}
\Eq{6IX18.5} leads to the following alternative form of \eqref{22XII19.112}:
\begin{eqnarray}
 \nonumber
  E_c[\phi,\mcC_{u,R}]
   & =  &
   \int_{\mcC_{u,R}}
   \Big( \frac 12
  \omega^\mu(\tphi, \Lie_X \tphi)
   +   \partial_{{\sigma}} \big(
    X^{[{{\sigma}}}
     \pi^{\mu]}  \tphi
     \big)
     \Big) dS_\mu
\\
 & = &
      \frac 12  \int_{\mcC_{u,R}}
        \Big( \partial_u \phi \, \partial_r \phi
   -    \phi \, \partial_r \partial_u \phi
     \Big)
       d\mu_{\mcC}
       \nonumber
\\
 &  &
       + \frac{1}{2}
        \int_{S_{u,R}}
   \phi \nabla^i \phi\sqrt{|\det g|}
   \partial_i\rfloor( dx^1\wedge dx^2\wedge dx^3)
    \label{27XII19.5-}
\\
 & = &
      \frac 12  \int_{\mcC_{u,R}}
        \Big( \partial_u \phi \, \partial_r \phi
   -    \phi \, \partial_r \partial_u \phi
     \Big)
       d\mu_{\mcC}
       \nonumber
\\
 &  &
       - \frac{1}{2}
        \int_{S_{u,R}}
   \phi (\partial_u \phi + \epsilon N^2 \partial_r \phi) \sqrt{|\det g|}
    \,
   dx^2\wedge dx^3
        \,.
         \phantom{xxxx}
    \label{27XII19.5}
\end{eqnarray}
A careful reader might justly worry about the convergence of the integrals, since a light cone is not a smooth manifold. The fact that all integrals in this section are well behaved for solutions of the wave equation which are smooth in a neighborhood of the light cone can be verified using \cite[Proposition~2.1]{ChJezierskiCIVP}.

As a check of equality of \eqref{27XII19.5-} and  \eqref{22XII19.112}, we note that the field equation for $\phi$  reads
\ptcheck{24VII20}
\begin{eqnarray}
 m^2  \sqrt{|\det g|} \phi
   & = &
    \sqrt{|\det g|}\, \Box \phi = \partial_\mu ( \sqrt{|\det g|} \nabla^\mu \phi)
    \nonumber
\\
 & = &
       -  \sqrt{|\det g|} \partial_u \partial_r \phi +
        \partial_i ( \sqrt{|\det g|} \nabla^i \phi)
  \label{27XII19.6}
\\
   & = &
       -  r^2\sqrt{\det \zzhTBW} \partial_u \partial_r \phi -
        \partial_r \big(
          r^2\sqrt{\det \zzhTBW}   (\partial_u \phi + \epsilon \lapseTB^2 \partial_r \phi)
          \big)
          \nonumber
\\
 &&
          +
        \partial_A (\sqrt{\det \zzhTBW}  \zzhTBW^{AB}\partial_B \phi) \nonumber
\\
   & = &
       -  2r \sqrt{\det \zzhTBW}\partial_r (r  \partial_u  \phi)  -
        \partial_r \big(
          r^2\sqrt{\det \zzhTBW}  \epsilon \lapseTB^2 \partial_r \phi
          \big)
          \nonumber
\\
 &&
          +
        \partial_A (\sqrt{\det \zzhTBW}  \zzhTBW^{AB}\partial_B \phi)
        \,.
  \label{27XII19.6b}
\end{eqnarray}
Equation~\eqref{27XII19.6} together with the divergence theorem can be used to replace the boundary term in \eqref{27XII19.5-} by a volume integral, indeed recovering \eqref{22XII19.112}.

The mass-flux formula \eqref{11IX18.14} becomes
\begin{eqnarray}
 \nonumber
   \frac{dE_c[\phi,\mcC_{u,R}]}{d u }
    &  \equiv &
  \frac{d\mcH (\mcC_{u,R}, \partial_u)}{d u }
    =
    -
      \int_{\partial \hyp_\tau}
    X^{[{{\sigma}}} \pi ^{\mu]} \partial_u\phi \, dS_{{{\sigma}} \mu}
\\
   &   =   &
        \int_{S_{u,R}}
    \nabla^{r} \phi \, \partial_u\phi \,  r^2 \, d\mu_{\zzhTBW}
     \nonumber
\\
   &   =  &
     -
        \int_{S_{u,R}}
   \big(
   (\partial_u\phi)^2 + \epsilon N^2 \partial_r \phi \, \partial_u \phi
   \big) \, r^2
    \, d\mu_{\zzhTBW}
    \,.
    \label{4VII20.23}
\end{eqnarray}
%

Let us momentarily assume that $\Lambda=0=m$. As is well known, there exists a large class of solutions of the wave equation on Minkowski spacetime with full asymptotic expansions, for large $r$,
\begin{equation}\label{4VI20.21}
  \phi(u,r, x^A) = \frac{\oone \phi(u,x^A)}{r}
  + \frac{\otwo \phi(u,x^A)}{r^2}
  + \frac{\othree \phi(u,x^A)}{r^3} + ...
\end{equation}
After passing to the limit $R\to \infty$, for such solutions \eqref{4VII20.23} becomes the usual energy-loss formula for the scalar field:
\begin{eqnarray}
   \frac{dE_c[\phi,\mcC_{u }]}{d u }
   &   =  &
     -
        \int_{S^2}
    (\partial_u\!\oone \phi)^2\, d\mu_{\zzhTBW}
    \,.
    \label{4VII20.24}
\end{eqnarray}

We now turn our attention to the de Sitter case. We first
 consider a massive scalar field, with the mass chosen so that the equation is conformally covariant,
\begin{equation}\label{5VII20.5}
  \Box_g \phi - \underbrace{\frac{(d-2)R(g)}{4 (d-1)}}_{=:m^2} \phi = 0
  \,,
\end{equation}
where $d$ is the dimension of spacetime and $R(g)$ is the scalar curvature of $g$. After a conformal transformation  $g\mapsto \Omega^2 g$ the field $\Omega^{d/2-1} \phi$ satisfies again \eqref{5VII20.5}, with $g$ there replaced by $\Omega^2 g$. This implies that solutions of \eqref{5VII20.5} with smooth initial data on a Cauchy surface in de Sitter spacetime behave asymptotically for large $r$, in spacetime dimension four, again as in \eqref{4VI20.21}.
We return to this in Section~\ref{App25VII20.1} below.

 For such solutions \eqref{27XII19.5} becomes
\begin{eqnarray}
 \nonumber
  E_c[\phi,\mcC_{u,R}]
 & = &
      \frac 12  \int_{\mcC_{u,R}}
        \Big( \partial_u \phi \, \partial_r \phi
   -    \phi \, \partial_r \partial_u \phi
     \Big)
       d\mu_{\mcC}
       \nonumber
\\
 &  &
       - \frac{1}{2}
        \int_{S^2}
   \oone \phi
        \big( \partial_u\!\oone \phi - \alpha^2 R  \oone \phi \big) \, d\mu_{\zzhTBW}
    + o(1)
        \,,
         \phantom{xxxx}
    \label{27XII19.519}
\end{eqnarray}
where the volume integral converges when passing with the radius $R$ to infinity, but the boundary integral diverges linearly in $R$ in general.
Indeed, given $\phi|_{\mcC_u}$ one can integrate \eqref{27XII19.6b} to \ptcheck{23VII20, checked with TS except for convergence at the origin }
 \ptcrr{AV says change $u$ to $U=ru$ and then this should be phg in this variable, but is not, a failed attempt was in the appendix}
\begin{eqnarray}
        \partial_u  \phi
        &= &
         \frac 1{2r}\int_{0}^{r}\frac{1}{\rho}
       \big(
        - \partial_r \big(
          r^2   \epsilon \lapseTB^2 \partial_r \phi
          \big)
           -  {m^2} r^2 \phi  +
         {\Delta_{\ringh}} \phi
        \big)\big|_{r=\rho} d\rho
        \nonumber
\\
 & = &
    \frac{ \partial_u  \oone \phi}{r}+  \frac{ 2 \alpha^2 \othree  \phi - \Delta_{\ringh} \oone \phi}{2 r^2} + \ldots
        \,,
  \label{27XII19.6bag}
\end{eqnarray}
%
%
where $\Delta_{\ringh}$ is the Laplace operator associated with the metric $\ringh$, and where we have used
\begin{equation}\label{23VII20}
  m^2 = 2 \alpha ^2
\end{equation}
in spacetime dimension four.
Here
 \ptcheck{24VII20}
\begin{equation}
        \partial_u  \oone \phi
        =
         \int_{0}^{\infty}\frac{1}{2\rho}
       \big(
        - \partial_r \big(
          r^2  \epsilon \lapseTB^2 \partial_r \phi
          \big)
           -  2 \alpha^2 r^2 \phi
          +
        {\Delta_{\ringh}} \phi
        \big)\big|_{r=\rho} d\rho
        \,,
  \label{8VII20.1}
\end{equation}
which shows that $\oone \phi$ will not be zero at later times in general even if it is initially.

As for \eqref{4VII20.23}, we find
\begin{eqnarray}
 \frac{dE_c[\phi,\mcC_{u,R}]}{d u }
   &   =  &
      -
        \int_{S^2}
        \bigg(
    (\partial_u\!\oone \phi)^2
     - \alpha^2 R \oone \phi \partial_u\!\oone \phi
    \bigg)
     \, d\mu_{\zzhTBW}
       + o(1)
    \,.
     \phantom{xxx}
    \label{4VII20.23a}
\end{eqnarray}
This diverges again when $R$ tends to infinity. However, we see that the divergent term in $E_c$ has a dynamics of its own, so that a \emph{renormalised energy} can be obtained by subtracting the divergent term in $E_c$ and passing with $R$ to infinity,
\begin{eqnarray}
  E [\phi,\mcC_{u }]
 & := &
      \frac 12  \int_{\mcC_{u }}
        \Big( \partial_u \phi \, \partial_r \phi
   -    \phi \, \partial_r \partial_u \phi
     \Big)
       d\mu_{\mcC}
       - \frac{1}{2}
        \int_{S^2}
   \oone \phi
         \partial_u\!\oone \phi   \, d\mu_{\zzhTBW}
        \,.
         \phantom{xxxxxx}
    \label{27XII19.520}
\end{eqnarray}
The renormalised energy satisfies again an energy-loss formula identical to \eqref{4VII20.24}, formally coinciding with that in Minkowski spacetime.

A similar behaviour is observed for the   massless scalar field. It follows from~\cite{VasydS} (cf.\ Section~\ref{App25VII20.1}) that scalar fields evolving out of smooth initial data on a Cauchy surface  have an asymptotic expansion of the form
\begin{equation}\label{4VI20.21a}
  \phi(u,r, x^A) = \ozero  \phi(u,r, x^A)
  +  \frac{\oone \phi(u,x^A)}{r}
  + \frac{\otwo \phi(u,x^A)}{r^2}
 \ldots
 \,,
\end{equation}
compare \eqref{25VII20.11}.

It turns out that the volume part of $E_c$ given by \eqref{27XII19.5} does not converge anymore as $R$ tends to infinity under \eqref{4VI20.21a}. Indeed, using \eqref{25VII20.11}-\eqref{27XII19.6bags} one finds
\begin{eqnarray}
  \lefteqn{
      \frac 12  \int_{\mcC_{u,R}}
        \Big( \partial_u \phi \, \partial_r \phi
   -    \phi \, \partial_r \partial_u \phi
     \Big)
       d\mu_{\mcC}
       }
       \nonumber & &
       \\
    &=&    -\frac 12  \int_{S^2}\bigg(
  \Big(
    \alpha^{2} (\oone \phi)^{2}-\ozero   \phi  \partial_u \oone \phi
  \Big) R
 \nonumber
\\
& &
+\zspaceD^{A}\Big( \oone \phi \zspaceD_{A}\ozero \phi- \ozero \phi  \zspaceD_{A} \oone \phi\Big) \ln R
 \bigg)
  \, d\mu_{\zzhTBW}
  +O(1) \, ,
  \phantom{xxx}
  \label{27VII20.ts223}
\end{eqnarray}
where $\zspaceD$ is the covariant derivative associated with the metric $\ringh$ and where $O(1)$ denotes terms which have a finite limit as $R$ tends to infinity.
Note that the logarithmic term  integrates-out to zero over $S^2$. This leads us to define the finite part, say $E_{\Vol}$, of the volume integral as
\begin{equation}\label{28VII20.101}
  E_{\Vol}:=
\lim_{R\to\infty}
 \bigg( \frac 12  \int_{\mcC_{u,R}}
        \Big( \partial_u \phi \, \partial_r \phi
   -    \phi \, \partial_r \partial_u \phi
     \Big)
       d\mu_{\mcC}
       + \frac R2  \int_{S^2}
  \big(
    \alpha^{2} (\oone \phi)^{2}-\ozero   \phi  \partial_u \oone \phi
  \big)
  \, d\mu_{\zzhTBW}
  \bigg)
  \,.
\end{equation}
One now finds
\ptcheck{28VII20, with TS}
  \begin{eqnarray}
  E_c [\phi,\mcC_{u,R }]
  & := & E_{\Vol}
  -\frac 12  \int_{S^2}\bigg(
  \Big(
    \alpha^{2} (\oone \phi)^{2}-\ozero \zspaceD_{A} \phi \zspaceD^{A} \ozero \phi
  \Big) R
 \nonumber
\\
 & &
  +\ozero \phi \oone \phi
  +\oone \phi \partial_{u} \oone \phi
  -3 \alpha^{2} \ozero \phi \othree \phi
    \nonumber
  \\
  & &
  \underbrace{
  +\oone \phi {\Delta_{\ringh}} \ozero \phi
  -\frac12 \ozero\phi {\Delta_{\ringh}} \oone \phi
    }_{=\frac12\oone \phi {\Delta_{\ringh}} \ozero \phi
    \ \mathrm{after\ integration\ by\ parts}}
    \bigg)
  \, d\mu_{\zzhTBW}
  +o(1) \, ,
  \phantom{xxx}
  \label{27VII20.ts2}
  \end{eqnarray}
  which continues to diverge   as $R$ tends to infinity in general. The associated energy flux formula reads
  \begin{eqnarray}
  \nonumber
  \frac{dE_c[\phi,\mcC_{u,R}]}{d u }
  &   =  &
  -
  \int_{S^2}
  \bigg(
    R \big(
     {\Delta_{\ringh}} \ozero \phi + \partial_{u} \oone \phi
     \big) \alpha^{2} \oone \phi
  \nonumber
  \\
  & &
  +\Big(
    \oone \phi
    -3\alpha^{2} \othree \phi
    - \frac12 {\Delta_{\ringh}} \oone \phi
    \Big) \alpha^{2} \oone \phi
  \nonumber
  \\
  & &
    +(\partial_{u} \oone \phi)^2
    + {\Delta_{\ringh}} \ozero \phi \partial_{u} \oone \phi
  \bigg)
  \, d\mu_{\zzhTBW}
  + o(1)
  \, .
  \phantom{xxxxx}
  \label{27VII20.ts1}
  \end{eqnarray}
  Using $\partial_u \ozero \phi = \alpha^2 \oone \phi$ (cf.\ \eqref{27XII19.6bags}), one finds (unsurprisingly) that the divergent term has a dynamics of its own, so that the finite renormalised energy, defined as
  \begin{eqnarray}
  \lefteqn{
  E  [\phi,\mcC_{u }]
  := E_{\Vol}
  }
  &&
  \nonumber
\\
 &&
  -\frac 12  \int_{S^2}\bigg( \ozero \phi \oone \phi
  +\oone \phi \partial_{u} \oone \phi
  -3 \alpha^{2} \ozero \phi \othree \phi
  +
   \frac12\oone \phi {\Delta_{\ringh}} \ozero \phi
    \bigg)
  \, d\mu_{\zzhTBW}
  \phantom{xxx}
  \label{27VII20.ts21}
  \end{eqnarray}
 has a finite and well-defined flux:
  \begin{eqnarray}
  \nonumber
  \frac{dE [\phi,\mcC_{u }]}{d u }
  &   =  &
  -
  \int_{S^2}
  \bigg( \Big(
    \oone \phi
    -3\alpha^{2} \othree \phi
    - \frac12 {\Delta_{\ringh}} \oone \phi
    \Big) \alpha^{2} \oone \phi
  \nonumber
  \\
  & &
    +(\partial_{u} \oone \phi)^2
    + {\Delta_{\ringh}} \ozero \phi \partial_{u} \oone \phi
  \bigg)
  \, d\mu_{\zzhTBW}
  \, .
  \phantom{xxxxx}
  \label{27VII20.tsasdf1}
  \end{eqnarray}
\subsubsection{Asymptotics of scalar fields on de Sitter spacetime}
 \label{App25VII20.1}

In order to understand the behaviour for large $r$, in Bondi coordinates,   of solutions of the massive or massless wave equation,
\begin{equation}\label{25VII20}
  \Box \phi = m^2 \phi
  \,,
\end{equation}
on de Sitter spacetime
it is most convenient to work using a foliation of  (part of) de Sitter spacetime  by flat submanifolds, so that the metric takes the form
 \begin{equation}\label{19VI13.201}
  g= -d\tau^2 + e^{2 \sqrt{\frac \Lambda 3}\tau} (dx^2 + dy^2 +dz^2)
  \,.
 \ee
Let
$\rho=\sqrt{x^2+ y^2 +z^2}$, where $(x,y,z)$ are as in \eqref{19VI13.201}, thus
 \begin{equation}\label{19VI13.202}
  g= -d\tau^2 + e^{2 \sqrt{\frac \Lambda 3}\tau}
   \big(
    d\rho^2 + \rho^2(d\theta^2 +\sin^2 \theta d\varphi^2)
     \big)
  \,.
 \ee
 Set
 \begin{equation}\label{23VII20.1}
   T:= \alpha^{-1}  e^{-\alpha\tau}
   \,,
   \quad \alpha = \sqrt{\frac{\Lambda}{3}}
    \,.
 \end{equation}

 We start our discussion with the case $m=0$.
 According to \cite{VasydS}, smooth solutions of the massless scalar wave equation on de Sitter spacetime extend through the conformal boundary $\{T=0\}$ as
 \begin{equation}\label{23VII20.7}
   \phi= f + T^3 \ln T \check f
   \,,
 \end{equation}
 where $f$ and $\check f$ are smooth functions of $(T,x,y,z)$. (By matching coefficients as below one finds in fact that $\check f\equiv 0$; in other words, $\phi$ extends smoothly across the conformal boundary.)
The coordinate transformation
\begin{equation}\label{3II19.5}
  r= \rho e^{\alpha \tau }
   \,,
   \quad
   t= \tau - \frac 1 { 2 \alpha} \ln (-1+ \rho^2 \alpha^2 e^{2 \alpha \tau }  )
   \,,
\end{equation}
brings \eqref{19VI13.202} to the form
\begin{equation}\label{4II19.1}
  g = - V dt^2 + \frac{dr^2}{V} + r^2 (d\theta^2 + \sin^2 \theta d\varphi^2)
  \,,
   \quad
    V = 1 - \alpha^2 {r^2}
   \,.
\end{equation}
In terms of the coordinate $u$ of \eqref{16I20.3},
\begin{equation}\label{23VII20.2}
  u =
    t + \frac{1}{2 \alpha} \ln \big(\frac{\alpha r -1}{\alpha r + 1} \big)
  \,,
\end{equation}
one finds
\begin{equation}\label{23VII20.4}
  u =
    \tau - \frac{1}{  \alpha} \ln ( \alpha r +  1)
    \quad
    \Longleftrightarrow
    \quad
    T= \frac{e^{-\alpha u}}{\alpha(1+\alpha r)}
  \,,
\end{equation}
as well as
\begin{equation}\label{23VII20.3}
  \rho = \frac{r e^{-\alpha u}}{1+\alpha r}
  \,.
\end{equation}
We thus have the expansions as $r\to\infty$, absolutely convergent for $r > \alpha^{-1}$,
\begin{eqnarray}
  \rho
   & = &
    \frac{  e^{-\alpha u}}{ \alpha  }
   \sum_{n=0}^\infty\left(-\frac{1}{ \alpha r}\right)^n
   = \frac{  e^{-\alpha u}}{ \alpha  }
   \left(1 -\frac{1}{ \alpha r} +  \frac{1}{ (\alpha r)^2} + \ldots\right)
   \,,
   \nn
\\
  T
   & = &
    \frac{  e^{-\alpha u}}{ \alpha^2 r }
   \sum_{n=0}^\infty\left(-\frac{1}{ \alpha r}\right)^n
   =    \frac{e^{-\alpha u}} \alpha
   \left( \frac{1}{ \alpha r} -  \frac{1}{ (\alpha r)^2} + \ldots\right)
   \,.
   \label{23VII20.5}
\end{eqnarray}
This shows that a Taylor-series in $T$, near $T=0$, for a function $f$,
\begin{equation}\label{23VII20.6a}
  f(T,\rho,\theta,\varphi) =
  f(0,\rho,\theta,\varphi) +
  \partial_T f(0,\rho,\theta,\varphi) T + \cdots
  \,,
\end{equation}
translates into a full asymptotic expansion in $1/r$, for large $r$:
\begin{equation}\label{23VII20.7a}
 \underbrace{f|_{(T=0,\rho=\frac{e^{-\alpha u}}{\alpha},\theta,\varphi)}}_{\ozero f (u,\theta,\varphi)} +
  \underbrace{
  \frac{
   e^{-\alpha u}(
     \partial_T  f- \partial_\rho f)|_{(T=0,\rho=\frac{e^{-\alpha u}}{\alpha},\theta,\varphi)}
    }{\alpha  } }_{\oone f(u,\theta,\varphi)}  \times  \frac{1 }{  r } + \cdots\,.
\end{equation}
Using
$$
 \ln (\alpha T) = - \alpha u -\ln (1+\alpha r) = \ln(\alpha r)
  \left( 1 + \frac{1}{\alpha r} + \ldots
  \right)
  - \alpha u
  \,,
$$
Equation~\eqref{23VII20.7a} translates to the following asymptotic expansion for solutions $\phi$ of the massless wave equation:
\begin{equation}\label{23VII20.8}
  \phi(u,r, x^A) = \ozero \phi(u,x^A)  + \frac{\oone \phi(u,x^A)}{r}
 + \frac{\otwo \phi(u,x^A) }{r^2}
   + \frac{\othree \phi(u,x^A)}{r^3}
   + \frac{ \othreeone \phi(u,x^A)\ln r}{r^3} +
    ...
  \,.
\end{equation}

Here some words of caution are in order. When considering the characteristic Cauchy problem for the wave equation with initial data on a light cone $\mcC(u_0)$, any function $\phi|_{\mcC(u_0)}$ can be used as initial data. In particular we can prescribe $\phi $ of the form \eqref{23VII20.8} at $u=u_0$ with arbitrary expansion functions $ \ok \phi$. However, some relations will have to be satisfied by  the coefficients so that $\phi$ extends as in \eqref{23VII20.7}. This can be seen by inserting the Taylor expansion \eqref{23VII20.6} in the massless wave equation in the coordinates of \eqref{19VI13.201} to find that solutions take the following form near $T=0$:
 \ptcheck{27VII20 with TS}
\begin{equation}\label{25VII20.2}
  \phi(T,x^i) = f(x^i) - \frac 12 \Delta_\delta  f(x^i) T^2 + \frac 16 \breve f(x^i) T^3 -
   \frac 18 \Delta_\delta ^2  f(x^i) T^4
   +
   \cdots
  \,,
\end{equation}
with arbitrary functions $f$ and $\breve f$, where $\Delta_\delta$ is the Laplacian on Euclidean $\R^3$, and
where no logarithmic terms occur, so that the function $\check f$ in \eqref{23VII20.7} is zero.

Using \eqref{25VII20.2}, the expansion \eqref{23VII20.8} becomes
 \ptcheck{27VII20 with TS}
\begin{eqnarray}
  \phi(u,r, x^A)
   & = &
   f\big(\frac{e^{-\alpha u}}{\alpha},x^A\big)
     - \frac{e^{-\alpha u}\partial_\rho f\big(\frac{e^{-\alpha u}}{\alpha},x^A\big)}{\alpha^2r}
      - \frac{\Delta_{\ringh}f\big(\frac{e^{-\alpha u}}{\alpha},x^A\big) }{2 (\alpha r)^2} +
    ...
    \nonumber
\\
 & = &
  \ozero \phi(u,x^A)
   + \frac{ \oone \phi(u, x^A) }{\alpha^2 r }   - \frac{\Delta_{\ringh}\ozero \phi(u,x^A)  }{2 (\alpha r)^2} +
    ...
  \,.\label{25VII20.11}
\end{eqnarray}
where $\ozero \phi(u,\cdot) =f\big(\frac{e^{-\alpha u}}{\alpha},\cdot\big)$ and $\othree \phi$ are arbitrary. Note that $\otwo \phi $ is determined uniquely by $\ozero \phi $.

As a consistency check, given $\phi|_{\mcC_u}$ of the form \eqref{25VII20.11} one can integrate \eqref{27XII19.6b} to determine $\partial_u \phi$ on $\mcC_u$:
 \ptcheck{ 20 VIII 20 by TS}
\begin{eqnarray}
        \partial_u  \phi
        &= &
         \frac 1{2r}\int_{0}^{r}\frac{1}{\rho}
       \big(%
        - \partial_r \big(
          r^2   \epsilon \lapseTB^2 \partial_r \phi
          \big)
          +
        {\Delta_{\ringh}} \phi
        \big)\big|_{r=\rho} d\rho
        \nonumber
\\
 & = &   \alpha^2   \oone \phi + \frac{\partial_u \oone \phi}{r } -
 \frac{\Delta_{\ringh}  \oone \phi }{2 r^2}  -
 \frac{(\Delta_{\ringh} +2) \otwo \phi }{ 4 r^3} + ...
        \,.
  \label{27XII19.6bags}
\end{eqnarray}
where
\begin{equation}
       \partial_u \oone \phi
        =
       \lim_{r\to\infty}
          \left(
          \frac 12
          \int_{0}^{r}\frac{1}{\rho}
       \big(
        - \partial_r \big(
          r^2   \epsilon \lapseTB^2 \partial_r \phi
          \big)
          +
        {\Delta_{\ringh}} \phi
        \big)\big|_{r=\rho} d\rho
         - \alpha^2   \oone \phi r
         \right)
        \,.
  \label{27XII19.6bagsxt}
\end{equation}
\ptcrr{inconclusive further blow up argument moved to VasyScalar.tex}
We finish by a short remark on the conformally covariant case. The corresponding wave equation in the metric \eqref{19VI13.202} becomes the massless Minkowskian wave equation in the coordinates $(T,x,y,z)$ for the function
\begin{equation}\label{24VII20.1}
  \hat \phi:= T^{-1} \phi
  \,.
\end{equation}
A Taylor expansion near $T=0$ of a solution $\hat \phi$ gives
\begin{equation}\label{24VII20.2a}
  \hat \phi(T,\rho,\theta,\varphi) = f(\rho,\theta,\varphi) + \hat f (\rho,\theta,\varphi) T + \frac{1}{2} \Delta_\delta f (\rho,\theta,\varphi) T^2 +
   \ldots
   \,,
\end{equation}
where $\Delta_\delta$ is the Laplace operator of the flat metric $\delta$ on $\R^3$, and where $f$ and $\hat f$ are arbitrary functions. Hence
\begin{eqnarray}
   \phi(u,r,\cdot)
     &=  &  f(\rho,\cdot)  T + \hat f (\rho,\cdot) T^2  + \frac{1}{2} \Delta_\delta f (\rho,\cdot) T^3 +
   \ldots
   \nonumber
\\
    & = &  \frac{e^{-\alpha u}}{1+\alpha r} f \big(\frac{r e^{-\alpha u}}{1+\alpha r },\cdot \big) + \hat f \big(\frac{r e^{-\alpha u}}{1+\alpha r},\cdot \big) \big(\frac{e^{-\alpha u}}{1+\alpha r}\big)^2
     \nonumber
\\
      &&+ \frac{1}{2} \Delta_\delta f \big(\frac{r e^{-\alpha u}}{1+\alpha r},\cdot \big) \big(\frac{e^{-\alpha u}}{1+\alpha r} )^3 +
   \ldots
   \nonumber
   \,,
\\
    & = &  \frac{e^{-\alpha u}}{\alpha^2 r} f \big(\frac{ e^{-\alpha u}}{ \alpha  },\cdot \big)
    \nonumber
\\
 &&
      - \frac{e^{-2\alpha u}}{ \alpha^4  r  ^2}
    \Big(
      \alpha e^{ \alpha u} f \big(\frac{ e^{-\alpha u}}{ \alpha  },\cdot \big)
      +  \partial_\rho  f \big(\frac{ e^{-\alpha u}}{ \alpha  },\cdot\big)
       -
    \hat f \big(\frac{  e^{-\alpha u}}{ \alpha  },\cdot \big)
      \Big)
 \nonumber
\\
 &&
+
   \ldots
   \,.
   \label{24VII20.2}
\end{eqnarray}
We see that the initial data for a solution which extends smoothly through the conformal boundary at infinity will have, in Bondi coordinates, arbitrary expansion coefficients $\oone \phi$ and $\otwo \phi$, with all the remaining expansion coefficients determined uniquely by these first two.

\subsection{Linearised gravity}
 \label{ss11IX18.2}

We apply the results of Section~\ref{ss11IX18.1} to vacuum general relativity with cosmological constant $\Lambda$, using the
background metric approach of \cite{ChAIHP}.
Thus
$$
 (\phi^A) \equiv  (g_{\mu\nu})
 \,.
$$

We note the usual ambiguity related to the question, how to differentiate with respect to a symmetric tensor field. When performing variations we resolve this by allowing $g_{\mu\nu}$ not to have any symmetries, with all geometric quantities such as the Christoffel symbols, the Ricci tensor, or the volume form defined using the symmetric part $g_{(\mu\nu)}$ of $g_{\mu\nu}$. The tensor field $g_{\mu\nu}$ is assumed to be symmetric in all final formulae and in all unrelated calculations.

In \cite{ChAIHP} the Lagrangian density is obtained by removing from the Hilbert one a divergence which is made covariant by using a background metric $\backg$. After allowing for a cosmological constant, in space-time dimension $d$ this leads to the Lagrangian (see \cite[Section~5.1]{CJK})
\begin{eqnarray}
\mcL  &  =  & \Kp^{\mu\nu}
 \Big[ \left(
    {\Gamma}^{{\noalpha}}_{{\nosigma}\mu} - {\Bgamma}^{{\noalpha}}_{{\nosigma}\mu}
  \right) \left( {\Gamma}^{{\nosigma}}_{{\noalpha}\nu}-
    {\Bgamma}^{{\nosigma}}_{{\noalpha}\nu} \right) - \left(
    {\Gamma}^{{\noalpha}}_{\mu\nu} - {\Bgamma}^{{\noalpha}}_{\mu\nu} \right)
  \left( {\Gamma}^{{\nosigma}}_{{\noalpha}{\nosigma}} -
    {\Bgamma}^{{\nosigma}}_{{\noalpha}{\nosigma}} \right)
 \nonumber
\\
 &&
 \phantom{\Kp^{\mu\nu} \Big[}
     + r_{\mu\nu}
    -\frac{2\Lambda} d g_{\mu\nu}
    \Big] \,,
 \label{12IX18.6}
\end{eqnarray}
where $r_{\mu\nu}$ is the Ricci tensor of $\backg$,
\be \label{12IX18.1}
 \Kp^{\mu\nu} :=
 \frac 1{16 \pi} \sqrt{-\det g } \ g^{\mu\nu} \,,
\ee
and where the $\Bgamma^{\noalpha}_{{\nobeta} \gamma}$'s are the Christoffel symbols of the background metric $\backg$.

Consider the field
$$
 (\delta \mypi_A{}^{\noalpha}):=  \Big(\dlambda{ \mypi_A{}^{\noalpha}}\Big) \equiv \Big(\dlambda {   \mypi^{{\nobeta}\gamma {\noalpha}}}\Big )  \equiv (\delta  \mypi^{{\nobeta}\gamma {\noalpha}} ) :=\left( \dlambda{}{\frac{\partial \mcL}{\partial (\partial _{\noalpha} g_{{\nobeta}\gamma})}}\right)
  \,.
$$
Viewing momentarily $\Gamma ^{\noalpha}_{\mu\nu}$ and $g_{\mu\nu}$ as independent variables we have
\begin{eqnarray}
 \nonumber 
  \delta \mcL &=& \Kp^{\mu\nu}
\Big[ \delta {\Gamma}^{{\noalpha}}_{{\nosigma}\mu}
  \left( {\Gamma}^{{\nosigma}}_{{\noalpha}\nu}-
    {\Bgamma}^{{\nosigma}}_{{\noalpha}\nu} \right)
     +   \left(
    {\Gamma}^{{\noalpha}}_{{\nosigma}\mu} - {\Bgamma}^{{\noalpha}}_{{\nosigma}\mu}
  \right)
  \delta {\Gamma}^{{\nosigma}}_{{\noalpha}\nu}
\\
 \nonumber
 &&
  -
    \delta
    {\Gamma}^{{\noalpha}}_{\mu\nu}
  \left( {\Gamma}^{{\nosigma}}_{{\noalpha}{\nosigma}}
    -
    {\Bgamma}^{{\nosigma}}_{{\noalpha}{\nosigma}} \right)
    -
    \left(
    {\Gamma}^{{\noalpha}}_{\mu\nu} - {\Bgamma}^{{\noalpha}}_{\mu\nu} \right)
  \delta {\Gamma}^{{\nosigma}}_{{\noalpha}{\nosigma}}
    \Big]  + \frac{\partial \mcL}{\partial g_{\mu\nu}} \delta g_{\mu\nu}
\\
 \nonumber
  &=&
  \Kp^{\mu\nu}
\Big[ 2
  \left( {\Gamma}^{{\nosigma}}_{{\noalpha}\nu}-
    {\Bgamma}^{{\nosigma}}_{{\noalpha}\nu} \right)
     \delta^\rho_\mu
%
%
\\
 &&
  -
  \left( {\Gamma}^{{\nobeta}}_{{\noalpha}{\nobeta}}
    -
    {\Bgamma}^{{\nobeta}}_{{\noalpha}{\nobeta}} \right)
  \delta^{\nosigma}_\mu \delta^\rho_\nu
    -
    \left(
    {\Gamma}^{{\nosigma}}_{\mu\nu} - {\Bgamma}^{{\nosigma}}_{\mu\nu} \right)
  \delta ^{{\rho}}_{{\noalpha} }
    \Big]
   \delta {\Gamma}^{{\noalpha}}_{{\nosigma}\rho}
    + \frac{\partial \mcL}{\partial g_{\mu\nu}} \delta g_{\mu\nu}
    \,.
    \phantom{xxxxx}
     \label{12IX18.5}
\end{eqnarray}
The point is that $\mcL$ depends upon the derivatives of the metric only through $ \delta {\Gamma}^{{\noalpha}}_{{\nosigma}\rho}$, so that the formula allows us to calculate $\mypi^{\b \c\a}$:
\begin{eqnarray}
 \nonumber
\mypi^{\beta\gamma \alpha}
&  =
 &
  \Kp^{\mu\nu}
\Big[
2
  \left( {\Gamma}^{{\nosigma}}_{{\noalpha}\nu}-
    {\Bgamma}^{{\nosigma}}_{{\noalpha}\nu} \right)
     \delta^\rho_\mu
%
%
\\
 &&
  -
  \left( {\Gamma}^{{\nobeta}}_{{\noalpha}{\nobeta}}
    -
    {\Bgamma}^{{\nobeta}}_{{\noalpha}{\nobeta}} \right)
  \delta^{\nosigma}_\mu \delta^\rho_\nu
    -
    \left(
    {\Gamma}^{{\nosigma}}_{\mu\nu} - {\Bgamma}^{{\nosigma}}_{\mu\nu} \right)
  \delta ^{{\rho}}_{{\noalpha} }
    \Big]
  \frac{ \partial {\Gamma}^{{\noalpha}}_{{{{\nosigma}}}\rho} }{\partial (\partial_\alpha g_{\beta\gamma})}
 \,.
 \label{11IX18.6-}
\end{eqnarray}
Denoting by $\znabla$ the covariant derivative associated with the background metric $\backg$, one has
\begin{equation}\label{26IX18.21}
  \Gamma ^\sigma _{\mu\nu}
  - \Bgamma ^\sigma _{\mu\nu} = \frac 12 g^{\sigma \rho}
   ( \znabla_\mu g_{\nu \rho}
   +
    \znabla_\nu g_{\mu \rho}
   -
    \znabla_\rho g_{\mu \nu  }
   )
   \,.
\end{equation}
A somewhat lengthy calculation allows one to rewrite \eq{11IX18.6-} as
\begin{eqnarray}
 \mypi^{\beta\gamma \alpha}
&  =
 &
 {\frac{1}{16 \pi} }
 \sqrt{|\det g|}
P^{ \a (\b \c) \d (\e \f) }
 \znabla_{ \d }   g_{\e \f }
 \,,
 \label{28IX18.1}
\end{eqnarray}
with
\bean
 P^{\a \b \c \d \e \f}
  & = &
  \frac 12
   \big(
    {\nobarg}^{\a \e} {\nobarg}^{\d \b} {\nobarg}^{\c \f}
    +
    {\nobarg}^{\a \e} {\nobarg}^{\f \b} {\nobarg}^{\c \d}
     -
       {\nobarg}^{\a
 \d} {\nobarg}^{\b \e} {\nobarg}^{\f \c}
  -
   {\nobarg}^{\a \b} {\nobarg}^{\c \d} {\nobarg}^{\e \f}
   \nonumber
\\
 &&
 \phantom{\frac 12 \big(}
    -
  {\nobarg}^{\b \c} {\nobarg}^{\a \e} {\nobarg}^{\f \d}
    +
     {\nobarg}^{\b \c} \nobarg^{\a \d} {\nobarg}^{\e \f}
  \big)
 \,.
\eeal{5VII18.12mine}

Note that most expressions that follow in this paper involve contractions of $P^{\a \b \c \d \e \f}$ with tensors which are symmetric both in the pairs $\b\c$ and $\e\f$, and in such expressions there is no need to symmetrise $P^{\a \b \c \d \e \f}$ as in \eqref{28IX18.1}, since  such a symmetrisation is  done automatically when the contraction is performed.

Incidentally, since $\mcL$ is quadratic in $\znabla g$, \eqref{5VII18.12mine} implies
\begin{equation}\label{28IX18.2}
  \mcL  =
  \frac 1 {32 \pi}  \sqrt{|\det g|}
  \big(
P^{ \a \b \c \d \e \f }
 \znabla_{ \a }   g_{\b \c }
 \znabla_{ \d }   g_{\e \f }
 + 2 g^{\mu\nu} ( r_{\mu\nu}
    -\frac{2\Lambda} d g_{\mu\nu}
     )
    \big)
  \,,
\end{equation}
which shows that it makes sense to require
\begin{equation}\label{5I20}
   P^{\a (\b \c) \d (\e \f)}
    =
   P^{ \d (\e \f) \a (\b \c)}
    \,.
\end{equation}
This last equation is not obvious by staring at \eqref{5VII18.12mine}, but can be checked by a direct calculation.

Recall that we are interested in the linearised theory. For this, it is clearly convenient to choose the background metric $\backg$ to be the metric $g$ at which we are linearising.
Denoting by $h_{\mu\nu}$ the linearised metric field, the Lagrangian $\wmcL$ for the linearised theory is thus
\begin{equation}\label{28IX18.2z}
  \wmcL  =
  \frac 1 {32 \pi}  \sqrt{|\det g|}
  \big(
P^{ \a \b \c \d \e \f }
 \znabla_{ \a }   h_{\b \c }
 \znabla_{ \d }   h_{\e \f }
 + Q(h)
     )
    \big)
  \,,
\end{equation}
where $Q$ arises from the quadratic terms in the Taylor expansion, at the background metric, of
\begin{equation}\label{6III20.2}
  F:=  \sqrt{|\det g|} g^{\mu\nu}
   \big( r_{\mu\nu}
    -\frac{2\Lambda} d g_{\mu\nu}
    \big)
    \,.
\end{equation}
We have,
 ignoring the usual issues related to the symmetry of $g_{\alpha\beta}$ as this will be taken care of by itself when calculating the Taylor expansion below,
\begin{eqnarray}
  \frac{\partial F}{\partial g_{\alpha\beta}}
   & = &
    -\sqrt{|\det g|}
   \big(
    g^{\alpha\mu}g^{\beta\nu}r_{\mu \nu}
    +
     \big(\Lambda - \frac {g^{\mu\nu}r_{\mu\nu}} 2) g^{\alpha\beta}
      \big)
   \,,
   \nonumber
\\
  \frac{\partial^2 F}{\partial g_{\rho\sigma}\partial g_{\alpha\beta}}
   & = & -\sqrt{|\det g|}\bigg(
   \frac 12
   \big(
    g^{\alpha\mu}g^{\beta\nu}r_{\mu \nu}
    +
     \big(\Lambda - \frac {g^{\mu\nu}r_{\mu\nu}} 2) g^{\alpha\beta}\big)g^{\rho\sigma}
   \nonumber
\\
 &&
  -
    (g^{\alpha\rho}g^{\mu\sigma}g^{\beta\nu}
    +
    g^{\alpha\mu}g^{\beta\rho}g^{\nu\sigma}
     )
      r_{\mu \nu}
    -
      \Lambda  g^{\alpha\rho}  g^{\beta\sigma}
      \nonumber
\\
 &&
 + \frac 12
     \big( g^{\mu\rho} g^{\nu\sigma}   g^{\alpha\beta}
 +
    g^{\mu\nu }g^{\alpha\rho} g^{\beta\sigma}
    )r_{\mu \nu}
   \bigg)
    \,.
\end{eqnarray}
%
Replacing $g$ by $g+h$ in the right-hand side of \eqref{6III20.2} we thus obtain the Taylor expansion, after replacing $r_{\mu\nu}$ by $R_{\mu\nu}$ in the result,
\begin{eqnarray}
\nonumber 
F &=& \sqrt{|\det g|}
\bigg[R - 2 \Lambda
- \big(R^{\alpha \beta}
+
\big(\Lambda -\frac R 2\big) g^{\alpha\beta}\big) h_{\alpha\beta}
\\
&&
-
\frac{1}{2}\bigg(
\big(R^{\alpha \beta} +\frac 12 \big(\Lambda - \frac R 2\big) g^{\alpha\beta}\big)g^{\rho \sigma}
\nonumber
\\
&&
-
2\big(R^{\alpha \rho} +\frac12 \big(\Lambda - \frac R 2\big) g^{\alpha\rho}\big)g^{\beta \sigma}
\bigg)h_{\alpha\beta}h_{\rho\sigma} \bigg]
+
O(h^3)
\,.
\end{eqnarray}
Hence
\begin{eqnarray}\label{6III20.5}
  Q(h)
  &=&
\bigg(
2\big(R^{\alpha \rho} +\frac12 \big(\Lambda - \frac R 2\big) g^{\alpha\rho}\big)g^{\beta \sigma}
\nonumber
\\
&&
\phantom{\bigg(}
-
\big(R^{\alpha \beta} +\frac 12 \big(\Lambda - \frac R 2\big) g^{\alpha\beta}\big)g^{\rho \sigma}
\bigg)h_{\alpha\beta}h_{\rho\sigma}
 \,.
 \label{6III20.21}
\end{eqnarray}
Assuming that the background satisfies the Einstein vacuum equations,
\begin{equation}\label{6III20.4}
  R_{\alpha \beta}
 +
  \big(\Lambda -\frac R 2\big) g_{\alpha\beta}
   =0
    \quad
     \Longleftrightarrow
     \quad
    R_{\alpha\beta} = \frac{2\Lambda}{d-2} g_{\alpha\beta}
   \,,
\end{equation}
Equation~\eqref{6III20.21} simplifies to
 \ptcheck{8III20}
\begin{equation}\label{6III20.6}
  Q(h)
  =
      \frac{2 \Lambda}{(d-2)} \bigg[g^{\alpha\rho}g^{\beta \sigma}h_{\alpha\beta}h_{\rho\sigma}-\frac12 (g^{\alpha\beta}
      h_{\alpha\beta})^2\bigg]
      \,.
\end{equation}

In view of \eq{28IX18.1}, the variation of $\mypi^{\b \c\a}$ at $g=\backg$ equals
\begin{eqnarray}
 \nonumber
\delta \mypi^{\beta\gamma \alpha}
&  =
 &
  \Kp^{\mu\nu}
\Big[ 2  \delta {\Gamma}^{{{{\nosigma}}}}_{{\noalpha}\nu}
     \delta^\rho_\mu
%
%
  -
  \delta{\Gamma}^{{\nobeta}}_{{\noalpha}{\nobeta}}
  \delta^{{{\nosigma}}}_\mu \delta^\rho_\nu
    -
    \delta
    {\Gamma}^{{{{\nosigma}}}}_{\mu\nu}
  \delta ^{{\rho}}_{{\noalpha} }
    \Big]
  \frac{ \partial {\Gamma}^{{\noalpha}}_{{{{\nosigma}}}\rho} }{\partial (\partial_\alpha g_{\beta\gamma})}
\\
 & = &
 {\frac{1}{16 \pi} }
 \sqrt{|\det g|}
P^{ \a (\b \c) \d (\e \f) }
 \znabla_{ \d } \delta  g_{\e \f }
 \,.
 \label{11IX18.6}
\end{eqnarray}
%
%

Given two solutions $\delta_i g$, $i=1,2$, of linearised Einstein equations,
the presymplectic current of vacuum Einstein gravity with a cosmological constant therefore reads
%
%
\be
 \omega^\a \equiv \omega^\a (\delta_1 g ,\delta_2 g)=
 {\frac{1}{16 \pi}
 \sqrt{|\det g|} }
 P^{ \a (\b \c) \d (\e \f) }
 \left( \delta_2 g_{ \b \c } {\znabla}_{ \d } \delta_1 g_{\e \f } - \delta_1 g_{\b \c } {\znabla}_{\d} \delta_2 g_{\e \f }\right)
  \,.
  \label{11IX18.4}
\ee

(Because of the symmetrisations occurring in \eq{11IX18.4}, one can use there instead an equivalent version of \eqref{5VII18.12mine} given by Wald and Zoupas in~\cite{WaldZoupas}:
\ptcheck{23IV20, with TS, that using this formula directly or a completely symmetrised version thereof, gives the same result for the boundary integral, not that this was really needed}
%
\bel{5VII18.12}
 P^{\a \b \c \d \e \f}_{WZ} = {\nobarg}^{\a \e} {\nobarg}^{\f \b} {\nobarg}^{\c \d} - \frac{1}{2} {\nobarg}^{\a
 \d} {\nobarg}^{\b \e} {\nobarg}^{\f \c} - \frac{1}{2} {\nobarg}^{\a \b} {\nobarg}^{\c \d} {\nobarg}^{\e \f} -
 \frac{1}{2} {\nobarg}^{\b \c} {\nobarg}^{\a \e} {\nobarg}^{\f \d} + \frac{1}{2}  {\nobarg}^{\b \c}
 \nobarg^{\a \d} {\nobarg}^{\e \f}
 \,.)
\ee
One can check that the part of the Lagrangian density which contains Christoffel symbols can be reduced to four terms. Indeed, we have
%
\begin{equation}
\label{2I20.t5}
P^{ \a \b \c \d \e \f }
\znabla{}_{ \a }   g_{\b \c }
\znabla{}_{ \d }   g_{\e \f } =
\widetilde{P}^{ \a \b \c \d \e \f }
\znabla{}_{ \a }   g_{\b \c }
\znabla{}_{ \d }   g_{\e \f }
\,,
\end{equation}
where
\begin{equation}
\widetilde{P}^{\alpha \beta \gamma \delta \epsilon \sigma}:= g^{\alpha \sigma} g^{\beta \epsilon} g^{\gamma \delta} -g^{\alpha \beta} g^{\gamma \delta} g^{\epsilon \sigma} +\frac{1}{2} g^{\alpha \delta} g^{\beta \gamma} g^{\epsilon \sigma}-\frac{1}{2} g^{\alpha \delta} g^{\beta \sigma} g^{ \gamma \epsilon} \, .
\end{equation}
Using \eqref{2I20.t5}, it follows  from \eqref{28IX18.2} that
\begin{equation}
 \mypi^{\beta\gamma \alpha}
=
{\frac{1}{32 \pi} }
\sqrt{|\det g|}\left(\widetilde{P}^{\alpha \beta \gamma \delta \epsilon \sigma}+\widetilde{P}^{\delta \epsilon \sigma \alpha \beta \gamma}\right)\znabla{}_{\delta} g_{\epsilon \sigma} \; .
\label{31X20.tn}
\end{equation}
Note that one of the terms constituting $\widetilde{P}^{\alpha \beta \gamma \delta \epsilon \sigma}$  is not invariant under exchange of the first three indices with the three last ones:
\begin{equation}
\widetilde{P}^{\alpha \beta \gamma \delta \epsilon \sigma}-\widetilde{P}^{\delta \epsilon \sigma \alpha \beta \gamma}=-g^{\alpha \beta} g^{\gamma \delta} g^{\epsilon \sigma}- \Big(-g^{\delta \epsilon} g^{\sigma \alpha} g^{\beta \gamma}\Big) \, ,
\end{equation}
which prevents us to express the canonical momenta in a simple form using $\widetilde{P}^{\alpha \beta \gamma \delta \epsilon \sigma}$.

 The following relations hold:
 \begin{eqnarray}
	 P^{\alpha (\beta \gamma) \delta (\epsilon \sigma)}
	 &=&
	 P^{\delta (\epsilon \sigma) \alpha (\beta \gamma)}
\,,
	 \\
	 P^{\alpha (\beta \gamma) \delta (\epsilon \sigma)}
	 &=&
	 P^{\alpha (\beta \gamma) \delta (\epsilon \sigma)}_{WZ}
\,,
	 \\
	 P^{\alpha (\beta \gamma) \delta (\epsilon \sigma)}
	 &=&
	 \frac{1}{2} \big(
	 	\widetilde{P}^{\alpha (\beta \gamma) \delta (\epsilon \sigma)}
	 	+
	 	\widetilde{P}^{\delta (\epsilon \sigma) \alpha (\beta \gamma)}
 	\big)
  \,.
 \end{eqnarray}
  \ptcheck{up to here in this subfile 24II20}

\subsubsection{Canonical energy of weak gravitational fields}
 \label{ss14IX18.1}

 \ptcrr{spacelike hypersurfaces, including passing to the limit to Scri on de Sitter, are already covered in Ashtekar, so no point of adding them? see also Jezierski APP 2008, around formula 5.9}
In the gravitational case \eq{6IX18.5+} reads
\begin{eqnarray}
\wmcH [\hyp, X]
 &
 = &
  \frac 12
  \left( \int_\hyp
  \omega^\mu(\delta g , \Lie_X \delta g)
       \, dS_\mu
    {-}
      \int_{\partial \hyp }
     \tpi ^{ \alpha\beta [\mu }
    X^{{{\sigma}}]} \delta g_{\alpha\beta}
     dS_{\sigma \mu}
      \right)
   .
      \phantom{xxxxxxx}
        \label{6IX18.5+asdf}
\end{eqnarray}
From \eq{6IX18.10} and \eq{11IX18.6} we find
\begin{eqnarray}
  (\tpi_A{}^\alpha (\delta \phi))
   &\equiv & (\tpi^{\beta\gamma\alpha} (\delta g) )
 =
 {\frac{1}{16 \pi} }
 \sqrt{|\det g|}
P^{ \a (\b \c) \d (\e \f) }
 {\znabla}_{ \d } \delta  g_{\e \f }
  \,.
      \phantom{xxxx}
  \label{12IX18.7}
\end{eqnarray}

If $\partial\hyp$ is a spacelike surface given by the equation
$$ \{u\equiv x^{0}=\const\,, \ r\equiv x^{\TSred 1}=\const'\}
\,,
$$
and if $X$ equals $\partial_u$, then   the boundary integral in \eq{6IX18.5+asdf} reads
\begin{eqnarray}
\lefteqn{-
 {\frac{1}{32 \pi} }
      \int_{\partial \hyp } X^{[\sigma  }
P^{ \mu] (\b \c) \d (\e \f) }
 {\znabla}_{ \d } \delta  g_{\e \f }
 \,
   \delta g_{\b\c}
   \,
 \sqrt{|\det g|}
 \,
     dS_{{{\sigma}}\mu}
      }
      &&
 \nonumber
\\
 & = &
 -
 {\frac{1}{32 \pi} }
      \int_{\partial \hyp }
P^{{\TSred r} (\b \c) \d (\e \f) }
 {\znabla}_{ \d } \delta  g_{\e \f }
   \, \delta g_{\b\c}
    \,
 \sqrt{|\det g|} \, dx^2\wedge dx^3
   \,.
      \phantom{xx}
        \label{13IX18.1}
\end{eqnarray}

If we denote by $h_{\mu\nu}$ the linearised metric field, the  ``generating equation'' \eq{13IX18.6} reads, again with $X=\partial_u$,
\begin{eqnarray}
\nonumber
 \lefteqn{
 \dlambda{  \wmcH {[\hyp,X]}  }
 =
\int_\hyp
 \big(
    \Lie_X  h_{\alpha\beta}  \dlambda{ \tpi^{\alpha \beta \mu}}   -
    \Lie_X \tpi^{\alpha \beta \mu}  \dlambda{ h_{\alpha\beta}}
    \big)
     dS_\mu
  }
  &&
\\
\nonumber
\\
 &&
  {-}
  {\frac{1}{16 \pi} }
      \int_{\partial \hyp }
       \underbrace{ X^{[\sigma  }
        P^{ \mu] (\b \c) \d (\e \f) }
     {\znabla}_{ \d } h_{\e \f }
    \,
     \dlambda{ h_{\b\c}}
     \,
     \sqrt{|\det g|}
     \,  dS_{{{\sigma}} \mu}
 }_
  {
P^{r (\b \c) \d (\e \f) }
 {\znabla}_{ \d } h_{\e \f }
 \,
   \dlambda{ h_{\b\c}}
   \, {\sqrt{|\det g|} \, dx^2\wedge dx^3}
  }
   \,.
     \label{13IX18.4}
\end{eqnarray}

\subsubsection{Energy flux}
 \label{ss14IX18.2}

In the setting just described, the linearised-fields version of the flux formula \eq{11IX18.14} takes the form
\begin{eqnarray}
\nonumber
   \frac{d\wmcH [\hyp_\tau, X]}{d\tau}
   & =
 &
  {-}
  {\frac{1}{16 \pi} }
      \int_{\partial \hyp } X^{[\sigma  }
P^{ \mu] (\b \c) \d (\e \f) }
 {\znabla}_{ \d } \delta  g_{\e \f }
\,\Lie_X
   \delta g_{\b\c}
   \,
 \sqrt{|\det g|}
 \,
     dS_{{{\sigma}}\mu}
\\
 & = &
   {-}
 {\frac{1}{16 \pi} }
  \int_{\partial \hyp_\tau}
P^{{\TSred r} (\b \c) \d (\e \f) }
 {\znabla}_{ \d }  \delta g_{\e \f }
   \,\Lie_X \delta  g_{\b\c}
   \, {\sqrt{|\det g|} \, dx^2\wedge dx^3}
    \,.
    \nonumber
\\
 &&
    \label{13IX18.3}
\end{eqnarray}
 \ptcrr{all boundary integrals look consistent to me so far}

\subsubsection{Gauge invariance}
 \label{ss14IX18.3}
%
%
%
%
%
%

It is shown in~\cite[Equations~(5.19)-(5.20)]{CJK} that for \emph{solutions of the field equations and for all vector fields $Y$} the current $\mcH^\mu[Y]$ takes the form  $\mcH^\mu=\partial_\alpha \ourU ^{\mu\alpha} + {\mycal G}^\mu$, where
\begin{eqnarray}
 \ourU^{\nu\lambda}&= &
{\ourU^{\nu\lambda}}_{\beta}Y^\beta - \frac 1{8\pi} \sqrt{|\det
g_{\rho\sigma}|} g^{\alpha[\nu}\delta^{\lambda]}_\beta
{Y^\beta}_{;\alpha} \,,\label{Fsup2newxx}
\\ {\ourU^{\nu\lambda}}_\beta &= & \displaystyle{\frac{2|\det
  \bmetric_{\mu\nu}|}{ 16\pi\sqrt{|\det g_{\rho\sigma}|}}}
g_{\beta\gamma}(e^2 g^{\gamma[\lambda}g^{\nu]\kappa})_{;\kappa}
\,,\label{14IY18.34}
\end{eqnarray}
where a semicolon denotes the covariant derivative of the
metric $\bmetric$, with
\begin{equation}\label{8I18.1}
    \textstyle
 e\equiv \frac{\sqrt{|\det
 g_{\rho\sigma}|}}{\sqrt{|\det\bmetric_{\mu\nu}|}}
  \,,
\end{equation}
and where ${\mycal G}^\mu$ does not depend upon the derivatives of $g$.

\ptcrr{not quite convincing arguments moved to recycling.tex}

Under our conditions, inspection of the analysis in~\cite[Section~5.1]{CJK} leads to the formula
\begin{eqnarray}
\int_\hyp  {\omega^\mu(\Lie_Yg,\delta g)}      \, dS_\mu
 & = &
 \dlambda{  \mcH_{\mathrm{boundary}} {[\hyp,Y]}  }
     \nonumber
\\
 & \equiv &
 \delta{  \mcH_{\mathrm{boundary}} {[\hyp,Y]}  }
   \,,
     \label{17IX18.2}
\end{eqnarray}
where
\begin{eqnarray}\label{14IY18.33}
 \mcH_{\mathrm{boundary}} {[\hyp,Y]}
   &
    := &
    \frac 12 \int_{\partial \hyp} \ourU^{\mu\nu} dS_{\mu\nu}
    \,.
\end{eqnarray}
A direct proof of \eqref{17IX18.2} will be provided shortly.
Thus, for all vector fields $Y$ \emph{which vanish together with their first derivatives at $\partial \hyp$} and for all variations $\delta g$ of the metric satisfying the linearised field equations it holds that
\begin{eqnarray}
\int_\hyp {\omega^\mu(\Lie_Yg,\delta g)}
   \,  dS_\mu =0
   \,,
     \label{13IY18.4a}
\end{eqnarray}
as already established by different arguments in~\cite{WaldLee,CrnkovicWitten,Friedman1978}. Since $
\Lie_Yg$ is the variation of the metric $g$ corresponding to infinitesimal coordinate-transformations, this is interpreted as the statement that \emph{the form obtained by integrating the presymplectic current is gauge-invariant}.

For our purposes the key significance of \eqref{13IY18.4a} is:

\begin{theorem}
  \label{T16IX18.1}
The total {Noether charge} $\wmcH [\hyp, X]$ of the linearised gravitational field associated with a compact hypersurface $\hyp$  with smooth boundary is invariant under the ``gauge transformation''
$$
 \delta g \mapsto \delta g + \Lie _Y g
$$
as long as the vector field $Y$ satisfies $Y= 0 ={\znabla} Y= [X,Y]={\znabla}([X,Y])$ at $\partial \hyp$.
\end{theorem}

\proof
Using \eqref{13IY18.4a} we have
\begin{eqnarray}
 \lefteqn{
 \int_\hyp
  \omega^\mu(\delta g +\Lie_Y  g, \Lie_X (\delta g +\Lie_Y  g))
       \, dS_\mu
      }
      &&
 \nonumber
\\
& = &\int_\hyp
  \omega^\mu(\delta g , \Lie_X (\delta g +\Lie_Y  g))
       \, dS_\mu
   +\underbrace{ \int_\hyp
  \omega^\mu (\Lie_Y  g , \Lie_X (\delta g +\Lie_Y  g))
       \, dS_\mu}_{=0}
 \nonumber
\\
& = &\int_\hyp
  \omega^\mu(\delta g , \Lie_X  \delta g )
       \, dS_\mu
   + \int_\hyp
  \omega^\mu(\delta g , {
   \underbrace{\Lie_Y \Lie_X  g + \Lie_{[X,Y]}
     g)
     }_{\equiv \Lie_X \Lie_Y g}
   }
       \, dS_\mu
 \nonumber
\\
& = &\int_\hyp
  \omega^\mu(\delta g , \Lie_X  \delta g )
       \, dS_\mu
   +\underbrace{\int_\hyp
  \omega^\mu(\delta g , \Lie_Y \Lie_X g )}_{=0}
   +\underbrace{\int_\hyp
  \omega^\mu(\delta g , \Lie_{[X,Y]} g )
       \, dS_\mu}_{=0}
   \,.
   \nonumber
\\
 &&
      \phantom{xxxxx}
        \label{17IX18.1}
\end{eqnarray}
The result follows now from \eqref{13IY18.4a}.
\qedskip

\begin{Remark}
  \label{R17IX18.1}
  {\rm
  There is an obvious version of Theorem~\ref{T16IX18.1} for non-compact $\hyp$'s, when suitable asymptotic conditions, ensuring the vanishing of the boundary integrals at the right-hand side of \eqref{17IX18.2}, are imposed on all objects involved.
  }
\qed
\end{Remark}

\ptcrr{give a formula for the boundary term in general?}
\subsubsection{Proof of \eqref{17IX18.2}}
The remainder of this section will be devoted to the proof of \eqref{17IX18.2}. For this it is convenient to write
\begin{equation}\label{23IX18.1}
\noomega^\mu := \frac{16 \pi}{\sqrt{|\det g |}} \omega^\mu
\,.
\end{equation}
The presymplectic form on a light cone $\mcC_u$, which we denote by  $ \Omega_{\mcC_u}$,  is obtained by integrating the presymplectic current:
\begin{eqnarray}
\Omega_{\mcC_u}(\delta_1g ,\delta_2 g)
& := &   \int_{\mcC_u} \omega^\mu (\delta_1g ,\delta_2 g)
\, dS_\mu
\nn
\\
&\equiv  & \frac 1 {16 \pi}   \int_{\mcC_u} \noomega^u (\delta_1g ,\delta_2 g) \underbrace{\sqrt{|\det \nobarg|} \, dr \, d^2 x^A}_{=:d\mu_{\mcC}}
\,,
\label{8VII18}
\end{eqnarray}
where the light cone is given by the equation $\{u=0\}$, and coordinatised by coordinates $(r,x^A)$.
Thus, to determine $  \Omega_{\mcC_u}(\delta_1g ,\delta_2 g)$, which in turn determines the volume part of the Noether charge \eq{6IX18.5+asdf}, we need to calculate $\noomega^u $.

We define
\begin{equation}\label{23XI19.3}
b^\mu(\delta_1 g, \delta_2 g):=   P^{ \mu (\b \c) \d (\e \f) }  \delta_1  g_{\b\c}
\nabla_{ \d }  \delta_2 g_{\e \f }
\,,
\end{equation}
so that the vector field $\noomega$ of \eqref{23IX18.1} equals
\begin{equation}\label{23XI19.2b}
\noomega^\mu (\delta_1 g,\delta_2 g) = b^\mu(\delta_2 g, \delta_1 g) -  b^\mu(\delta_1 g, \delta_2 g)
\,.
\end{equation}
We consider the following gauge transformations
\begin{eqnarray}
\delta_{1} g \to \delta_{1} g + \Lie_{\xi_{1}}g \, ,
\\
\delta_{2} g \to \delta_{2} g + \Lie_{\xi_{2}}g \, .
\end{eqnarray}
A gauge transformation of the vector field $\noomega$ of \eqref{23IX18.1} leads to
\begin{eqnarray}
\noomega^{\alpha}(\delta_{1} g + \Lie_{\xi_{1}} g ,\delta_{2} g + \Lie_{\xi_{2}}g )&=&b^{\alpha} (\delta_2 g,\delta_1 g) -   b^\alpha(\delta_1 g, \delta_2 g)
\nonumber
\\
& &
    +[
    b^{\alpha} (\delta_2 g, \Lie_{\xi_{1}} g) -   b^\alpha(\Lie_{\xi_{1}} g, \delta_2 g)
    ]
\nonumber
\\
& &
   -[
        b^{\alpha}(\delta_1 g, \Lie_{\xi_{2}} g) -   b^\alpha(\Lie_{\xi_{2}} g, \delta_1 g)
    ]
\nonumber
\\
& &
  +[
b^{\alpha} (\Lie_{\xi_{2}} g, \Lie_{\xi_{1}} g) -   b^\alpha(\Lie_{\xi_{1}} g, \Lie_{\xi_{2}} g)
]
\, .
\phantom{xxxxxx}
\label{24X20.t1}
\end{eqnarray}

In order to avoid a notational confusion between fields such as $\delta g^{\mu\nu}$, understood as a variation of $g^{\mu\nu}$, and $g^{\mu\alpha}g^{\nu\beta}\delta g_{\alpha\beta}$, as before we will write $h_{\mu\nu}$ for $\delta g_{\mu\nu}$. It is convenient introduce
\begin{equation}
\tshbar_{\mu \nu}:= h_{\mu \nu} - \frac12 g_{\mu \nu} g^{\alpha \beta} h_{\alpha \beta} \, .
\end{equation}
Indices on $h_{\mu \nu}$ and $\tshbar_{\mu \nu}$ are of course raised and lowered with the metric $g$.
Each term in square brackets in \eqref{24X20.t1} can be rewritten using the identity
\begin{eqnarray}
\lefteqn{
[
b^\alpha(h, \Lie_{\xi } g) -   b^\alpha(\Lie_{\xi } g, h)
]
=
}
& &
\nonumber
\\
& &     \nabla_{\beta}\Big[
        \big(
            g^{\beta\gamma}{\delta^{\alpha}}_{\sigma} \tshbar_{\gamma\delta}
            -
            g^{\alpha\gamma}{\delta^{\beta }}_{\sigma}\tshbar_{\gamma  \delta}
        \big) \nabla^{\delta}{\xi }^{\sigma}
    \Big]
    -  \nabla_{\beta} \Big[\nabla_{\delta}U^{\alpha \beta}{}_{\gamma}{}^{  \delta}   \xi^{\gamma} \Big]
\nonumber
\\
& &
    +2 R_{\delta \gamma} \xi^{\gamma} \tshbar^{\alpha \delta}
       -
   g^{\alpha \beta} (
    2 \delta R_{  \beta\gamma}
    -  g^{\nu \rho} \delta R_{\nu \rho}g_{\beta \gamma}
    )
    \xi^{\gamma}
\, ,
\label{24X20.t2}
\end{eqnarray}
where
\begin{equation}
U^{\alpha \beta \gamma \delta}
 :=
  g^{\beta \delta}{\tshbar^{\alpha \gamma}}
   +g^{\alpha \gamma}{\tshbar^{\beta \delta}}
    -g^{\beta \gamma}{\tshbar^{\alpha \delta}}
     -g^{\alpha \delta}{\tshbar^{\beta \gamma}}
     \, .
\end{equation}
The tensor $U^{\alpha \beta \gamma \delta}$ fulfills $U^{\mu \lambda \nu \kappa}=U^{[\mu \lambda][\nu \kappa]}=U^{\nu \kappa \mu \lambda} \, .$

Note that   the last  line  in \eqref{24X20.t2} vanishes on a background which  satisfies the vacuum Einstein equations \eqref{6III20.4}  and for metric perturbation satisfying the linearised vacuum Einstein equations:
\ptcrr{the lhs checked by JH in the appendix C}
\begin{equation}
\label{26X20.t1}
\underbrace{
    \nabla^{\mu} \nabla_{\alpha} \tshbar_{\beta \mu}
    +\nabla^{\mu} \nabla_{\beta} \tshbar_{\alpha \mu}
    -\nabla_{\mu} \nabla^{\mu} \tshbar_{ \alpha \beta}
    -g_{\alpha \beta} \nabla^{\kappa} \nabla^{\lambda} \tshbar_{\kappa \lambda}
}_{2 \delta R_{\alpha \beta} - g_{\alpha \beta} g^{\nu \rho} \delta R_{\nu \rho}}
=
2 \Lambda \tshbar_{\alpha\beta}
\, ,
\end{equation}
Indeed, \eqref{26X20.t1} is equivalent to the linearised Einstein equations
\begin{equation}
\label{31X20.t0}
\delta \big(G_{\mu \nu}+\Lambda g_{\mu\nu} \big)
=0 \, ,
\end{equation}
when $R_{\mu \nu}=\Lambda g_{\mu \nu}$ holds; see Appendix~\ref{s3XI20.1}.

In order to show \eqref{24X20.t2}  we start by noting that, by definition of $b^\alpha$, we have
\begin{eqnarray}
  b^\alpha(h, \Lie_{\xi } g)&=&   \big(
    g^{\beta\gamma}{\delta^{\alpha}}_{\sigma} \tshbar_{\gamma\delta}
    -
    g^{\alpha\gamma}{\delta^{\beta }}_{\sigma}\tshbar_{\gamma  \delta}
    \big)  \nabla_{\beta} \nabla^{\delta}{\xi }^{\sigma}
\nonumber
\\
& &
    +R^{\alpha}{}_{\delta \beta \sigma} \xi^{\beta} \tshbar^{\delta \sigma}
    +R_{\beta\delta} \xi^{\beta} \tshbar^{\delta \alpha}
    \,.
     \label{27X20.1p}
\end{eqnarray}
In order to find $b^\alpha(\Lie_{\xi } g, h)$, we calculate $\nabla_{\beta}\nabla_{\delta}U^{\alpha \beta \gamma  \delta}$ and use  \eqref{26X20.t1} to obtain
\tscomr{Commutation without bar here! Update: it does not matter.}
%
\begin{eqnarray}
 \nonumber
 \lefteqn{
\nabla_{\beta}\nabla_{\delta}U^{\alpha \beta \gamma  \delta}
 }
 \nonumber
\\
   & = &
   g^{\beta \delta}  \nabla_{\beta}  \nabla_{\delta} {\tshbar^{\alpha \gamma}}
   +g^{\alpha \gamma}  \nabla_{\beta}  \nabla_{\delta}{\tshbar^{\beta \delta}}
    -g^{\beta \gamma}  \nabla_{\beta}  \nabla_{\delta}{\tshbar^{\alpha \delta}}
     -g^{\alpha \delta}  \nabla_{\beta}  \nabla_{\delta}{\tshbar^{\beta \gamma}}
     \nonumber
     \\
      & = & -\big(2 \delta R_{\mu \beta} - g_{\mu \beta} g^{\sigma \rho} \delta R_{\sigma \rho}\big) g^{\mu \alpha} g^{\gamma \beta}
    -R^{\alpha}{}_{\beta}{}^{\gamma}{}_{\delta} \tshbar^{\beta \delta}
    + R_{\delta}{}^{\gamma} \tshbar^{\alpha \delta} \, .
    \phantom{xxx}
\end{eqnarray}
%
Thus
\begin{eqnarray}
 \nonumber 
\lefteqn{
  \nabla_{\beta} \Big[\nabla_{\delta}U^{\alpha \beta}{}_{\gamma}{}^{  \delta}   \xi^{\gamma} \Big]
 }
\\
 & = &
 \nabla_{\beta}\nabla_{\delta}U^{\alpha \beta}{}_{\gamma}{}^{  \delta}
     \xi^\gamma
 +
 ( \nabla_{\delta}U^{\alpha \beta}{}_{\gamma}{}^{  \delta} )  \nabla_{\beta}   \xi^{\gamma}
  \nonumber
\\
 &= &
  -\big(2 \delta R_{\mu \nu} - g_{\mu \nu} g^{\sigma \rho} \delta R_{\sigma \rho}\big) g^{\mu \alpha} \xi^\nu
  \nonumber
  \\
  & &
  -R^{\alpha}{}_{\beta \gamma \delta} h^{\beta \delta} \xi^\gamma
  + R_{\delta\gamma} h^{\alpha \delta} \xi^\gamma
  +\big(
   \nabla_{\delta}U^{\alpha \beta}{}_{\gamma}{}^{  \delta}
    \big )  \nabla_{\beta}   \xi^{\gamma}
  \nonumber
\\
 &= &
-\big(2 \delta R_{\mu \nu} - g_{\mu \nu} g^{\sigma \rho} \delta R_{\sigma \rho}\big) g^{\mu \alpha} \xi^\nu
-R^{\alpha}{}_{\beta \gamma}{}_{\delta} h^{\beta \delta} \xi^\gamma
+ R_{\delta\gamma} h^{\alpha \delta} \xi^\gamma
\nonumber
\\
& &
  +
 \underbrace{
  \big(
  g^{\beta \delta} \nabla_{\delta}{\tshbar^{\alpha}{}_{ \gamma}}
   +\delta^{\alpha}{}_{\gamma} \nabla_{\delta}{\tshbar^{\beta \delta}}
    -\delta^{\beta}{}_{ \gamma} \nabla_{\delta}{\tshbar^{\alpha \delta}}
     -g^{\alpha \delta}  \nabla_{\delta} {\tshbar^{\beta}{}_{\gamma}}
     \big)  \nabla_{\beta}  \xi^{\gamma}
     }_{=:I }
    \,.
    \nonumber
\end{eqnarray}
Next
\begin{eqnarray}
\nonumber
\lefteqn{
 \nabla_{\beta}\Big[
        \big(
            g^{\beta\gamma}{\delta^{\alpha}}_{\sigma} \tshbar_{\gamma\delta}
            -
            g^{\alpha\gamma}{\delta^{\beta }}_{\sigma}\tshbar_{\gamma  \delta}
        \big) \nabla^{\delta}{\xi }^{\sigma}
    \Big]
    }
    &&
\\
 \nonumber
 & =&
 \underbrace{
        \big(
            g^{\beta\gamma}{\delta^{\alpha}}_{\sigma} \nabla_{\beta}  \tshbar_{\gamma\delta}
            -
            g^{\alpha\gamma}{\delta^{\beta }}_{\sigma} \nabla_{\beta} \tshbar_{\gamma  \delta}
        \big) \nabla^{\delta}{\xi }^{\sigma}
        }_{=: II}
  \,,
\\
 & &
 +
        \big(
            g^{\beta\gamma}{\delta^{\alpha}}_{\sigma} \tshbar_{\gamma\delta}
            -
            g^{\alpha\gamma}{\delta^{\beta }}_{\sigma}\tshbar_{\gamma  \delta}
        \big)  \nabla_{\beta} \nabla^{\delta}{\xi }^{\sigma}
  \,.
\end{eqnarray}
A calculation gives
\begin{eqnarray}
 b^\alpha(\Lie_{\xi } g, h)
  &=&
  I  - II
   \label{27X20.2p}
\end{eqnarray}
Subtracting  \eqref{27X20.2p} from \eqref{27X20.1p} gives \eqref{24X20.t2}

To finish the argument it remains to compare  \eqref{17IX18.2} with   \eqref{24X20.t2}. The variation of the boundary Hamiltonian \eqref{14IY18.33}
 reads
\begin{eqnarray}
\lefteqn{
\delta \mcH_{\mathrm{boundary}} {[\hyp,Y]}
=
\frac 12 \int_{\partial \hyp} \delta \ourU^{\mu\nu} dS_{\mu\nu}
}
&&
\nonumber
\\
&=&
\frac 12 \int_{\partial \hyp} \Big[
 \delta {\ourU^{\alpha \beta}}_{\gamma}Y^{\gamma}
- \frac 1{8\pi} \sqrt{|\det
    g_{\rho\kappa}|} g^{\mu [\alpha}\dgbar{}_{\mu \gamma} \delta^{\beta]}_{\sigma}
Y^{\sigma ;\gamma}
\Big] dS_{\alpha \beta} \, .
\phantom{xxx}
\label{3XI20.t1}
\end{eqnarray}
The linearisation of \eqref{14IY18.34} gives
\begin{eqnarray}
\frac{16 \pi}{\sqrt{|\det g_{\rho\kappa}|}} Y^{\gamma} {\ourU^{\alpha \beta}}_{\gamma}
&=&
    \nabla_{\delta} \Big(
    g^{\beta \gamma}{\tshbar^{\alpha \delta}}
    +g^{\alpha \delta}{\tshbar^{\beta \gamma}}
    -g^{\beta \delta}{\tshbar^{\alpha \gamma}}
    -g^{\alpha \gamma}{\tshbar^{\beta \delta}}
    \Big)
    Y_{\gamma}
    \, ,
    \phantom{xxxxxxx}
\\
\frac{16 \pi}{\sqrt{|\det g_{\rho\kappa}|}} Y^{\gamma} {\ourU^{\alpha \beta}}_{\gamma}
&=&
    -\nabla_{\delta}U^{\alpha \beta}{}_{\gamma}{}^{  \delta}   \xi^{\gamma}
\, ,
\end{eqnarray}
which shows that $\delta \mcH_{\mathrm{boundary}}$  equals the second line in \eqref{24X20.t2}.\tscomr{Section has been re-read.}
\subsection{Adding matter fields}
 \label{s15XI19.1}
We consider now Einstein equations interacting with matter fields. The fields $\phi^A$ under consideration   take the form
\begin{equation}\label{15XII19.2}
  \phi^A = (g_{\mu\nu}, \TSphi^\TSa)
  \,,
\end{equation}
where $\TSphi^\TSa$ are matter fields. We write the Lagrangian in the form
 \begin{equation}\label{15XII19.1}
   \mcL = \mcL_g + \mcL_m
 \end{equation}
where $\mcL_g$ is the Lagrangian \eqref{12IX18.6} and $\mcL_m$ is the Lagrangian describing matter fields, which is assumed to depend upon the metric but not its derivatives. The examples of main interest in the current context would be the Einstein-Maxwell equations, as well as the equations for gravitating elastic bodies.\\
Assuming that $\TSphi^\TSa$ and $\partial_{\mu} \TSphi^{\TSa}$ are independent fields, the variation of $\mcL$ reads
 \begin{equation} \label{15XII19.ts1}
\delta \mcL =  \delta \mcL_H + \frac{\partial \mcL_m}{\partial\left(\partial_{\mu}{\TSphi^\TSa}\right) } \delta \left(\partial_{\mu}{\TSphi^\TSa} \right) + \frac{\partial \mcL_m}{\partial{\TSphi^\TSa} } \delta \TSphi^\TSa + \frac{\partial \mcL_m}{\partial g_{\mu\nu}} \delta g_{\mu\nu}
\,,
\end{equation}
where $\delta \mcL_g$ is given by (\ref{12IX18.5}).

 The momenta $\mypi_A$ split into gravitational and matter parts, with the gravitational momenta given by \eqref{28IX18.1}, and the matter ones  defined as before:
\begin{eqnarray}
&
\displaystyle
	\TSpi_\TSa{}^\mu  :=
		\frac{\partial  \mcL_m}{\partial \TSphi^\TSa{}_\mu }
		\,,
		\quad
	\TSpi_\TSa  :=
		\frac{\partial   \mcL_m}{\partial \TSphi^\TSa  }
		\,,
&
\\
&
\displaystyle
	\TSpi_\TSa{}^\mu{}_\TSb{}^\nu :=
		\frac{\partial^2  \mcL_m}{\partial \TSphi^\TSa{}_\mu\partial \TSphi^\TSb{}_\nu }
		\,,
	\
	\TSpi_\TSa{}^\mu{}_\TSb  :=
		\frac{\partial^2  \mcL_m}{\partial \TSphi^\TSa{}_\mu\partial \TSphi^\TSb  }
		\,,
	\
	\TSpi_{\TSa \TSb}  :=
		\frac{\partial^2  \mcL_m}{ \partial \TSphi^\TSa \partial \TSphi^\TSb  }
		\,.
&
\label{15XI19.ts2}
\end{eqnarray}
The Hamiltonian density of linearised fields equals now
\begin{equation}
\wmcH [\hyp, X]=\wmcH_{g} [\hyp, X]+\wmcH_{m} [\hyp, X] \,.
\end{equation}
According to (\ref{6IX18.5+}), the contribution to the total energy arising from the linearised matter fields equals
\begin{equation}
\nonumber
\wmcH_m [\hyp, X]
=
\frac 12
\left( \int_\hyp
\omega^\mu(\TStphi, \Lie_X \TStphi)
\, dS_\mu
  -
   \int_{\partial \hyp }
X^{[{{\sigma}}}
\TStpi_\TSa{}^{\mu]}  \TStphi^\TSa
dS_{{{\sigma}}\mu}
\right)
\,,
\phantom{xx}
\label{15XI19.ts3}
\end{equation}
where $\omega^\mu(\TStphi, \Lie_X \TStphi)= \Lie_X \TStphi^\TSa \, \TStpi_\TSa{}^\mu -
\Lie_X \TStpi_\TSa{}^\mu  \TStphi^\TSa $. From (\ref{11IX18.14}), the energy flux formula for matter fields is equal to
\begin{equation}\label{15XI19.ts4}
\frac{d\mcH_m (\hyp_\tau, X)}{d\tau}
 =
   -
    \int_{\partial \hyp_\tau}
X^{[{{\sigma}}} \TSpi_\TSa{}^{\mu]} \Lie_X\TSphi^\TSa dS_{{{\sigma}} \mu}
\,.
\end{equation}
%





\subsubsection{The presymplectic form on null hypersurfaces}
 \label{s20X20.1}

\subsubsection{General Gauge}

While we are mainly interested in families of light cones in this work, the calculations that follow apply to any null hypersurfaces.
Hence we consider a set of coordinates which is adapted to a space-time foliation by null hypersurfaces. The null generator of  each surface is proportional to $\partial_{r}$. We will write the space-time metric in adapted coordinates as
\begin{eqnarray}
g
= g_{uu}  du^2-2 e^{2\beta}du\,dr + 2 g_{uA} dx^A
+g_{AB} dx^A dx^B
\, .
\label{28V20.t1}
\end{eqnarray}
In terms of coordinate components, the field $b^{u}$ defined in \eqref{23XI19.3} takes the form
 \ptcheck{end May 2020; maple calculations of $b^u$ and $b^r$ at the level of four dimensional covariant derivatives checked together with TS , copy from Maple to tex checked  7VI20}
\begin{eqnarray}
b^u(\delta_1 g,\delta_2 g)
&=&
\frac{ e^{-4 \beta}\delta_{1}g_{u r} }{2}g^{A B}
	\left(
		\nabla_{r}\delta_{2} g_{A B}-2 \nabla_{A} \delta_{2}g_{r B}
	\right)
\nonumber
\\
& &+\frac{e^{-2 \beta}\delta_{1} g_{A B}}{2 }
	\Big[
		  g^{A B} \left(
			g^{C D}\left(
				\nabla_{C}\delta_{2} g_{r D}- \nabla_{r} \delta_{2} g_{C D}
			\right)
			+e^{-2 \beta} \nabla_{r} \delta_{2} g_{u r}
		\right)
\nonumber
\\
& &
 + g^{A C} g^{B D} \left(\nabla_{r} \delta_{2} g_{CD}- 2 \nabla_{C}\delta_{2}g_{r D} \right) \Big]
 \,.
\label{28V20.t2}
\end{eqnarray}
While we use the notation \eqref{28V20.t1} for the components of the metric at which the variations are taking place, most of the time  we simply use $\delta g_{\mu\nu}$ for the variations, which are assumed to satisfy
$$
 \delta g_{rr}=\delta g_{rA}=0
 \,.
$$
Note that $ \nabla_{A} \delta_{2}g_{r B}$ will not be zero in general even though $ \delta g_{rA} $ vanishes.
Writing-out the Christoffel symbols, we find
 \ptcheck{8VII20, together with TS, copy from Maple but not checked the Maple file}
\begin{eqnarray}
b^{u}(\delta_1 g,\delta_2 g)
&=&
\frac{ e^{-6 \beta}\delta_1 g_ {u r}}{2 } g^{AB}
	\left(
		\delta_2 g_ {u r} \partial_{r} g_{AB}+e^{2 \beta} \partial_r \delta_{2}g_{AB}
	\right)
\nonumber
\\
& &
 + e^{-4 \beta}\delta_{1}g_{AB} \left\{
	g^{AB} \left[
		\frac12 \partial_{r} \delta_{2} g_{u r}
		-\delta_{2} g_{u r}\left(
			\frac14 g^{CD}\partial_{r} g_{CD}
			+\partial_{r} \beta
		\right)		
	\right.
\right.
	\nonumber
	\\
& & + \left.
		\frac{e^{2 \beta}}{4}\left(
			\delta_{2}g_{CD}g^{CE}g^{DF} \partial_{r} g_{EF}-2 g^{CD} \partial_{r} \delta_{2}g_{CD}
		\right)
	\right]
\nonumber
\\
& &
+ \left.
	\frac{g^{AC} g^{BD}}{2} \left(e^{2 \beta} \partial_{r} \delta_{2} g_{CD}+ \delta_{2} g_{u r} \partial_{r} g_{CD}   \right)
\right\}
 \,.
\label{5VI20.t1}
\end{eqnarray}
Antisymmetrising over $\delta_1g$ and $\delta_2g$ to obtain the field  $\noomega^u$  of \eqref{23IX18.1}, the first term at the right-hand side drops out, which is the only obvious simplification.
If we assume moreover the Bondi condition
\begin{equation}\label{8VII20.12}
 g^{AB}\delta g_{AB} = 0
 \,,
\end{equation}
we find
\begin{eqnarray}
\noomega^u (\delta_1 g,\delta_2 g)
&=&
\frac{ e^{-4 \beta}}{2 } \bigg[\delta_2 g_ {u r} g^{AB} \partial_r \delta_{1}g_{AB}-  \delta_1 g_ {u r} g^{AB} \partial_r \delta_{2}g_{AB}
\nonumber
\\
& &
\phantom{\frac{ e^{-4 \beta}}{2 }}
+  g^{AC} g^{BD}\big ( \delta_{2}g_{AB}(e^{2 \beta} \partial_{r} \delta_{1} g_{CD}+ \delta_{1} g_{u r} \partial_{r} g_{CD}   )
\nonumber
\\
& &
\phantom{\frac{ e^{-4 \beta}}{2 }}
\left.
-\delta_{1}g_{AB} (e^{2 \beta} \partial_{r} \delta_{2} g_{CD}+ \delta_{2} g_{u r} \partial_{r} g_{CD}
 )\big)
\right]
\,.
\phantom{xxxxx}
\label{9VI20.t1}
\end{eqnarray}
Using
\begin{equation*}
\delta_2 g_ {u r} g^{AB} \partial_r \delta_{1}g_{AB}-\delta_{1}g_{AB} \delta_{2} g_{u r} g^{AC} g^{BD} \partial_{r} g_{CD}
=\delta_2 g_ {u r} \partial_{r} \left(g^{AB}\delta_{1}g_{AB}\right)=0
\end{equation*}
we obtain
\begin{equation}
\noomega^u (\delta_1 g,\delta_2 g)=\frac{ e^{-2 \beta}}{2 } g^{AC} g^{BD} \left(\delta_{2}g_{AB} \partial_{r} \delta_{1} g_{CD}-\delta_{1}g_{AB} \partial_{r} \delta_{2} g_{CD}\right)
\,.
 \label{2I20.t3}
\end{equation}
Explicit formulae for the field $b^r$ can be found in Appendix~\ref{s15VII20.t1}, see also \eqref{29IV20.t1}.

\ptcrr{b0reduction.tex commented out now}

\newcommand{\sgn}{\mathop {\rm sgn}\nolimits}
\subsubsection{The nonlinear theory}

Let us denote by $K$  a family of future directed generators of $\mcN$. We choose the orientation of  Bondi coordinates so that $K=f \partial_{r}$ where $f>0$.
For each point $p \in \mcN$ the tangent space $T_p \mcN$ may be quotiented by the subspace spanned by $K$. This quotient space $T_p \mcN/K$ carries a non-degenerate Riemannian metric $h$ and,
therefore, is equipped with a volume form $\omega$. Consider a two-form ${\bf L}$ which is equal to the pull-back of $\omega$ from the quotient space $T_p \mcN /K$ to $T_p \mcN $
\[ \pi : T_p \mcN  \longrightarrow T_p \mcN /K  \, , \quad {\bf L}:=\pi^*\omega \, . \]
We choose a one-form  $\alpha$   on $N$, such that $<K,\alpha>
\equiv 1$, and define a   three-form ${\bf v}_K$  as the product
\[ {\bf v}_K = \alpha \wedge {\bf L} \, .\]
Note that ${\bf v}_K$ does not depend upon $\alpha$ because $K\wedge {\bf L} =0$.
We can write
\begin{equation}
 {\bf v}_K =v_{K} dr\wedge dx^2\wedge dx^3\,,
 \ \mbox{ where } \ v_K=\frac{\sqrt{\mathrm{det} \, g_{AB}}}{f}.
\end{equation}
We define the following vector density
\begin{equation}
	\TSlambden=v_{K} K \equiv \sqrt{\mathrm{det} \, g_{AB}} \partial_r \, ,
\end{equation}
which is equivalent to the equation
\begin{equation}
{\bf L} = \TSlambden^a \left( \partial_a \  \rfloor \  dr \wedge
dx^2 \wedge dx^3 \right) \, .
\end{equation}
Following \cite{Jezierski:2003hh}, we  define the tensor density 
\begin{equation}\label{Q-fund}
{{\TSred \NullQ}^a}_b (K) := -s \left\{ v_K \left( \nabla_b K^a - \delta_b^a
\nabla_c K^c \right) + \delta_b^a \partial_c {\TSred \TSlambden^c} \right\}
\, ,
\end{equation}
where
$$s:=\sgn g^{u r}=\pm 1
\,,
$$
thus
$s$
equals minus one if both $\partial_u$ and $\partial_r$ are causal and consistently time-oriented.
(In the case where $\partial_u$ changes type, as happens for light cones in the de Sitter metric when $\partial_u$ is taken to be a timelike Killing vector near the vertex of the cone, the value of $s$ is determined near the vertex and extended to the light cone by continuity.)
Choosing
$$
 K:={\rm e}^{-2\beta} \partial_r
$$
%
and assuming that the variations of the metric preserve the Bondi form of the metric (including the determinant condition), the pre-symplectic form can be rewritten as~\cite{Jezierski:2003hh}
\begin{eqnarray}
\Omega_{\mcN} (\delta_1g, \delta_2 g)
& = &
\frac 1 {16 \pi}
\int_{\mcN}
(
\delta_1 \NullQ  ^{AB}   \delta_2 g_{AB}
-
\delta_2 \NullQ  ^{AB}   \delta_1 g_{AB}
)
\,
d^3 x
\,,
 \phantom{xxxx}
\label{27IX18.2}
\end{eqnarray}
where  $\NullQ $ is given by
\begin{eqnarray}
\NullQ ^{A B} &:=& g^{BC} \NullQ^A{}_C(K) \equiv \frac s2 {\sqrt{\mathrm{det} \, g_{EF}}} \, g^{AC}g^{BD}     \partial_r g_{CD}
\label{27IX18.2++}
\,.
\end{eqnarray}
The reader is warned that the field
${\NullQ ^{a}}_{b}$ is not invariant under rescalings of the null generator $K$.
Linearising  \eqref{27IX18.2}-\eqref{27IX18.2++}  provides another derivation of \eqref{2I20.t3}. Compare~\cite{KorbiczTafel}.
\label{pc19VIII20.2}

\ptcrr{further material in the file symplectiOnNull, commented out}

\ptcrr{lots of explicit calculations, made obsolete by mathematica, removed but kept in AsymptoticallyMinkowskianOldWithExplictiCalculations.tex}

\ptcrr{WithLambda removed}

\section{Bondi gauge
\\
{\small\em by PTC, JH,  TS}
}
\label{C26X20.2}

In this section we show how to put a linearised metric perturbation in Bondi gauge, and analyse the gauge freedom remaining.
%

\subsection{Coordinate transformations,  gauge freedom}
\label{s8XII19.1}

Linearised gravitational fields are defined up to a gauge transformation
\begin{equation}\label{4XII19.12}
  h\mapsto h+ \Lie_\TSxip g
\end{equation}
determined by a vector field $\TSxip$.
The aim of this section is to analyse the gauge transformations which bring a smooth linearised solution $h$ of the vacuum Einstein equations to the Bondi gauge. We will assume that, near the conformal boundary at future infinity, the linearised solution behaves as if it arose from a one-parameter of smoothly conformally compactifiable solutions near the de Sitter metric. Thus we take a background of the form
\begin{equation}\label{4XII19.11}
  g= \epsilon \lapseTB^2 du^2 - 2 du \, dr + r^2 \zzhTBW _{AB}dx^A dx^B
  \,,
\end{equation}
where $\lapseTB$ depends only upon $r$, with $\epsilon \in \{\pm1\}$, and where
$\partial_u \zzhTBW _{AB}=0=\partial_r\zzhTBW _{AB} $.

It turns out that the transformation to Bondi gauge introduces singularities at the vertex. For this reason in this section in formulae where ambiguities might arise, \emph{and only these formulae}, we will write $\barh$ for the metric perturbation in the original manifestly smooth gauge (where ``reg'' stands for ``regular'') in the cone-adapted coordinates $(u,r,x^A)$, and we will write $\hBo$ for the metric in the Bondi gauge. For instance, in order not to overburden the  notation we will continue to write $h_{tt}$, $h_{ti}$ and $h_{ij}$ instead of $\barh_{tt}$, $\barh_{ti}$ and $\barh_{ij}$ in the original manifestly smooth coordinates $(t,x^i)$, since the metric in the Bondi coordinates will only be considered in the $(u,r,x^A)$--coordinate system.

 The ``infinitesimal coordinate transformations'' \eqref{4XII19.12} should transform the metric perturbation to the Bondi gauge:
\begin{eqnarray}
\Lie_{\TSxip} g_{\TSr \TSr} + \barh_{\TSr \TSr}&=&0 \,,
 \label{4II20.3}
  \\
   \label{4II20.4}
\Lie_{\TSxip} g_{\TSr A} + \barh_{\TSr A}&=&0   \,,
\\
 g^{AB} (\Lie_{\TSxip} g_{AB} + \barh_{AB} )
  &  =  &
  0
   \label{4XII19.14}
 \, .
\end{eqnarray}
The last condition deserves a justification. For this, consider a one-parameter family of metrics, say $\lambda \mapsto g(\lambda)$ in Bondi coordinates.  (In the current case of interest $\lambda$ is the flow parameter along the vector field $\zeta$, but the argument applies to any such family.) We then have
\begin{equation}\label{4II20.1}
  g(\lambda)_{22} g(\lambda)_{33} - g_{23}(\lambda)^2 = r^4 \sin^2 \theta
  \,.
\end{equation}
Differentiating with respect to $\lambda$ one finds
\begin{equation}\label{4II20.2}
  g^{AB}h_{AB}\equiv g^{AB} \frac{d g_{AB}}{d\lambda}= 0
  \,.
\end{equation}
After performing a gauge transformation \eqref{4XII19.12}, in the new gauge we must likewise have
\begin{equation}\label{4II20.2a}
  g^{AB}(h_{AB} + \Lie_{\TSxi} g_{AB}) =0
  \,,
\end{equation}
which 
explains \eqref{4XII19.14}.

The conditions \eqref{4II20.3}-\eqref{4II20.4} are equivalent to:
\ptcheck{20II20 and corrected by JH 26V and rechecked 31VIII20; in normal coordinates the source terms are zero on the initial light cone...}
\begin{eqnarray}
\partial_{\TSr} \TSxip^{\TSu}  - \frac 12 \barh_{\TSr \TSr}&=&0
\,,
 \nonumber \\
\partial_{\TSr}\TSxip^{A}-  \frac{\ringh^{AB}}{\TSr^2}
 \left( \partial_{B} \TSxip^{\TSu}
  - \barh_{\TSr B}
   \right)
   &=&
    0
 \,, \nonumber
\end{eqnarray}
which is solved by
\ptcheck{20II20 and correction JH 26V and rechecked 31VIII20; note that there might be a convergence problem in a general smooth gauge, better to integrate from e.g. 1 as done in the CMHS paper}
\begin{eqnarray}
\TSxip^{\TSu}(u,r,x^A)&=&\TSxi^{\TSu}(\TSu,x^A) + \frac 12  \int_0^r \barh_{rr} (u,s,x^A) \, ds
 \,,
  \label{3XII19.t1}
\\
 \TSxip^{B}(u,r,x^A) &=&
   \TSxi^{B} (\TSu,x^A)
  -\ringh^{BC}
  \left(
   \frac{1}{\TSr}  \partial_{C} \TSxi^{\TSu}(\TSu,x^A)
\right.
 \nonumber
\\
  &&
	+ \int_{r_0}^r  \frac 1 {s^{2}} \left( \barh_{rC}(u,s,x^A)
    \right.
\nonumber
\\
&&
    \left.
    \left.
     - \frac 12    \int_0^s\partial_{C}  \barh_{rr}(u,\rho, x^A)\,  d\rho\right) ds
       \right)
    \,,
   \phantom{xxxxxx}
   \label{3XII19.t2}
\end{eqnarray}
for some fields $\TSxi^{u} (\TSu,x^A)$, $\TSxi^{B} (\TSu,x^A)$,
and where $r_0$ can be chosen conveniently according to the context.

\ptcrr{the stuff in normal coordinates near the tip went to NearTheTipNormalCoordinates.tex
}

In order to address \eqref{4XII19.14} we will use  the symbol
$$
 \TSoLie_\TSxip
$$
to denote   Lie-derivation in the $x^A$-variables
with respect to the vector field $\TSxip^A\partial_A$. We have
%
\begin{eqnarray}
 \TSr^{-2} \ringh^{AB} \Lie_{\TSxip} g_{AB}
&=&
	\TSr^{-2} \ringh^{AB} \left( 2 \TSr \TSxip^{\TSr} \ringh_{AB}
	+
	 \TSxip^{C}\partial_{C}g_{AB}
	+
	2 \TSxip^{C}{}_{,(A}g_{B) C}
	\right)
  \nonumber
\\
&=&
	\TSr^{-2} \ringh^{AB} \left(
	2 \TSr \TSxip^{\TSr} \ringh_{AB}
	+
	\TSr^2 \TSoLie_{\TSxip} \ringh_{AB}
	\right)
 \nonumber
\\
&=&
\TSr^{-2} \left(
	4 \TSr \TSxip^{\TSr}
	+
	2 \TSr^2 \zspaceD_{B} \TSxip^{B}
	\right)
  \nonumber
\\
    &=&
     -  \TSr^{-2} \ringh^{AB} \barh_{AB}
     \, ,
      \label{3XII19.t3}
\end{eqnarray}
where in the last line we used \eqref{4XII19.14}. Hence
\begin{equation}\label{20VII20.1}
    \TSxip^{\TSr}
	= -\frac 1	2 \TSr  \zspaceD_{B} \TSxip^{B}
     -  \frac 14 \TSr^{-1} \ringh^{AB} \barh_{AB}
     \,.
\end{equation}

We will denote by
$$\TSxip[\barh] \equiv \TSxip|_{\TSxi^A=\TSxi^u =0}
$$
the part of $\TSxip$ which depends explicitly upon $h$,   and write
\begin{equation}
\TSxip^{\mu}=
\TSxip^{\mu} [\barh] +
 \mathring \TSxip^{\mu}
\,.
 \label{4II20.10a}
\end{equation}
We note that $\mathring \TSxip$ still contains a part which depends upon $h$, as needed to satisfy the asymptotic boundary conditions. This is discussed in more detail in Section~\ref{ss5X20.2}.  The remaining part of $\mathring \TSxip$ describes asymptotic symmetries, we return to this in Section~\ref{ss5XI20.5}.

\subsubsection{Small $r$}
 \label{ss5XI20}

An analysis of the behavior of the metric at the tip of the light cone is in order. For definiteness consider a smooth  metric perturbation of Minkowski, Anti de Sitter or de Sitter spacetime.
After transforming to Bondi coordinates of the background metric we have for small $r$
\begin{eqnarray}
 \nonumber
 &
  \barh_{rr} = O(1)
  \,,
  \quad
  \barh_{rA} = O(r )
  \,,
  \quad
  \barh_{AB} = O(r^2)
  \,,
  &
\\
 &
  \barh_{ur} = O(1)
  \,,
  \quad
  \barh_{uA} = O(r )
  \,,
  \quad
  \barh_{uu} = O(1)
  \,.
   &
  \label{12IX20.1}
\end{eqnarray}

Equation \eqref{3XII19.t1} gives,  for small $r$,
\begin{eqnarray}
&
  \TSxip [\barh]^u = O(r)
  \,.
   &
    \label{23IX20.3}
\end{eqnarray}

Now, there could be a $1/r$ term  for small $r$ in the integral defining $\zeta^A$, which could lead to logarithmic terms. To see that there is a cancellation, we note that
\begin{eqnarray}
 \barh_{rr}
&= &
 h_{tt}
+ 2 h_{ti}\frac{x^i}{r}
+ h_{ij}\frac{x^i x^j}{r^2}
\,,
\\
  \barh_{rA}
&= &
\big(
 h_{tj}
+
 h_{ij} \frac{x^i}{r}
\big)
\frac{\partial x^j}{\partial x^A}
 \,,
\\
 \barh_{ur}
&= &
h_{tt}
+h_{ti}\frac{x^i}{r}
 \,.
\end{eqnarray}
Then
 \ptcheck{30IX20, missing terms added by TS}
%
\begin{eqnarray}
\nonumber
\lefteqn{
    \barh_{rC} - \frac 12 \int_{0}^{r }\partial_C \barh_{rr} dr
}
\\
&=&
    \big(
    h_{tj}
    +
    h_{ij} \frac{x^i}{r}
    \big)
    \frac{\partial x^j}{\partial x^A}
    - \frac 12 \int_0^r\partial_C \big(
     h_{tt}
    + 2 h_{ti}\frac{x^i}{r}
    + h_{ij}\frac{x^i x^j}{r^2}
    \big) dr
\nonumber
\\
&&
\nonumber
\\ &=&
\int_0^r \bigg[
     \frac{d}{dr } \bigg(
          \big(
            h_{tj}
            +
            h_{ij} \frac{x^i}{r}
          \big) \frac{\partial x^j}{ \partial x^C}
    \bigg)
- \frac 12 \partial_C \big(
     h_{tt}
     + 2 h_{ti}\frac{x^i}{r}
     + h_{ij}\frac{x^i x^j}{r^2}
\big)
\bigg] dr
\nonumber
\\ &=&
\int_0^r \bigg(
(\partial_k h_{t j}) \frac{x^k}{r} \frac{\partial x^j}{ \partial x^C}
+\frac{1}{r} h_{t j} \frac{\partial x^j}{ \partial x^C}
- \frac 12 (\partial_k h_{tt}) \frac{\partial x^{k}}{\partial x^C}
\nonumber
\\
&&
\phantom{\int_0^r \bigg()}
- (\partial_k h_{tj}) \frac{x^{j}}{r} \frac{\partial x^{k}}{\partial x^C}
-\frac{1}{r} h_{t j} \frac{\partial x^j}{ \partial x^C}
\bigg) dr
\nonumber
\\
&&
+
\int_0^r \bigg(
(\partial_k h_{ij}) \frac{x^k }{r }\frac{x^i }{r } \frac{\partial x^j}{ \partial x^C}
+   h_{ij} \frac{x^i }{r^2 } \frac{\partial x^j}{ \partial x^C}
- \frac 12 \partial_ C\big( h_{ij } \frac{x^ix^j }{r^2 }
\big)
\bigg) dr
\nonumber
\\ &=&
\int_0^r \bigg(
2(\partial_{[k} h_{ j] t}) \frac{x^k}{r} \frac{\partial x^j}{ \partial x^C}
- \frac 12 (\partial_k h_{tt}) \frac{\partial x^{k}}{\partial x^C} \bigg) dr
\nonumber
\\
&&
+
\int_0^r
\bigg(
    (\partial_k h_{ij})
    \bigg(
    \frac{x^k }{r }
    \frac{\partial x^j}{ \partial x^C}
    - \frac 12  \frac{x^j }{r }
    \frac{\partial x^k }{\partial x^C }\bigg)\frac{x^i }{r }
\bigg)dr
\nonumber
\\
&=&
O(r^2)    \,,
\end{eqnarray}
and \eqref{3XII19.t2} gives, again for small $r$,
\begin{eqnarray}
 \TSxip^{B}(u,r,x^A) &=&
   \TSxi^{B} (\TSu,x^A)
  -\ringh^{BC}
  \bigg[
   \frac{1}{\TSr}  \partial_{C} \TSxi^{\TSu}(\TSu,x^A)
 \nonumber
\\
  &&
	+
 \frac 12 \partial_k  h_{ij}\big|_{r=0}
    \bigg(
    \frac{x^k }{r}
     \frac{\partial x^j}{ \partial x^C}
   - \frac 12 \frac{x^j }{r}
    \frac{\partial x^k }{\partial x^C }\bigg)\frac{x^i }{r }
     \bigg]
 +    O(r^2)
   \phantom{xxxxxx}
   \nonumber
\\
 &   = &
    O(r^{-1})
   \,,
   \label{23IX20.6}
\end{eqnarray}
so that
\begin{equation}
   \TSxip [h]^A =
    O(r)
   \,.
\end{equation}
Equations~\eqref{3XII19.t1}-\eqref{3XII19.t2} together with \eqref{20VII20.1} lead to, again for small $r$,
\begin{eqnarray}
&
   \TSxip [h]^r = O(r )
  \,,
   &
   \label{3IX20.11}
\\
&
  \TSxi^u = O(1)
  \,,
  \quad
  \TSxi^A = O(1 )
  \,,
  \quad
  \zeta^u = \TSxi^u +  O(r)
  \,,
  &
\\
 &
 \displaystyle
  \zeta^A =  \TSxi^{A} (\TSu,x^B)
  - \frac{1}{\TSr} \zspaceD ^A
  \TSxi^{\TSu}(\TSu,x^B) + O(r)
  \,.
   &
    \label{14VII20.3a}
\end{eqnarray}
Inserting \eqref{3XII19.t2} in \eqref{20VII20.1} we find
\begin{equation}
 \mathring \zeta^r=
  \frac{\TSzlap \TSxi^{\TSu}}{2} - \frac{\TSr \zspaceD_{B} \TSxi^{B}}{2}
   = O(1)
  \, ,
  \quad
   \TSxip ^r = \frac 12  \Delta_\ringh \xi^u + O(r ) = O(1)
  \,,
 \label{4II20.10b}
\end{equation}
where $\TSzlap$ is a Laplace operator associated with the metric $\ringh_{AB}$.

The behaviour of various derivatives   should be clear from the above.

We emphasise that the original behaviour \eqref{12IX20.1} of $h$ near the vertex will not be true in general for the metric coefficients in Bondi coordinates.
For instance:
\tscomr{Checked red part until 12.20 24IX20}
\begin{eqnarray}
  \hBo_{uA}
   & = &
  \barh_{uA} + \mcL_\zeta g_{uA} =
  \barh_{uA} + \partial_A\zeta ^\mu \eta_{u\mu}
  + \partial_u\zeta ^\mu \eta_{\mu A}
  \nonumber
\\
  & = &
  \barh_{uA} +
   \partial_A(\epsilon N^2 \zeta ^u - \zeta ^r)
  + r^2 \ringh_{AB} \partial_u\zeta ^B
  \nonumber
\\
  & = &
-
   \zspaceD_A (1+ \frac 12 \Delta_\ringh) \xi ^u
  + O(r)
  \,
  \nonumber
\\
  & = &
   O(1)
  \,,
  \label{24IX20.1}
  \\
  \hBo_{ur}
   &= &
  \barh_{ur} + \mcL_\zeta g_{ur} =
  \barh_{ur}
  - \partial_u \zeta^u
  + \epsilon N^2 \partial_r \zeta^u
  - \partial_r \zeta^r
  \nonumber
\\
&=&
  \barh_{ur}
  - \partial_u \xi^u
  - \frac{1}{2 } \barh_{rr}
  + \frac{1}{2} \zspaceD^A
   \xi_A
    \nonumber
\\
 &&
  +
   \frac{1}{4} \partial_r(r^{-1} \ringh^{AB}\barh_{AB})
    +
    O(r)
    \,,
    \label{24IX20.23a}
\\
 {
      \hBo_{uu}}
    & = &
 {\redc \barh_{uu} + \mcL_\zeta g_{uu} =
  \barh_{uu}
  + \epsilon \TSxip^{\TSr} \partial_r N^2
	+2 \partial_u(\epsilon N^2 \zeta^u - \zeta^r)
}
  \nonumber
\\
&=&  \barh_{uu}
 -
    (2 + \Delta_\ringh) \partial_u\xi ^u
  + O(r)
    \,,
    \label{24IX20.23x}
\\
  \hBo_{AB}
    & = &
  \barh_{AB} + \mcL_\zeta g_{AB} =
  \barh_{AB}
  +2 \TSr \TSxip^{\TSr} \ringh_{AB}
	+
	\TSr^2 \TSoLie_{\TSxip} \ringh_{AB}
  \nonumber
\\
&=&
 \TSr
  (\Delta_\ringh \xi^u \ringh_{AB}
	- 2
	 \zspaceD_A\zspaceD_B \xi^u
)
+ O(r^2)
 \,.
  \label{24IX20.23}
\end{eqnarray}
%
%
\tscomr{Checked red part until here 28IX20}

Summarising, for small $r$,
\ptcheck{23IX20}
\begin{equation}
   \hBo_{uu}= O(1)
   \,,
    \quad
  \hBo_{ur} =  O(1) \,,
  \quad
  \hBo_{uA} = O(1)
   \,,
\quad
  \hBo_{AB} =O(r)
   \,.
   \label{14VII20.5}
\end{equation}

For further reference we emphasise that
\begin{eqnarray}
  \zspaceD^A \hBo_{uA}|_{r=0} =
 -
    \frac 12 (\Delta_\ringh+ 2) \Delta_\ringh \xi ^u
  \,.
  \label{24IX20.21}
\end{eqnarray}

Using
\begin{eqnarray}\label{18IX20.50}
 &
\mathring{\gamma}^{AB}\frac{\partial x^{i}}{\partial x^{A}}\frac{\partial x^{j}}{\partial x^{B}}
=
(\delta^{ij} - \frac{x^{i}x^{j}}{r^{2}})r^{2}
&
\\
  &
 \Delta_\ringh x^i = -2 x^i
  \,,
  \quad
 0 = \Delta_\ringh (\underbrace{x^ix^i}_{r^2}) = 2 \zspaceD^A(x^i \zspaceD_A x^i)
  \,,
 &
  \label{21IX20.2}
\\
 &
 \displaystyle
 \zspaceD^C
  \bigg(
     \frac{x^{j} }{r^2}\frac{\partial x^{i}}{\partial x^C}
     \bigg)
      =
     \frac{ \zspaceD^C x^{j} }{r^2}\frac{\partial x^{i}}{\partial x^C}
     +
     \frac{x^{j} }{r^2} \Delta_\ringh x^{i}
     =
      \delta^{ij} - \frac{3 x^i x^j}{r^2}
      \,,
     &
\\
 &
 \displaystyle
 \zspaceD^C
  \bigg(
     \frac{x^{(j} }{r^2}\frac{\partial x^{i)}}{\partial x^C}
     \bigg) =
      \delta^{ij} - \frac{3 x^i x^j}{r^2}\,,
      \quad
      \Delta_\ringh \frac{x^i x^j}{r^2}= 2 \delta^{ij} -6
       \frac{x^i x^j}{r^2}
      \,,
     &
  \label{21IX20.3asd}
\end{eqnarray}
 \ptcheck{checked up to here by JH 26IX20}
 \tscomr{Checked red part until here 28IX20}
Equation \eqref{24IX20.23a} can be rewritten as
\tscomr{Checked with JH up to $h_{tt}$ and $h_{ti}$ terms.}
\begin{eqnarray}
\hBo_{ur}
& = &
\barh_{ur}|_{r=0}
- \partial_u \xi^u
+ \frac{1}{2} \zspaceD^A
\xi_A
\nonumber
\\
&&
- \frac12 \Big(
     h_{tt}|_{r=0}
    + 2 h_{ti}|_{r=0}\frac{x^i}{r}
    + h_{ij}|_{r=0}\frac{x^i x^j}{r^2}
\Big)
\nonumber
\\
&&
+
\frac{1}{4} \partial_r
\bigg(
\underbrace{
    r^{-1} \ringh^{AB}h_{ij}\frac{\partial x^{i}}{\partial x^{A}}\frac{\partial x^{j}}{\partial x^{B}}
}_{r             h_{ij}   \big(\delta^{ij} -\frac{x^ix^j}{r^2}\big)      }
\bigg)
+
O(r)
\nonumber
\\
&=&
\barh_{ur}|_{r=0}
- \partial_u \xi^u
+ \frac{1}{2} \zspaceD^A \xi_A - \frac12 \Big(
h_{tt}|_{r=0}
+ 2 h_{ti}|_{r=0}\frac{x^i}{r}
\Big)
\nonumber
\\
& &
+\frac14 h_{ij}|_{r=0}  \big(\delta^{ij} -\frac{x^ix^j}{r^2}\big)
- \frac12 h_{ij}|_{r=0}\frac{x^ix^j}{r^2}
+
O(r)
\nonumber
\\
&=&
\frac12 h_{tt}|_{r=0}
- \partial_u \xi^u
+ \frac{1}{2} \zspaceD^A \xi_A
\nonumber
\\
& &
+\frac14 h_{ij}|_{r=0}  \big(\delta^{ij} -3\frac{x^ix^j}{r^2}\big)
+
O(r)
\,.
\label{24IX20.11a}
\end{eqnarray}
\subsubsection{From smooth to Bondi}
 \label{ss5X20.2}

In view of the formulae so far, one proceeds as follows. Given a smooth linearised metric perturbation $h$ we perform  the ``infinitesimal coordinate transformation'' $\TSxip[h]$ as defined above to obtain a metric in Bondi gauge, still denoted by $h$ but occasionally by $\hBo$. It still remains to take care of the boundary conditions. In particular,  after the above, the asymptotic field $\ozero h_{AB}(u,\cdot)$ will not necessarily vanish, regardless of whether or not $\ozero h_{AB}(u,\cdot)$ was zero before the $\TSxip[h]$-transformation.
In order to remedy this we take the  $u$-parameterised family of covector fields  $\TSxi_B(u ,\cdot)$ to be any family of solutions, smooth in $u$, of the equations (cf., e.g., \cite[Th\'eor\`eme~3.4]{Boucetta1999})
\begin{equation}\label{6III20.1}
  \zspaceD_A \TSxi_B + \zspaceD_B \TSxi_A - \zspaceD^C \TSxi_C \ringh_{AB} + \ozero \zhTB_{AB}(u,\cdot)- \frac12
   \ozero \zhTB{}^{C}{}_{C}(u,\cdot) \ringh_{AB}
    =0
  \,,
\end{equation}
where
$$
 \ozero \zhTB_{\mu\nu}:= \lim_{r\to\infty} r^{-2}h_{\mu\nu}
  \,.
$$
These solutions will be denoted by $\xi^A[h]$.

It follows from the first line of \eqref{24IX20.23} that the gauge-transformed fields $\ozero \zhTB_{AB}(u,\cdot) $ will  vanish.

Equation~\eqref{6III20.1} determines $\TSxi_A[h](u,x^B)$ up to a $u$-dependent family of conformal Killing vectors of the round two-dimensional sphere; we will return to this freedom shortly.

We let $\TSxi^u[h]$ be a solution of (compare \eqref{24IX20.23a} and \eqref{24IX20.11a})
\begin{eqnarray}
	 \partial_{\TSu}  \TSxi^{\TSu}[h](\TSu,x^A)
 &=&
 \big(\barh_{ur}
  + \frac{1}{2 }\epsilon N^2 \barh_{rr}
  + \frac{1}{2} \zspaceD^A
   \xi_A [h]
   \big)\big|_{r=0}
    \nonumber
\\
\nonumber
&=&
\frac12 \barh_{tt}|_{r=0}
+\frac14 \barh_{ij}|_{r=0}  \big(\delta^{ij} - 3 \frac{x^ix^j}{r^2}\big)
+ \frac{1}{2} \zspaceD^A \xi_A[h] \,,
\nonumber
\\
& &
 \label{5XII19.1n}
\end{eqnarray}
with smooth initial data $\TSxi^u(u_0,\cdot)$ for some $u_0$.
In vacuum (cf.\ \eqref{28XI19.4aa} below) this leads to a gauge-transformed field for which
$$h_{ur} \equiv 0
 \,.
$$

This procedure leads to  a metric perturbation satisfying all Bondi gauge conditions together with
\begin{equation}\label{28V20.1}
 \ozero \zhTB_{AB}\equiv 0 \equiv h_{ur}
 \,,
\end{equation}
where the second equality requires $T_{ur}\equiv 0$.

\subsection{Residual gauge}
 \label{ss5XI20.3}

The freedom of choosing the vector field $\ringzeta $ of \eqref{4II20.10a}
 describes the freedom to perform coordinate transformations preserving the Bondi form of the metric.
These will be referred to as \emph{residual gauge-transformations}.
An example is provided by the vector field $\xi[h]$ just defined.

Under these $\ringzeta $-transformations,
the linearised metric components acquire the following terms:
 \ptcheck{20VII20 notation cleaned up, corrected; previously checked    PTC and TS on 10VII20; first two lines rechecked 31VIII20 }
\begin{eqnarray}
h_{uu} : \  \Lie_{\ringzeta } g_{uu}
 &=&\ringzeta^r \epsilon\partial_r \lapseTB^2
	+2 \epsilon\lapseTB^2 \partial_{\TSu} \ringzeta^{\TSu}
	-2 \partial_{\TSu}\ringzeta^{\TSr}
\nonumber
\\
	&=&
	 -(2+\TSzlap ) \partial_{u} \xi^{u}
	               +\TSr (\zspaceD_{B} \partial_{\TSu}\TSxi^{B}+\alpha^{2} \TSzlap \TSxi^{\TSu})
 \nonumber
\\
 &&
+\alpha^{2}r^{2}(2\partial_{\TSu}\TSxi^{\TSu} - \zspaceD_{B}\TSxi^{B})
\,,
\label{4XII19.t1}
\\
h_{ur}: \  \Lie_{\ringzeta } g_{ur}
 &=&- \partial_{\TSu}  \ringzeta^{\TSu}- \partial_{\TSr}  \ringzeta^{\TSr}
\nonumber
\\
&=&
	- \partial_{\TSu}  \TSxi^{\TSu}
 +
   \frac{ \zspaceD_{B} \TSxi^{B} }{2}
 \,,
\label{4XII19.t2}
\\
h_{uA} : \  \Lie_{\ringzeta } g_{uA}
&=&
	\TSr^2 \ringh_{AB}\partial_{\TSu}{  \mathring \TSxip^{B}}
	-  \partial_{A} \ringzeta ^{\TSr}
    + \epsilon\lapseTB^2 \partial_{A} \ringzeta ^{\TSu}
\nonumber
\\
&=&
	- \frac{1}{2}\partial_{A} \big[
 \big(\TSzlap \TSxi^{\TSu}+ 2 \xi^{u}\big)
	+ {r }(   \zspaceD_{B} \TSxi^{B}-2\partial_{u} \xi^{u})
\big]
\nonumber
\\
& &
	+ r^{2}(\ringh_{AB}\partial_{\TSu} \TSxi^{B}+\alpha^{2}\partial_{A}\xi^{u})
 \,,
\label{4XII19.t3}
\\
h_{AB} : \  \Lie_{\ringzeta } g_{AB}
&=&
	2 \TSr \ringzeta^{\TSr} \ringh_{AB}
	+
	\TSr^2 \TSoLie_{\ringzeta } \ringh_{AB}
\nonumber
\\
 &=&
  \TSr^2
   (  \TSoLie_{\TSxi} \ringh_{AB}
    - \ringh_{AB}  \zspaceD_{C} \TSxi^{C}
    )
\nonumber
\\
& &
	- \TSr  ( 2 \zspaceD_{A}\zspaceD_{B}   \TSxi^{\TSu}- \ringh_{AB} 		   \TSzlap \TSxi^{\TSu} )
	 \,.
 \label{20II20.1}
\end{eqnarray}

\ptcrr{commented out the part about conformal KV's which was  only correct if we exclude $\xi[h]$ from $\mathring \zeta$, so either exclude, or say here that we have already carried out $\zeta[h]$ and $\xi[h]$; but then one could as well move this already to the asymptotic symmetries subsection...}
The residual gauge transformations are thus defined by a $u$-parameterised family of vector fields $\TSxi^A(u,\cdot)$ on $S^2$ together with
\begin{equation}\label{5XII19.1a}
	 \partial_{\TSu}  \TSxi^{\TSu}(\TSu,x^A) =
\frac{ \zspaceD_{B} \TSxi^{B}(\TSu, x^{A})}{2}
 \,,
\end{equation}
and \eqref{4II20.10b}.
Explicitly:
 \ptcheck{consistent with MK calculations up to here according to TS on 11VII20, corrected for misleading notation 29VIII}
\begin{eqnarray}
 \nonumber 
 \mathring \zeta &=&
  \left(
   \int
\frac{ \zspaceD_{B} \TSxi^{B}(u, x^{A})}{2} du
 + \TSoxi{}^u(x^ A)
  \right)\partial_u +
  \frac 12 \left(
  {\TSzlap \TSxi^{\TSu}}  -  {\TSr \zspaceD_{B} \TSxi^{B}}
  \right) \partial_r
   \\
    && + \big(\TSxi^B(u,x^A) - \frac{1}{\TSr} \zspaceD ^B
  \TSxi^{\TSu}(\TSu,x^A) \big)\partial_B
     \,,
      \label{20II20.9}
\end{eqnarray}
with an arbitrary function $\TSoxi^u(x^A)$.

\subsubsection{Asymptotic symmetries}
 \label{ss5XI20.5}

 \ptcheck{4XII19 up to here and rechecked 20II20, old version confirmed JH on 24IV20}

Unless explicitly indicated otherwise, in the remainder of this work we suppose that the metric has been transformed to Bondi form with $\ozero \zhTB_{AB} \equiv 0$.  The residual gauge transformation which preserve this condition will take the form \eqref{20II20.9} with, at each $u$, $ \TSxi^B(u, x^A)\partial_B  $ being a conformal Killing vector field of $\ringh$.

The conformal Killing vectors of $S^2$ are related to the Lorentz group, and the remaining freedom in $\TSxi^u$ corresponds to translations and supertranslations.

All these gauge transformations are interpreted as governing  \emph{asymptotic symmetries}, defined here as transformation which preserve the Bondi gauge as well
as the asymptotic  fall off condition for linearised fields.

In the asymptotically flat case (thus $\Lambda=0$) the fields $\TSxi^A$ become $u$-independent by (consistently) requiring in addition that $\ozero \zhTB_{uA} \equiv 0$.

However, when $\Lambda>0$, the requirement  $\ozero \zhTB_{uA} \equiv 0$ is not consistent with \eqref{28V20.1} for general metric perturbations considered so far. Indeed, it follows from  \eqref{4XII19.t3} that  under  the $\TSxi$-gauge transformations, the $r^2$-terms in   $h_{uA}$ transform as
\begin{equation}\label{20II20.7}
    { \ozero \zhTB_{uA}}(\TSu,x^A)
     \mapsto
      { \ozero \zhTB_{uA}}(\TSu,x^A)
     + \ringh_{AB}\partial_{\TSu} \TSxi^{B} (\TSu,x^A)
	+ 
\alpha^2 \partial_{A} \TSxi^{\TSu}(\TSu, x^{A})
 \,,
\end{equation}
keeping in mind that $\epsilon =1$ for large $r$
While this shows that  we can always choose $\partial_u\TSxi^B$ so that the gauge-transformed field $  { \ozero \zhTB_{uB}}$ vanishes identically, such a choice will not be compatible with the requirement that $ \ozero \zhTB_{AB}\equiv 0$ in general when $\Lambda \ne 0$.
Indeed, we have the transformation law
\begin{equation}\label{6III20.1asb}
  \ozero \zhTB_{AB}(u,\cdot)  \mapsto
  \big(
   \zspaceD_A \TSxi_B + \zspaceD_B \TSxi_A - \zspaceD^C \TSxi_C \ringh_{AB} + \ozero \zhTB_{AB}
   \big)
    (u,\cdot)
  \,.
\end{equation}
One could therefore choose  $\TSxi^B$   so that
\begin{equation}\label{6III20.1asbc}
\ozero \zhTB_{AB}(u_0,\cdot) \equiv 0 \equiv    \ozero \zhTB_{uB}(u.\cdot)
 \,,
\end{equation}
but there does not seem to be any reason why $\ozero \zhTB_{AB}(u ,\cdot)$ should then  be zero in general for $u\ne u_0$.

In the  gauge \eqref{6III20.1asbc} the  canonical energy on $\mcC_{u,R}$ diverges  as $R^3$
\ptcheck{20VIII10 with TS: the volume integral diverges as $R$ because of leading order cancellation, this comes from boundary term}
 when $R$ tends to infinity for $u\ne u_0$, while $R^2$ is replaced by $R$ when  \eqref{28V20.1} holds. This   property makes the gauge $\ozero \zhTB_{AB}\equiv 0$ more attractive from our perspective.

\subsubsection{Rigid transport of sections of $\scrip$}
 \label{ss5XI20.5a}

When $\Lambda \ne 0$,  a prescription to reduce the set of asymptotic symmetries has been presented in \cite{ChHS}. For the sake of completeness we reproduce the construction here.

 The Hodge-Kodaira decomposition of one-forms on $S^2$ shows that there exist functions $\hdpot(u)$ and $\hrotpot(u)$ on $S^2$ such that
\begin{equation}\label{20II20.7a}
     \ozero \zhTB_{uA}(\TSu,\cdot)=\mathring{D}_{A}\hdpot(u)+{\varepsilon_{A}}^{B} \mathring{D}_{B}\hrotpot(u)
 \,,
\end{equation}
where ${\varepsilon_{B}}^{C}$ is the two-dimensional Levi-Civita tensor.
We can similarly write $\xi_B$ as
\begin{equation}
\xi_{B}(u,\cdot)=\mathring{D}_{B}\iota(u)+{\varepsilon_{B}}^{C} \mathring{D}_{C} \upsilon(u)
 \,,
\label{1VII20.t4a}
\end{equation}
where the functions $\iota(u)$ and $\upsilon(u)$ are linear combinations of $\ell=1$ spherical harmonics (see Appendix~\ref{App2VII20}).
Equation \eqref{20II20.7} can be rewritten as
\begin{eqnarray}\label{20II20.7ab}
  \hdpot(u)
     & \mapsto &
     \hdpot(u)
     + \partial_u \iota(u)
      +
\alpha^2   \TSxi^{\TSu}(\TSu, \cdot)
 \,,
\\\label{20II20.7ac}
    \hrotpot(u)   & \mapsto &
      \hrotpot(u)
     + \partial_u \upsilon(u)
 \,.
\end{eqnarray}
Let $P_1$ denote the $L^2(S^2)$-orthogonal projection on the space of $\ell=1$ spherical harmonics.
 We can arrange that $ P_1  \big(\hrotpot  \big)$ vanishes by solving the linear ODE
\begin{eqnarray}\label{20II20.7abcd} &
      \partial_u \upsilon
      =
      -
      P_1  \big(\hrotpot  \big)
 \,,
  &
\end{eqnarray}
which leaves the freedom of choosing $\upsilon(u_0)$.

Next, using \eqref{20II20.7ab} and \eqref{5XII19.1a} we obtain
\begin{eqnarray}
  \partial_u\hdpot(u)
     & \mapsto &
    \partial_u \hdpot(u)
     + \partial^2_u \iota(u)
      +
\alpha^2   \partial_u\TSxi^{\TSu}(\TSu, \cdot)
\nonumber
\\
      & = &
      \partial_u \hdpot(u)
     + \partial^2_u \iota(u)
    -
 \alpha^2  \iota(u)
 \,.
  \label{20II20.7acc}
\end{eqnarray}
We can arrange that $\partial_u \big(P_1(\hdpot)\big)$ vanishes by solving the equation
 \ptcheck{11VII20 by TS}
\begin{eqnarray}  &
     \partial^2_u \iota
     -
\alpha^2   \iota
 = P_1 \big( \partial_u \hdpot
     \big)
 \,.
  & \label{2VII20.10}
\end{eqnarray}
Equation~\eqref{20II20.7ab} shows that $ P_1\hdpot $ will vanish if
\begin{eqnarray}  &
    \partial_u \iota(u_0)
      + P_1\big(
       \hdpot(u_0)
        +
\alpha^2   \TSxi^{\TSu}(\TSu_0, \cdot)
     \big)= 0
 \,.
  & \label{2VII20.10asdf}
\end{eqnarray}

There remains  the freedom of choosing $\iota(u_0)$,
with the solutions of the homogeneous equation \eqref{2VII20.10} taking the form
\begin{equation}\label{11VII20.1}
  \iota(u, \cdot) = e^{\alpha u} \iota_+(\cdot) + e^{-\alpha u} \iota_-(\cdot)
  \,,
\end{equation}
where $\iota_\pm$ are linear combinations for $\ell=1$ spherical harmonics.

Summarising, we can achieve a \emph{rigid transport} of the Bondi coordinates from one sphere to the other by requiring that the potentials $\hdpot$ and $\hrotpot$ of \eqref{1VII20.t4a} satisfy
\begin{equation}\label{2VII10.11}
  P_1(\hdpot)\equiv 0 \equiv P_1(\hrotpot)
  \,.
\end{equation}
We will refer to \eqref{2VII10.11} as the \emph{rigid transport condition}.

Under the rigid transport conditions, we have the freedom of choosing
$\big(\iota(u_0),   \partial_u \iota (u_0),  \upsilon (u_0)\big)$, which is related to the freedom of rotating and boosting the initial light cone $\mcC_{u_0}$, and of choosing  $ \TSxi^u(u_0,\cdot)$, which is the equivalent of the supertranslations that arise in the case $\Lambda =0$, subject to the constraint
\begin{eqnarray}  &
    \partial_u \iota(u_0)
      + \alpha^2 P_1\big(
  \TSxi^{\TSu}(\TSu_0, \cdot)
     \big)= 0
 \,.
  & \label{2VII20.10asdfb}
\end{eqnarray}

After imposing \eqref{2VII10.11}, the residual gauge transformations  which also preserve  the rigid transport condition \eqref{2VII10.11}  take the form \eqref{20II20.9}
with an arbitrary function $\TSoxi^u(x^A)$, and
where  $ \TSoxi^B (u,x^A)\partial_B  $ is the angular part of a Killing vector field of de Sitter spacetime as in \eqref{1VII20.t4a}, thus $\upsilon$ is a $u$-independent linear combination of $\ell=1$ spherical harmonics, the potential $\iota$ takes the form \eqref{11VII20.1}, with $\partial_u \iota(u_0)$ satisfying \eqref{2VII20.10asdfb}.

\section{The linearised Einstein equations
\\
{\small\em by PTC, JH, MM, TS}
}
\label{C26X20.3}

\ptcrr{all frame stuff for de Sitter moved to deSitterFrame}

\subsection{Linearised metric perturbations  in Bondi coordinates}
 \label{s9VII20}

Let $\mcN$ be a null hypersurface given by $u=\mathrm{const}$.
We will use \emph{Bondi-type coordinates} and a \emph{Bondi parameterisation} of the metric on $\mcN$:
\begin{eqnarray}
g_{\alpha \beta}dx^{\alpha}dx^{\beta}
 &  =  &-\frac{V}{r}e^{2\beta} du^2-2 e^{2\beta}dudr
  \nonumber
\\
 &&
   +r^2\zhTBW_{AB}\Big(dx^A-U^Adu\Big)\Big(dx^B-U^Bdu\Big)
    \, .
     \label{30XI19.100}
\end{eqnarray}
Here it is also assumed that $\det \gamma_{AB}$ takes a canonical, $r$- and $u$-independent value, namely
\begin{equation}\label{5XI20.p1}
\det \gamma = \sin^2 \theta
 \,.
\end{equation}
\ptcrr{stuff about frames removed, not needed anymore}
 In spacetime dimension four,  the Euler--Lagrange equations for the Lagrangian \eqref{28IX18.2z}  with $\znabla$ replaced by $\nabla$ are
   \ptcheck{13VIII20 together}
    \begin{eqnarray}
    \label{20VIII20.t0}
    \lefteqn{
    \nabla^{\mu} \nabla_{\alpha} \delta g_{\beta \mu}
    +\nabla^{\mu} \nabla_{\beta} \delta g_{\alpha \mu}
    +g_{\alpha \beta}\nabla_{\kappa}\nabla^{\kappa} \delta {g_{\lambda}}^{\lambda}
    -\nabla_{\mu} \nabla^{\mu} \delta g_{ \alpha \beta}
    } &&
     \nonumber \\
     &&
    -g_{\alpha \beta} \nabla^{\kappa} \nabla^{\lambda} \delta g_{\kappa \lambda}
    -\nabla_{\alpha} \nabla_{\beta} \delta {g^{\kappa}}_{\kappa}
    =  2 \Lambda ( \delta g_{\alpha\beta} - \frac{1}{2}  \delta g^\kappa{}_\kappa g_{\alpha\beta} )
     \, .
    \end{eqnarray}
   Assuming $\delta {g^{\kappa}}_{\kappa}=0$, which is the case in the Bondi gauge, we obtain
  \begin{equation}
      \mcE_{\alpha \beta}
      :=
      \nabla^{\mu} \nabla_{\alpha} \delta g_{\beta \mu}
      +\nabla^{\mu} \nabla_{\beta} \delta g_{\alpha \mu}
      -\nabla_{\mu} \nabla^{\mu} \delta g_{ \alpha \beta}
      -g_{\alpha \beta} \nabla^{\kappa} \nabla^{\lambda} \delta g_{\kappa \lambda} -   2 \Lambda  \delta g_{\alpha\beta}
      =0 \, .
      \label{20VIII20.t1}
  \end{equation}
  Taking a trace we find
  \begin{equation}
      \nabla^{\kappa} \nabla^{\lambda} \delta g_{\kappa \lambda}
      =0 \, ,
  \end{equation}
which simplifies \eqref{20VIII20.t1} further to
  \begin{equation}
      \nabla^{\mu} \nabla_{\alpha} \delta g_{\beta \mu}
      +\nabla^{\mu} \nabla_{\beta} \delta g_{\alpha \mu}
      -\nabla_{\mu} \nabla^{\mu} \delta g_{ \alpha \beta}  -   2 \Lambda  \delta g_{\alpha\beta}
      =0 \, .
      \label{20VIII20.t1asf}
  \end{equation}
These equations are still unpleasant enough so that it appears simpler to instead linearise the equations as written down in~\cite{MaedlerWinicour}, and we will do so. Nevertheless,  to  be on the safe side, we have checked,  using Maple, that the set of equations $ {\mcE^{u}}_{r}={\mcE^{u}}_{u}={\mcE^{u}}_{A}=0$ is equivalent to the linearised  equations  (\ref{28XI19.4aa}), (\ref{28XI19.5}) and (\ref{30XI19.11}) below, obtained from the equations in~\cite{MaedlerWinicour}.

Our asymptotic conditions on the linearised perturbations of the metric will be modelled on the asymptotic behaviour of the full solutions of the Einstein vacuum field equations with positive cosmological constant and with smooth initial Cauchy data on $S^3$, as constructed by Friedrich in~\cite{Friedrich}. The resulting  spacetimes  have a smooth conformal completion with a (necessarily spacelike) boundary at (timelike) infinity $\scrip$ (denoted by $I^+$ by some authors).  It is shown in~\cite[Section~2.1]{ChIfsits} that, given a foliation of $\scrip$ by a function $y^{  0}$ (in our case, this foliation will be provided by the intersections of a family of null light cones emanating from a world-line in spacetime), there exists a neighborhood of $\scrip$ on which  the metric takes the Bondi form \eqref{30XI19.100},
where $u_\Bo|_{\scrip} = y^{ 0}$.
Now, Bondi \emph{et al.} consider the case $\Lambda=0$ and assume
\bel{eq:bc}
	\lim_{r_{\Bobo} \to \infty}  U^A_\Bo = 0
	\,, \quad
	\lim_{r_{\Bobo} \to \infty} \beta  = 0
	\,, \quad
	\lim_{r_{\Bobo} \to \infty} \left( r_{\Bobo}^{-2} \og^\Bo_{AB} \right) = \zzhTBW _{AB}
	\,,
\ee
where $\zzhTBW ^\Bo_{AB} $ is the standard metric on $S^2$.
%
%
It follows e.g.\ from~\cite[Section~2.1]{ChIfsits} that the last equation in \eq{eq:bc} is justified under the hypothesis of existence of a smooth conformal completion at infinity regardless of the value of $\Lambda$. However, it is not clear at all whether the first two equations \eq{eq:bc} can be assumed to hold \emph{for all retarded times} in general: When $\Lambda<0$ this is part of asymptotic conditions which one is free to impose, and which are usually imposed in this context, but which one might  \emph{not} want to impose in \emph{some} situations. When $\Lambda=0$ these conditions can be realised by choosing the function $y^{0}$ suitably.  However, when $\Lambda>0$  there is little doubt that all three conditions in \eq{eq:bc} can be simultaneously satisfied for all retarded times by a \emph{restricted class of metrics} only. In the linearised theory this will be clear from the calculations that follow.

We have mentioned above the results of Friedrich on the spacelike relativistic Cauchy problem, as they guarantee existence of a large class of vacuum spacetimes with a positive cosmological constant, near the de Sitter or Anti de Sitter spacetime, with a smooth conformal completion at Scri. As such, in our context it is more natural to think of the characteristic rather than the spacelike Cauchy problem (cf., e.g., \cite{ChPaetz}). In this context, in the linearised theory we are free to prescribe arbitrarily the angular part $h_{AB} dx^A dx^B$ of the linearised metric perturbation on the light cone, with the remaining fields, and their asymptotics, determined by these free data and the residual gauge conditions to which we return in Section~\ref{s8XII19.1}. This follows quite generally from the analysis in e.g.\ \cite{ChPaetz}, but can also be deduced directly from the considerations that we are about to present. Friedrich's results just mentioned guarantee, e.g. by taking data induced on light cones from his solutions, that there exists a large class of free data $h_{AB} dx^A dx^B$ on the initial light cone with an evolution which is smoothly conformally compactifiable at $\scrip$, and we restrict our attention to such data.

Writing interchangeably
$$
 \mbox{$h_{\mu\nu}$ for $\delta g_{\mu\nu}$, and $\zhTB_{\mu\nu}$ for $r^{-2}h_{\mu\nu}$,}
$$
we thus assume the following large-$r$ expansion
\begin{eqnarray}
  h_{AB}
   & = &
     r^2 \underbrace{\omtwo h_{AB}}_0
   + r  \omone h_{AB}
   +   \ozero  h_{AB}
   + r^{-1} \oone h_{AB}
   + o(r^{-1})
    \big)
    \nonumber
\\
   & \equiv  &  r^2
   \big(
     \underbrace{\ozero \zhTB_{AB}}_0
   + r^{-1} \oone \zhTB_{AB}
   + r^{-2} \otwo  \zhTB_{AB}
   + r^{-3} \othree \zhTB_{AB}
   + o(r^{-3})
    \big)
   \,,
   \phantom{xxx}
   \label{18IV20.1}
\end{eqnarray}
where the expansion tensors $\omone h_{AB}\equiv\oone \zhTB_{AB}$, etc., are independent of $r$.

A comment on our hypothesis that $\ozero \zhTB_{AB}\equiv 0$ is in order. As discussed in detail in Section~\ref{s8XII19.1},
 after transforming the metric perturbation to the Bondi form  we can always use the remaining coordinate freedom to achieve $\ozero \zhTB_{AB}=0$. An alternative possibility is   $\ozero \zhTB_{uA}=0$. A key fact is that $\ozero \zhTB_{AB}=0$ and $\ozero \zhTB_{uA}=0$ cannot be achieved simultaneously in general.  As already mentioned, the  condition $\ozero \zhTB_{uA}=0$ (in which case $\ozero \zhTB_{AB}\ne 0$ in general)  leads to energy integrals on balls of radius $R$ which diverge as $R^3$, and therefore we have opted for $\ozero \zhTB_{AB}=0$ which leads to a slower divergence.
\ptcheck{20VIII20 with TS}

Smooth compactifiability of the  solution guarantees that there will be no logarithmic terms in the asymptotic expansion, which in turn requires
\begin{equation}\label{9VII20.1}
  \otwo  \zhTB_{AB} \equiv 0
   \,,
\end{equation}
as follows from our calculations below (compare~\cite{ChIfsits}). However, we will not assume \eqref{9VII20.1} at this stage, to be able to track down the role of this term in the equations that follow.

We will derive precise information on the asymptotic behaviour of the remaining linearised fields  using the characteristic constraint equations in Bondi coordinates, which read~\cite{MaedlerWinicour}
\ptcheck{20II20, and actually the first equation checked by TS in Maple and agrees on 1VIII20 and the second as well on 3III20}
\begin{equation}
         \label{eq:beta_eq}
          \partial_r \beta  = \frac{r}{16}\zhTBW^{AC}\zhTBW^{BD} (\partial_r \zhTBW_{AB})(\partial_r \zhTBW_{CD}) + 2\pi r    T_{rr} \,,
         \end{equation}
\begin{eqnarray}
          &&  \partial_r \left[r^4 e^{-2\beta}\zhTBW_{AB}(\partial_r U^B)\right]
             =   2r^4\partial_r \Big(\frac{1}{r^2}\spaceD_A\beta  \Big)
                 \nonumber \\ &&\qquad
                 -r^2\zhTBW^{EF} \spaceD_E (\partial_r \zhTBW_{AF})
                  +16\pi r^2    T_{rA}
                  \,,
                            \label{eq:UA_eq}
           \end{eqnarray}
as well as \eqref{eq:V_eqn} below.
Here $\spaceD_A$ is the covariant derivative
of the 2-metric $\zhTBW_{AB}$.

\subsubsection{$h_{ur}$}
 \label{ss29VII20.1}

Since the right-hand side of \eqref{eq:beta_eq} is quadratic in $\partial_r\gamma_{AB}$, assuming a vacuum spacetime everywhere, after linearising we find
\begin{equation}\label{28XI19.4aa}
 \partial_r \delta \beta =0  \quad
   \Longleftrightarrow
   \quad
   \delta \beta = \delta \beta (u,x^A)
  \,.
\end{equation}
Hence we can use the $\TSxi$-gauge transformation \eqref{4XII19.t2} to obtain
\begin{equation}\label{28XI19.4a}
   \delta \beta \equiv 0
 \quad
 \Longleftrightarrow
 \quad
  \delta g_{ur } \equiv 0
  \,.
\end{equation}

\subsubsection{$h_{uA}$}
 \label{ss29VII20.2}

The linearisation of \eqref{eq:UA_eq} at the de Sitter metric gives now, in vacuum,
\begin{eqnarray}
          &&  \partial_r \left[r^4  \partial_r(r^{-2} \delta  g_{uA})\right] =
                    r^2
                    \zspaceD_E
                     \left ( \zzhTBW ^{EF}\partial_r \left(r^{-2}\delta g_{AF}\right)\right)
                 \,.
                            \label{28XI19.5}
           \end{eqnarray}
%
%
Let $\psi_A$ denote the right-hand side of the last equation,
\begin{eqnarray}
         &
                 \psi_A:=  r^2
                    \zspaceD_E
                     \left ( \zzhTBW ^{EF}\partial_r \left(r^{-2}\delta g_{AF}\right)\right)
                 \,.
                  &
                            \label{28XI19.7}
           \end{eqnarray}
\ptcheck{18IV20 up to here, with TS }
Equation~\eqref{18IV20.1} gives, for large $r$,
\ptcheck{21IV20  with TS }
\begin{equation}\label{18IV20.2}
  \psi_A =
   -   \zspaceD^B  \oone \zhTB_{AB}
   -2  r^{-1}\zspaceD^B  \otwo  \zhTB_{AB}
   -3  r^{-2} \zspaceD^B  \othree \zhTB_{AB}
   + {\TSblue o(r^{-2}) }
   \,.
\end{equation}
while for small $r$ we have, from \eqref{24IX20.23},
 \ptcheck{1IX20 with TS, and on 30IX for new asymptotics }
\begin{equation}\label{18IV20.2a}
  \psi_A =
   \zspaceD^B (
	 2     \zspaceD_{A}\zspaceD_{B}   \TSxi^{\TSu}
		 - \TSzlap \TSxi^{\TSu}
   \ringh_{AB} )
   +  O(r   )
   \,.
\end{equation}
Here $\TSxi$ is the gauge field of \eqref{3XII19.t1}, cf.\ \eqref{4XII19.12}.

As $\psi_A$ tends to a non-zero covector field as $r$ goes to infinity in general, it turns out to be convenient to write the general solution of \eqref{28XI19.5} as
	\begin{eqnarray}
		r^{-2} \delta g_{u A}
		& =  &
 \Ichi_A(u,x^B)
 +  \frac{\Ipsi_A(u,x^B)}{r^3 }
+
		\int_{1}^{r}\left[\frac{1}{\rho^{4}} \int_{0}^{\rho} \psi_{A}(s) d s\right] d \rho
 \nonumber
\\
		& =  &
 \Ichi_A(u,x^B)
  + \frac{\Ipsi_A(u,x^B)}{r^3 }
+
		\underbrace{
 \int_{1}^{\infty}\left[\frac{1}{\rho^{4}} \int_{0}^{\rho} \psi_{A}(s) d s\right] d \rho
 }_{=:\mathring \Ichi_A}
 \nonumber
\\
&&
  -\int_{r}^{\infty} \left[\frac{1}{\rho^{4}} \int_{0}^{\rho} \psi_{A}(s) d s\right] d \rho
  \,,
  \label{21IV20.1}
	\end{eqnarray}
with  fields $\Ichi_A$ and $\Ipsi_A$ depending upon the arguments indicated.
Here $\psi_A(s)$ stands for $\psi_A(u,s,x^A)$. It follows from \eqref{4XII19.t3}  that the requirement, that $\delta g$ is obtained by an infinitesimal coordinate transformation from a metric perturbation which is smooth near the vertex of the cone, enforces boundedness of $r^{-2} \delta g_{u A}$.
This, together with \eqref{18IV20.2a} shows that
\begin{equation}\label{14VI18.7}
\Ipsi_A \equiv 0
\,.
\end{equation}
The field $\Ichi_A$ has a gauge character and can be determined by imposing convenient conditions at infinity, as follows from the results in  Section~\ref{ss5XI20.3}.

 We have for large $\rho$
\begin{eqnarray}
  \int_{0}^{\rho} \psi_{A}(s) d s
  & =  &
     -   \zspaceD^B  \oone \zhTB_{AB} \,  \rho
   -2 \zspaceD^B  \otwo  \zhTB_{AB} \,   \ln \rho
    -3  \IntPsi_A
   \nonumber
\\
 &&
   +3  \rho^{-1} \zspaceD^B  \othree \zhTB_{AB}
   + {\TSblue o(\rho^{-1}) }
  \,,
  \label{21IV20.2}
	\end{eqnarray}
where
\begin{equation}\label{21IV20.1000}
  \IntPsi_A:=  - \frac 13 \lim_{r\to\infty}
  \bigg(
   \int_{0}^{r} \psi_{A}(s) d s
 +     \zspaceD^B  \oone \zhTB_{AB} \, r
   +2  \zspaceD^B  \otwo  \zhTB_{AB}\, \ln r
   \bigg)
   \,.
\end{equation}
This leads to the following expansion, for large $r$,
\begin{eqnarray}
-\int_{r}^{\infty} \left[\frac{1}{\rho^{4}} \int_{0}^{\rho} \psi_{A}(s) d s\right] d \rho
 & = &    \frac 12  \zspaceD^B  \oone \zhTB_{AB} \, r^{-2}
  + \frac 2 9 \zspaceD^B  \otwo  \zhTB_{AB} \,  (3  \ln r + 1) r^{-3}
   \nonumber
\\
 &&
   +  \IntPsi_A  r^{-3}
   - \frac 34  \zspaceD^B  \othree \zhTB_{AB} r^{-4}
   + {  o(r^{-4}) }
  \,,
  \phantom{xxxxx}
  \label{21IV20.6}
	\end{eqnarray}
resulting in%
\footnote{We take this opportunity to note that there is a misprint in the power of the log term in Equation~(4.33) in~\cite{ChIfsits}; the correct power is $r^{-3}$, as here. There are also some powers of $\nu^0$  missing in this equation, which is innocuous since $\nu^0$ equals 1 in situations of interest.}
 \ptcheck{2V, confirmed by JH }
	\begin{eqnarray}
		{\zhTB_{uA}} & := &
 r^{-2} \delta g_{u A}
 =
 \Ichi_A  + \mathring \Ichi_A
  + \frac 12  \zspaceD^B  \oone \zhTB_{AB} \, r^{-2}
   \nonumber
\\
 &&
 +  \big(
 \mathring \Ipsi_A   + \frac 2 9 \zspaceD^B  \otwo  \zhTB_{AB} \,  (3  \ln r + 1)
   \big)
    r^{-3}
   \nonumber
\\
 &&
   - \frac 34  \zspaceD^B  \othree \zhTB_{AB} \, r^{-4}
   + {  o(r^{-4}) }
  \,.
  \label{21IV20.13}
	\end{eqnarray}
\ptcrr{lots of stuff moved to EnergydeSitter}

\subsubsection{$\partial_u h_{AB}$}
 \label{ss29VII20.3}

 We continue with an analysis of  the asymptotics of $\partial_u h_{AB}$, which can be determined from \cite[Equation~(32)]{MaedlerWinicour}. Denoting
the traceless symmetric part of a tensor on the sphere by
$$
 TS[\cdot]
 \,,
$$
we have
\begin{eqnarray}\label{eq:ev_eqn}
     \lefteqn{
     TS\Big[
      r\partial_r [r  (\partial_u \zhTBW_{AB})]
     	 - \frac{1}{2}  \partial_r[ rV  (\partial_r \zhTBW_{AB})]
          -2e^{\beta} \spaceD_A \spaceD_B e^\beta
          }
          &&
           \nonumber
\\
         && + \zhTBW_{CA} \spaceD_B[ \partial_r (r^2U^C) ]
          - \frac{1}{2} r^4 e^{-2\beta}\zhTBW_{AC}\zhTBW_{BD} (\partial_r U^C) (\partial_r U^D)
          \nonumber \\
       &&
       +
               \frac{r^2}{2}  (\partial_r \zhTBW_{AB}) (\spaceD_C U^C )
              +r^2 U^C \spaceD_C (\partial_r \zhTBW_{AB})
                \nonumber \\
       &&
        -
	r^2 (\partial_r \zhTBW_{AC}) \zhTBW_{BE} (\spaceD^C U^E -\spaceD^E U^C)
       -8\pi e^{2\beta}T_{AB}
       \Big]
        =0.
\end{eqnarray}
Linearising around the de Sitter background one obtains
\begin{equation}\label{30XI19.12}
       r\partial_r [r  (\partial_u  \zhTB_{AB})]
     	 + \frac{\epsilon}{2}  \partial_r[ r^2 \lapseTB^2  (\partial_r \zhTB_{AB})]
        - TS
         \big[ \zspaceD_A\big( \partial_r (r^2\zhTB_{uB}) \big)\big]
       = 8\pi TS[
       \delta T_{AB}
       ]
        \,.
\end{equation}
%
Integrating in vacuum, we find for large $r$
\ptcheck{30IX20 for new asymptotics at small r}
\begin{eqnarray}
 \partial_u  \zhTB_{AB} (r,\cdot)
 & = &
 - \frac 1 r \int_{0}^r \frac{1}{s}
 \Big(
     	   \frac{\epsilon}{2}  \partial_r[r^2 \lapseTB^2  (\partial_r  \zhTB_{AB})]
         -  TS
         \big[ \zspaceD_A\big( \partial_r (r^2\zhTB_{uB}) \big)\big]
         \Big)(s,\cdot) ds
          \nonumber
\\
 &= &
   \partial_u  \ozero \zhTB_{AB}(\cdot) +
    \frac {  \partial_u  \oone \zhTB_{AB}(\cdot) }{r } +
    \frac {  \partial_u  \othree \zhTB_{AB}(\cdot) }{r^3}
    + o(r^{-3})
        \,,
        \label{30XI19.13}
\end{eqnarray}
where
 \ptcheck{15 IV 20; confirmed by JH 2V20 \\ -- \\ ptc:  corrected 6VI20; Last equation   added  by TS 9 II 2020}
\begin{eqnarray}\label{30XI19.14}
  \partial_u  \ozero \zhTB_{AB}(\cdot)
    & = &
    \alpha^2 \oone \zhTB_{AB} +
     \big( \zspaceD_A   { \ozero \zhTB_{uB}}  +  \zspaceD_B   { \ozero \zhTB_{uA}} - \zspaceD^C \ozero \zhTB_{u C} \zh_{AB}
     \big)
         \,,
\\
  \partial_u  \oone \zhTB_{AB}(\cdot)
    & = &
    -
   \lim_{R\to\infty}\bigg(  \int_{0}^{R}\frac{1}{s}
 \Big(
     	   \frac{\epsilon}{2}  \partial_r[ r^2\lapseTB^2  (\partial_r  \zhTB_{AB})]
         -  TS
         \big[ \zspaceD_A\big( \partial_r (r^2\zhTB_{uB}) \big)\big]
         \Big)(s,\cdot) ds
         \nonumber
\\
 &&
 + \partial_u  \ozero \zhTB_{AB}(\cdot) R \bigg)
         \,,
 \label{8XII19.2}
 \\
 \partial_u  \othree \zhTB_{AB}(\cdot)
 & = &
     \alpha^2 \ofour \zhTB_{AB} +
 \big( \zspaceD_A   { \othree \zhTB_{uB}}  +  \zspaceD_B   { \othree \zhTB_{uA}} - \zspaceD^C \othree \zhTB_{u C} \zh_{AB}
 \big)
 \,,
\end{eqnarray}
and note that the limit in \eqref{8XII19.2} exists and is finite under the current conditions.
Also note that \eqref{30XI19.13} has no $r^{-2}$ terms, which shows that the expansion coefficients $\otwo \zhTB_{AB}$  are constants of motion.

The choice of asymptotic gauge $  \ozero \zhTB_{AB} \equiv 0$ implies   $\partial_u  \ozero \zhTB_{AB}\equiv 0$, equivalently
\begin{equation}\label{4II20.7}
  \zspaceD_A   { \ozero \zhTB_{uB}}  +  \zspaceD_B   { \ozero \zhTB_{uA}} -  \zspaceD_C \ozero \zhTB_u{}^C \zh_{AB}
  = - \alpha^2 \oone \zhTB_{AB}
  \,.
\end{equation}
%



\subsubsection{$h_{uu}$}
 \label{ss29VII20.4}

The $V$ function occurring in the Bondi form of the metric solves the equation~\cite{MaedlerWinicour}
\ptcheck{1VII20 some calculations done by JH on this, and checked with maple by TS on 3VIII20}
 \begin{eqnarray}
            &&
                2 e^{-2\beta} \partial_r V
                =
                \mathscr{R}
                -2\zhTBW^{AB}  \Big[\spaceD_A \spaceD_B \beta
                + (\spaceD_A\beta) (\spaceD_B \beta)\Big]\nonumber\\
                &&\qquad
               +\frac{e^{-2\beta}}{r^2 }\spaceD_A \Big[ \partial_r (r^4U^A)\Big]
                -\frac{1}{2}r^4 e^{-4\beta}\zhTBW_{AB}(\partial_r U^A)(\partial_r U^B)
                \nonumber\\&&\qquad
                  - 8\pi  \Big[ \zhTBW^{AB}T_{AB}-r^2 T^\alpha{}_\alpha\Big]
                   -   {2 \Lambda r^2}\,,
                   \label{eq:V_eqn}
           \end{eqnarray}
where $\mathscr{R}$ is the Ricci scalar of the conformal  two-metric $\zhTBW_{AB}$.
Assuming $\delta \beta \equiv 0$,
the linearised version of \eqref{eq:V_eqn} reads
 \ptcheck{20VII20, the leading term in $h_{uA}$ at the rhs with \cite{ChMS}}
              \begin{equation}\label{30XI19.11}
                2   \partial_r \delta V
                =
                \delta \mathscr{R}
              - \frac{1}{r^2 }\zspaceD^A \Big[ \partial_r (r^4\zhTB_{uA})\Big]
                  + 8\pi \delta \Big[ \zhTBW^{AB}T_{AB}-r^2 T\ud{a}{a}\Big]
                   \,.
           \end{equation}

Let $\mathring R_{AB}$ denote the Ricci tensor of the metric $\ringh_{AB}$. As $h_{AB}$ is $\ringh$-traceless we have
 \ptcheck{31VIII20, with two books}
\begin{eqnarray}
\nonumber 
 r^2\delta \mathscr{R}
  & = &
   -\zspaceD^A\zspaceD_A( \ringh^{BC }h_{BC})+\zspaceD^A\zspaceD^B h_{AB}-\mathring R^{AB}h_{AB}
\\
  & =  &
  \zspaceD^A\zspaceD^B h_{AB}
  \,.
\end{eqnarray}
This leads to, in vacuum,
              \begin{equation}\label{2VI20.1}
                2   \partial_r \delta V
                =
                \zspaceD^A
                 \big(
                  \zspaceD^B \zhTB_{AB}
              -\frac{1}{r^2 } \big[ \partial_r (r^4\zhTB_{uA })\big]
               \big)
               \,.
           \end{equation}
It follows from \eqref{21IV20.13} that the $r^{-1}$ terms in the large-$r$ expansion of the right-hand side of \eqref{2VI20.1} cancel out, and that for large $r$ we have
\begin{equation}\label{10VII20.1}
  2   \partial_r \delta V = \zspaceD^A
                 \big( -4  r
              \ozero \zhTB_{uA }
              -  r^{-2} \othree \zhTB_{uA }
        		 + o (r^{-2})
               \big)
               \,.
\end{equation}
Hence
\ptcheck{30IX20  with the new asymptotics at small r, see also comments in one of the paragraphs below}
\begin{eqnarray}
 \nonumber
  \delta V
   &  = &
   \frac 12  \zspaceD^A
   \big(\int_0^r
                 \big(
                  \zspaceD^B \zhTB_{AB}
              -\frac{1}{r^2 } \big[ \partial_r (r^4\zhTB_{uA })\big]
               \big)
                \big |_{r=\rho} d\rho
                \big)
\\
 & = &
 	-\zspaceD^{A}
 \big(
  r^2  \ozero \zhTB_{uA }
  -
   \nu_A
 	-
  \frac{1}{2 r} \othree \zhTB_{uA }
 	+ o(1/r)
 	\big)
  \phantom{xxx}
 \label{9VII20.3}
\end{eqnarray}
as $r$ goes to infinity,
where we used that $\delta V|_{r=0}=0$ vanishes when transforming to Bondi coordinates a field which was originally smooth near the tip of the light cone;
compare \eqref{4XII19.t1}.
Here $\nu_A= \nu_A(u,x^B)$ equals
\begin{equation}
  \nu_A
  =
   \lim_{r\to\infty}
  \big( \frac 12  \int_0^r
                \big( \zspaceD^B
                 \zhTB_{AB}
              -\frac{1}{r^2 } \big[ \partial_r (r^4\zhTB_{uA })\big]
                \big)\big |_{r=\rho} d\rho
+
  r^2  \ozero \zhTB_{uA }
  \big)
              \,.
 \label{9VII20.3a}
\end{equation}

We note that the individual terms in the integrands of \eqref{9VII20.3} and \eqref{9VII20.3a} have, for small $r$, potentially dangerous $1/r$ terms  which could lead to a logarithmic divergence of the integral near $r=0$. However, these terms have to cancel out for fields obtained by a coordinate transformation from metric perturbations which are smooth near the vertex. This can be seen by a direct calculation from \eqref{4XII19.t3} and \eqref{20II20.1}. A simpler argument is to notice that a pure gauge field satisfies the linearised field equations, and that there are no logarithmic terms in the pure gauge fields \eqref{4XII19.t1}-\eqref{20II20.1}.
\ptcrr{Stuff MovedToConsistencyCheck}

\subsubsection{The remaining Einstein equations}
 \label{ss29VII20.5}

The remaining Einstein equations are irrelevant for our analysis in this paper, in the following sense (see~\cite{MaedlerWinicour}):

\begin{enumerate}
  \item The $uu$ part of the Einstein equations is an equation involving $\partial_u V$, which did not occur in the equations above and therefore cannot put further constraints on the expansion coefficients so far.
      \item Similarly for the $uA$ part of the Einstein equations, which involves $\partial_u U^A$.
  \item The trace part of the angular part of the Einstein equations is automatically satisfied in Bondi coordinates once the remaining equations are satisfied.

\end{enumerate}

We present detailed derivations for completeness.


The equation
$\mcE_{u u}=0$ reads
 \ptcheck{13 VIII 20  with TS whole  Maple file with Lambda =0 and 14VIII with Lambda }
\jhbr{rechecked, added 6XI20, precise version of $\mcE_{uu}$}
\begin{eqnarray}
\lefteqn{
    \frac{1}{ r^2}\bigg[2\Big(
      \partial_{u}
    +(\alpha^2 r^2-1) \partial_{r}
    -\frac{1}{r}
    \Big) \zspaceD^{A} {h}_{u A}
    -  \zspaceD^{A} \zspaceD_{A} {h}_{u u}
}
& & \nonumber \\
&&
-(\alpha^2 r^2-1) \bigg(\frac{\zspaceD^{A} \zspaceD^{B} {h}_{A B}}{r^2}\bigg)
-2 r \partial_{u} {h}_{u u}
-2 (\alpha^2 r^2-1) \partial_{r}(r {h}_{u u})
\bigg]
=0
\, . \phantom{xxxxxx}
\label{13VIII20.t3n}
\end{eqnarray}
We insert the asymptotic expansion of $h_{\mu\nu}$ into \eqref{13VIII20.t3n}, obtaining
\begin{eqnarray}
 4\alpha^{2}r\big(\zspaceD^{A}\omtwo h_{uA}-\omone h_{uu}\big) + 2\partial_{u}\big(\zspaceD^{A}\omtwo h_{uA}-\omone h_{uu}\big)
 & &
 \nonumber
 \\
 +\frac{1}{r} \big(
    4 \omone h_{uu}- 6 \zspaceD^{A}\omtwo h_{u A}-\TSzlap \omone h_{uu} - \alpha^2 D^{A} D^{B} \omone h_{A B}
 \big)
 & &
 \nonumber
 \\
 \frac{2}{r^2} \big(
    \partial_{u} \zspaceD^{A} \ozero h_{u A} -\alpha^2 \zspaceD^{A} \oone h_{u A} - \partial_{u} \oone h_{u u}+ \alpha^2 \otwo h_{u u}
 \big)
 &=& o \Big(\frac{1}{r^2} \Big) \, .
 \phantom{xxxx}
 \label{10XI20.t1a}
\end{eqnarray}
Hence
\begin{eqnarray}
  &
    \alpha^{2} \big(\zspaceD^{A}\omtwo h_{uA}-\omone h_{uu}\big)
    = 0
    \,,
    &
    \label{11XI20.1}
\\
     &
     \partial_{u}\big(\zspaceD^{A}\omtwo h_{uA}-\omone h_{uu}\big)
      = 0
      \,,
 &
    \label{11XI20.2}
 \\
 &
    4 \omone h_{uu}- 6 \zspaceD^{A}\omtwo h_{u A}-\TSzlap \omone h_{uu} - \alpha^2 D^{A} D^{B} \omone h_{A B}
  = 0
  \,,
    \label{11XI20.3}
 &
 \\
 &
    \partial_{u} \zspaceD^{A} \ozero h_{u A} -\alpha^2 \zspaceD^{A} \oone h_{u A} - \partial_{u} \oone h_{u u}+ \alpha^2 \otwo h_{u u}= 0
     \, , &
 \label{10XI20.t1}
\end{eqnarray}
and note that \eqref{11XI20.1} implies \eqref{11XI20.2} only if $\alpha \ne 0$; however,  with our boundary conditions, the latter is trivially satisfied when $\alpha$ vanishes.
Now,  \eqref{9VII20.3} gives
\begin{equation}
   \otwo h_{u u} = - \frac 12  \zspaceD^{A} \oone h_{u A} \, ,
 \label{10XI20.t2}
\end{equation}
while from \eqref{21IV20.13} we obtain
\begin{equation}
\zspaceD^{A} \ozero h_{u A}=\frac12 \zspaceD^{A} \zspaceD^{B} \omone h_{A B}
 \,.
\label{10XI20.t3}
\end{equation}
Substituting \eqref{10XI20.t2} and \eqref{10XI20.t3} into  \eqref{10XI20.t1} we are led to
\begin{equation}
\partial_{u} \oone h_{u u}= \frac12 \partial_{u}  \zspaceD^{A} \zspaceD^{B} \omone h_{A B} +  3\alpha^2 \otwo h_{u u}
\label{10XI20.t4}
\end{equation}

Suppose, first, that $\alpha \ne 0$.
Inserting \eqref{11XI20.1} into  \eqref{11XI20.3} gives
\begin{equation}
\zspaceD^{A} \zspaceD^{B} \omone h_{A B}=- \frac{1}{\alpha^2} \big(\TSzlap \omone h_{uu} +2 \omone h_{uu}\big)
\label{10XI20.t5}
\end{equation}
Combining \eqref{10XI20.t4} and \eqref{10XI20.t5} leads to
\begin{eqnarray}
 \displaystyle
   \partial_{u} {\oone h}_{u u}
  & = &
   - \frac1 { 2 \alpha^2 }
(\TSzlap + 2) \partial_u \omone h_{uu}
+
  3  \alpha^2  {\otwo h}_{uu}
 \nonumber
\\
  & = &
    \frac 12    \zspaceD^{A} \zspaceD^{B}
    \partial_{u} \omone h_{AB}
+
  3  \alpha^2  {\otwo h}_{uu}
 \,.
\label{7XI20.2a}
\end{eqnarray}

Similar manipulations show that \eqref{7XI20.2a} remains valid for $\alpha=0$:
 \ptcrr{
    ts:checked}
\begin{eqnarray}
 \displaystyle
   \partial_{u} {\oone h}_{u u}
  & = &
    \frac 12    \zspaceD^{A} \zspaceD^{B}
    \partial_{u} {\omone h}_{AB}
 \,,
\label{7XI20.2b}
\end{eqnarray}
which is the linearisation of the usual equation for the time-evolution of the mass aspect function (cf.,\ e.g.~\cite[Equation~(5.102)]{CJK}).

 \bigskip


We continue with an analysis of the equations $\mcE_{u A}=0$, which read
\begin{eqnarray}
0
& = &
\frac{1}{ r^2} \Bigg[
\zspaceD^{B}\zspaceD_{A}{h}_{uB}
-\zspaceD^{B}\zspaceD_{B}{h}_{uA}
+\partial_{u} \zspaceD^{B}{h}_{A B}
\nonumber
\\
&&
\blue{-}r^{2}\left(
\left(1-r^{2}\alpha^{2}\right)\partial_{r}^{2}h_{uA}
+ 2\alpha^{2}h_{uA}-r^{2}\partial_{r}\partial_{u}\left(\frac{h_{uA}}{r^{2}}\right)
+ \partial_{r}\zspaceD_{A}{h}_{uu}
\right)\Bigg]
\,.
\nonumber
\\
&&
\label{9XI20.t1}
\end{eqnarray}
An asymptotic expansion of the above equation takes the form
\begin{eqnarray}
    \mcE_{uA}
&=&
    \zspaceD^{B}\zspaceD_{A} \omtwo h_{uB}
    -\zspaceD^{B}\zspaceD_{B} \omtwo h_{uA}
    - 2 \omtwo h_{uA}
    -2\alpha^{2} \ozero h_{uA}
    -\zspaceD_{A}\omone h_{uu}
\nonumber
\\
& &
+ \frac{1}{r}\partial_{u}\big(\zspaceD^{B} \omone h_{AB} -2 \ozero h_{uA} \big)
\nonumber
\\
& &
    +\frac{1}{r^2} \big(
         \zspaceD^{B} \zspaceD_{A} \ozero h_{u B}
         -\zspaceD^{B} \zspaceD_{B} \ozero h_{u A}
         - 3 \partial_{u} \oone h_{u A}
         +4 \alpha^2 \otwo h_{u A}
         +\zspaceD_{A} \oone h_{u u}
    \big)
\nonumber
\\
& &
    + o\left(r^{-2}\right)\,.
\end{eqnarray}
Hence
\begin{eqnarray}
&
        \zspaceD^{B}\zspaceD_{A} \omtwo h_{uB}
    -\zspaceD^{B}\zspaceD_{B} \omtwo h_{uA}
    - 2 \omtwo h_{uA}
    -2\alpha^{2} \ozero h_{uA}
    -\zspaceD_{A}\omone h_{uu}
    =
    0
    \, ,
    \phantom{xx}
    \label{11Xi20.t1}
&
\\
&
    \partial_{u}\big(\zspaceD^{B} \omone h_{AB} -2 \ozero h_{uA} \big)
    =
    0
    \, ,
    \label{11Xi20.t2}
&
\\
&
    \zspaceD^{B} \zspaceD_{A} \ozero h_{u B}
    -\zspaceD^{B} \zspaceD_{B} \ozero h_{u A}
    - 3 \partial_{u} \oone h_{u A}
    +4 \alpha^2 \otwo h_{u A}
    +\zspaceD_{A} \oone h_{u u}
    =
    0
    \, .
    \phantom{xxxxx}
    \label{11Xi20.t3}
&
\end{eqnarray}
Using the identity
\begin{equation}
    \zspaceD^{B}\zspaceD_{A} \omtwo h_{uB}=\zspaceD_{A}\zspaceD^{B} \omtwo h_{uB} +\omtwo h_{uA}
     \, ,
    \label{11XI20.t3a}
\end{equation}
\eqref{11Xi20.t1} becomes
\begin{equation}
    -\big(
        \TSzlap
        + 1
    \big) \omtwo h_{uA}
    +\zspaceD_{A} \zspaceD^{B} \omtwo h_{uB}
    -2\alpha^{2} \ozero h_{uA}
    -\zspaceD_{A}\omone h_{uu}
    =
    0
     \,.
    \label{11Xi20.t4}
\end{equation}
For $\alpha \neq 0$, substituting \eqref{11XI20.1} we obtain
\begin{equation}
      -\big(
  \TSzlap
  + 1
  \big) \omtwo h_{uA}
  -2\alpha^{2} \ozero h_{uA}
  =
  0
  \,.
\end{equation}
 Regardless of whether $\alpha$ vanishes or not, we similarly substitute \eqref{11XI20.t3a} into
\eqref{11Xi20.t3} to obtain an evolution equation for $\partial_{u} \oone h_{u A}$
\begin{equation}
  3 \partial_{u} \oone h_{u A}
     =
      \zspaceD_{A} \zspaceD^{B}  \ozero h_{u B}
    -\big(
        \TSzlap
        +
        1
    \big)\ozero h_{u A}
    +2 \ozero h_{u A}
+4 \alpha^2 \otwo h_{u A}
+\zspaceD_{A} \oone h_{u u}
 \,.
 \label{14XI20.2}
\end{equation}
Recall that $\ozero h_{uA}= \zspaceD^B \omone h_{AB}/2$ and $\otwo h_{uA}= -3/4 \zspaceD^B \oone h_{AB}$, both  by \eqref{21IV20.13}, which allows us to rewrite \eqref{14XI20.2} as
\tscomr{Agrees with JH's \eqref{14XI20.1}.}
 \begin{eqnarray}
 3 \partial_{u} \oone h_{u A}
 &=&
\frac{1}{2} \zspaceD_{A} \zspaceD^{B}  \zspaceD^{C} \omone h_{CB}
+ \frac{1}{2}\zspaceD^B \omone h_{AB}
-\frac{1}{2} \TSzlap \zspaceD^B \omone h_{AB}
\phantom{xxx}
\nonumber
\\
& &
 -3 \alpha^2 \zspaceD^B \oone h_{AB}
 +\zspaceD_{A} \oone h_{u u}
 \,.
 \end{eqnarray}
The asymptotically flat case is obtained by setting $\alpha=0$, compare~\cite[Equation~(5.103)]{CJK}.

\ptcrr{MM's formulae for crosschecking in the file expansions.tex, commented out}

An alternative analysis of the equations $\mcE_{uu}=0$ and $\mcE_{uA}=0$ can be found in Appendix~\ref{A12XI20.1}.

\ptcrr{Jahanur's analysis of the remaining equations went to JahanurRemainingEquations.tex}
\subsection{An example: the Blanchet-Damour solutions}
 \label{s11IX20}

An interesting class of linearised solutions of the linearised Einstein equations with $\Lambda =0$ has been introduced in~\cite{BlanchetDamour}. We use these  solutions to provide an explicit example of the behaviour both at the origin and at infinity of a vacuum metric perturbation in Bondi coordinates. In particular we calculate the news tensor for the Blanchet-Damour metrics, see \eqref{22IX20.6}.

In~\cite{BlanchetDamour}  the solutions are presented in harmonic coordinates, where the trace-reversed tensor
$$
\bar h_{\mu\nu} := h_{\mu\nu} - \frac{1}{2 } \eta^{\alpha\beta} h_{\alpha\beta} \eta_{\mu \nu}
$$
satisfies
\begin{equation}\label{11IX20.p1}
  \Box_\eta \bar h_{\mu\nu} = 0
  \,,
  \quad
   \partial_\mu \bar h^{\mu\nu} = 0
   \,.
\end{equation}
Here $\eta$ is the Minkowski metric, taken to be $-(dx^0)^2+(dx^1)^2+(dx^2)^2+(dx^3)^2$ in the coordinates of \eqref{11IX20.p1}, and $\Box_\eta$ the associated wave operator.

Given a collection of smooth functions $I_{ij}:\R\to \R$ such that $I_{ij}=I_{ji}$, the tensor field
\mmr{checked on 19IX20}
\begin{eqnarray}
 \bar h_{tt} &=& \partial_i \partial_j
   \big(
    \frac{I_{ij}(t-r) - I_{ij}(t+r)}{r}
    \big)
    \nonumber
\\ &=&
   \big(
   \ddot I_{ij}(t-r) -  \ddot I_{ij}(t+r)
    \big)
     \frac{x^i x^j }{r^3}
    + O(r^{-2})
    \,,
       \label{11IX20.p2a}
\\
  \bar h_{ti} &=&  \partial_j
   \big(
     \frac{\dot I_{ij}(t-r) - \dot I_{ij}(t+r)}{r}
      \big)
    \nonumber
\\
 &=&
   -\big(
   \ddot I_{ij}(t-r) +  \ddot I_{ij}(t+r)
    \big)
     \frac{x^j }{r^2}
    + O(r^{-2})
    \,,
\\
 \bar  h_{ij} &=&   \frac{\ddot I_{ij}(t-r) - \ddot I_{ij}(t+r)}{r}
  \,,
   \label{11IX20.p2}
\end{eqnarray}
where each dot represents a derivative with respect to the argument of $I_{ij}$,
is a smooth tensor field on Minkowski spacetime solving \eqref{11IX20.p1}.

Since the operators appearing in \eqref{11IX20.p1} commute with partial differentiation, further solutions can be constructed by applying $\partial_{\mu_1}\cdots \partial_{\mu_\ell}$ to $\bar h_{\mu\nu}$, and by applying Poincar\'e transformations.

Transforming to the cone-adapted coordinates $(u=t-r,r,x^A)$ one has
\begin{equation}\label{11X20.p6}
  \partial_u = \partial_t
  \,,
  \quad
  \partial_r = \partial_t + \frac{x^i}{r}\partial_i
  \,,
  \quad
  \partial_A = \frac{\partial x^i}{\partial x^A} \partial_i\
  \,.
\end{equation}
It follows that
\mmr{checked on 19IX20, $h{ti}$, $h_{uu}$, trace, $h_{rr}$, }
\ptcheck{19IX20}
\tscomr{\purple Checked $h_{uu}, {h^{\mu}}_{\mu}$ below up to $I_{ij}(u)$ terms and its derivatives.}
\begin{eqnarray}
   h_{ti} &=& \bar h_{ti}  =  \partial_j
   \big(
     \frac{\dot I_{ij}(t-r) - \dot I_{ij}(t+r)}{r}
      \big)
    \nonumber
\\
 &=&
   -\big(\frac{
   \ddot I_{ij}(u) +  \ddot I_{ij}(u+2r)}{r}
   +
    \frac{  \dot I_{ij}(u) - \dot I_{ij}(u+2r)}{ r^2}
    \big)
    \frac{  x^j}{r}
    \,,
    \nonumber
    \\
  \bar h_{tt}
   &= & \mbox{``$\partial_i h_{ti}$  where in $h_{ti}$ the functions $\dot I_{ij}$ have been replaced by $I_{ij}$''}
   %
     \nonumber
\\
 \nonumber
&   = &
    \frac{\big(\ddot I_{ij}(u) - \ddot I_{ij}(u+2r)\big)}{ r}
       \frac{x^i x^j}{r^2}
     +
    \frac{\dot I_{ij}(u)+ \dot I_{ij}(u+2r)}{r^2}
    \big(3 \frac{x^i x^j}{r^2} - \delta^{ij}
   \big)
\\
 \nonumber
&   &
      +
    \frac{  I_{ij}(u) -  I_{ij}(u+2r)}{ r^3}
    \big(3 \frac{x^i x^j}{r^2} - \delta^{ij}
   \big)
   \,,
\\
  h_{uu}
   &= &
    h_{tt} = \bar h_{tt} + \frac 12 \bar h^{\mu}{}_\mu = \frac 12
    \big(
      \bar h_{tt} +\bar h^{i}{}_i
      \big)
    \nonumber
\\
 \nonumber
&   = &
    \frac{ \ddot I_{ij}(u) - \ddot I_{ij}(u+2r) }{2 r}
      \big( \delta^{ij} + \frac{x^i x^j}{r^2}
       \big)
     +
    \frac{\dot I_{ij}(u)+ \dot I_{ij}(u+2r)}{2r^2}
    \big(3 \frac{x^i x^j}{r^2} - \delta^{ij}
   \big)
\\
 \nonumber
&   &
      +
    \frac{  I_{ij}(u) -  I_{ij}(u+2r)}{2 r^3}
    \big(3 \frac{x^i x^j}{r^2} - \delta^{ij}
   \big)
       \,,
\\
  \bar h^\mu {}_{\mu }
   & = &
    - \bar h_{tt} +  \frac{\ddot I_{ii}(t-r) - \ddot I_{ii}(t+r)}{r}
     \nonumber
\\
   & = &
    \frac{\ddot I_{ij}(u) - \ddot I_{ij}(u+2r)}{r}
     \big(
      \delta^{ij} - \frac{x^i x^j}{r^2}
      \big)
     \nonumber
\\
   & &
           -
    \frac{\dot I_{ij}(u)+ \dot I_{ij}(u+2r)}{r^2}
    \big(3 \frac{x^i x^j}{r^2} - \delta^{ij}
   \big)
 \nonumber
\\
&   &
      -
    \frac{  I_{ij}(u) -  I_{ij}(u+2r)}{ r^3}
    \big(3 \frac{x^i x^j}{r^2} - \delta^{ij}
   \big)
       \,,
       \nonumber
\\
  h_{rr}
   &= &
   \bar h_{rr}
   =
    \bar h_{tt} + 2 \bar h_{ti}\frac{x^i}{r}
    +  \bar h_{ij}\frac{x^i x^j}{r^2}
   \nonumber
\\
&   = &
   -
    \frac{4\ddot I_{ij}(u+2r)}{ r}  \frac{x^i x^j}{r^2}
      \nonumber
\\
 &&
    -\frac{\dot I_{ij}(u) }{  r^2}
     \big(
      \delta^{ij} -  \frac{x^i x^j}{r^2}
      \big)
       -
    \frac{\dot I_{ij}(u+2r)}{  r^2}
     \big(
       \delta^{ij} -  5 \frac{x^i x^j}{r^2}
      \big)
      \nonumber
\\
 \nonumber
&   &
      +
    \frac{  I_{ij}(u) -  I_{ij}(u+2r)}{  r^3}
    \big(3 \frac{x^i x^j}{r^2} -  \delta^{ij}
   \big)
   \,,
       \nonumber
\\
  h_{rA}
   &= &
   \bar h_{rA}
    =
   \big(
   \bar h_{tj}
   +
   \bar h_{ij} \frac{x^i}{r}
   \big)
    \frac{\partial x^j}{\partial x^A}
     \nonumber
\\
 & = & \Big(\partial_j
   \big(
     \frac{\dot I_{ij}(t-r) - \dot I_{ij}(t+r)}{r}\big)
      + \frac{\ddot I_{ij}(t-r) - \ddot I_{ij}(t+r)}{r}
     \frac{x^j }{r}
      \Big) \frac{\partial x^i}{\partial x^A}
      \nonumber
\\
&   = &
    - \big(
      \frac{2\ddot I_{ij}(u+2r) }r
      + \frac{\dot I_{ij}(u) - \dot I_{ij}(u+2r)}{r^2}
      \big)
     \frac{x^j }{r}\frac{\partial x^i}{\partial x^A}
       \,,
   \label{11IX20.p4}
\end{eqnarray}
 \ptcheck{20X20: TS: Checked $h_{rr},h_{r A}$ below up to $I_{ij}(u)$ terms and its derivatives\\ -- \\ $h_{rr}$ also by MM}
\begin{eqnarray}
  h_{AB}
   &= &
   \bar h_{AB}  - \frac 12 \bar h^{\mu}{}_\mu  r^2 \ringh_{AB}
    =
   \bar h_{ij}
    \frac{\partial x^i}{\partial x^A}
    \frac{\partial x^j}{\partial x^B}- \frac 12 \bar h^{\mu}{}_\mu  r^2 \ringh_{AB}
    \nonumber
\\
 & = &
   \big(
      \frac{\ddot I_{ij}(u) - \ddot I_{ij}(u+2r)}{r}
      \big)
    \frac{\partial x^i}{\partial x^A}
    \frac{\partial x^j}{\partial x^B}
    \nonumber
\\
 &&
- \frac 12
    \bigg(
    \frac{\ddot I_{ij}(u) - \ddot I_{ij}(u+2r)}{r}
     \big(
      \delta^{ij} - \frac{x^i x^j}{r^2}
      \big)
     \nonumber
\\
   & &
           -
    \frac{\dot I_{ij}(u)+ \dot I_{ij}(u+2r)}{r^2}
    \big(3 \frac{x^i x^j}{r^2} - \delta^{ij}
   \big)
 \nonumber
\\
&   &
      -
    \frac{  I_{ij}(u) -  I_{ij}(u+2r)}{ r^3}
    \big(3 \frac{x^i x^j}{r^2} - \delta^{ij}
   \big)\bigg)
       r^2  \ringh_{AB}
     \nonumber
\\
   & = &
    \big(
      \frac{\ddot I_{ij}(u) - \ddot I_{ij}(u+2r)}{r}
      \big)
    \frac{\partial x^i}{\partial x^A}
    \frac{\partial x^j}{\partial x^B}
      \nonumber
\\
 &&
    -
    \frac{\ddot I_{ij}(u) - \ddot I_{ij}(u+2r)}{2r}
     \big(
      \delta^{ij} - \frac{x^i x^j}{r^2}
      \big)
        r^2 \ringh_{AB}
       + O(r^{-2})
      \,.
 \label{15X20.1}
\end{eqnarray}
Finally
(with all expansions in the last equation and below for large $r$),
  \tscomr{\purple Checked $h_{ur}, h_{u A}$ below up to $I_{ij}(u)$ terms and its derivatives. I have lower expansion terms also.}
  \jhbr{all perturbations checked on 23IX20}
\begin{eqnarray}
  h_{ur}
   &= &
    \bar h_{ur} + \frac 12 \bar h^{\mu}{}_\mu =
    \bar h_{tt}  + \bar h_{ti} \frac{x^i}{r}+ \frac 12 \bar h^{\mu}{}_\mu
    \nonumber
\\
   & = &
     \partial_i \partial_j
   \big(
    \frac{I_{ij}(t-r) - I_{ij}(t+r)}{r}
    \big)
    + \partial_j
   \big(
     \frac{\dot I_{ij}(t-r) - \dot I_{ij}(t+r)}{r}
      \big) \frac{x^i}{r}
    \nonumber
\\
&&
-\partial_i \partial_j
   \big(
    \frac{I_{ij}(t-r) - I_{ij}(t+r)}{2r}
    \big) +  \frac{\ddot I_{ii}(t-r) - \ddot I_{ii}(t+r)}{2r}
     \nonumber
\\
&   = &
    \frac{\ddot I_{ij}(u) - \ddot I_{ij}(u+2r)}{2 r}
     \big(
      \delta^{ij} + \frac{x^i x^j}{r^2}
      \big)
      -\frac{\ddot I_{ij}(u) + \ddot I_{ij}(u+2r)}{r} \frac{x^i x^j}{r^2}
       + O(r^{-2})
       \,,
       \nonumber
\\
  h_{uA}
   &= &
    \bar h_{uA}   =
 \bar h_{ti} \frac{\partial x^i}{\partial x^A}
    \nonumber
\\
   & = & \partial_j
   \big(
     \frac{\dot I_{ij}(t-r) - \dot I_{ij}(t+r)}{r}
      \big)  \frac{\partial x^i}{\partial x^A}
    \nonumber
\\
&   = &
      -\frac{\ddot I_{ij}(u) + \ddot I_{ij}(u+2r)} r \frac{\partial x^i}{\partial x^A}  \frac{x^j}{r}
       + O(r^{-2})
        \,.
        \label{11IX20.p5}
\end{eqnarray}
%
%


To continue, for simplicity we assume that all the  $I_{ij}$'s vanish for sufficiently large arguments, and that $r$ is so large that all the $I_{ij}(u+2r)$'s vanish.
The vector field $\TSxi$ which brings the metric perturbation to the Bondi form is (cf.\ Section~\ref{s8XII19.1})
%
%
\begin{eqnarray}
\TSxip^{\TSu} &=&\TSxi^{\TSu}(\TSu,x^A)
    +\frac{\dot I_{ij}(u) }{2 r}
     \big(
      \delta^{ij} -  \frac{x^i x^j}{r^2}
      \big)
      \nonumber
\\
&   &
      -
    \frac{  I_{ij}(u)
     }{4 r^2}
    \big(3\frac{x^i x^j}{r^2} -  \delta^{ij}
   \big)
       \,,
  \label{18IX20.6}
\\
 \TSxip^{B}  &=&
   \TSxi^{B} (\TSu,x^A)
  +
  \zspaceD^{B}
  \bigg[
     -\frac{1}{\TSr}   \TSxi^{\TSu}(\TSu,x^A)
      +
    \frac{   I_{ij}(u)  }{4r^3}
      \frac{x^i x^j}{r^2}
   \bigg]
   \, ,
  \phantom{xxxx}
  \\
r \zspaceD_B \TSxip^{B}  &=&
   r \zspaceD_B \TSxi^{B} (\TSu,x^A)
 -
    \Delta_{\ringh} \TSxi^{\TSu}(\TSu,x^A)
     +
    \frac{   I_{ij}(u)  }{4 r^2}
\Delta_{\ringh}
        \bigg(
      \frac{x^i x^j}{r^2}
      \bigg)
   \, ,
  \phantom{xxxx}
\\
    \TSxip^{\TSr}
	&=&
 -\frac 1	2 \TSr  \zspaceD_{B} \TSxip^{B}
     -  \frac 14 \TSr^{-1} \ringh^{AB} h_{AB}
     \nonumber
\\
	&=&
 -\frac 1	2 \TSr  \zspaceD_{B} \TSxip^{B}
     -  \frac {r h_{ij}}{4}
     \bigg(
      \delta^{ij} - \frac{x^ix^j}{r^2}
     \bigg)
     \nonumber
\\
	&=&
 -\frac 1	2 \TSr  \zspaceD_{B} \TSxip^{B}
     -  \frac {\ddot I_{ij}(u)}{4}
     \bigg(
      \delta^{ij} - \frac{x^ix^j}{r^2}
     \bigg) + \frac{r\bar h ^\mu {}_\mu}4
     \nonumber
\\
	&=&
 -\frac 1	2 \TSr  \zspaceD_{B} \xi^{B}
 + \frac 12 \Delta_\ringh \xi^u
           -
    \frac{\dot I_{ij}(u) }{4r}
    \big(3 \frac{x^i x^j}{r^2} - \delta^{ij}
   \big)
     \,.
 \label{16X20.5}
\end{eqnarray}

To obtain the mass aspect function one needs the gauge-transformed $h_{uu}$
%
\begin{eqnarray}
 h_{uu} &\to&  h_{uu}+ \mcL_\zeta \eta_{uu}
  =h_{uu}- 2 \partial_{\TSu}
 \big( \zeta^{\TSu} +\zeta^r
 \big)
 \,.
\label{11IX20.p11}
\end{eqnarray}
After applying a $u$-derivative to \eqref{16X20.5},
  we will obtain a solution for which $h_{uu}$ tends to zero as $r$ tends to infinity in Bondi gauge if and only if
\begin{equation}\label{19IX20.61}
   (\Delta_\ringh +2)\partial_u \xi^u =  0
  \,,
  \quad
  \zspaceD_A\partial_u \xi^A = 0
  \,,
\end{equation}

In the calculations that follow the identities  \eqref{18IX20.50}  are useful;  we repeat them here for the convenience of the reader:
\begin{eqnarray}\label{18IX20.50q}
 &
\mathring{\gamma}^{AB}\frac{\partial x^{i}}{\partial x^{A}}\frac{\partial x^{j}}{\partial x^{B}}
=
(\delta^{ij} - \frac{x^{i}x^{j}}{r^{2}})r^{2}
 \,,
 \qquad
 \zspaceD^A(x^i \zspaceD_A x^i) = 0
  \,,
 &
  \label{21IX20.2b}
\\
 &
 \displaystyle
 \zspaceD^C
  \bigg(
     \frac{x^{(j} }{r^2}\frac{\partial x^{i)}}{\partial x^C}
     \bigg) =
      \delta^{ij} - \frac{3 x^i x^j}{r^2}\,,
      \quad
      \Delta_\ringh \frac{x^i x^j}{r^2}= 2 \delta^{ij} -6
       \frac{x^i x^j}{r^2}
      \,.
     &
  \label{21IX20.3}
\end{eqnarray}
%
One then finds that  in the linearised Bondi gauge the original field $h_{uu}$ becomes
\begin{eqnarray}
\nonumber
 h_{uu} &\to&
    h_{uu}
    - \frac{ \ddot I_{ij}(u) }{2 r}
     \big(3 \delta^{ij} - 5\frac{x^i x^j}{r^2}
     \big)
      +O(r^{-2})
 \nonumber
\\       &=&
    \frac{ \ddot I_{ij}(u) }{ r}
     \big(
     \frac{3 x^i x^j}{r^2} - \delta^{ij}
     \big)
      +O(r^{-2})
     \,.
\label{11IX20.p110}
\end{eqnarray}
Let $\delta \mu$ denote the linearised mass aspect function, thus $h_{uu}= 2 \delta \mu/r + O(r^{-2})$. We see that
\ptcheck{21IX20 by MM and TS, corrected 22IX20}
\jhbr{checked 24IX20}
\begin{eqnarray}
  \delta \mu   &=&  \frac{\ddot I_{ij}(u) }2
     \big(
     \frac{3 x^i x^j}{r^2} - \delta^{ij}
     \big)
     \,.
\label{21IX20.1}
\end{eqnarray}

The function $\chi$ of \eqref{13VIII20.t1} can be calculated by inspecting the asymptotic behaviour, for large $r$, of the functions appearing there. One finds
\begin{eqnarray}
 \chi & = & 4 \delta \mu
  -
   \frac{1}{2}  (\TSzlap +2)\TSzlap\TSxi^{\TSu}
   \nonumber
\\
    &= &
     2\ddot I_{ij}(u)
     \big(
     \frac{3 x^i x^j}{r^2} - \delta^{ij}
     \big)
  -
   \frac{1}{2}  (\TSzlap +2)\TSzlap\TSxi^{\TSu}
     \,.
\label{22IX20.1}
\end{eqnarray}
It then follows from \eqref{14VIII20.3} that
\begin{equation}\label{22IX20.2}
 \zspaceD^A {h}_{uA}
  \big|_{r=0} =  2 \ddot I_{ij}(u)
     \big(
     \frac{3 x^i x^j}{r^2} - \delta^{ij}
     \big)
     -
   \frac{1}{2}  (\TSzlap +2)\TSzlap\TSxi^{\TSu}
  \,.
\end{equation}

This equation looks surprising at first sight, since the left-hand side is zero for a smooth tensor field in the $(u,r,x^A)$ coordinates. However, \eqref{22IX20.2} makes it clear that the linearised Bondi gauge introduces a singular behaviour of the gauge-transformed metric perturbation at $r=0$. This singularity can be removed on some chosen light cone, say $\mcC_{u_0}$, but cannot be removed for all $u$ by a residual gauge transformation (in other words, asymptotic symmetry) in general.

Recall the formula for the gauge-transformed $h_{AB}$:
\begin{eqnarray}
h_{AB} &\to&
 h_{AB}+  2 r \zeta^r \ringh_{AB}
	+ \TSr^2 \TSoLie_{\zeta} \ringh_{AB}
  \,,
 \label{20II20.p1asdf}
\end{eqnarray}
where $\TSoLie_{\zeta}$ denotes the two-dimensional Lie derivative with respect to the field $\zeta^A\partial_A$. This leads to
\begin{eqnarray}
h_{AB} &\to&
    \big(
      \frac{\ddot I_{ij}(u) - \ddot I_{ij}(u+2r)}{r}
      \big)
    \frac{\partial x^i}{\partial x^A}
    \frac{\partial x^j}{\partial x^B}
    \nonumber
\\
 &&
    - \ringh^{CD} \big(
      \frac{\ddot I_{ij}(u) - \ddot I_{ij}(u+2r)}{2r}
      \big)
    \frac{\partial x^i}{\partial x^C}
    \frac{\partial x^j}{\partial x^D}
    \ringh_{AB}
       + O(r^{-3})
  \,,
  \phantom{xx}
 \label{22IX20.5}
\end{eqnarray}
from which the news tensor $\partial_u \omone h_{AB}$ is readily obtained:
\begin{eqnarray}
\partial_u \omone h_{AB} =
       \dddot I_{ij}(u)
       \left(
    \frac{\partial x^i}{\partial x^A}
    \frac{\partial x^j}{\partial x^B}  - \frac 12 \ringh^{CD}
    \frac{\partial x^i}{\partial x^C}
    \frac{\partial x^j}{\partial x^D}
    \ringh_{AB}
    \right)
  \,.
 \label{22IX20.6}
\end{eqnarray}

\section{The energy of weak gravitational fields
\\
{\small\em by PTC, JH, TS}
}
\label{C26X20.4}

\subsection{Trautman-Bondi mass}
 \label{s9VII20.2}

When $\Lambda=0$, the Trautman-Bondi mass is determined from the function $V$ appearing in the metric. Based on the variational arguments so far, one expects that its linearised-theory equivalent will be determined by the second variation of $V$. Our aim in this section is to derive a formula for this second variation.

Recall that the one-parameter families of vacuum metrics $g_{\mu\nu}(\lambda, \cdot)$ which define the variations considered here satisfy
%
\begin{equation}\label{6VI20.1}
  e^{2\beta} |_{\lambda=0}= 1   \,,
   \quad  V |_{\lambda=0}= r - \frac{\Lambda r^3}{3}  \,,
   \quad
    \partial_r V |_{\lambda=0} = 
    1 -{\Lambda}r^2  \,.
\end{equation}
Using \eqref{6VI20.1} and our remaining asymptotic conditions, the second variation equation is obtained by differentiating \eqref{eq:V_eqn} twice with respect to $\lambda$ and  reads (cf.\ \eqref{19VIII20.2} below)
\ptcheck{30IX20 again and again 26II21 }
              \begin{eqnarray}
            \lefteqn{
                  2   \partial_r \delta^2 V
                   { + }
                    4 ( \Lambda r^2 -1) \delta^2 \beta
                =
                \delta^2 \mathscr{R}
               - \frac{1}{r^2 }\zspaceD^A \Big[ \partial_r (r^4\delta \zhTB_{u A})\Big]
                }
                &&
                \nonumber
                \\
                 && -2 \ringh^{AB} \zspaceD_{A} \zspaceD_{B} \delta^2 \beta
                - r^4 \ringh^{AB}(\partial_r \zhTB_{uA})(\partial_r \zhTB_{uB})
                 - 8\pi \delta^2 \Big[ \zhTBW^{AB}T_{AB}-r^2 T\ud{a}{a}\Big]
                 \nonumber
                 \\
                 & &
                  +  \frac{2}{r^2}\zspaceD^A \Big[ \partial_r (r^4 \ringh^{BD} \zhTB_{A B}  \zhTB_{u D})\Big]
                   \,.
                   \label{2VI20.2}
           \end{eqnarray}
We need the second variation of \eqref{eq:beta_eq}, which reads
%
\begin{equation}
         \label{eq:beta_eq2}
          \partial_r  \delta^2\beta  =   \frac{r}{8}\ringh^{AC}\ringh^{BD} \partial_r \zhTB_{AB}\partial_r \zhTB_{CD} + 2\pi r   \delta^2  T_{rr} \,,
         \end{equation}
and which can be explicitly integrated  after {requiring} that $\delta^2\beta$ {goes to zero as $r$ tends to infinity}:
\begin{equation}
         \label{eq:beta_eq3}
          \delta^2\beta  = - \int_ r^\infty
           \big( \frac{r}{8}\ringh^{AC}\ringh^{BD} \partial_r \zhTB_{AB}\partial_r \zhTB_{CD} + 2\pi r   \delta^2  T_{rr}
            \big) dr
              \,.
         \end{equation}
(We note that this is negative if we impose the dominant energy condition.)
In vacuum and for large $r$ we obtain the expansion
\begin{equation}\label{28VIII20.1}
  \delta^2 \beta = -
          \frac{1}{16r^2 }\ringh^{AC}\ringh^{BD}
            \oone \zhTB _{AB} \big(  \oone   \zhTB _{CD}
           + \frac{3}{ r^2} \othree   \zhTB _{CD}
           \big )
          + O (r^{-5})
           \,,
\end{equation}
while for small $r$ we have, using \eqref{24IX20.23},
\begin{equation}\label{28VIII20.1123}
  \delta^2 \beta =
          -\frac{1}{8 r^2 }
	 \zspaceD^A\zspaceD^B \xi^u
  ( 2
	 \zspaceD_A\zspaceD_B \xi^u -\Delta_\ringh \xi^u \ringh_{AB}
	)
          + O (r^{-1})
           \,.
\end{equation}
We see that $\delta^2\beta$ diverges badly at the origin unless $\xi^u$ is a linear combination of $\ell=0$ and $\ell=1$ spherical harmonics. Now, as explained in Section~\ref{C26X20.2}, the field $\xi^u$ is determined by the linearised metric up to the choice of a $u$-independent initial datum. Hence, given a light cone we can always find a gauge so that the leading-order singularity above vanishes. Equation \eqref{24IX20.23} shows that in this gauge the integrand in \eqref{eq:beta_eq3} is bounded and $\delta^2 \beta$ is finite everywhere.

In the remainder of this section we assume a vacuum metric perturbation, and a gauge so that $\delta^2\beta$ is bounded on the light cone under consideration.

We rewrite \eqref{2VI20.2} as
%
              \begin{eqnarray}
            \lefteqn{
                  2   \partial_r \delta^2 V
                =
                \delta^2 \mathscr{R}
               - \frac{1}{r^2 }\zspaceD_A \Big[ \partial_r (r^4\delta \zhTB_u{}^A)\Big] - {\TSgreen 4}( \Lambda r^2 -1) \delta^2 \beta
                }
                &&
                \nonumber
                \\
                 &&- 2 \ringh^{AB} \zspaceD_{A} \zspaceD_{B} \delta^2 \beta
                - r^4 \ringh^{AB}\partial_r \zhTB_{uA} \partial_r \zhTB_{uB}
                  +   \frac{2}{r^2}\zspaceD^A \Big[
                     \partial_r (r^4 \ringh^{BD} \zhTB_{A B}  \zhTB_{u D})
                     \Big]
                   \,.
                   \nonumber
\\
                   &&
                   \label{2VI20.2a}
           \end{eqnarray}

Recall that for $\Lambda=0$ the Trautman-Bondi mass $m_{\TB}$ is defined as~\cite{BBM,T,Tlectures}
\begin{equation}\label{23II21.4}
  m_\TB = -\frac{1}{8\pi}\int_{S^2} \ozero V d\mu_{\ringh}
  \,,
\end{equation}
where $\ozero V$ is the $r$-independent coefficient in an asymptotic expansion of $V$. It was proposed in~\cite{ChIfsits} to  use this definition for $\Lambda\ne 0$, which was motivated by Cauchy-problem considerations.

Before proceeding further, recall that a one-parameter family of metrics $\lambda\mapsto g_{\mu\nu}(\lambda)$ in Bondi gauge satisfies the condition
\begin{equation}\label{19VIII20.1}
  \sqrt{\det g_{AB}(\lambda)} = \sqrt{\det g_{AB}(0)}
  \,.
\end{equation}
This implies in particular
\begin{equation}\label{19VIII20.2}
  \frac{d\sqrt{\det g_{AB}(\lambda)}}{d\lambda} =  0=  \frac{d^2 \sqrt{\det g_{AB}(\lambda)}}{d\lambda^2}
  \,.
\end{equation}
Equivalently,
\begin{equation}\label{19VIII20.3}
  \delta  \sqrt{\det g_{AB} } =   0= \delta^2  \sqrt{\det g_{AB} }
  \,.
\end{equation}
It then follows from the Gauss-Bonnet theorem that
\begin{equation}\label{2VI20.5}
  \delta \int_{S^2} \mathscr{R} \, d\mu_\gamma = 0
   = \delta^2 \int_{S^2} \mathscr{R} \, d\mu_\gamma
   \,,
\end{equation}
Integrating \eqref{2VI20.2a} over a truncated cone
$$
 \mcC_{u,R} = \mcC_u \cap \{0\le r \le R\}
$$
one obtains
\begin{eqnarray}
\lefteqn{
               \delta^2 m_\TB(\mcC_{u,R})
                : =
                  -\frac{1}{8\pi}\int_{r=R} \delta^2  V d\mu_{\ringh}
                  }
                  &&
 \nonumber
\\
 & = &                  \frac{1}{16 \pi } \int_{\mcC_{u,R}}
              \left(
               r^4  \ringh^{AB}\partial_r \zhTB_{uA} \partial_r \zhTB_{uB}
               - {\TSgreen 4}(1- \Lambda r^2 ) \delta^2 \beta
               \right)
                 \, dr \, d\mu_{\ringh}
                   \,.
                    \phantom{xxx}
                   \label{2VI20.4}
\end{eqnarray}
When $\Lambda \le 0$ we have $\delta^2\beta \le 0$,  which proves positivity of $\delta^2 m_\TB(\mcC_{u,R})$, and of its limit $\delta^2 m_\TB(\mcC_{u })$ with $R\to\infty$.

The notation in \eqref{2VI20.4} is somewhat misleading, as the limit as $R$ goes to infinity of \eqref{2VI20.4} will only reproduce the Trautman-Bondi mass if this limit converges.
 This is the case when $\Lambda\le 0$ (compare \eqref{26X20.2}, Section~\ref{Cs26X20.1}). However, when $\Lambda >0$ the integrand for large $r$ behaves as
              \begin{equation}\label{28VIII20.31}
                - \frac{\Lambda r}4
         \ringh^{AC}\ringh^{BD}
            \oone \zhTB _{AB}  \oone   \zhTB _{CD}
          + O (r^{-1})
                   \,,
           \end{equation}
so that
\begin{eqnarray}
                   \label{23II21.11}
               \lefteqn{
                 \delta^2 m_\TB(\mcC_{u})
             :=
                   -\frac{1}{8\pi}\int_{S^2} \delta^2  \ozero V d\mu_{\ringh}
                  }
                  &&
\\
 & =
           \displaystyle \lim_{R\to\infty}
          \frac{1}{16 \pi } \Big(
           &
           \int_{\mcC_{u,R}}
           \displaystyle
              \left(
               r^4  \ringh^{AB}\partial_r \zhTB_{uA} \partial_r \zhTB_{uB}
               - {\TSgreen 4}(1- \Lambda r^2 ) \delta^2 \beta
               \right)
                 \, dr \, d\mu_{\ringh}
                 \nonumber
\\
 &&
                +
                 \frac{\Lambda R}4 \int_{S_R}
         \ringh^{AC}\ringh^{BD}
            \oone \zhTB _{AB}  \oone   \zhTB _{CD}
              \,
                d\mu_{\ringh}
            \
            \Big)
                   \,.
                   \label{28VIII20.32}
           \end{eqnarray}
\subsection{Boundary integrand and $\Lambda$}
\label{s8VII20.1}

We assume now that the background takes the form
\begin{equation}
\nobarg_{\a \b} dx^\a dx^\b = \epsilon N^2 du^2-2du \, dr
+ r^2
\underbrace{(d\theta^2+\sin^2 \theta d\phi^2)}_{=:\zzhTBW }
\,,
\end{equation}
which is compatible both with the Minkowski and the de Sitter metrics.
We suppose that the metric perturbations $\delta g_{\mu\nu}$ satisfy the Bondi gauge conditions:
\begin{equation}\label{8VII20.13}
 \delta g_{rr}=0 =  \delta g_{rA}=
 g^{AB}\delta g_{AB}
 \,.
\end{equation}
This implies
\begin{equation}
\nabla_{ \mu}\delta g_{r r}=0 =  \nabla_{u}\delta g_{r A} =
\nabla_{r}\delta g_{r A}
\,,
\label{13I20.t1}
\end{equation}
and
\ptcheck{31VII20 on the maple files with TS}
\begin{eqnarray*}
	&
	\displaystyle
	\nabla_{u} \delta g_{u u} = \partial_{u}(\delta g_{u u})
 -2 \epsilon \lapseTB^{'} \lapseTB \left(\delta g_{u u}+ \epsilon\lapseTB^2 \delta g_{u r}  \right) \,,
	\quad
	\nabla_{u}\delta g_{u r} =\partial_{u}(\delta g_{u r})  \,,
	&
	\\
	&
	\nabla_{u}\delta g_{AB} =\partial_{u}(\delta g_{A B})\,,
	\qquad
	\nabla_{u}\delta g_{u A} =\partial_{u}(\delta g_{u A})-  \epsilon \lapseTB^{'} \lapseTB \delta g_{u A}  \,,
	&
	\\
	&
	\displaystyle
	\nabla_{r} \delta g_{u u} =
	\partial_{r}(\delta g_{u u}) + 2  \epsilon \lapseTB^{'} \lapseTB \delta g_{u r}
	\,, \quad
	\nabla_{r} \delta g_{u r} = \partial_{r} ( \delta g_{u r})
	\,,
	&
	\\
	&
	\nabla_{r}\delta g_{AB} =\partial_{r}(\delta g_{AB})-\frac{2}{r}\delta g_{AB}\,,
	\qquad
	\nabla_{r}\delta g_{u A} =\partial_{r}(\delta g_{u A})- \frac{1}{r} \delta g_{u A}  \,,
	&
	\\
	&
	\displaystyle
	\nabla_A \delta g_{u u} =
	\partial_A (\delta g_{u u})
	\,, \quad
	\nabla_A \delta g_{u r} = \partial_{A} (\delta g_{u r})-\frac{1}{r} \delta g_{u A}
	&
	\\
	&
	\nabla_{A}\delta g_{u B} =\zspaceD_{A}(\delta g_{u B})-
	\frac{1}{r} g_{AB} \left(\delta g_{u u}+  \epsilon\lapseTB^2 \delta g_{u r}\right) \,,
	&
	\\
	&
	\nabla_{A}\delta g_{BC} = \zspaceD _{A} ( \delta g_{BC} ) -\frac{2}{r} g_{A(B} \delta g_{C) u}
	\,,
	\quad
	\nabla_{A}\delta g_{r B} =-\frac{1}{r}\left( \delta g_{AB}+ g_{AB} \delta g_{u r} \right)
	\,. &
\end{eqnarray*}
Under the above assumptions, the functional $b^r(\delta_1 g,\delta_2 g)$ given by \eqref{1IX20.40}, Appendix~\ref{s15VII20.t1} below  becomes
%
%
\begin{eqnarray}
\nonumber 
\lefteqn{b^{\TSr} (\delta_1 g,\delta_2 g)
	=-\frac{1}{r^2}\delta_1g_{\TSu \TSr} \zzhTBW^{A B} \nabla_{B}(\delta_2g_{\TSu B}
+\epsilon \lapseTB^2 \delta_2g_{\TSr B})}
\\
&&
\nonumber
+ \frac{1}{2r^4} \zzhTBW^{A C} \zzhTBW^{B D}\delta_1g_{C D}
\big( \epsilon\lapseTB^2 \nabla_{\TSr}\delta_2g_{AB}
+
\nabla_{\TSu}\delta_2g_{AB}
\\
&&
\nonumber
-2\nabla_A(\epsilon\lapseTB^2 \delta_2g_{\TSr B}
+
 \delta_2g_{\TSu B}
)
\big)
\\
&&
+  \frac{1}{r^2} \zzhTBW^{AC}\delta_1 g_{\TSu C} \left( \nabla_{\TSr} \delta_2 g_{\TSu A}
+   \frac1{2r^2} \zzhTBW^{B D} \nabla_{A} \delta_2 g_{B D}
\right)
\,.
\phantom{xxx}
\label{13I20.t2}
\end{eqnarray}
\ptcheck{29IV20 together in TS's Maple file  and again 31VII20 with Maple and by hand, N depends only upon $r$, angular metric spherical, $g_{ur}=-1$}
After some further algebra one obtains
\begin{eqnarray}
\nonumber 
\lefteqn{b^{\TSr} (\delta_1 g,\delta_2 g)
	=\frac{1}{r}\delta_1g_{\TSu \TSr}
	 \left[
		2(1+2 \epsilon\lapseTB^2) \delta_2g_{\TSu \TSr}
		-
		\frac{1}{r} \zzhTBW^{C D} \zspaceD_C\delta_2 g_{u D}
	\right]}
\\
&&
\nonumber
+  \frac{1}{2r^4} \zzhTBW^{A C} \zzhTBW^{B D}\delta_1g_{C D}
\big( \epsilon \lapseTB^2 \partial_{\TSr}\delta_2g_{AB}
+
\partial_{\TSu}\delta_2g_{AB}
-
2 \zspaceD_A\delta_2 g_{u B}
\big)
\\
&&
+\frac{1}{r^2} \delta_{1}g_{\TSu A} \zzhTBW^{A B}
	\left(
	\partial_{\TSr}\delta_{2}g_{\TSu B}
	-
	\frac{2}{r}\delta_{2}g_{\TSu B}
	\right) \, .
	\phantom{xxx}
	\label{29IV20.t1}
\end{eqnarray}
Upon antisymmetrisation the first and last terms above drop out:
 \ptcheck{8VII20 with TS}
\begin{eqnarray}
\nonumber 
 \lefteqn{
   \noomega^{\TSr} (\delta_1 g,\delta_2 g)
  \equiv
b^{\TSr} (\delta_2 g,\delta_1 g)-
b^{\TSr} (\delta_1 g,\delta_2 g)
}
\\
 & = &
	\frac{1}{r^2} \zzhTBW^{C D}
\left(\delta_1g_{\TSu \TSr}
		  \zspaceD_C\delta_2 g_{u D}
-
	 \delta_2g_{\TSu \TSr}
		   \zspaceD_C\delta_1 g_{u D}
\right)
 \nonumber
\\
&&
\nonumber
+  \frac{1}{2r^4} \zzhTBW^{A C} \zzhTBW^{B D}\Big(\delta_2g_{C D}
\big( \epsilon \lapseTB^2 \partial_{\TSr}\delta_1g_{AB}
+
\partial_{\TSu}\delta_1g_{AB}
-
2 \zspaceD_A\delta_1 g_{u B }
 \big)
\\
&&
\nonumber
- \delta_1g_{C D}\big(  \epsilon \lapseTB^2 \partial_{\TSr}\delta_2g_{AB}
+
\partial_{\TSu}\delta_2g_{AB}
-
2 \zspaceD_A\delta_2 g_{u B}
\big)
 \Big)
\\
&&
+\frac{1}{r^2} \zzhTBW^{A B}
	\left(
 \delta_{2}g_{\TSu A}
	\partial_{\TSr}\delta_{1}g_{\TSu B}
	-
 \delta_{1}g_{\TSu A}
	\partial_{\TSr}\delta_{2}g_{\TSu B}
	\right) \, .
	\phantom{xxx}
	\label{29IV20.t1sb}
\end{eqnarray}

\subsubsection{Large $r$}
 \label{ss2XI20.2}
We use the following asymptotics for the linearised metric components, as  derived in Section~\ref{s9VII20}  under the conditions there:
%
\begin{eqnarray}
\delta g_{u r} &\equiv &    0 \ \, \equiv \ \, \delta g_{r A}  \ \, \equiv \ \,\delta g_{rr} \,,
\label{15I20.t1}
\\
\label{15I20.t2}
\delta g_{AB} &=& \delta\gmone_{A B} \rBo
 + o(1)
\,,
\\
\label{15I20.t3}
\delta g_{u A} &=&  \delta\gmtwo_{u A} \rBo^2 + \delta\gzero_{u A} + o(1)
\,,
\\
\delta g_{u u} &=&
\delta\gmone_{u u} r
+\frac{\delta\gone_{u u}}{\rBo }
+ o(\rBo^{-1})
\,,
\label{15I20.t4}
\end{eqnarray}
one finds
 \ptcheck{17IV20. checked with maple together with TS the leading order term, and on 24 IV 20 the zeroth order, wrong indices corrected; decay indices corrected 29IV20, confirmed by JH; rechecked in Minkowski with TS and Maple on 31VII20}
%
\begin{eqnarray}
\lefteqn{
	 r^2 b^{r}(\delta_1 g,\delta_2 g)
	\,
	=
	}
&&
\nonumber
\\
\nonumber
&&
r
\Big[
	  -2 \zzhTBW^{AB} \delta_1 \gmtwo_{u A} \delta_2 \gzero_{u B}
\nonumber
\\
&&
\phantom{+ 	r \Big[}
	+
	\zzhTBW^{AC} \zzhTBW^{BD}
	\Big(
		\frac{\alpha^2}{2   } \delta_1\gmone_{AB}
		\delta_2\gmone_{CD} -    \delta_1\gmone_{AB}
		\zspaceD_C\delta_2 \gmtwo_{u D}
	\Big)
\Big]
\nonumber
\\
&&
	  -3 \zzhTBW^{AB}\delta_1\gmtwo_{u A} \delta_2\gone_{u B}
 +
   \frac{1}{2}
	\zzhTBW^{AC} \zzhTBW^{BD} \delta_1\gmone_{AB} \partial_{u} \delta_2\gmone_{CD}
+ o(1)
\,.
 \phantom{xxx}
\label{15I20.t5}
\end{eqnarray}
\ptcrr{corrected 8VII20; this might propagate? confirmed by JH 22VII20}
Hence
\begin{eqnarray}
\lefteqn{
	 r^2 \omega^{r}(\delta_1 g,\delta_2 g)
	=
 r^2 \big(
  b^{r}(\delta_2 g,\delta_1 g)
 - b^{r}(\delta_1 g,\delta_2 g)
  \big)
	}
&&
\nonumber
\\
\nonumber
&=&
r
\Big[
  2 \zzhTBW^{AB} (  \delta_1 \gmtwo_{u A} \delta_2 \gzero_{u B}
 -  \delta_2 \gmtwo_{u A} \delta_1 \gzero_{u B} )
\\
\nonumber
&& +
	\zzhTBW^{AC} \zzhTBW^{BD}
	\Big(    \delta_1\gmone_{AB}
		\zspaceD_C\delta_2 \gmtwo_{u D}
 -  \delta_2\gmone_{AB}
		\zspaceD_C\delta_1 \gmtwo_{u D}
	\Big)
\Big]
\nonumber
\\
&&
  +3 \zzhTBW^{AB}
 (  \delta_1\gmtwo_{u A} \delta_2\gone_{u B}
 -\delta_2\gmtwo_{u A} \delta_1\gone_{u B}
  )
\nonumber
\\
&&	+
 \frac{1}{2}
	\zzhTBW^{AC} \zzhTBW^{BD}
(
\delta_2\gmone_{AB} \partial_{u} \delta_1\gmone_{CD}
-
\delta_1\gmone_{AB} \partial_{u} \delta_2\gmone_{CD})
+ o(1)
\,.
 \phantom{xxxxxx}
\label{15I20.t5st}
\end{eqnarray}

\ptcrr{an inconclusive calculation commented out}


\subsection{The energy and its flux}
 \label{s8VII20.2}
\ptcheck{14VII20: dangerous  macro  delta , produces now $\zhTB$,  but it might lead to a wrong sign in some previous equations, crosschecked globally on the part which was showing up in the version on that day...}
We return now to a linearised vacuum gravitation field in Bondi gauge,
\begin{eqnarray}
   h
    & \equiv &
      h_{\mu\nu}dx^\mu dx^\nu = h_{uu} du^2 + 2 h_{uA} dx^A du + h_{AB} dx^A dx^ B
 \nonumber
\\
 &     =: &
     r^2\big(
      \zhTB_{uu} du^2 +    \underbrace{\zh_{AB}  \deltaU^B}_{=:\deltaU_A}   dx^ B du +     \zhTB_{AB} dx^A  dx^ B
      \big)
     \,.
     \label{30XI19.21}
\end{eqnarray}
Let, as before, $\mcC_{u,R}$ denote a light cone $\mcC_u$ truncated at radius $R$, and let $E_c[h,\mcC_{u,R}]$ denote the canonical energy contained in $\mcC_{u,R}$, defined using the vector field $\partial_u$:
\begin{equation}\label{30XI19.3}
  E_c[h,\mcC_{u,R}]:= \wmcH [\mcC_{u,R}, \partial_u]
  \,.
\end{equation}

Recall that the Bondi gauge introduces singular behaviour at the tip of the light cone in general. When integrating formulae such as \eqref{6IX18.5} over $\hyp$ one obtains a ``boundary integral at $r=0$''. One can keep track of this but the resulting formulae do not seem to be very enlightening, so in what follows we can, and will, choose a Bondi gauge so that the (freely specifiable) gauge vector field $\xi$ in \eqref{3XII19.t1}, which is part of the  transformation which takes the metric from a smooth gauge to a Bondi gauge, satisfies
\begin{equation}\label{2XI20,1}
	 \zspaceD_A\zspaceD_B \xi^u =
 \frac 12  \Delta_\ringh \xi^u \ringh_{AB}
\end{equation}
at \emph{a fixed light cone $\mcC_u$ under consideration}.
(Compare \eqref{13XI20.p1}-\eqref{26VI20.2b} below and  \eqref{24IX20.23}.)
Equivalently, $\xi^u$ is a linear combination of $\ell=0$ and $\ell=1$ spherical harmonics.

We note that this choice is tied to the chosen light cone $\mcC_u$, and  will \emph{not} be satisfied by nearby light cones in general. This turns out to be irrelevant for the calculations in this section.

We further note that the Bondi gauge is mainly relevant for us for the analysis of the field at large distances. One can sweep under the carpet the problem of the singularity at the origin by using Bondi coordinates for large $r$, and any other coordinates for small $r$. It follows from Section~\ref{ss14IX18.3} that the total energy does not depend upon the choice of the coordinates for small $r$.  The volume integral will then not take the simple form presented here for ``non-Bondi'' values of $r$, but this would again be irrelevant from the point of view of the large-$r$ analysis of the fields.

Putting together our calculations so far, namely \eqref{6IX18.5+asdf}-\eqref{13IX18.1}, \eqref{23IX18.1}, \eqref{23XI19.3}, \eqref{2I20.t3}  and \eqref{15I20.t5} we find:
%
%
\begin{eqnarray}
 \lefteqn{
  E_c[h,\mcC_{u,R}]
 =
     }
     &&
       \nonumber
\\
 \nonumber
 &= &
  \frac{1}{64 \pi}
    \int_{\mcC_{u,R}}   {\nobarg}^{BE } {\nobarg}^{FC}
    \big(
    \partial_u h  _{BC } \partial_{ r } h_{EF }
    -
     h_{BC } \partial_{ r } \partial_u h_{EF }
     \big)\, d\mu_{\mcC}
\\
\nonumber
 &&
       +
         {\frac{R}{64  \pi} }
      \int_{S^2}
      \Big(   4\zh^{AB} \ozero \zhTB_{uA}  \otwo  \zhTB_{uB}
\\
\nonumber
&&
           - \zh^{AB}\zh^{CD} (\alpha^2 \oone \zhTB_{AC}
             \oone \zhTB_{BD}
               -
                  2  \oone \zhTB_{AC}
          \zspaceD_B \ozero \deltaU{}_{D}
          )
    \Big)
     d \mu_{\zzhTBW }
     \nonumber
\\
 &&{\purple
	+
	{\frac{1}{64 \pi} }
	\int_{S^2}
	\Big(
	   6\zh^{AB} \ozero \zhTB_{uA}  \othree  \zhTB_{uB}
	-
    \zh^{AB}\zh^{CD} \oone \zhTB_{AC} \partial_{u}
	\oone \zhTB_{BD}
	\nonumber   }
\\
&&
\phantom{+  {\frac{\alpha R}{64 \pi} }
	\int_{S^2}
	\Big(}
{\purple - \zh^{AB}\zh^{CD}(
	 {\alpha^2} \oone \zhTB_{AC}
	\otwo \zhTB_{BD}
 - 2  \otwo \zhTB_{AC}
	\zspaceD_B \ozero \deltaU{}_{D}
)
	\Big)
	d \mu_{\zzhTBW }}
\nonumber
\\
 &&
 + o(1)
  \,.
  \label{30XI19.t1}
\end{eqnarray}

The associated energy flux formula  follows from \eqref{11IX18.14}  together with  the equations just listed:
 \ptcheck{29IV20. signs of some terms corrected together with TS, equation retypeset}
%
\begin{eqnarray}
\nonumber
\lefteqn{
	\frac{d E_c[h,\mcC_{u,R}]}{du}
	= -\frac{1}{32 \pi} \int_{S_R} r^2 b^r(\partial_uh, h) \, d\mu_{\ringh}
=
}
&& \\
&&{
	{\frac{R}{32 \pi} }
	\int_{S^2}
	\Big( 4
	{  \ringh{}^{AB} \otwo \zhTB_{uA} \partial_{u} \ozero \zhTB_{uB}}
	       }
\nonumber
\\
&&
\phantom{- {\frac{\alpha R}{16 \pi} }
	\int_{S^2}
	\Big(}
{-
	 \zh^{AB}\zh^{CD} (
	{\alpha^2}  \oone \zhTB_{AC} \partial_{u} \oone \zhTB_{BD}
	-  2 \partial_{u} \oone \zhTB_{AC}	\zspaceD_B  \ozero \deltaU{}_{D}
)
	\Big)
	d \mu_{\zzhTBW }}
\nonumber
\\
&&{
	-
	{\frac{1}{32 \pi} }
	\int_{S^2}
	\Big(
	{ 	-  6 \ringh^{AB} \othree \deltaU{}_{A} \partial_{u}\ozero \deltaU{}_{B}}
	+  \zh^{AB}\zh^{CD} \partial_{u} \oone \zhTB_{AC}
	\partial_{u} \oone \zhTB_{BD}
	\nonumber   }
\\
&&
\phantom{+  {\frac{\alpha R}{16 \pi} }
	\int_{S^2}
	\Big(}
{ +
	 \zh^{AB}\zh^{CD}
(
  {\alpha^2} \oone \zhTB_{AC}	 \partial_{u}\otwo \zhTB_{BD}
	-
    2 \partial_{u} \otwo \zhTB_{AC}	\zspaceD_B  \ozero \deltaU{}_D
    )
	\Big)
	d \mu_{\zzhTBW }}
\nonumber
\\
&&
\phantom{+  {\frac{\alpha R}{16 \pi} }
	\int_{S^2}
	\Big(}
+ o(1)
\,.
\label{30XI19.5}
\end{eqnarray}

Let us first analyse the convergence,  as $R$ tends to infinity, of the volume integral in \eqref{30XI19.t1}.

Consider the part of the boundary integral in \eqref{30XI19.t1} which diverges linearly with $R$; up to a multiplicative coefficient $R/(64\pi)$ this term equals
\begin{equation}
      \int_{S^2}
      \Big({\TSgrn  4\zh^{AB} \ozero \zhTB_{uA}  \otwo \zhTB_{uB} }
           -\zh^{AB}\zh^{CD} (\alpha^2 \oone \zhTB_{AC}
             \oone \zhTB_{BD}
              -   2  \oone \zhTB_{AC}
          \zspaceD_B \ozero \deltaU{}_{D}
          )
    \Big)
           \,  d \mu_{\zzhTBW }
  \,.
  \label{30XI19.t1abc}
\end{equation}
Using \eqref{21IV20.13}, the first term in \eqref{30XI19.t1abc} can be rewritten as
 \ptcheck{2V20 by JH}
\begin{eqnarray}
       -  4
      \int_{S^2}
      {\TSgrn \zh^{AB} \ozero \zhTB {}_{uA} \otwo \zhTB{}_{uB}}
           \,  d \mu_{\zzhTBW }
          & = &
         - 2
      \int_{S^2}
      \TSgrn   \zh^{AB} \ozero \zhTB {}_{uA} \zspaceD^C \oone \zhTB_{BC}
           \,  d \mu_{\zzhTBW }
           \nonumber
\\
 &
            =
            &
   2
      \int_{S^2}
      \TSgrn   \zh^{AB} \zspaceD^C \ozero \zhTB {}_{uA}  \oone \zhTB_{BC}
           \,  d \mu_{\zzhTBW }
  \,,
  \label{21IV20.21}
\end{eqnarray}
which cancels-out the last term in  \eqref{30XI19.t1abc}.

Incidentally, from \eqref{4II20.7} and the symmetries of $\oone \zhTB_{AC}$ we obtain
\begin{eqnarray}
\nonumber
    \lefteqn{
        -   2\int_{S^2}
              \zh^{AB}\zh^{CD} \oone \zhTB_{AC}
          \zspaceD_B \ozero \deltaU{}_{D}
           \,  d \mu_{\zzhTBW }
          }
          &&
\\
  &&
    =
    -    \int_{S^2}
              \zh^{AB}\zh^{CD}  \oone \zhTB_{AC}
         ( \zspaceD_B \ozero \zhTB_D  +  \zspaceD_D   { \ozero \zhTB_{uB}} -  \zspaceD_E \ozero \zhTB{}^E \zh_{BD}
         ) d \mu_{\zzhTBW }
         \nonumber
\\
  &&
    =
    \alpha^2     \int_{S^2}
              \zh^{AB}\zh^{CD}  \oone \zhTB_{AC} \oone \zhTB_{BD}
         \,  d \mu_{\zzhTBW }
  \,,
  \label{21IV20.11}
\end{eqnarray}
so all three terms in \eqref{30XI19.t1abc} are equal up to signs. Thus \eqref{30XI19.t1} can be rewritten as
\begin{eqnarray}
 \lefteqn{
  E_c[h,\mcC_{u,R}]
 =
     }
     &&
       \nonumber
\\
 \nonumber
 &= &
  \frac{1}{64 \pi}
    \int_{\mcC_{u,R}}   {\nobarg}^{BE } {\nobarg}^{FC}
    \big(
    \partial_u h  _{BC } \partial_{ r } h_{EF }
    -
     h_{BC } \partial_{ r } \partial_u h_{EF }
     \big)\, d\mu_{\mcC}
\\
\nonumber
   &&
       -   {\frac{\alpha^2 R}{64  \pi} }
      \int_{S^2} \zh^{AB}\zh^{CD}\oone \zhTB_{AC}
             \oone \zhTB_{BD}
             \,
	d\mu_{\ringh}
     \nonumber
\\
 &&
	-
	{\frac{1}{{\TSgrn 64} \pi} }
	\int_{S^2}  \zh^{\TSgrn AB}
	\Big( \zh^{\TSgrn CD} \oone \zhTB_{AC} \partial_{u}
	\oone \zhTB_{BD}
-
	{   6  \ozero \zhTB {}_{uA} \othree \zhTB{}_{uB}}
	\Big)
	d\mu_{\ringh}
\nonumber
\\
&&
+ o(1)
\,.
\label{30VIII20.-1}
\end{eqnarray}

We finish this section by a remark concerning the vanishing of $\ozero \zhTB_{AB}$. If this were not the case,
the insertion of the expansion \eqref{30XI19.13} into the volume integral in \eqref{30XI19.t1} would lead to a divergent leading-order behaviour:
 \ptcheck{30IX20; volume integral convergent at the origin, there is a cancellation; 6I21 the formula rechecked with mathematica and with TS}
%
\begin{eqnarray}
  \lefteqn{   \int_{\mcC_{u,R}}   {\nobarg}^{BE } {\nobarg}^{FC}
    \big(
     h_{BC } \partial_{ r } \partial_u h_{EF }
     -
    \partial_u h_{BC } \partial_{ r } h_{EF }
     \big)\, d\mu_{\mcC}
     }
     &&
     \nonumber
\\
 && =   R  \int_{S^2}  \ringh^{BE } \ringh^{FC} \big(
     \oone \zhTB_{BC } \partial_{u } \ozero \zhTB_{EF }
     -
    \partial_u \oone \zhTB_{BC } \ozero \zhTB_{EF }
     \big) d\mu_{\zzhTBW }
     \nonumber
\\
 && +  2\ln(  R)  \int_{S^2}  \ringh^{BE } \ringh^{FC} \big(
     \otwo \zhTB_{BC } \partial_{u } \ozero \zhTB_{EF }
     -
    \partial_u \otwo \zhTB_{BC } \ozero \zhTB_{EF }
     \big) d\mu_{\zzhTBW }
     \nonumber
\\
 &&
      + O (1)
      \,.
      \label{8XII19.5}
\end{eqnarray}
Further, the boundary term in the energy would diverge as $R^2$. As already pointed out, this provides the justification while it is natural to choose a gauge in which $\ozero \zhTB_{AB}$ vanishes.

\subsubsection{Energy-loss revisited}

In this section we rederive the energy-loss formula by a somewhat more direct calculation.
    For this we integrate over $\mcC_{u,R}$ the identity
\begin{equation}
 \label{26VI20.2a}
  \partial_u \omega^u = -\partial_i \omega^i
\end{equation}
to obtain
\begin{equation}
 \label{26VI20.2b}
 \frac{d}{du} \int_{\mcC_{u,R}}  \omega^u r^2 dr  \,d\mu_\ringh =
 -\int_{S_{u,R}} \omega^r r^2 \,d\mu_\ringh
 + \lim_{\epsilon\to 0}
  \int_{S_{u,\epsilon}} \omega^r r^2 \,d\mu_\ringh
 \,.
\end{equation}
One way to guarantee  the vanishing of the last integral in \eqref{26VI20.2b} is to choose a gauge so that
\begin{equation}\label{13XI20.p1}
   \partial_r h_{AB}|_{r=0} = 0
   \,.
\end{equation}
For simplicity  this gauge choice will be made in the rest of this section.

From \eqref{2I20.t3}
and
\eqref{15I20.t5st}
this is equivalent to
%
\begin{eqnarray}
\lefteqn{
  \frac 12 \frac{d}{du} \int_{\mcC_{u,R}} g^{AB}g^{CD}\left(\delta_2 g_ {AC}\partial_{\TSr} \delta_{1}g_{BD}-\delta_1 g_ {AC}\partial_{\TSr} \delta_{2}g_{BD}\right)
  r^2 dr  \,d\mu_\ringh
  }
 \nonumber
\\
 &=
  \displaystyle
 -\int_{S_{u,R}}\!\!\!\!\!
 &
\Big[
r
\big[
  2 \zzhTBW^{AB} (  \delta_1 \gmtwo_{u A} \delta_2 \gzero_{u B}
 -  \delta_2 \gmtwo_{u A} \delta_1 \gzero_{u B} )
 \nonumber
\\
\nonumber
&& +
	\zzhTBW^{AC} \zzhTBW^{BD}
	\big(    \delta_1\gmone_{AB}
		\zspaceD_C\delta_2 \gmtwo_{u D}
 -  \delta_2\gmone_{AB}
		\zspaceD_C\delta_1 \gmtwo_{u D}
	\big)
\big]
\nonumber
\\
&&
  +3 \zzhTBW^{AB}
 (  \delta_1\gmtwo_{u A} \delta_2\gone_{u B}
 -\delta_2\gmtwo_{u A} \delta_1\gone_{u B}
  )
\nonumber
\\
&&	+
	\zzhTBW^{AC} \zzhTBW^{BD} \big( \frac{1}{2}
(
\delta_2\gmone_{AB} \partial_{u} \delta_1\gmone_{CD}
-
\delta_1\gmone_{AB} \partial_{u} \delta_2\gmone_{CD})
\nonumber
\\
&&
	\phantom{+\zzhTBW^{AC} \zzhTBW^{BD} \Big(}
	 + \frac{\alpha^2}{2   }
(
  \delta_2\gzero_{AB}
		\delta_1\gmone_{CD}
-
  \delta_1\gzero_{AB}
		\delta_2\gmone_{CD}
)
\nonumber
\\
&&
	\phantom{+\zzhTBW^{AC} \zzhTBW^{BD} \Big(}
	+
	\delta_1\gzero_{AB}
	\zspaceD_C\delta_2\gmtwo_{u D}
	-
	\delta_2\gzero_{AB}
	\zspaceD_C\delta_1\gmtwo_{u D}
	\big)
 \Big]
 \,d\mu_\ringh
 \nonumber
\\
 &
 \phantom{=}
+ o(1)
 \,. &
\label{15I20.t5sast}
\end{eqnarray}
Letting $\delta_1g_{\mu\nu}= h_{\mu\nu}$, $\delta_2g_{\mu\nu}= \partial_u h_{\mu\nu}$,  using
$$\ozero \zhTB_{AB}\equiv 0 \equiv \otwo \zhTB_{AB}
$$
(similarly for the $u$-derivatives),
and keeping in mind that $\breh:= r^{-1} h$ and $\zhTB=r^{-2} h$, this becomes
\begin{eqnarray}
\lefteqn{
 \frac 12
 \frac{d}{du} \int_{\mcC_{u,R}} \zzhTBW^{AB}\zzhTBW^{CD}\left(\partial_u \breh_ {AC}\partial_{\TSr} \breh_{BD}-\breh_ {AC}\partial_{\TSr} \partial_u \breh_{BD}\right)
   dr  \,d\mu_\ringh
  }
 \nonumber
\\
 &=
  \displaystyle
 -\int_{S_{u,R}}
\Big[\!\!\!\!\!
 &
 r\big[
  2 \zzhTBW^{AB} (  \ozero \zhTB_{u A} \partial_u \otwo \zhTB_{u B}
 -  \partial_u \ozero \zhTB_{u A}   \otwo \zhTB _{u B} )
 \nonumber
\\
\nonumber
&& +
	\zzhTBW^{AC} \zzhTBW^{BD}
	\big(    \oone  \zhTB_{AB}
		\zspaceD_C\partial_u \ozero \zhTB_{u D}
 -  \partial_u \oone \zhTB_{AB}
		\zspaceD_C\ozero \zhTB_{u D}
	\big)
 \big]
\nonumber
\\
&&
 +3 \zzhTBW^{AB}
 \big(
  \ozero \zhTB_{u A} \partial_u \othree\zhTB_{u B}
 -
  \partial_u \ozero \zhTB_{u A} \othree \zhTB_{u B}
  \big)
\nonumber
\\
&&	+
	 \frac{1}{2}
 \zzhTBW^{AC} \zzhTBW^{BD}
\big(
\partial_u \oone \zhTB_{AB} \partial_{u} \oone  \zhTB_{CD}
-
\oone  \zhTB_{AB} \partial^2_{u}  \oone \zhTB_{CD}
\big)
 \Big]
 \,d\mu_\ringh
 \nonumber
\\
 &
 \phantom{=}
+ o(1)
\,. &
\label{3VII20.1n}
\end{eqnarray}
In order to obtain $ {d E_c[h,\mcC_{u,R}]}/{du}$ from this formula, we analyse the  surface terms in the definition of canonical energy \eqref{30XI19.t1}:
%
%
\begin{eqnarray}
B_{E_{c}}
:=
&&
\nonumber
R
\int_{S^2}
\Big(  2 \zh^{AB} \ozero \deltaU{}_{A} \otwo \deltaU{}_{B}
-\frac{1}{2}\zh^{AB}\zh^{CD} (
     \alpha^2  \oone \zhTB_{AC}
    \oone \zhTB_{BD}
    -   2  \oone \zhTB_{AC}
    \zspaceD_B \ozero \zhTB_{BD}
)
\Big)
d \mu_{\zzhTBW }
\nonumber
\\
&&
+
\int_{S^2}
\Big(
3 \zh^{AB} \ozero \deltaU{}_{A} \omthree \deltaU{}_{B}
-\frac{1}{2}\zh^{AB}\zh^{CD} (
    \oone \zhTB_{AC} \partial_{u}\oone \zhTB_{BD}
)
\Big)
d \mu_{\zzhTBW }
\nonumber
\\
&&
+ o(1)
\,.
\end{eqnarray}
Explicitly computing $ {d{B}_{E_c}}/{du}$, adding   both sides to
\eqref{3VII20.1n}, after multiplying by $1/(32 \pi)$ we recover the mass
loss formula \eqref{30XI19.5}, keeping in mind that $\otwo \zhTB_{AB}=0 $ when no logarithms occur.

%
In the case $\Lambda=0$ things simplify as then $\alpha=\ozero \zhTB_{u B} =0$. One can integrate by parts in the first line above so that the left-hand side of \eqref{3VII20.1n} reads instead
\begin{eqnarray}
\frac{d}{du}
\left( \int_{\mcC_{u,R}} \zzhTBW^{AB}\zzhTBW^{CD} \partial_u \breh_ {AC}\partial_{\TSr} \breh_{BD} dr  \,d\mu_\ringh
-  \frac 12 \int_{S_{u,R}}  \zzhTBW^{AB}\zzhTBW^{CD} \breh_ {AC}  \partial_u \breh_{BD}
\,d\mu_\ringh
\right)
\,.
\nonumber
\\
\label{3VII20.2}
\end{eqnarray}
From \eqref{3VII20.1n} one obtains
\begin{eqnarray}
\lefteqn{
 \frac{d}{du}
  \left( \int_{\mcC_{u,R}} \zzhTBW^{AB}\zzhTBW^{CD} \partial_u \breh_ {AC}\partial_{\TSr} \breh_{BD} dr  \,d\mu_\ringh
   -  \frac 12 \int_{S_{u,R}}  \zzhTBW^{AB}\zzhTBW^{CD} \breh_ {AC}  \partial_u \breh_{BD}
     \,d\mu_\ringh
   \right)
  }
 \nonumber
\\
 &=
  \displaystyle
 -
	 \frac{1}{2}\int_{S_{u,R}}
\Big[\!\!\!\!\!
 &
 \zzhTBW^{AC} \zzhTBW^{BD}
\big(
\partial_u \oone \zhTB_{AB} \partial_{u} \oone  \zhTB_{CD}
-
\oone  \zhTB_{AB} \partial^2_{u}  \oone \zhTB_{CD}
\big)
 \Big]
 \,d\mu_\ringh
 \nonumber
\\
 &
 \phantom{=}
 + o(1)
 \,. &
\label{3VII20.3}
\end{eqnarray}
Equivalently
\begin{eqnarray}
\lefteqn{
 \frac{d}{du}
  \left( \int_{\mcC_{u,R}} \zzhTBW^{AB}\zzhTBW^{CD} \partial_u \breh_ {AC}\partial_{\TSr} \breh_{BD} dr  \,d\mu_\ringh
   -   \int_{S_{u,R}}  \zzhTBW^{AB}\zzhTBW^{CD} \breh_ {AC}  \partial_u \breh_{BD}
     \,d\mu_\ringh
   \right)
  }
 \nonumber
\\
 &
 & \phantom{xxxxxxx} =
	-\int_{S_{u,R}}
 \zzhTBW^{AC} \zzhTBW^{BD}
\partial_u \oone \zhTB_{AB} \partial_{u} \oone  \zhTB_{CD}
 \,d\mu_\ringh + o(1)
\,,
\label{3VII20.4}
\end{eqnarray}
or
\begin{eqnarray}
\lefteqn{
 \frac{d}{du}
  \Big( \int_{\mcC_{u,R}} \zzhTBW^{AB}\zzhTBW^{CD}\left(\partial_u \breh_ {AC}\partial_{\TSr} \breh_{BD}-\breh_ {AC}\partial_{\TSr} \partial_u \breh_{BD}\right)
   dr  \,d\mu_\ringh
   }
&&
 \nonumber
\\
&&
   -   \int_{S_{u,R}}  \zzhTBW^{AB}\zzhTBW^{CD} \breh_ {AC}  \partial_u \breh_{BD}
     \,d\mu_\ringh
   \Big)
 \nonumber
\\
 &
 & \phantom{xxxxxxx} =
	-2 \int_{S_{u,R}}
 \zzhTBW^{AC} \zzhTBW^{BD}
\partial_u \oone \zhTB_{AB} \partial_{u} \oone  \zhTB_{CD}
 \,d\mu_\ringh + o(1)
\,.
\phantom{xxx}
\label{3VII20.5}
\end{eqnarray}
Passing with $R$ to infinity, after multiplying by $1/(64 \pi)$
the right-hand side becomes (up to a multiplicative factor) the right-hand side of the familiar Trautman-Bondi mass loss formula. Hence the expression which is differentiated at the left-hand side must be the Trautman-Bondi energy, up to the addition of a time-independent  functional of the fields.

Now, it is known that there are no such functionals in the nonlinear theory which are gauge-independent and
 which vanish when $h_{\mu\nu}$ vanishes by \cite{CJM}, and it is clear that the argument there carries over to the linearised theory. However, gauge-independence of the functional being differentiated in \eqref{3VII20.5}, namely
\begin{eqnarray}
\lefteqn{  \int_{\mcC_{u }} \zzhTBW^{AB}\zzhTBW^{CD}\left(\partial_u \breh_ {AC}\partial_{\TSr} \breh_{BD}-\breh_ {AC}\partial_{\TSr} \partial_u \breh_{BD}\right)
   dr  \,d\mu_\ringh
   }
&&
 \nonumber
\\
&&
   \phantom{xxxxxxx}
   -  \lim_{R\to\infty}
      \int_{S_{u,R}}  \zzhTBW^{AB}\zzhTBW^{CD} \breh_ {AC}  \partial_u \breh_{BD}
     \,d\mu_\ringh
     \,,
\label{3VII20.6}
\end{eqnarray}
is not clear at this stage. In Section~\ref{s1IX20.1} below we settle the issue by providing a direct proof of the equality of $\hat E_c$ with half of the quadratisation of the Trautman-Bondi mass. When $\Lambda$ vanishes this could have been anticipated by the results of \cite{CJK}, where it is shown that the canonical energy for the nonlinear field is the Trautman-Bondi mass. However, such statements  involve a careful choice  of boundary terms so that the correspondence is not automatic. And no such statement for $\Lambda >0$ has been established so far in any case.

\ptcrr{obsolete material moved to check2}
\subsection{Renormalised energy}
 \label{s9XI20.1}

Equation \eqref{30XI19.5} shows that the divergent term in $E_c$ has a dynamics of its own, evolving separately from the remaining part of the canonical energy.
It is therefore natural to introduce a \emph{renormalised canonical energy}, say $\hat E_c[h,\mcC_{u,R }]$, by removing the divergent term in \eqref{30VIII20.-1}.
After having done this, we can pass to the limit $R\to\infty$ to obtain:
\begin{eqnarray}
 \lefteqn{
 \hat E_c[h,\mcC_{u }]
 : =
 }
 &&
 \nonumber
\\
 \nonumber
 &&
  \frac{1}{64 \pi}
    \int_{\mcC_{u }}   {\nobarg}^{BE } {\nobarg}^{FC}
    \big(
    \partial_u h_{BC } \partial_{ r } h_{EF }
    -
     h_{BC } \partial_{ r } \partial_u h_{EF }
     \big)\,r^2   \,dr \sin \theta\, d\theta\, d\varphi
\\
 &&
	-
	{\frac{1}{64 \pi} }
	\int_{S^2}
	  \zh^{AB} \Big(\zh^{CD}\oone \zhTB_{AC} \partial_{u}
	\oone \zhTB_{BD}
-
	{   6  \ozero \zhTB {}_{uA} \othree \zhTB{}_{uB}}
	\Big)
	\, \sin \theta\, d\theta\, d\varphi
  \,.
  \label{8XII19.31}
\end{eqnarray}
This is our first main result here, and is our proposal how to calculate the total energy contained in a light cone of a weak gravitational wave on a de Sitter background.

The flux equation for the renormalised energy $\hat E_c$ coincides with the one obtained by dropping the term linear in $R$ in \eqref{30XI19.5} and passing again to the limit $R\to \infty$:
%
%
\begin{eqnarray}
\lefteqn{
	\frac{d \hat E_c[h,\mcC_{u,R}]}{du}
	=
}
&&
 \nonumber
\\
&&
	-
	{\frac{1}{32 \pi} }
	\int_{S^2}
	\zh^{AB}\Big(\zh^{CD}
	 	\partial_{u} \oone \zhTB_{AC}
		\partial_{u} \oone \zhTB_{BD}
-
	{   6 \othree \zhTB{}_{uA}} \partial_u	 \ozero \zhTB {}_{uB}
	\Big)
	\, \sin \theta\, d\theta\, d\varphi
\,.
\nonumber
\\
&&
\phantom{xx}
\label{8XII19.32}
\end{eqnarray}
This is our key new formula.
When $\Lambda =0$   we recover the weak-field version of the usual Trautman-Bondi mass loss formula
since, as already pointed out, $\ozero\zhTB_{uA}\equiv 0$ in the  asymptotically Minkowskian case.
Hence the last term in \eqref{8XII19.32} shows how the cosmological constant affects the flux of energy emitted by a gravitating astrophysical system.

\subsection{Energy and gauge transformations}
 \label{s9XI20.2}

We have seen by general considerations that the energy integral is invariant under gauge transformations, up to boundary terms. It is instructive to rederive this result for the residual gauge transformations by a direct calculation. As a byproduct we obtain the explicit form of the boundary terms arising.


We focus attention on the volume part  of the energy integral:
\begin{equation}
  E_{\Vol}[h](u)
  :=
     \frac{1}{64 \pi}
    \int_{\mcC_u}  {\nobarg}^{BE } {\nobarg}^{FC}
    \big(
     \partial_u h_{BC } \partial_{ r } h_{EF }
       -
       h_{BC } \partial_{ r } \partial_u h_{EF }
    \big)\, d\mu_{\mcC}
     \,.
  \label{6VII18.1a}
\end{equation}
It is convenient to define a new field
\begin{equation}\label{31V20.1}
  \breh_{\mu\nu} := r^{-1} h_{\mu\nu} \equiv r \zhTB_{\mu\nu}
  \,.
\end{equation}
In terms of $\breh$ the integral $E_{\Vol}$ takes the form
\begin{equation}
  E_{\Vol}[h](u)
    =
      \frac{1}{64 \pi}
    \int_{\mcC_u}  \ringh^{BE } \ringh^{FC}
    \big(
     \partial_u \breh_{BC } \partial_{ r } \breh_{EF }
       -
       \breh_{BC } \partial_{ r } \partial_u \breh_{EF }
    \big)dr \,d\mu_\ringh
     \,.
  \label{6VII18.1bn}
\end{equation}

Under asymptotic symmetries we have (cf.\ \eqref{20II20.1})
\bea
 \breh_{AB} & \mapsto &    \breh_{AB}
  + r \big(
 \underbrace{\mathring{D}_{A} \xi_{B}
+ \mathring{D}_{B} \xi_{A}- \mathring{\gamma}_{AB}\mathring{D}^{C} \xi_{C}
 }_{0}
 \big)
  \nonumber
\\
 && +
 \big( \mathring{\gamma}_{AB}
\Delta _{\mathring{\gamma}} -2 \mathring{D}_{A}\mathring{D}
_{B}
 \big)  \xi^{u}
\,,
\label{31V20.5}
\eea
where the   term linear in $r$ vanishes since $\TSxi^A(u,\cdot)$ is a conformal Killing vector of $S^2$. Whence
\bea
\label{31V20.2}
 \partial_u\breh_{AB} & \mapsto &    \partial_u\breh_{AB}
   +
 \big( \mathring{\gamma}_{AB} \Delta _{\mathring{\gamma}} -2 \mathring{D}_{A}\mathring{D} _{B} \big)\partial_u\xi^{u}(u, x^{A})
\,,
\\
  \partial_r\breh_{AB} & \mapsto &   \partial_r\breh_{AB}
  \,.
\eea
Now, for each $u$ the function $\partial_u\TSxi^u$ is a linear combination of $\ell=0$ and $\ell=1$ spherical harmonics (see Appendix~\ref{App2VII20}), which implies that the terms involving  $\partial_u\TSxi^u$ in $\partial_u \breh_{AB}$ vanish. We conclude that both $\partial_u h_{AB} \equiv r \partial_u \breh_{AB}$ and $\partial_r(r^{-1} h_{AB})\equiv \partial_r\breh_{AB}$ are invariant under asymptotic symmetries.

Since the measure  $ d\mu_\ringh = \sqrt{\det \ringh}\, dx^2\wedge dx^3$ is $r$-independent, integration by parts in \eqref{6VII18.1a} gives
\begin{eqnarray}
  E_{\Vol}[h](u)
  & = &  \frac{1}{32 \pi}
    \int_{\mcC_u}  \ringh^{BE } \ringh^{FC}
     \partial_u \breh_{BC } \partial_{ r } \breh_{EF }\,dr \,  d\mu_{\ringh}
     \nonumber
\\
 &&
      +\lim_{\epsilon\to0}\frac{1}{64 \pi}
       \int_{r=\epsilon} \ringh^{BE } \ringh^{FC}
       \breh_{BC }   \partial_u  \breh_{EF }
     \,d\mu_\ringh
     \nonumber
\\
 &&
       -  \lim_{R\to\infty}\frac{1}{64 \pi}
       \int_{r=R} \ringh^{BE } \ringh^{FC}
       \breh_{BC }   \partial_u  \breh_{EF }
     \,d\mu_\ringh
     \,,
      \phantom{xxxxx}
  \label{6VII18.1basdf}
\end{eqnarray}
and note that the  second line above does not vanish   in a general Bondi gauge, since then both $\breh_{AB}$ and $\partial_u\breh_{AB}$ are of order one for small $r$ by \eqref{31V20.5}.
So
%
\begin{eqnarray}
\lefteqn{
  E_{\Vol}[h](u)
  =
    \frac{1}{32 \pi}
    \int_{\mcC_u}  \ringh^{BE } \ringh^{FC}
     \partial_u \breh_{BC } \partial_{ r } \breh_{EF }\,dr \,  d\mu_{\ringh}
     }
     &&
     \nonumber
\\
 &&
       +\frac{1}{64 \pi}
       \int_{S^2} \ringh^{BE } \ringh^{FC}
       \big(
       (  \breh_{BC }   \partial_u   \breh_{EF })|_{r=0}
       -
       \ozero \breh_{BC }   \partial_u \ozero \breh_{EF }
       \big)
     \,d\mu_\ringh
     \,,
      \phantom{xxxxx}
  \label{6VII18.1bm}
\end{eqnarray}
where we assumed that the metric $h_{\mu\nu}$ has the usual asymptotic expansion for large $r$ as considered elsewhere in this paper.

From what has been said the volume integral in \eqref{6VII18.1bm} is  invariant under residual gauge transformations, so that we have
\begin{eqnarray}
 \lefteqn{
  E_{\Vol}[h](u)
    \mapsto
      E_{\Vol}[h](u)
      }
      &&
      \nonumber
\\
 &&
       + \frac{1}{64 \pi}
       \int_{S^2}
          \ringh^{AB } \ringh^{CD}
          \big(
            \partial_u   \zhTB_{AC }|_{r=0}
            -
            \partial_u \oone \zhTB_{AC }
            \big)
 \big( \mathring{\gamma}_{BD}
\Delta _{\mathring{\gamma}} -2 \mathring{D}_{B}\mathring{D}
_{D}
 \big)  \xi^{u}
     \,d\mu_\ringh
     \,.
      \phantom{xxxxxxx}
  \label{6VII18.1bacn}
\end{eqnarray}
%
The second line in \eqref{6VII18.1bacn} vanishes when $\xi^u$ is a linear combination of $\ell=0$ and $\ell=1$ spherical harmonics.

%

It easily follows that the canononical energy is invariant under such residual gauge transformations.

\ptcrr{Some calculations of JH on gauge moved to JHonGauge.tex}
\subsection{$\hat E_c=\frac{1}{2}\delta^2 m_\TB$}
 \label{s1IX20.1}

In this section we show by a direct calculation
that our renormalised energy  $\hat E_c$ coincides with the Trautman-Bondi mass of the linearised theory.
As before we fix a light cone $\mcC_u$ and use the gauge \eqref{2XI20,1} on $\mcC_u$.

We start with \eqref{2VI20.4}, which we repeat here for the convenience of the reader:
%
              \begin{equation}
               \delta^2 m_\TB(\mcC_{u,R}) =
                \frac{1}{16 \pi } \int_{\mcC_{u,R}}
              \Big(
               r^4  \ringh^{AB}\partial_r \zhTB_{uA} \partial_r \zhTB_{uB}%
 {\TSgreen -4}(1- \Lambda r^2 ) \delta^2 \beta
               \Big)
                 \, dr \, d\mu_{\ringh}
                   \,.\label{2VI20.4a}
              \end{equation}
%
We integrate by parts on the $\delta^2 \beta$ term and use the vacuum version of \eqref{eq:beta_eq2},
%
\begin{equation}
         \label{23VIII20.1}
          \partial_r  \delta^2\beta  =   \frac{r}{8}\ringh^{AC}\ringh^{BD} \partial_r \zhTB_{AB}\partial_r \zhTB_{CD}    \,,
         \end{equation}
to obtain
\ptcheck{28VIII20 and before with TS but not for the boundary integral at the origin, which might be missing, but there is not much point of doing this if the equation for the quadratised mass is wrong}
              \begin{eqnarray}
               \delta^2 m_\TB(\mcC_{u,R})
                &
                =
                &
                \frac{1}{16 \pi } \int_{\mcC_{u,R}}
              \Big(
               r^4  \ringh^{AB}\partial_r \zhTB_{u A}\partial_r \zhTB_{uB}
               \nonumber
\\
 &&
              {\TSgreen +}  \frac{3 r^2-  {\Lambda r^4}  }{6}\ringh^{AC}\ringh^{BD}
               \partial_r \zhTB_{AB}\partial_r \zhTB_{CD}
               \Big)
                 \, dr\,
                d\mu_{\ringh}
               \nonumber
\\
 &&
             {\TSgreen -}  \frac{1}{4  \pi }\int_{S_R}  r
             \big(1-  \frac{\Lambda r^2}3\big )
               \delta^2 \beta
              \,
                d\mu_{\ringh}
                   \,.\label{23VIII20.2}
           \end{eqnarray}
The first volume integral can be handled using \eqref{28XI19.5},
\begin{eqnarray}
          &&  \partial_r \left[r^4  \partial_r(r^{-2} \delta  g_{uA})\right] =
                    r^2
                    \zspaceD_E
                     \left ( \zzhTBW ^{EF}\partial_r \left(r^{-2}\delta g_{AF}\right)\right)
                 \,.
                            \label{23VII20.5a}
           \end{eqnarray}
Equivalently,
\begin{eqnarray}
          &&    \partial_r \left[r^4  \partial_r \zhTB_{uA} \right]   =
                    r^{ 2}
\partial_r \zspaceD ^{ F}\zhTB_{AF}
                 \,.
                            \label{23VII20.6}
           \end{eqnarray}
Hence
\ptcheck{28VIII20 and before with TS}
\begin{eqnarray}
              \lefteqn{
                \frac{1}{16 \pi } \int_{\mcC_{u,R}}
               r^4  \ringh^{AB}\partial_r \zhTB_{u A}\partial_r \zhTB_{uB}
                \,
                { d\mu_{\ringh} \, dr }
                }
                &&
               \nonumber
\\
                &= &
                 \frac{1}{16 \pi } \int_{S_R}
               r^4  \ringh^{AB}  \zhTB_{u A} \partial_r \zhTB_{uB}\, d\mu_{\ringh}
               - \frac{1}{16 \pi } \int_{\mcC_{u,R}}
               \ringh^{AB} \zhTB_{uA} \partial_r (r^4 \partial_r \zhTB_{u B})
                \,
                { d\mu_{\ringh} \, dr }     \nonumber
\\
                &= &
                 \frac{1}{16 \pi } \int_{S_R}
               r^4  \ringh^{AB}  \zhTB_{u A} \partial_r \zhTB_{uB}\, d\mu_{\ringh}
               - \frac{1}{16 \pi } \int_{\mcC_{u,R}}
               r^{ 2}\ringh^{AB} \zhTB_{uA}
 \partial_r\zspaceD ^{ F}\zhTB_{BF}
                \,
                { d\mu_{\ringh} \, dr }
  \,.
                \nonumber
\\
 &&
 \label{23VIII20.7}
\end{eqnarray}
Equation~\eqref{23VIII20.2} can thus be rewritten as
\ptcheck{28VIII20 and before with TS}
              \begin{eqnarray}
               \delta^2 m_\TB(\mcC_{u,R})
                &
                =
                &
               -  \frac{1}{16 \pi } \int_{\mcC_{u,R}}
              \Big( r^{ 2}\ringh^{AB} \zhTB_{uA}
    \partial_r \zspaceD ^{ F}\zhTB_{BF}
               \nonumber
\\
 &&
              {\TSgreen -}  \frac{3 r^2-  {\Lambda r^4}  }{{6}}
              \ringh^{AC}\ringh^{BD} \partial_r \zhTB_{AB}\partial_r \zhTB_{CD}
               \Big)
                 \, dr\,
                d\mu_{\ringh}
               \nonumber
\\
 &&
             +  \frac{1}{16  \pi }\int_{S_R}
             \Big(
                r^4  \ringh^{AB}  \zhTB_{u A} \partial_r \zhTB_{uB}
             -
               4 r
             \big(1-  \frac{\Lambda r^2}3\big )
               \delta^2 \beta
               \Big)
              \,
                d\mu_{\ringh}
                   \,.
                    \nonumber
\\
 &&                   \label{23VIII20.6}
           \end{eqnarray}

Denoting by
$ E_V[h,\mcC_{u,R}]$
the volume term in \eqref{30XI19.t1} and
$$
 \breve h_{AB} := r^{-1} h_{AB}
 \,,
$$
after another integration  by parts we  find
\ptcheck{28VIII20 and before with TS; checked 7XI20 that the singularity of the time derivative does not affect the convergence of the integral, nor introduces new boundary terms at r=0; jh says it is ok}
\begin{eqnarray}
 \nonumber
  E_V[h,\mcC_{u,R}]
 &:= &
  \frac{1}{64 \pi}
    \int_{\mcC_{u,R}}   {\nobarg}^{BE } {\nobarg}^{FC}
    \big(
    \partial_u h  _{BC } \partial_{ r } h_{EF }
    -
     h_{BC } \partial_{ r } \partial_u h_{EF }
     \big)\, d\mu_{\mcC}
\\
\nonumber
 & = &
  \frac{1}{64 \pi}
    \int_{\mcC_{u,R}}   \ringh^{BE } \ringh^{FC}
    \big(
    \partial_u \breve h  _{BC } \partial_{ r } \breve h_{EF }
    -
     \breve h_{BC } \partial_{ r } \partial_u \breve h_{EF }
     \big)\, { d\mu_{\ringh} \, dr }
\\
\nonumber  &= &
  - \frac{1}{32 \pi}
    \int_{\mcC_{u,R}}
     {\ringh}^{BE } {\ringh}^{FC} \breve h_{BC } \partial_{ r } \partial_u \breve h_{EF }
     \, dr \,  d\mu_{\ringh}
\\
 &&
       +
  \frac{1}{64 \pi}
    \int_{S_R}  r^{-2} {\ringh}^{BE } \ringh^{FC}
    \partial_u h  _{BC }  h_{EF }
    \, d\mu_{\ringh}
  \,.
  \label{23VIII20.3}
\end{eqnarray}
Inserting the vacuum version of \eqref{30XI19.12},
\ptcheck{28VIII20 and before with TS}
\begin{equation}\label{23VIII20.4}
       \partial_r  \partial_u  \breve h_{EF}
     	 + \frac{\epsilon}{2r  }  \partial_r[ r^2 \lapseTB^2  \partial_r \zhTB_{EF}]
        - TS
         \big[\frac{1}{r } \zspaceD_E\big( \partial_r (r^2\zhTB_{uF}) \big)\big]
       =  0
        \,,
\end{equation}
into \eqref{23VIII20.3} we are led to
\ptcheck{28VIII20 and before with TS}
\begin{eqnarray}
 \lefteqn{
  E_V[h,\mcC_{u,R}]
     }
     &&
       \nonumber
\\
\nonumber  &= &
    \frac{1}{32 \pi}
    \int_{\mcC_{u,R}}
     {\ringh}^{BE } {\ringh}^{FC} \zhTB_{BC }
      \Big(
     \frac{\epsilon}{\green 2}   \partial_r[ r^2 \lapseTB^2
     \partial_r \zhTB_{EF}]
        -
          \zspaceD_E\big( \partial_r (r^2\zhTB_{uF}) \big)
         \Big)
     \, { d\mu_{\ringh} \, dr }
 \nonumber
\\
 &&       +
  \frac{1}{64 \pi}
    \int_{S_R}  r^{-2} {\ringh}^{BE } \ringh^{FC}
    \partial_u h  _{BC }  h_{EF }
    \, d\mu_{\ringh}
       \nonumber%
\\
\nonumber  &= &
 -
   \frac{1}{32 \pi}
    \int_{\mcC_{u,R}}
     {\ringh}^{BE } {\ringh}^{FC}
      \Big(
          \frac{\epsilon}{2  }  r^2 \lapseTB^2  \partial_r \zhTB_{BC }  \partial_r \zhTB_{EF}
      +
        \zhTB_{BC }  \zspaceD_E\big( \partial_r (r^2\zhTB_{uF}) \big)
         \Big)
     \, d\mu_{\ringh}\, dr
 \nonumber
\\
 &&       +
  \frac{1}{64 \pi}
    \int_{S_R}   {\ringh}^{BE } \ringh^{FC}
   h  _{BC }
     \big( \partial_u  \zhTB_{EF } +  {\epsilon}    \lapseTB^2  \partial_r \zhTB_{EF}
     \big)
    \, d\mu_{\ringh}
  \,.
  \label{23VIII20.9}
\end{eqnarray}
The last integral in the before-last line can be integrated by parts over the angles to become, using \eqref{23VIII20.6} to pass from  \eqref{23VIII20.99} to \eqref{23VIII20.8},
\ptcheck{28VIII20 and before with TS}
\begin{eqnarray}
\lefteqn{
  -\frac{1}{32 \pi}
    \int_{\mcC_{u,R}}
     {\ringh}^{BE } {\ringh}^{FC} \zhTB_{BC }
         \zspaceD_E\big( \partial_r (r^2\zhTB_{uF}) \big)
         \Big)
     \, d\mu_{\ringh}\, dr
     }
 \nonumber
\\
 &=&
    \frac{1}{32 \pi}
    \int_{\mcC_{u,R}}
      {\ringh}^{FC} \zspaceD^B \zhTB_{BC }
          \partial_r (r^2\zhTB_{uF})
     \, d\mu_{\ringh}\, dr
 \nonumber
\\
 &=&
    - \frac{1}{32 \pi}
    \int_{\mcC_{u,R}}
       \ringh^{FC}
          r^2\zhTB_{uF}
          \partial_r \zspaceD^B \zhTB_{BC }
     \, d\mu_{\ringh}\, dr
 \nonumber
\\
 & &
     +
      \frac{1}{32 \pi}
    \int_{S_R }r^2
      {\ringh}^{FC}
    \zhTB_{uF}
     \zspaceD^B \zhTB_{BC}
     \, d\mu_{\ringh}
                \label{23VIII20.99}
\\
 &=&
   \frac 12
               \delta^2 m_\TB(\mcC_{u,R})
 \nonumber
\\
 &&
            {\TSgreen -}\frac{1}{32 \pi}
    \int_{\mcC_{u,R}} \underbrace{\frac{3 r^2-  {\Lambda r^4}  }{6}}_{-\epsilon r^2 \lapseTB^2/2}
    \ringh^{AC}\ringh^{BD} \partial_r \zhTB_{A B}\partial_r \zhTB_{C D}
                 \, dr\,
                d\mu_{\ringh}
               \nonumber
\\
 &&
             -  \frac{1}{32  \pi }\int_{S_R}
             \Big(
               r^4  \ringh^{AB}  \zhTB_{u A} \partial_r \zhTB_{uB}
              -
                4r
             \big(1-  \frac{\Lambda r^2}3\big )
               \delta^2 \beta
 \nonumber
\\
 & &
    \phantom{xxxxxxxxx}
     -  r^2  {\ringh}^{FC}
    \zhTB_{uF}
     \zspaceD^B \zhTB_{BC}
               \Big)
              \,
                d\mu_{\ringh}
  \,.
  \label{23VIII20.8}
\end{eqnarray}
Hence
 \ptcheck{27VIII with TS, and again 28VIII20; the volume term must cancel out and all boundary terms should be identical, otherwise the time derivative will not fit}
\begin{eqnarray}
\lefteqn{
  E_V[h,\mcC_{u,R}]
   =   \frac 12
               \delta^2 m_\TB(\mcC_{u,R})
                }
&&
\nonumber
\\
 &&
             -  \frac{1}{64  \pi }\int_{S_R}
             \Big(
              2 r^4  \ringh^{AB}  \zhTB_{u A} \partial_r \zhTB_{uB}
              -
                8 r
             \big(1-  \frac{\Lambda r^2}3\big )
               \delta^2 \beta
 \nonumber
\\
 & &
     -  2 r^2{\ringh}^{FC}
    \zhTB_{uF}
     \zspaceD^B \zhTB_{BC} -  {\ringh}^{BE } \ringh^{FC}
   h  _{BC }
     \big( \partial_u  \zhTB_{EF } + {\epsilon}    \lapseTB^2  \partial_r \zhTB_{EF}
     \big)
               \Big)
              \,
                d\mu_{\ringh}
  \,.
    \nonumber
\\
 &&
  \label{23VIII20.98}
\end{eqnarray}
Now, using \eqref{6IX18.5+asdf}, \eqref{13IX18.1}, \eqref{23XI19.3}, \eqref{29IV20.t1}, the total canonical energy equals
%
\begin{eqnarray}
 \lefteqn{
E_c [\mcC_{u,R},h]
 :=
  \frac 12
  \left( \int_{\mcC_{u,R}}
  \omega^\mu(h , \Lie_X h)
       \, dS_\mu
    -
      \int_{S_R}
     \tpi ^{ \alpha\beta [\mu }
    X^{{{\sigma}}]} h_{\alpha\beta}
     dS_{\sigma \mu}
      \right)
      }
      &&
 \nonumber
\\
 &= &
  E_V[h,\mcC_{u,R}]
   -
 {\frac{1}{32 \pi} }
      \int_{S_R}
P^{{\TSred r} (\b \c) \d (\e \f) }h_{\b\c}
 {\znabla}_{ \d } h_{\e \f }
   \,
   r^2
   d\mu_{\ringh}
 \nonumber
\\
 &= &
  E_V[h,\mcC_{u,R}]
   -
 {\frac{1}{32 \pi} }
      \int_{S_R} b^r (h,h)
       \,
   r^2
   d\mu_{\ringh}
 \nonumber
\\
 \nonumber
 &= &
  E_V[h,\mcC_{u,R}]
  \nonumber
\\
\nonumber
 &&
   -
 \frac{1}{32 \pi}
      \int_{S_R}
       \Big(
   \frac{1}{2r^4} \zzhTBW^{A C} \zzhTBW^{B D}h_{C D}
 \big( \epsilon \lapseTB^2 \partial_{\TSr}h_{AB}
 +
 \partial_{\TSu}h_{AB}
 -
 2 \zspaceD_Ah_{u B}
 \big)
\\
 \nonumber
 &&
 + h_{\TSu A} \zzhTBW^{A B}
 \underbrace{ \frac{1}{r^2}
	\big(
	\partial_{\TSr}h_{\TSu B}
	-
	\frac{2}{r}h_{\TSu B}
	\big)
  }_{\partial_r \zhTB_{uB}}
 \Big)
       \,
   r^2
   d\mu_{\ringh}
\\
 \nonumber
 &= &
   \frac 12
               \delta^2 m_\TB(\mcC_{u,R})
\\
 &&
             -  \frac{1}{64  \pi }\int_{S_R}
             \Big(
              2 r^4  \ringh^{AB}  \zhTB_{u A} \partial_r \zhTB_{uB}
               -
                8 r
             \big(1-  \frac{\Lambda r^2}3\big )
               \delta^2 \beta
 \nonumber
\\
 & &
     -  2 r^2{\ringh}^{FC}
    \zhTB_{uF}
     \zspaceD^B \zhTB_{BC} -  {\ringh}^{BE } \ringh^{FC}
   h  _{BC }
     \big( \partial_u  \zhTB_{EF } +  \epsilon    \lapseTB^2  \partial_r \zhTB_{EF}
     \big)
               \Big)
              \,
                d\mu_{\ringh}
                 \nonumber
\\
\nonumber
 &&
   -
 \frac{1}{32 \pi}
      \int_{S_R}
       \Big(
   \frac{1}{2 } \zzhTBW^{A C} \zzhTBW^{B D} { \zhTB} _{C D}
 \big( \epsilon \lapseTB^2 \partial_{\TSr}h_{AB}
 +
 \partial_{\TSu}h_{AB}
 -
 2 \zspaceD_Ah_{u B}
 \big)
\\
 &&
 +
   r^2 h_{\TSu A} \zzhTBW^{A B}
	\partial_{\TSr}\zhTB _{\TSu B}
 \Big)
   \,
   d\mu_{\ringh}
 \nonumber
\\
\nonumber
&=&
   \frac 12
\delta^2 m_\TB(\mcC_{u,R})
\\
&&
-  \frac{1}{64  \pi }\int_{S_R}
\Big(
4 r^4  \ringh^{AB}  \zhTB_{u A} \partial_r \zhTB_{uB}
-
8 r
\underbrace{\big(1-  \frac{\Lambda r^2}3\big )}_{-\epsilon N^2}
\delta^2 \beta
\nonumber
\\
& &
+  {\ringh}^{BE } \ringh^{FC} \epsilon    \lapseTB^2
\big(
\zhTB  _{BC }  \partial_r h_{EF}
-h  _{BC }  \partial_r \zhTB_{EF}
\big)
\Big)
\,
d\mu_{\ringh}
 \,.
\label{25II21.t1}
\end{eqnarray}
We emphasise that no asymptotic conditions have been used in the analysis in this section  so far.

To continue we invoke the asymptotic conditions used in the preceding sections. From the expansion \eqref{28VIII20.1} of $\delta^2\beta$ one obtains
\begin{eqnarray}
\lefteqn{
E_c [\mcC_{u,R},h] =
   \frac 12
               \delta^2 m_\TB(\mcC_{u,R})
            }
        & &
               \nonumber
\\
 &&
            -   \frac{1}{64  \pi }\int_{S_R}
              \Big(
              \frac{3 \alpha^2}2 r  {\ringh}^{BE } \ringh^{FC}
  \oone \zhTB_{BC } \oone \zhTB_{EF}
          +   4 r^4  \ringh^{AB}  \zhTB_{u A} \partial_r \zhTB_{uB}
               \Big)
              \,
                d\mu_{\ringh}
    \nonumber
\\
  &&
  + o (1)
    \,   .
        \label{26VIII20.1a}
\end{eqnarray}
Using \eqref{28VIII20.32} this becomes
 \ptcheck{1IX20 with Maple by TS and by hand; for Minkowski this is ok now, but  de Sitter? the first boundary  term doubles the third}
\begin{eqnarray}
 \nonumber
 \lefteqn{
 E_c [\mcC_{u,R},h]
  =
   \frac 12
               \delta^2 m_\TB(\mcC_{u })
               }
               &&
\\
 &&
             +   \frac{R}{64  \pi }\int_{S_R}
             \Big(
     - \frac{3 \alpha^2}2   {\ringh}^{BE } \ringh^{FC}
  \oone \zhTB_{BC } \oone \zhTB_{EF}
           +8    \ringh^{AB} \ozero \zhTB_{u A} \otwo \zhTB_{uB}
               \Big)
              \,
                d\mu_{\ringh}
    \nonumber
\\
  &&
 -
                \frac{\Lambda R}{128 \pi }\int_{S_R}
         \ringh^{AC}\ringh^{BD}
            \oone \zhTB _{AB}  \oone   \zhTB _{CD}
              \,
                d\mu_{\ringh}
  + o (1)
    \,   .
        \label{26VIII20.1ab}
\end{eqnarray}
The last term in the second line can be integrated by parts as in \eqref{21IV20.21}-\eqref{21IV20.11} to obtain
%
\begin{eqnarray}
 \nonumber
 \lefteqn{
 E_c [\mcC_{u,R},h]
  =
   \frac 12
               \delta^2 m_\TB(\mcC_{u })
               }
               &&
\\
  &&
  -
                \frac{\alpha^2 R}{64 \pi }\int_{S_R}
         \ringh^{AC}\ringh^{BD}
            \oone \zhTB _{AB}  \oone   \zhTB _{CD}
              \,
                d\mu_{\ringh}
  + o (1)
    \,   .
        \label{26VIII20.1ac}
\end{eqnarray}
We see that the renormalised energy $\hat E_c$ coincides with one half of the quadratisation of the Trautman-Bondi mass.

\subsection{Energy in the asymptotically block-diagonal gauge}

So far we have assumed a gauge where the leading order corrections to the metric are encoded in the off-diagonal components $h_{uA}$ of the metric perturbation.
As such, the residual gauge freedom of Section~\ref{ss5XI20.3}  allows us to pass between two natural choices of asymptotic gauge:
\begin{eqnarray}
\mathrm{(I)} & & \ozero \zhTB_{A B} \equiv 0 \, ,
 \quad
 \ozero \zhTB_{uA} \neq 0 \, ,
 \\
\mathrm{(II)} & & \ozero \zhTB_{A B} \neq 0 \, ,
\quad
\ozero \zhTB_{uA} \equiv 0 \, .
\end{eqnarray}
Both choices are possible. However,  as already emphasised in Section \ref{ss5XI20.5}, when $\Lambda>0$  the asymptotic gauge  $\ozero \zhTB_{uA} \equiv 0$ cannot be attained simultaneously with $\ozero \zhTB_{A B} \equiv 0$ for all $u$ for general metric perturbations considered in this work.

We wish to revisit our analysis in the asymptotically block-diagonal  gauge (I).
Thus in the remainder of this section we assume
\begin{eqnarray}
\delta\gmtwo_{A B} &\neq& 0 \, , \\
\delta\gmtwo_{u A } &\equiv& 0 \, .
\end{eqnarray}

\subsubsection{The energy and its flux}
\label{s18II21.1}

Repeating the analysis of Section~\ref{s9VII20}, one finds:
\begin{enumerate}
    \item $\delta\gmtwo_{A B} \neq 0$ does not change the expansion  \eqref{21IV20.13} of $g_{u A}$. In particular $\delta\gmone_{u A}$ remains related with the log terms in the asymptotic expansion of $g_{AB}$.
    \item When $\delta\gmtwo_{A B} \neq 0$,  Equation \eqref{9VII20.3} becomes
    \begin{eqnarray}
    \nonumber
    \delta V
    &  = &
    \frac 12  \zspaceD^A
    \big(\int_0^r
    \big(
    \zspaceD^B \zhTB_{AB}
    -\frac{1}{r^2 } \big[ \partial_r (r^4\zhTB_{uA })\big]
    \big)
    \big |_{r=\rho} d\rho
    \big)
    \\
    & = &
     \zspaceD^{A}
    \big(
     \frac{1}{2} r \zspaceD^{B} \ozero \zhTB_{AB }
    +
    {\hat \nu }_A
    +
    \frac{1}{2 r} \othree \zhTB_{uA }
    + o(1/r)
    \big)
                \,.
    \phantom{xxxxxx}
    \label{21XII20.t1}
    \end{eqnarray}
    %
   \begin{eqnarray}
   {\hat \nu}_A
    &\!\!\!\!=   \lim_{r \to \infty} &\!\!\!\!\!
   \Big( \frac{1}{2}\int_0^r
    \big(
    \zspaceD^B \zhTB_{AB}
    -\frac{1}{r^2 } \big[ \partial_r (r^4\zhTB_{uA })\big]
    \big)
    \big |_{r=\rho} d\rho
    \nonumber
\\
 &&
    -\frac{r}{2} \zspaceD^{B} \ozero \zhTB_{AB}
    \Big)
    \,.
   \end{eqnarray}
   This differs from \eqref{9VII20.3a}, but does not affect the mass because of the divergence structure of the right-hand side of \eqref{21XII20.t1}.
\end{enumerate}

The asymptotics of the linearised metric perturbations reads
\begin{eqnarray}
\delta g_{u r} &\equiv &    0 \ \, \equiv \ \, \delta g_{r A}  \ \, \equiv \ \,\delta g_{rr} \,,
\label{18XII20.t1}
\\
\label{18XII20.t2}
\delta g_{AB} &=&\delta\gmtwo_{A B}\rBo^2 + \delta\gmone_{A B} \rBo + o(1)
\,,
\\
\label{18XII20.t3}
\delta g_{u A} &=&
\delta\gzero_{u A} + o(1)
\,,
\\
\delta g_{u u} &=&
 \delta\gzero_{u u}
+\frac{\delta\gone_{u u}}{\rBo }
+ o(\rBo^{-1})
\,.
\label{18XII20.t4}
\end{eqnarray}
%

Let us denote by $\hat E_{c,II}$ the renormalised canonical energy in the  block-diagonal gauge. To determine $\hat E_{c,II}$ we use  \eqref{25II21.t1} with the asymptotically block-diagonal boundary conditions.
One checks that the asymptotic behaviour \eqref{28VIII20.1} of   $\delta^2 \beta$ remains unchanged at the order needed; the analysis of the remaining terms in  \eqref{25II21.t1} is likewise straightforward, leading to
\begin{equation}
\hat E_{c,II}
=
\Big(
\frac 12
\delta^2 m_\TB(\mcC_{u }) -\frac{1}{16 \pi }
\int_{S^2}
\ringh^{AC }\ringh^{BD }\big(\frac{\Lambda}{3} \othree \zhTB_{AB} \ozero \zhTB_{CD}
-\oone \zhTB_{AB}  \ozero \zhTB_{CD}\big)
\, d \mu_{\zzhTBW}
\Big)
\,.
\label{25II21.t2}
\end{equation}

In order to make clear the comparison with our previous calculations,  let us denote by $\hat E_{c,I}$ the renormalised energy $\hat E_c$ calculated in the asymptotically off-diagonal gauge (I):
\begin{equation}\label{23II21.21}
\hat E_{c,I} \equiv \hat E_{c}
\,.
\end{equation}
Now, $\delta^2 \ozero V$ is gauge-independent (cf.\ \eqref{4XII19.t1}), and therefore so is $\delta^2 m_\TB(\mcC_{u })$. And we have seen that this last quantity coincides with $2 \hat E_c$. We conclude that
\begin{equation}\label{23II21.3}
\hat E_{c,II} =\hat E_{c,I }
- \frac{1}{16 \pi }
\int_{S^2}
\ringh^{AC }\ringh^{BD }
\big(\frac{\Lambda}{3} \othree \zhTB_{AB} \ozero \zhTB_{CD}
-\oone \zhTB_{AB}  \ozero \zhTB_{CD}\big)
\, d \mu_{\zzhTBW}
\,.
\end{equation}

An alternative,  direct  way of calculating $\hat E_{c,II}$ invokes the boundary integrand $b^r$ in \eqref{29IV20.t1}, which now reads:
\begin{eqnarray}
\lefteqn{
    r^2 b^{r}(\delta_1 g,\delta_2 g)
    \,
    =
}
&&
\nonumber
\\
\nonumber
& &\zzhTBW^{AC}\zzhTBW^{BD} \Big[r^3 \alpha^2 \delta_{1} \gmtwo_{A B} \delta_{2} \gmtwo_{C D}
\\
\nonumber
& &+ \frac{1}{2} r^2 \big(
        \delta_{1} \gmtwo_{A B} \partial_{u}\delta_{2} \gmtwo_{C D}
        +\alpha^2 \delta_{1} \gmtwo_{A B} \delta_{2} \gmone_{C D}
        +2 \alpha^2 \delta_{1} \gmone_{A B} \delta_{2} \gmtwo_{C D}
    \big)
\\
& &
    + \frac{1}{2} r \big(
        \delta_{1} \gmtwo_{A B} \partial_{u} \delta_{2} \gmone_{C D}
        +\delta_{1} \gmone_{A B} \partial_{u}\delta_{2} \gmtwo_{C D}
\nonumber
\\
& &
        \phantom{ + \frac{1}{2} r \big(}
        +\alpha^2 \delta_{1} \gmone_{A B}\delta_{2} \gmone_{C D}
        -2\delta_{1} \gmtwo_{A B} \delta_{2} \gmtwo_{C D}
    \big)
\nonumber
\\
\nonumber
& &
    +\frac{1}{2} \big(
        \delta_{1} \gmone_{A B} \partial_{u}\delta_{2} \gmone_{C D}
        -\delta_{1}\gmtwo_{A B} \delta_{2} \gmone_{C D}
        -2 \delta_{1}\gmone_{A B} \delta_{2} \gmtwo_{C D}
\\
& &
    \phantom{+\frac{1}{2} \big(}
    +2\alpha^2 \delta_{1} \gone_{A B}\delta_{2} \gmtwo_{C D}
    -\alpha^2 \delta_{1} \gmtwo_{A B}\delta_{2} \gone_{C D}
    -2 \delta_{1} \gmtwo_{A B} \zspaceD_C \delta_{2}   \gzero_{u D}
    \big)
\Big]
 \nonumber
\\
 &&
 + o(1)
 \,.
 \label{11II21.1}
\end{eqnarray}
%

Proceedings as in the derivation of \eqref{30XI19.t1}, after discarding the divergent terms both in the volume (compare \eqref{8XII19.5}) and boundary integrals, the renormalised canonical energy in the block-diagonal gauge (II) is found to be
 \begin{eqnarray}
 \lefteqn{
 \hat E_{c,II}[h,\mcC_{u }]
 : =
 }
 &&
 \nonumber
\\
 \nonumber
 &&
  \Big[
   \frac{1}{64 \pi}
    \int_{\mcC_{u }}   {\nobarg}^{BE } {\nobarg}^{FC}
    \big(
    \partial_u h_{BC } \partial_{ r } h_{EF }
    -
     h_{BC } \partial_{ r } \partial_u h_{EF }
     \big)\,r^2   \,dr \sin \theta\, d\theta\, d\varphi
      \Big]\Big|_{\mathrm{ren}}
\\
 \nonumber
 &&
	-
	{\frac{1}{64 \pi} }
	\int_{S^2}
	  \zh^{AB} \zh^{CD} \Big(\oone \zhTB_{AC} \partial_{u}
	\oone \zhTB_{BD}
-
	{     \ozero \zhTB {}_{AC} \oone \zhTB{}_{BD}}
	-\ozero \zhTB{}_{AC} \mathring{D}_{B} \mathring{D}^{F} \oone \zhTB{}_{FD}
	\\
	&&
	-2 \oone \zhTB{}_{AC} \ozero \zhTB{}_{BD}
	+\alpha^{2}(2 \othree \zhTB{}_{AC} \ozero \zhTB{}_{BD}- \ozero \zhTB{}_{AC} \othree \zhTB{}_{BD})
	\Big)
	\, \sin \theta\, d\theta\, d\varphi
  \,.
  \label{22XII20.1}
\end{eqnarray}
(Here one could use  \eqref{24X20.t2} to analyse the change of the volume integral under changes of gauges to isolate the divergent terms.)

The energy flux for the renormalised energy $\hat E_{c,II}[h,\mcC_{u }] $ is
\begin{eqnarray}
\lefteqn{
	\frac{d \hat E_{c,II}[h,\mcC_{u }]}{du}
	=
}
&&
 \nonumber
\\ \nonumber
&&
	-
	{\frac{1}{32 \pi} }
	\int_{S^2}
	\zh^{AB}\zh^{CD}
	 	\Big( \partial_{u} \oone \zhTB_{AC}
		\partial_{u} \oone \zhTB_{BD}
-
	{   2  \partial_{u}\oone \zhTB{}_{AC}}  \ozero \zhTB {}_{BD}
	\\ \nonumber
	&&
	+\alpha^{2}\big(2 \partial_{u} \othree \zhTB{}_{AC}\ozero \zhTB{}_{BD}-\oone \zhTB{}_{AC}\oone \zhTB{}_{BD}- \oone \zhTB{}_{AC} \mathring{D}_{B}\mathring{D}^{F} \oone \zhTB{}_{FD}  \big)
	\\
	&&
	-\alpha^{4} \oone \zhTB{}_{AC} \othree \zhTB{}_{BD}
	\Big)
	\, \sin \theta\, d\theta\, d\varphi
\,,
\label{22XII20.2}
\end{eqnarray}
where we have used \eqref{11II21.1} and $\partial_{u} \ozero \zhTB{}_{AB}=\alpha^{2}\oone \zhTB{}_{AB}$.

\subsubsection{Comparing with~\cite{compere2019lambda}}
 \label{s23II21/1a}

Comp\`ere et al.~\cite{compere2019lambda,Compere} proposed a version of  mass  in the non-linear theory, using the asymptotically block-diagonal gauge.
In~\cite[Equation~(2.39)]{compere2019lambda} they define
\begin{align}
M^{(\Lambda)} &=  M + \frac{1}{16} (\partial_u + l)(C_{CD}C^{CD})
 \,,\label{eq:hatMa}
\end{align}
where $M$ is an integration constant which appears when analysing the characteristic constraint equations, $l$ is a gauge field which can be set to zero   and $C_{AB}$ is, essentially, our field $\oone \zhTB_{AB}$.
They propose to define a mass, which we will denote by  $E^{(\Lambda)}$, as
\begin{equation}\label{18II21.5a}
  E^{(\Lambda)} :=
   \frac{1}{4\pi}
    \int  M^{(\Lambda)}  \,d \mu_{\zzhTBW}
  \,.
\end{equation}
(Strictly speaking, a different multiplicative constant factor is probably used in \cite{compere2019lambda}.)
Using our notation, the quadratised version of \eqref{eq:hatMa} reads
\begin{align}
\delta^2 M^{(\Lambda)} &=   \label{eq:hatMab}
  -\frac 12 \delta^2  \ozero V + \frac{1}{8}  \partial_u  (\ringh^{AB}\ringh^{CD}\oone \zhTB_{AC}\oone \zhTB_{BD})\,.
\end{align}
Integrating over $S^2$ we find:
\begin{eqnarray}
\delta^2 E^{(\Lambda)} & := &
 \frac{1}{4\pi} \int_{S^2} \delta^2 M^{(\Lambda)}
        \, d \mu_{\zzhTBW}
\\
& = &  \delta^2 m_\TB(\mcC_{u }) +
  \frac{1}{16 \pi}
  \int_{S^{2}}  \ringh^{AE }\ringh^{FB }\oone \zhTB_{E F} \partial_{u} \oone \zhTB_{AB}  \,d \mu_{\zzhTBW}
  \label{7VII21.t1}
\\
& = &  2 \hat E_{c,I} +
  \frac{1}{16  \pi}
  \int_{S^{2}}  \ringh^{AE }\ringh^{FB }\oone \zhTB_{E F} \partial_{u} \oone \zhTB_{AB}  \,d \mu_{\zzhTBW}
  \label{7VII21.t2}
\\
        \nonumber
\\  & = & 2  \hat E_{c,II} +
  \frac{1}{16 \pi}
  \int_{S^{2}}  \ringh^{AE }\ringh^{FB }\oone \zhTB_{E F} \partial_{u} \oone \zhTB_{AB}  \,d \mu_{\zzhTBW}
  \nonumber
\\
 &&
      +\frac{1}{8 \pi }  \int_{S^2}
         \ringh^{AC }\ringh^{BD }\big(\frac{\Lambda}{3}
         \othree \zhTB_{AB} \ozero \zhTB_{CD}
       -\oone \zhTB_{AB}  \ozero \zhTB_{CD}\big)
        \, d \mu_{\zzhTBW}
   \,,
    \phantom{xxx}
    \label{23II21.2ab}
\end{eqnarray}
where we used
\begin{eqnarray}
 \hat E_{c,I }
 & = &
   \frac 12
               \delta^2 m_\TB(\mcC_{u }) =
 - \frac{1}{16 \pi} \int_{S^2} \delta^2 \ozero V
        \, d \mu_{\zzhTBW}
               \,,
\\
  \hat E_{c,II}
 & = &
 \hat E_{c,I}
 \nonumber
\\
&&
  -\frac{1}{16 \pi }
        \int_{S^2}
         \ringh^{AC }\ringh^{BD }
         \big(\frac{\Lambda}{3} \othree \zhTB_{AB} \ozero \zhTB_{CD}
       -\oone \zhTB_{AB}  \ozero \zhTB_{CD}\big)
        \, d \mu_{\zzhTBW}
   \,.
    \phantom{xxxx}
    \label{23II21.2abc}
\end{eqnarray}

For completeness we note the following flux formula
\begin{eqnarray}
    \lefteqn{
        \partial_u  \int_{S^{2}} \delta^2 M^{(\Lambda)} d \mu_{\zzhTBW }
        =
 - \frac{1}{24} \int_{S^{2}}   \oone \zhTB_{E F} \ringh^{AE } \ringh^{FB }
  \Big\{
   -\Lambda^2\othree \zhTB_{A B}}
& &
\nonumber \\
& &
 -3 \partial^{2}_{u} \oone \zhTB_{AB} - \Lambda \Big[\zspaceD_{(A} \zspaceD^{C}\oone \zhTB_{B) C} - \oone \zhTB_{AB}\Big] \Big \}d \mu_{\zzhTBW}
\nonumber
\\
&=& \frac{1}{8} \int_{S^{2}} \partial_{u} \Big[\oone \zhTB_{E F} \ringh^{AE }\ringh^{FB }\partial_{u} \oone \zhTB_{AB} \Big] d \mu_{\zzhTBW }
-\frac{1}{8}\int_{S^{2}} \ringh^{AC} \ringh^{BD} \Big(\partial_{u} \oone \zhTB_{AB} \partial_{u} \oone \zhTB_{CD} \nonumber \\
& &-\frac{\Lambda}{3}(\oone \zhTB_{AB} \mathring{D}_{C}\mathring{D}^{F} \oone \zhTB_{FD}-\oone \zhTB_{AB} \oone \zhTB_{CD})
-\frac{\Lambda^{2}}{3} \oone \zhTB_{AB} \othree \zhTB_{CD}\Big)
 d \mu_{\zzhTBW} \, .
\phantom{xxxxxx}
\label{18II21.7}
\end{eqnarray}
\subsection{$\Lambda<0$}
 \label{Cs26X20.1}
 {
All our results so far are  independent of the sign of $\Lambda$: it suffices to replace $\alpha^2$ by $-\alpha^2$ wherever relevant.

Now, it should be pointed out that there is  a key conceptual difference arising from the causal character of the boundary at infinity. While in the $\Lambda\ge0 $ case the data on the initial light cone determine the remainder of the evolution uniquely, this is not the case anymore for $\Lambda <0$, where we have the freedom to add  \emph{and control} boundary data on $\scri$ (compare~\cite{Friedrich:aDS,HLSW}).

In the nonlinear theory the  ``pre-holographic''} approach is to require that the conformal boundary at infinity is the same as that of anti de Sitter spacetime. At a linearised level, this translates to the requirement that the evolution preserves the condition
\begin{equation}\label{26X20.1}
\omtwo h_{\mu\nu} \equiv 0
  \,,
\end{equation}
up to gauge.
This, in turn, leads to corner conditions on the data at the intersection of the initial light cone with the conformal boundary at infinity. For instance, choosing a gauge so that \eqref{26X20.1} holds, Equation~\eqref{4II20.7}  implies
\begin{equation}\label{26X20.2}
  \oone \zhTB_{AB} \equiv 0
  \,,
\end{equation}
which together with \eqref{26X20.1} shows that the canonical energy
$ E_c [\mcC_{u },h]$ is finite, $u$-independent, and equals the Trautman-Bondi mass, without the need for  renormalisation.
{


A natural option is to relax \eqref{26X20.1} by allowing the asymptotic data (but \emph{not} the whole linearised solution) to be stationary. For this we reanalyze the linearised vacuum Einstein equation under the assumption that $X=\partial_{u}$ is a Killing field for a leading terms in metric perturbation. In other words, we assume
\begin{eqnarray}
\Lie_{X} \omtwo h_{\mu\nu} \equiv 0 \, .
 \label{6IV21.21}
\end{eqnarray}
In the off-diagonal gauge this the same as
\begin{equation}
\partial_{u} \omtwo h_{u A} = 0 = \omtwo h_{AB}\, .
\label{31III21.t1}
\end{equation}
%
Equation \eqref{11XI20.2} then gives
\begin{equation}
\partial_{u}\omone h_{uu}
= 0
\, .
\label{31III21.t2}
\end{equation}
Applying a $u$-derivative to  \eqref{11Xi20.t1} and \eqref{11Xi20.t3} we obtain
\begin{eqnarray}
&
\partial_{u} \ozero h_{u A} = 0  \, ,
&
\label{31III21.t4}
\\
&
4 \alpha^2 \partial_{u} \otwo h_{u A}
+\partial_{u}\zspaceD_{A} \oone h_{u u}
- 3 \partial_{u}^2 \oone h_{u A}
=
0
\, .
&
\end{eqnarray}
Under \eqref{31III21.t1} the mass-loss formula \eqref{8XII19.32} reduces, in view of \eqref{31III21.t4}, to the one known from the $\Lambda=0$ case:
\begin{eqnarray}
	\frac{d \hat E_c[h,\mcC_{u,R}]}{du}
	=
	-
	{\frac{1}{32 \pi} }
	\int_{S^2}
	\zh^{AB} \zh^{CD}
	 	\partial_{u} \oone \zhTB_{AC}
		\partial_{u} \oone \zhTB_{BD}
	\, \sin \theta\, d\theta\, d\varphi
\,.
\label{3III21.1}
\end{eqnarray}

For completeness we list some further simplifications which arise in the asymptotics of the field. Equation
  \eqref{11Xi20.t1} gives
\begin{equation}
\partial_{u} D^{A} \omone h_{A B}
=
0
\, .
\label{31III21.t3}
\end{equation}
Using  \eqref{10XI20.t4} and \eqref{31III21.t3}
 one obtains the following formula for the evolution of the linearised mass aspect function (not to be confused with the evolution of the quadritised mass aspect, which is relevant for \eqref{3III21.1}):
\begin{equation}
\partial_{u} \oone h_{u u}=  3\alpha^2 \otwo h_{u u} \, .
 \label{3III21.21}
\end{equation}
Equations \eqref{31III21.t4} and \eqref{10XI20.t1} lead to
\begin{equation}
    \partial_{u} \oone h_{u u}
    =  \alpha^2 \otwo h_{u u}
    -\alpha^2 \zspaceD^{A} \oone h_{u A} \, .
\label{31III21.t5}
\end{equation}
Comparing \eqref{3III21.21} with \eqref{31III21.t5} we find
\begin{equation}
    \otwo h_{u u} = \frac 12
    \zspaceD^{A} \oone h_{u A} \, .
\label{31III21.t5a}
\end{equation}
 \ptcheck{20VII20 notation cleaned up, corrected; previously checked    PTC and TS on 10VII20; first two lines rechecked 31VIII20 }
The asymptotic symmetries of interest are now those  preserving \eqref{31III21.t1}.
Equations \eqref{20II20.1}, \eqref{31III21.t2}-\eqref{31III21.t4} and \eqref{31III21.t3} show that these satisfy
\begin{eqnarray}
\partial_u h_{uu} : \qquad  0
	&=&
	               \partial_u (
 \zspaceD_{B} \partial_{\TSu}\TSxi^{B}+\alpha^{2} \TSzlap \TSxi^{\TSu})
  \label{3III21.28}
\\
 &=&
  \partial_u(2\partial_{\TSu}\TSxi^{\TSu} - \zspaceD_{B}\TSxi^{B})
\,,
\label{4XII19.t1a}
\\
\partial_u h_{uA} :
 \qquad  0
&=&
	 \partial_{A}
 \big(\TSzlap \partial_u\TSxi^{\TSu}+ 2 \partial_u\xi^{u}\big)
  \label{3III21.24}
\\
 & = &
	\partial_u \Delta_\ringh(   \zspaceD_{B} \TSxi^{B}-2\partial_{u} \xi^{u})
  \label{3III21.25}
\\
&= &
	\partial_u(\ringh_{AB}\partial_{\TSu} \TSxi^{B}+\alpha^{2}\partial_{A}\xi^{u})
 \,,
\label{4XII19.t3a}
\\
\partial_uh_{AB} :\qquad
  0
 &=&
  \partial_u
   (  \TSoLie_{\TSxi} \ringh_{AB}
    - \ringh_{AB}  \zspaceD_{C} \TSxi^{C}
    )
	 \,.
 \label{3III21.22}
\end{eqnarray}
Equations~\eqref{4XII19.t1a} and \eqref{3III21.25} are automatically satisfied by \eqref{5XII19.1a}.
Equation~\eqref{3III21.24} shows that $\partial_u \TSxi^u$ is a linear combination of $\ell=0$ and $\ell=1$ spherical harmonics, and thus its gradient is a conformal Killing vector on $S^2$. Equation~\eqref{4XII19.t3a} gives
\begin{equation}
	 \partial_{\TSu}^2 \TSxi^{A}= - \alpha^2\zspaceD^{A}\partial_u\xi^{u}
 \,,
\label{3III21.26}
\end{equation}
and \eqref{3III21.28} follows automatically by taking the divergence of \eqref{3III21.26}.
By \eqref{3III21.22} the vector field $\partial_u \TSxi^A$ is also a conformal vector field on $S^2$.

By \eqref{31V20.2} the right-hand side of
\eqref{3III21.1} is gauge-invariant under the current asymptotic symmetries. We conclude that
$$\hat E_c[h,\mcC_{u}] \equiv \hat E_{c,I}[h,\mcC_{u}]
$$
is gauge-invariant up do the addition of a functional which is $u$-independent.

Note that the flux of energy seems to be of more interest than the energy itself, since energy is always defined up to an additive constant anyway, so gauge-invariance of the flux is the key for a physically meaningful quantity.

Let us return to \eqref{7VII21.t1}-\eqref{7VII21.t2}:
\begin{eqnarray}
\delta^2 E^{(\Lambda)}
& = &  \delta^2 m_\TB(\mcC_{u }) +
  \frac{1}{16 \pi}
  \int_{S^{2}}  \ringh^{AE }\ringh^{FB }\oone \zhTB_{E F} \partial_{u} \oone \zhTB_{AB}  \,d \mu_{\zzhTBW}
   \phantom{xxx}
   \label{6IV21.18}
\\
& = &  2 \hat E_{c,I} +
  \frac{1}{16  \pi}
  \int_{S^{2}}  \ringh^{AE }\ringh^{FB }\oone \zhTB_{E F} \partial_{u} \oone \zhTB_{AB}  \,d \mu_{\zzhTBW}
   \,.
    \phantom{xxx}
    \label{6IV21.11}
\end{eqnarray}
The first equation is quite general, while the second holds in a Bondi gauge which is ``as regular at the origin as can be''. But since $d\hat E_{c,I} /du$ is gauge-independent, the flux of $\hat E_{c,I} $ coincides with that  of $\delta^2 m_\TB(\mcC_{u })/2$.

Next, consider the explicit integral term in  \eqref{6IV21.18}. Under a residual gauge transformation we have,  by \eqref{20II20.1} and by the above,
\begin{eqnarray}
  \oone \zhTB_{AB} &\mapsto &
   \oone \zhTB_{AB} - 2 \zspaceD_{A}\zspaceD_{B}   \TSxi^{\TSu}
    +
     \ringh_{AB} 		   \TSzlap \TSxi^{\TSu}
 \nonumber
\\
  \partial_{u} \oone \zhTB_{AB} &\mapsto & \partial_{u}
   (\oone \zhTB_{AB} - 2 \zspaceD_{A}\zspaceD_{B}   \TSxi^{\TSu}
    +
     \ringh_{AB} 		   \TSzlap \TSxi^{\TSu} )
 \nonumber
\\
 &  &
 =
         \partial_{u} \oone \zhTB_{AB}
	 \,.
 \label{6IV21.12}
\end{eqnarray}
Hence
\begin{eqnarray}
 \lefteqn{
  \frac{1}{16  \pi}
  \int_{S^{2}}  \ringh^{AE }\ringh^{FB }\oone \zhTB_{E F} \partial_{u} \oone \zhTB_{AB}  \,d \mu_{\zzhTBW}
  }
  &&
   \nonumber
\\
 & \mapsto &
  \frac{1}{16  \pi}
  \int_{S^{2}}  \ringh^{AE }\ringh^{FB }(\oone \zhTB_{E F}- 2 \zspaceD_{E}\zspaceD_{F}   \TSxi^{\TSu} ) \partial_{u} \oone \zhTB_{AB}  \,d \mu_{\zzhTBW}\nonumber
\\
 &  & =
   \frac{1}{16 \pi}
  \int_{S^{2}}
  \big(\oone \zhTB{}^{AB}- 2\TSxi^{\TSu} \zspaceD^{A}   \zspaceD^{B}
   \big)
   \partial_{u} \oone \zhTB_{AB}
   \,d \mu_{\zzhTBW}
 \nonumber
\\
   & & =  \frac{1}{16  \pi}
  \int_{S^{2}}  \ringh^{AE }\ringh^{FB }\oone \zhTB_{E F} \partial_{u} \oone \zhTB_{AB}  \,d \mu_{\zzhTBW}
   \,.
    \label{6IV21.14}
\end{eqnarray}
Since $ \delta^2 m_\TB(\mcC_{u }) $ is invariant under all residual gauge-transformations, we conclude that
both $ \delta^2 m_\TB(\mcC_{u }) $ and
$
\delta^2 E^{(\Lambda)}$ are invariant under asymptotic symmetries preserving \eqref{6IV21.21}. But their fluxes differ. Which of the two fluxes is more relevant for specific physical applications requires further justification.

}

\section*{Appendices}
 \addcontentsline{toc}{section}{Appendices}
 \renewcommand{\thesection}{\Alph{section}}
 \setcounter{section}{0}

\bigskip

\ptcrr{Christoffels and Bianchi commented out}
\section{The conformal Killing operator on $S^2$}
 \label{App2VII20}

Consider the conformal Killing equation with a source on a round two-dimensional sphere,
\begin{equation}
\mathring{D}_{A} \xi_{B}
+ \mathring{D}_{B} \xi_{A}- \mathring{\gamma}_{AB}\mathring{D}^{C} \xi_{C}=  H_{AB} \, .
\label{1VII20.t1}
\end{equation}
Its properties have been extensively analysed by Boucetta in \cite{Boucetta1999}, see also Appendix A in \cite{CzajkaJezierski} and Appendix E in \cite{JJpeeling}). The CKV operator, defined by the left-hand side of (\ref{1VII20.t1}), has a  six-dimensional kernel, see (\ref{1VII20.t5}). Solutions of  (\ref{1VII20.t1}) are not unique and exist if and only if the symmetric traceless tensor  $H_{AB}$ is orthogonal to the kernel.

 We calculate the kernel of the CKV operator
\begin{eqnarray}
0&=&\mathring{D}_{A} \xi_{B}
	+ \mathring{D}_{B} \xi_{A}- \mathring{\gamma}_{AB}\mathring{D}^{C} \xi_{C}
	\nonumber
	\\
 &=&\mathring{D}^{A}\mathring{D}_{A} \xi_{B}
 + \mathring{D}^{A}\mathring{D}_{B} \xi_{A}- \mathring{\gamma}_{AB}\mathring{D}^{A}\mathring{D}^{C} \xi_{C}
 \nonumber
 \\
 &=&\mathring{D}^{A}\mathring{D}_{A} \xi_{B}+\xi_{C}{{R^{C}}_{AB}}^{A}
 \nonumber
 \\
 &=&(\Delta_{\ringh}+1)\xi_{B} \, .
 \label{1VII20.t2}
\end{eqnarray}
In the last line, we have used
\begin{equation}
{{R^{C}}_{AB}}^{A} =
	{R^{C}}_{B} =
	{\delta^{C}}_{B} \, .
	\label{1VII20.t3}
\end{equation}

We wish to show that only a gradient of a ``dipole function'', i.e.\ a combination of $\ell=1$ spherical harmonics, together with a co-gradient of a dipole function fulfill  (\ref{1VII20.t2}). According to the Hodge-Kodaira theorem, each one-form can be expressed as the sum of an exterior derivative, a coderivative and a harmonic form. Furthermore, it is  well-known that there are no harmonic one-forms on $S^{2}$. This leads to the following decomposition of $\xi_{B}$:
\begin{equation}
\xi_{B}=\mathring{D}_{B}\iota+{\varepsilon_{B}}^{C} \mathring{D}_{C} \upsilon
 \,,
\label{1VII20.t4}
\end{equation}
where ${\varepsilon_{B}}^{C}$ is the two-dimensional Levi-Civita tensor. Inserting this into (\ref{1VII20.t2}) one obtains
\begin{eqnarray}
0&=&
(\Delta_{\ringh}+1)(\mathring{D}_{B}\iota+{\varepsilon_{B}}^{C} \mathring{D}_{C} \upsilon)
\nonumber
\\
&=&
\mathring{D}_{B}(\Delta_{\ringh}+2)\iota+{\varepsilon_{B}}^{C} \mathring{D}_{C}(\Delta_{\ringh}+2)\upsilon \, .
\label{1VII20.t5a}
\end{eqnarray}
Taking a divergence and codivergence of \eqref{1VII20.t5a} gives
\begin{eqnarray}
 \Delta_{\ringh}(\Delta_{\ringh}+2)\iota= 0 =  \Delta_{\ringh}(\Delta_{\ringh}+2)\upsilon \, .
\label{1VII20.t5}
\end{eqnarray}
Basic facts about spherical harmonics show that the kernel of $ \Delta_{\ringh}(\Delta_{\ringh}+2)$ is a mono-dipole function. Since a constant part of $\iota$ or $\upsilon$ does  not contribute to the co-vector $\xi_{B}$, we conclude that conformal Killing vectors of $S^2$ can be uniquely written in the form \eqref{1VII20.t4} where $\iota$ and $\upsilon$ are linear combinations of $\ell=1$ spherical harmonics.
\section{The field $b^{\alpha}$}
 \label{s15VII20.t1}
\subsection{Alternative representation}
 \label{s24VIII20.t1}

Recall that
\begin{equation}\label{31VII20.11}
  b^\alpha(\delta_1 g, \delta_2 g):=   P^{ \alpha (\b \c) \d (\e \f) }  \delta_1  g_{\b\c}
 \nabla_{ \d }  \delta_2 g_{\e \f }
   \,,
\end{equation}
where
\bean
 P^{\a \b \c \d \e \f}
  & = &
  \frac 12
   \big(
    {\nobarg}^{\a \e} {\nobarg}^{\d \b} {\nobarg}^{\c \f}
    +
    {\nobarg}^{\a \e} {\nobarg}^{\f \b} {\nobarg}^{\c \d}
     -
       {\nobarg}^{\a
 \d} {\nobarg}^{\b \e} {\nobarg}^{\f \c}
  -
   {\nobarg}^{\a \b} {\nobarg}^{\c \d} {\nobarg}^{\e \f}
   \nonumber
\\
 &&
 \phantom{\frac 12 \big(}
    -
  {\nobarg}^{\b \c} {\nobarg}^{\a \e} {\nobarg}^{\f \d}
    +
     {\nobarg}^{\b \c} \nobarg^{\a \d} {\nobarg}^{\e \f}
  \big)
 \,.
\eeal{31VII20.12}

In Bondi gauge for $g$ and $\delta g$ we have
\begin{equation}\label{31VII20.3}
  g^{\alpha\beta} \delta g_{\alpha\beta}
  =
   2 g^{ur} \delta g_{ur} + \underbrace{g^{AB} \delta g_{AB}}_{0}
  \,.
\end{equation}
For  solutions of interest here it holds that $\delta g_{ur}\equiv 0$, which leads us to consider the gauge
\begin{equation}\label{31VII20.1}
  g^{\alpha\beta} \delta g_{\alpha\beta} \equiv  0
  \,,
\end{equation}
hence also
\begin{equation}\label{31VII20.2}
  g^{\alpha\beta}\nabla_\sigma  \delta g_{\alpha\beta} \equiv 0
  \,.
\end{equation}
After insertion in \eqref{31VII20.11}  the last three terms of  \eqref{31VII20.12}  drop out when \eqref{31VII20.1}-\eqref{31VII20.2} hold, and we obtain
\begin{eqnarray}
  b^\alpha(\delta_1 g, \delta_2 g) & = &    P^{ \alpha (\b \c) \d (\e \f) }  \delta_1  g_{\b\c}
 \nabla_{ \d }  \delta_2 g_{\e \f }
  \nonumber
\\
 & = &
 \frac 12
   \big(
    {\nobarg}^{\a \e} {\nobarg}^{\d \b} {\nobarg}^{\c \f}
    +
    {\nobarg}^{\a \e} {\nobarg}^{\f \b} {\nobarg}^{\c \d}
     -
       {\nobarg}^{\a
 \d} {\nobarg}^{\b \e} {\nobarg}^{\f \c}\big) \delta_1  g_{\b\c}
 \nabla_{ \d }  \delta_2 g_{\e \f }
  \nonumber
\\
 & = &
\frac 12
\big(
2{\nobarg}^{\a \e} {\nobarg}^{\d \b} {\nobarg}^{\c \f}
-
{\nobarg}^{\a
    \d} {\nobarg}^{\b \e} {\nobarg}^{\f \c}\big) \delta_1  g_{\b\c}
\nabla_{ \d }  \delta_2 g_{\e \f }
   \,.
   \label{31VII20.4}
\end{eqnarray}
This formula can be used as a starting point for a simpler calculation of a $u$+$r$+angles decomposed form of the fields $b^\alpha$ and $\omega^\alpha$.

\subsection{The field $b^r$}
\newcommand{\nozspaceD}{{D}}

We assume the  Bondi form of the metric
\begin{eqnarray}
g_{\alpha \beta}dx^{\alpha}dx^{\beta}
&  =  &-\frac{V}{r}e^{2\beta} du^2-2 e^{2\beta}dudr
\nonumber
\\
&&
+g_{AB}\Big(dx^A-U^Adu\Big)\Big(dx^B-U^Bdu\Big)
\, ,
\label{18XI20.t5}
\end{eqnarray}
with variations satisfying
\begin{equation}\label{1IX20.41}
  \delta g_{rr} = \delta g_{rA} = 0
  \,.
\end{equation}
The metric allows a more general form of two-dimensional metric than the one used in \eqref{30XI19.100}. In other words, we use general  coordinates adapted to a family of null hypersurfaces $\{u=\const\}$. In particular we do \emph{not}  assume the trace condition $g^{AB}\delta g_{AB} =0$ and that  the determinant  $\det g_{AB}$ takes a canonical form. Using the definition \eqref{23XI19.3}, we then find
\begin{eqnarray}
b^{r}(\delta_1 g,\delta_2 g)
&=&
	e^{-4 \beta}\delta_{1} g_{u r} \bigg[
		g^{A B} \big(
			\frac12 \nabla_{u}\delta_{2}g_{A B}
			- \nabla_{A}\delta_{2}g_{u B}
			+\frac{V}{r} \nabla_{A}\delta_{2}g_{r B}
		\big)		
\nonumber
\\
& &
			+\frac12 \big(
		U^{A} g^{BC}-2 U^{B}g^{AC}
		\big) \nabla_{A}\delta_{2} g_{BC}
	\bigg]
	- \frac{e^{-4 \beta}}{2} \delta_{1}g_{u u} g^{AB} \nabla_{r} \delta 2 g_{A B}
\nonumber
\\
& &
	+e^{-4 \beta}\delta_{1} g_{u A} \bigg[
		\big(
			g^{A B}U^{C} - U^{A} g^{B C}
		\big) \nabla_{r} \delta_{2}g_{B C}
		+ g^{A B}\big(
			\nabla_{r} \delta_{2}g_{u B } - \nabla_{u} \delta_{2}g_{r B }
		\big)
\nonumber
\\
& &
		+g^{A B} \bigg(
			\frac{e^{2 \beta}}{2} g^{C D} \nabla_{B} \delta_{2}g_{C D }
			- U^{C} \nabla_{C} \delta_{2}g_{r B}
		\bigg)
	\bigg]
\nonumber
\\
& &
	+\frac{e^{-2 \beta}}{2} \delta_{1} g_{A B} \bigg \{
		g^{A B} \bigg[
			e^{-2 \beta} \bigg(
			 U^{C} U^{D} \big(
				\nabla_{C} \delta_{2}g_{r D } - \nabla_{r} \delta_{2}g_{C D }
			\big)
\nonumber
\\
& &
			+\nabla_{u} \delta_{2}g_{u r }
			-
			\nabla_{r} \delta_{2}g_{u u }
			+U^{C} \big(
				\nabla_{C} \delta_{2}g_{u r} -2 \nabla_{r} \delta_{2}g_{u C } +\nabla_{u} \delta_{2}g_{r C }	
			\big)
			\bigg)			
\nonumber
\\
& &	
		+g^{C D} \bigg(
		\nabla_{C} \delta_{2}g_{u D}
		-\nabla_{u} \delta_{2}g_{C D}
			+\frac{V}{r} \big(
				\nabla_{r} \delta_{2}g_{C D}-\nabla_{C} \delta_{2}g_{r D}
			\big)	
		\bigg)
		\bigg]
\nonumber
\\
& &
	+2 e^{-2 \beta} U^{A} \bigg[
		g^{B C} \big(
				\nabla_{r} \delta_{2}g_{u C}
				- \nabla_{u} \delta_{2}g_{r C}
				+U^{D} \nabla_{r} \delta_{2}g_{C D}
		\big)
\nonumber
\\
& &	
	-g^{B D} U^{C} \nabla_{C} \delta_{2}g_{r D}
	- \frac{1}{2}U^{B} g^{CD} \nabla_{r} \delta_{2}g_{C D}
	\bigg]
\nonumber
\\
& &
		+g^{A C} g^{B D} \bigg(
			\nabla_{u} \delta_{2}g_{C D} - 2 \nabla_{C} \delta_{2}g_{u D}
			- \frac{V}{r} \big(
				\nabla_{r} \delta_{2}g_{C D} - 2 \nabla_{C} \delta_{2}g_{r D}
			\big)	
		\bigg)
\nonumber	
\\
& &
		+\big(
			U^{A} g^{B D} g^{C E}
			-g^{A C} g^{B D} U^{E}
		\big)\nabla_{C} \delta_{2}g_{D E}
	\bigg \} \; .
\phantom{xxxxx}
\label{15VII20.t1}
\end{eqnarray}
Letting $\nozspaceD$ be the covariant derivative associated with the metric $g_{AB}$, we find
\begin{eqnarray}
b^{r}(\delta_1 g,\delta_2 g)
&=&
\frac{e^{-4 \beta}}{4} \delta_{1} g_{ur} \bigg\{\delta_{2}g_{A B}\bigg[
			g^{AC} g^{BD}\big(
				\partial_{u} g_{C D}-\frac{2 V}{r} \partial_{r} g_{C D}
			\big)
		\bigg]
\nonumber
\\
&&
	+ 2 g^{A B} \partial_{u} \delta_{2}g_{A B}+ 2 \left(U^{A} g^{BC}- 2 U^{B}g^{AC}\right) \nozspaceD_{A} \delta_{2}g_{BC}
\nonumber
\\
&&
	+4 \left(\delta_{2} g_{ur}e^{-2 \beta} \nozspaceD_{A}U^{A}
 -g^{AB} \nozspaceD_{A}\delta_{2}g_{u B} \right)
	\bigg \}
\nonumber
\\
& &+\frac{e^{-2 \beta}}{2} \delta_{1} g_{AB} \bigg\{
	g^{AB} \bigg[
			2 \delta_{2} g_{C D} U^{C} g^{D E} \nozspaceD_{E} \beta
\nonumber
\\
& &+ g^{CD}\left(
		\frac{V}{r} \partial_{r}\delta_{2}g_{C D} -\partial_{u}\delta_{2}g_{C D}+\nozspaceD_{C} \delta_{2}g_{u D}+\delta_{2}g_{u C} \nozspaceD_{D} \beta
	\right)	
\nonumber
\\
& &
	+e^{-2 \beta} \bigg( U^{C} \big(
	\delta_{2}g_{uD}g^{D E} \partial{r}g_{C E} -2 \partial_{r}\delta_{2}g_{uC}
	- \delta_{2}g_{C D}\partial_{r}U^{D}
\nonumber
\\
& &
	+\delta_{2}g_{u r} \nozspaceD_{C}\beta-U^{D} \partial_{r} \delta_{2}g_{C D}
	\big)
	+\nabla_{u}\delta_{2}g_{u r}-\nabla_{r}\delta_{2}g_{u u}
\nonumber
\\
& &
		 -\delta_{2} g_{u r} \nozspaceD_{C}U^{C} - \frac{1}{2} \delta_{2} g_{u C} \partial_{r} U^{C} +\frac{3}{2} e^{-2 \beta} \delta_{2}g_{u r}U^{C}g_{C D} \partial_{r} U^{D}
\nonumber
\\
& &
+U^{C} \big(
 \nozspaceD_{C} \delta_{2} g_{u r}
 -2\delta_{2} g_{u r} \nozspaceD_{C} \beta
\nonumber
\\
& &
 +\frac12 e^{-2 \beta} \delta_{2} g_{u r} U^{E} \partial_{r} g_{E C}
-\frac12 \delta_{2} g_{u E} g^{E F} \partial_{r} g_{F C}
\big)
	 \bigg)
	\bigg]
\nonumber
\\
& &
+	g^{A C} g^{B D} \bigg[ \partial_{u} \delta_{2} g_{C D}
-\frac{V}{r} \partial_{r} \delta_{2} g_{C D}
-2 \nozspaceD_{C}\delta_{2}g_{uD}
- U^{E} \nozspaceD_{C}\delta_{2}g_{D E}
\nonumber
\\
& &
+\frac{e^{-2 \beta}}{2}  \partial_{r}g_{CD} \bigg( 3 U^{E}\delta_{2}g_{u E}
+U^{E} U^{F} \delta_{2}g_{E F}\bigg) \bigg]	+U^{A} g^{B C} g^{D E} \nozspaceD_{D}\delta_{2}g_{C E}
\nonumber
\\
& &
	+\frac{e^{-2 \beta}}{2}U^{A} \bigg[
			\big(
				U^{B}g^{C E} -2 g^{BC} U^{E} - U^{C}g^{BE}
			\big)g^{D F} \partial_{r} g_{E F} \delta_{2}g_{C D}
\nonumber
\\
& &
		+2\big(
		2 U^{D}g^{BC} -U^{B}g^{C D}
		\big) \partial_{r} \delta_{2} g_{C D}
\nonumber
\\
& &
		+4g^{B C} \partial_{r} \delta_{2} g_{u C}
		- g^{B C} g^{D E} \partial_{r} g_{C D} \delta_{2} g_{u E}
	\bigg]
\nonumber
\\
& &
+ \frac{e^{-2 \beta}}{2} \bigg( g^{A C} g^{B F} U^{D} U^{E} \delta_{2} \partial_{r}g_{E F} g_{C D} + g^{A E} g^{B D} U^{C} \partial_{r} g_{C D } \delta_{2} g_{u E} \bigg) \bigg\}
\nonumber
\\
& &
	+\frac{e^{-4 \beta}}{4} \delta_{1}g_{u A} \bigg[
		\big(
			2 U^{A} g^{B C} g^{D E}  \delta_{2} g_{C E}
\nonumber
\\
& &
			-3(U^{E} \delta_{2} g_{C E}+\delta_{2} g_{u C}) g^{A B} g^{C D}
		\big)\partial_{r} g_{B D}
\nonumber
\\
& &
    	+2 g^{AB} \big(
            2 \partial_{r} \delta_{2} g_{u B}
        \big)
	-4 e^{-2 \beta} \partial_{r} U^{A} \delta_{r} g_{u r}
\nonumber
\\
& &
   	+4\big(
        g^{AC}U^{B}-U^{A} g^{B C}
    \big)\partial_{r} \delta_{2} g_{B C}
\bigg]
\nonumber
\\
& &
	+\frac{e^{-4 \beta}}{4} \delta_{1} g_{u u} g^{A C} \left( g^{B D} \partial_{r} g_{AB} \delta_{2} g_{C D} - 2 \partial_{r} \delta_{2} g_{A C}\right) \; ,
	\label{15VII20.t2}
\phantom{xxxxxxxxxxxxxxxxx}
\end{eqnarray}
where
\begin{eqnarray}
 \nabla_{u}\delta_{2}g_{u r}
 &=&
 \partial_{u} \delta_{2}g_{u r}
  + \delta_{2}g_{u r} \big[
    U^{A} \nozspaceD_{A} \beta
    -2 \partial_{u} \beta
    -\frac{e^{-2 \beta}}{2} U^{A} \partial_{r}\big(
        g_{A B} U^{B}
    \big)
 \big]
 \nonumber
 \\
  & &
  +\delta_{2}g_{u A} \big[
    \frac12 g^{AB} \partial_{r} \big(
        g_{B C} U^{C}
    \big)
    -e^{2 \beta} g^{A B} \nozspaceD_{B} \beta
  \big]
 \, ,
 \phantom{xxxxxxxxxxxxxxx}
 \\
 \nabla_{r}\delta_{2}g_{u u}
 &=&
    \partial_{r} g_{u u}
    +\delta_{2}g_{u A} \bigg[
        g^{A B} \partial_{r}\big(
         g_{B C} U^{C}
        \big)
    -2 e^{2 \beta} g^{A B} \nozspaceD_{B} \beta
    \bigg]
 \nonumber
 \\
 & &
    +\delta_{2}g_{u r} \bigg[
        +2 U^{A} \nozspaceD_{A} \beta
        +e^{-2 \beta} U^{A}g_{A B} \partial_{r} U^{B}
\nonumber
\\
& &
        +2 \frac{V}{r} \partial_{r} \beta
        - \partial_{r} \bigg(\frac{V}{r}\bigg)
    \bigg]
 \, .
\end{eqnarray}
Assuming $U^{A}=0$ and $g^{A B} \delta g_{AB}=0$,  equation  (\ref{15VII20.t1}) simplifies to
\begin{eqnarray}
b^{r}(\delta_1 g,\delta_2 g)
&=&
e^{-4 \beta}\delta_{1} g_{u r} \bigg[
g^{A B} \big(
\frac12 \nabla_{u}\delta_{2}g_{A B}
- \nabla_{A}\delta_{2}g_{u B}
+\frac{V}{r} \nabla_{A}\delta_{2}g_{r B}
\big)
\bigg]		
\nonumber
\\
& &
- \frac{e^{-4 \beta}}{2} \delta_{1}g_{u u} g^{AB} \nabla_{r} \delta_2 g_{A B}
\nonumber
\\
& &
+e^{-4 \beta}\delta_{1} g_{u A} \bigg[
g^{A B}\big(
\nabla_{r} \delta_{2}g_{u B } - \nabla_{u} \delta_{2}g_{r B }
\big)
\nonumber
\\
& &
+g^{A B} \bigg(
\frac{e^{2 \beta}}{2} g^{C D} \nabla_{B} \delta_{2}g_{C D }
\bigg)
\bigg]
\nonumber
\\
& &
+\frac{e^{-2 \beta}}{2} \delta_{1} g_{A B}
g^{A C} g^{B D} \bigg(
\nabla_{u} \delta_{2}g_{C D} - 2 \nabla_{C} \delta_{2}g_{u D}
\nonumber
\\
& &
- \frac{V}{r} \big(
\nabla_{r} \delta_{2}g_{C D} - 2 \nabla_{C} \delta_{2}g_{r D}
\big)	
\bigg)
 \,.
\label{15VII20.t3}
\end{eqnarray}
Likewise we have the following version of  (\ref{15VII20.t2}):
\begin{eqnarray}
b^{r}(\delta_1 g,\delta_2 g)
&=&
\frac{e^{-4 \beta}}{4} \delta_{1} g_{ur} \bigg\{\delta_{2}g_{A B}\bigg[
g^{AC} g^{BD}\left(
\partial_{u} g_{C D}-\frac{2 V}{r} \partial_{r} g_{C D}
\right)
\bigg]
\nonumber
\\
&&
+ 2 g^{A B} \partial_{u} \delta_{2}g_{A B}
-4g^{AB} \nozspaceD_{A}\delta_{2}g_{u B}
\bigg \}
\nonumber
\\
& &+\frac{e^{-2 \beta}}{2} \delta_{1} g_{AB} g^{A C} g^{B D} \bigg[ + \partial_{u} \delta_{2} g_{C D} -\frac{V}{r} \partial_{r} \delta_{2} g_{C D} -2 \nozspaceD_{C}\delta_{2}g_{uD}
\bigg]	
\nonumber
\\
& &
+\frac{e^{-4 \beta}}{4} \delta_{1}g_{u A} \bigg[
-3\delta_{2} g_{u C} g^{A B} g^{C D} \partial_{r} g_{B D}
+2 g^{AB} \big(
2 \partial_{r} \delta_{2} g_{u B}
\big)
\bigg]
\nonumber
\\
& &
+\frac{e^{-4 \beta}}{4} \delta_{1} g_{u u} g^{A C} \left( g^{B D} \partial_{r} g_{AB} \delta_{2} g_{C D} - 2 \partial_{r} \delta_{2} g_{A C}\right) \; .
 \label{1IX20.40}
\end{eqnarray}
%
\section{Linearised curvature}
 \label{s3XI20.1}
  \jhbr{Checked 10XI20}

For completeness we derive here a formula for the linearised Ricci tensor, as used in the main text. While the calculation is standard, our final formulae \eqref{3XI20.1} and \eqref{3XI20.2} are nonstandard in that they do not involve any explicit curvature tensors at the right-hand sides.
\begin{eqnarray}
\delta \Gamma^{\mu}{}_{\nu \lambda}
&=&
\frac{1}{2} g^{\mu \kappa}
\big(
    \nabla_{\lambda}\delta g_{\kappa \nu}
    +\nabla_{\nu}\delta g_{\kappa \lambda}
    -\nabla_{\kappa}\delta g_{\lambda \nu}
\big)
\, ,
\\
\delta R^{\alpha}{}_{\beta \mu \nu}
&=&
\nabla_{\mu} \delta \Gamma^{\alpha}{}_{\beta \nu}
-\nabla_{\nu}\delta \Gamma^{\alpha}{}_{\beta \mu}
\, ,
\\
2 g_{\mu \alpha}\delta R^{\alpha}{}_{\nu \sigma \rho}
&=&
    \nabla_{\sigma} \nabla_{\rho}\delta g_{\mu \nu}
    -\nabla_{\rho} \nabla_{\sigma}\delta g_{\mu \nu}
    +\nabla_{\sigma} \nabla_{\nu}\delta g_{\mu \rho}
\nonumber
\\
& &
    -\nabla_{\sigma} \nabla_{\mu}\delta g_{\nu \rho}
    -\nabla_{\rho} \nabla_{\nu}\delta g_{\mu \sigma}
    +\nabla_{\rho} \nabla_{\mu}\delta g_{\nu \sigma}
\, ,
\nonumber
\\
&=&
    \nabla_{\sigma}\nabla_{\nu}\delta g_{\mu \rho}
    -\nabla_{\sigma}\nabla_{\mu}\delta g_{\nu \rho}
    -\nabla_{\rho}\nabla_{\nu}\delta g_{\mu \sigma }
    +\nabla_{\mu}\nabla_{\rho}\delta g_{\nu \sigma }
\nonumber
\\
& &
    -R^{\alpha}{}_{ \mu \sigma \rho} \delta g_{\alpha \nu}
    -R^{\alpha}{}_{\nu \sigma \rho} \delta g_{\alpha \mu}
    -R^{\alpha}{}_{\nu \rho \mu} \delta g_{\alpha \sigma}
    -R^{\alpha}{}_{\sigma \rho \mu} \delta g_{\alpha \nu}
    \, ,
    \phantom{xxxxxx}
\\
2 \delta R_{\nu \rho}
&=&
    \nabla^{\sigma}\nabla_{\nu}\delta g_{\rho \sigma }
    +\nabla^{\sigma}\nabla_{\rho}\delta g_{\sigma \nu}
    -\nabla^{\sigma}\nabla_{\sigma}\delta g_{\nu \rho}
    -g^{\mu \sigma}\nabla_{\nu}\nabla_{\rho}\delta g_{\sigma \mu}
    \,,
    \label{3XI20.1}
\\
2 \delta R_{\alpha \beta} - g_{\alpha \beta} g^{\nu \rho} \delta R_{\nu \rho}
&=&
    \nabla^{\mu} \nabla_{\alpha} \delta \bar g_{\beta \mu}
    +\nabla^{\mu} \nabla_{\beta}\delta \bar g_{\alpha \mu}
    -\nabla_{\mu} \nabla^{\mu} \delta \bar g_{ \alpha \beta}
    -g_{\alpha \beta} \nabla^{\kappa} \nabla^{\lambda} \delta \bar g_{\kappa  \lambda}
    \, ,
    \phantom{xxxx}
\,,
 \label{3XI20.2}
\end{eqnarray}
where $\delta \bar g_{\alpha\beta} = \delta  g_{\alpha\beta} - \frac 12 g^{\mu \nu}\delta   g_{\mu\nu}  g_{\alpha\beta} $.

The linearisation of the Einstein tensor   reads
\begin{equation}
\delta G_{\alpha \nu}
=
\delta R_{\alpha \nu}
- \frac{1}{2} g_{\alpha \nu} g^{\sigma \rho} \delta R_{\sigma \rho} -\frac{1}{2} R \delta g_{\alpha \nu}
+\frac{1}{2}g_{\alpha \nu}\delta g_{\sigma \rho} R^{\sigma \rho}
\label{18XI20.t1}
\end{equation}
In order to compare the linearised Einstein equations,
\begin{equation}
\delta \big( G_{\mu \nu} + \Lambda g_{\mu \nu} \big)=0 \, ,
\label{18XI20.t2}
\end{equation}
with the Euler--Lagrange  equations, we substitute \eqref{18XI20.t1} into \eqref{18XI20.t2} and regroup the terms as follows
\begin{equation}
    \delta R_{\alpha \nu}
- \frac{1}{2} g_{\alpha \nu} g^{\sigma \rho} \delta R_{\sigma \rho}
=
\underbrace{
    \delta G_{\alpha \nu}
    +\frac{1}{2} R \delta g_{\alpha \nu}
    -\frac{1}{2}g_{\alpha \nu}\delta g_{\sigma \rho} R^{\sigma \rho}
}_{\Lambda \tshbar_{\alpha\nu} }
\, .
\label{18XI20.t3}
\end{equation}
Using \eqref{3XI20.2}, the last equations are equivalent to \eqref{26X20.t1} when the background satisfies  $R_{\mu \nu}=\Lambda g_{\mu \nu}$ holds. Furthermore, under this last assumption,  equation  \eqref{18XI20.t3} coincides with \eqref{20VIII20.t0}.

\section{The remaining Einstein equations revisited}
 \label{A12XI20.1}

In this appendix we derive some further consequences of the linearised Einstein equations.

Adding, in vacuum, the equation $\zspaceD^A \times $(\ref{28XI19.5}),
\begin{equation}
            \partial_r \left[r^4  \partial_r(r^{-2} \zspaceD^A {h}_{uA})\right] =
                    r^2
                     \left[ \partial_r \left(r^{-2}
                    \zspaceD^A \zspaceD^B {h}_{AB}\right)\right]
                 \,,
                            \label{14VIII20.1}
           \end{equation}
to $-r^2\partial_r \times $(\ref{2VI20.1}),
              \begin{equation}\label{14VIII20.2}
                2  r^2  \partial_r^2 (r {h}_{uu})
                =
              - r^2 \partial_r
               \big[r^{-2}
               \big(
  \zspaceD^A\zspaceD^B {h}_{AB}
              -   \partial_r (r^2 \zspaceD^A {h}_{uA} \big)
              \big]
                   \,,
           \end{equation}
one finds
\begin{equation}\label{14VIII20.3}
  \partial_r\big( r^2   \partial_r {h}_{uu} {\purple-} \zspaceD^A {h}_{uA}
  \big) = 0
  \,.
\end{equation}
Integrating, we find that there exists a function $\chi(u,x^{C})$ such that
 \ptcheck{13VIII20 with TS on the blackboard, sign corrected 15VIII; regardless of whether or not the argument that follows holds, as far as I can see $\chi$ is zero for the BD solutions before we transformed them to the Bondi gauge, the question is, what does the $\zeta[h]$ transformation do to them?}
\begin{equation}
  \zspaceD^{A} {h}_{u A}
  =
   r^2 \partial_{r}  {h}_{u u}
   +\chi(u,x^{C})
  \,.
  \label{13VIII20.t1}
\end{equation}
We evaluate this equation at $r=0$ for a globally smooth linearised field ${h}_{\mu\nu}$. Let $\zeta[h]$ denote the gauge vector field which brings $h_{\mu\nu}$ to the Bondi gauge;
compare \eqref{3XII19.t1}-\eqref{3XII19.t2} with $\xi$ given by \eqref{6III20.1}-\eqref{5XII19.1n}.
By  \eqref{24IX20.1}  and \eqref{24IX20.23x} one obtains
\begin{equation}\label{2IX20.1}
  \chi = ( \zspaceD^{A} {h}_{u A}-
   r^2 \partial_{r}  {h}_{u u})|_{r=0}
  =
  -
   \frac{1}{2}  (\TSzlap +2)\TSzlap\TSxi^{\TSu}
  \,.
\end{equation}
Keeping in mind that $\xi^u$ is freely specifiable \emph{at a chosen value of $u$}, but that $\partial_u\xi^u$ is essentially determined by the solution (cf.\ \eqref{2IX20.2} below), Equation \eqref{2IX20.1} shows that the function $\chi$ can be made to vanish by applying an asymptotic symmetry at a chosen retarded time $u$, and that the vanishing of $\chi$ at some value of $u$ will \emph{not} be preserved by evolution in general.
We will exploit  the possibility of setting $\chi$ to zero at the chosen light cone later,
 but for the moment we remain in a general gauge compatible with our setting so far.

Equations \eqref{2IX20.1} and \eqref{5XII19.1n}
 show that
\begin{eqnarray}
  \partial_u \chi
  & = &
  - \frac{1}{2}  (\TSzlap +2)\TSzlap \partial_u \TSxi^{\TSu}
  \nonumber
\\
 & = &
  -\frac{1}{2}  (\TSzlap +2)\TSzlap
 \big[
\frac12 h_{tt}|_{r=0}
+\frac14 h_{ij}|_{r=0}  \big(\delta^{ij} - 3 \frac{x^ix^j}{r^2}\big)
  \nonumber
\\
 &   &
+ \frac{1}{2} \zspaceD^A \xi_A[h]
 \big]
  \,.
   \label{2IX20.2}
\end{eqnarray}
where we have used the fact that, by \cite[Appendix~A.1]{CzajkaJezierski},
when $ \TSxi^{B}(\TSu, x^{A})$ is a conformal vector field its divergence is a linear combination of   $\ell=1$ spherical harmonics.

We continue by inserting \eqref{13VIII20.t1} into the right-hand side of (\ref{2VI20.1}):
 \ptcheck{13VIII20 with TS on the blackboard, including Lambda,  corrected 14VIII20 and again 21VIII; all equations in the current draft (6IX20) confirmed by MM}
\begin{equation}
 \zspaceD^{A} \zspaceD^{B} {h}_{A B} =
  r^4 \partial^{2}_{r} {h}_{u u}
  +2 r^3 \partial_{r} {h}_{u u}
  -2r^2 {h}_{u u}
    { +}2 r \chi(u, x^C)  \, .
   \label{13VIII20.t2}
\end{equation}

Recall that the equation $r^2 \mcE_{u u}=0$ reads
 \ptcheck{13 VIII 20  with TS whole  Maple file with Lambda =0 and 14VIII with Lambda }
\jhbr{rechecked, added 6XI20, precise version of $\mcE_{uu}$}
\begin{eqnarray}
\lefteqn{2\Big(
      \partial_{u}
    +(\alpha^2 r^2-1) \partial_{r}
    -\frac{1}{r}
    \Big) \zspaceD^{A} {h}_{u A}
    -  \zspaceD^{A} \zspaceD_{A} {h}_{u u}
}
& & \nonumber \\
&&
-(\alpha^2 r^2-1) \bigg(\frac{\zspaceD^{A} \zspaceD^{B} {h}_{A B}}{r^2}\bigg)
-2 r \partial_{u} {h}_{u u}
 \nonumber \\
 &&
-2 (\alpha^2 r^2-1) \partial_{r}(r {h}_{u u})
=0
\, . \phantom{xxxxxx}
\label{13VIII20.t3}
\end{eqnarray}
Substituting  (\ref{30XI19.11}) into (\ref{13VIII20.t3}) one obtains
\begin{eqnarray}
    \Big(
    2 r \partial_{u}
    + (\alpha^2 r^3-r) \partial_{r}
    -2 \alpha^2r^2
    \Big) \zspaceD^{A} {h}_{u A}
& & \nonumber \\
 - r \zspaceD^{A} \zspaceD_{A} {h}_{u u}-2 r^2 \partial_{u} {h}_{u u}
&=&
0
\, . \phantom{xxxxx}
\label{17VIII20.t1}
\end{eqnarray}
We insert the asymptotic expansion of $h_{\mu\nu}$ into \eqref{17VIII20.t1}, obtaining
\begin{eqnarray}
r^2 \Big(
2 \alpha^2 {\oone h}_{uu}
- 2 \alpha^2 \chi
-(\TSzlap + 2) \omone h_{uu}
\Big)
&&
\nonumber
\\
+ r \big(
6 \alpha^2  {\otwo h}_{uu}
- 4 \partial_{u} {\oone h}_{u u}
+2 \partial_{u} \chi
\big)
&=&
O (1)
 \,,
  \phantom{xxx}
\label{3IX20.t1}
\end{eqnarray}
thus both the coefficient of $r$ and that of $r^2$ in the left-hand side have to vanish:
\begin{eqnarray}
 &
 \displaystyle
 \chi =  {\oone h}_{uu}
-\frac{1}{
2 \alpha^2 }(\TSzlap + 2) \omone h_{uu}
\,,
&
\label{7XI20.1}
\\
 &
  \partial_{u} \chi
   =
 2 \partial_{u} {\oone h}_{u u}
 - 3 \alpha^2  {\otwo h}_{uu}
 \,.
  &
\label{3IX20.t1x}
\end{eqnarray}

Assume momentarily that we are in Minkowski spacetime.   From~\eqref{3IX20.t1}  we obtain
%
\begin{equation}\label{13VIII20.p2}
 \partial_u \chi = 2
  \partial_u {\oone h}_{uu}
\,.
\end{equation}
An asymptotic expansion
of the right-hand side of \eqref{13VIII20.t2} gives
\begin{equation}
\zspaceD^{A} \zspaceD^{B} {h}_{A B}
 = 2r   (\chi- {\oone h}_{uu} ) +O(1)  \, .
\label{13VIII20.p4}
\end{equation}
Differentiating \eqref{13VIII20.p4} with respect to $u$ one  finds
\ptcrr{all equations derived in Maple by TS and checked with Mathematica by MM on 3IX20, and this does not give anything new?}
\begin{equation}
    \partial_u {\oone h}_{uu}
   =   \frac 12    \zspaceD^{A} \zspaceD^{B}
    \partial_{u} \oone \zhTB_{AB}
  \,.
   \label{13VIII20.p5}
\end{equation}
This is the linearised version of the usual formula for the time evolution of the mass aspect function.

When $\Lambda \ne 0$, we can  insert \eqref{7XI20.1} into \eqref{13VIII20.p4} to obtain a \emph{corner condition}
\begin{equation}
     (\TSzlap + 2) \omone h_{uu}
   =
   -\alpha^2 \zspaceD^{A} \zspaceD^{B} \oone {\zhTB}_{A B}
   \, .
\label{13VIII20.p4x}
\end{equation}
Differentiating \eqref{7XI20.1} with respect to $u$ and inserting into \eqref{3IX20.t1x} gives the following evolution equation for $\oone h_{uu}$:
\begin{eqnarray}
 \displaystyle
   \partial_{u} {\oone h}_{u u}
  & = &
   - \frac1 { 2 \alpha^2 }
(\TSzlap + 2) \partial_u \omone h_{uu}
+
  3  \alpha^2  {\otwo h}_{uu}
 \nonumber
\\
  & = &
    \frac 12    \zspaceD^{A} \zspaceD^{B}
    \partial_{u} \oone \zhTB_{AB}
+
  3  \alpha^2  {\otwo h}_{uu}
 \,.
\label{7XI20.2}
\end{eqnarray}
%

For further reference we note the following. In~\cite{JJGRG} Jezierski has introduced two gauge-invariant scalars describing linearised gravitational fields on spherically symmetric backgrounds.
These scalars are closely related to the fields $  \zspaceD^B h_{uB}$ and $\varepsilon^{A B} \zspaceD_{A} h_{uB}$. The time evolution of these last fields is governed by the divergence and co-divergence of \eqref{9XI20.t1}:
\begin{eqnarray}
0&=&\Big[
\partial_{u} \partial_{r}
+ (\alpha^2 r^2 -1) \partial_{r} ^2
-\frac{2}{r}\partial_{u}
-2 \alpha^2
\Big] \zspaceD^{A} {h}_{u A}
 \nonumber \\
 & &
+\frac{1}{r^2} \partial_{u} \zspaceD^{A} \zspaceD^{B} {h}_{A B}
- \partial_{r} \zspaceD^{A} \zspaceD_{A}{h}_{u u}
    \,,
\\
0
 & = &
    \Big[
        (\alpha^2 r^2-1) r^2 \partial^{2}_{r}
        +r^2 \partial_{u} \partial_{r}
        -2 r \partial_{u}
        -2 \alpha^2 r^2
    \Big]\varepsilon^{A B} \zspaceD_{A} h_{uB}
    \phantom{xx}
\nonumber
\\
  &&   +\partial_{u} \big(
        \varepsilon^{A B} \zspaceD_{A} \zspaceD^{C} h_{C B}
    \big)
    -\TSzlap \varepsilon^{A B} \zspaceD_{A} h_{uB}
    \, .
  \phantom{xxxxx}
\end{eqnarray}
\ptcrr{Todo commented out 2XI20}

 \addcontentsline{toc}{section}{References}
\bibliographystyle{amsplain}

\bibliography{LinearizedBondiSachsWithLambdaNew-minimal,%
../references/newbiblio,%
../references/reffile,%
../references/bibl,%
../references/prop,%
../references/newbib,%
../references/netbiblio,%
../references/hip_bib,%
../references/newbiblio2,%
../references/besse,%
../references/howard,%
../references/bartnik,%
../references/myGR,%
../references/Energy,%
../references/Chrusciel,%
../references/dp-BAMS,%
../references/prop2,%
../references/besse2%
}
\end{document}